\documentclass[12pt,PhD]{phdthesis}
\usepackage {color, amsfonts, amsmath, amsthm, amssymb}
\usepackage {hyperref}
\usepackage {epsf, epsfig}
\usepackage {graphicx}
\usepackage {verbatim}
\usepackage{latexsym}
\usepackage{bm}
\usepackage[english]{babel}

\makeatletter

\newcommand{\Rmnum}[1]{\expandafter\@slowromancap\romannumeral #1@}
\makeatother

\begin{document}

\title{COSMOLOGICAL SIGNATURES OF BRANE INFLATION}

\author{Taichi Kidani}

\dept {Institute of Cosmology and Gravitation}

\principaladvisor{Dr. Kazuya Koyama}

\secondsupervisor {Prof. David Wands}

\secondreader {Dr. Gianmassimo Tasinato} 



\copyrightfalse


\tablespagefalse

\beforeabstract

\prefacesection{Abstract}

Cosmology motivated by string theory has been studied extensively in the recent literature. String theory is promising because it has interesting features such as unifying gravity, electromagnetic, weak and strong nuclear forces. However, even the energy scale of the experiments at the Large Hadron Collider ($\sim$TeV) is too low to detect any strong evidence for string theory. The energy scale of inflation can be above $\sim10^{9}$ TeV. Therefore, it is expected to find some signature of string theory in cosmology. 

String theory predicts ten space-time dimensions. In the brane world scenario, our four dimensional Universe is confined onto the higher dimensional object called the Brane in the ten dimensional space time. The Dirac-Born-Infeld (DBI) inflation is based on this idea. DBI inflation predicts a characteristic statistical feature in the Cosmic Microwave Background (CMB) temperature anisotropies. In this thesis, we study the predictions of the DBI inflation models on the CMB temperature anisotropies. 

In chapter \ref{chapter:inflationintro}, the idea of inflation in the early stage of the Universe is introduced after explaining why we need inflation in addition to the standard Big Bang scenario. At the end of this chapter, we introduce the CMB observables that quantify the statistical properties of the CMB anisotropies. 

In chapter \ref{chapter:generalperturbations}, we introduce the cosmological perturbation theory for general multi-field inflation including DBI inflation. After studying the linear perturbation theory, we introduce the higher order perturbations that produce the non-Gaussianities. The analytic formulae for the CMB observables that are valid in cases with the effective single field dynamics around horizon crossing are summarised at the end of this chapter. 

In chapter \ref{chapter:stringinflation}, the idea of DBI inflation is introduced. Some analytic predictions for the CMB observables are given in a simple single field DBI inflation model. After introducing the microphysical constraint that excludes the single field DBI inflation, we show that this constraint can be significantly relaxed if the trajectory in the field space is bent in multi-field DBI inflation models. 

In chapter \ref{chapter:spinflation}, we study the specific two-field DBI inflation model with a potential that is derived in string theory. The potential contains only the leading order term ignoring all other possible corrections in string theory. After studying how curves in the trajectories in the field space affect the CMB observable, we show that this model is excluded by observation in the regime where the analytic formulae introduced in chapter \ref{chapter:generalperturbations} are valid. At the end of this chapter, we discuss the cases where we cannot use the analytic formulae and discuss possible implications.  

In chapter \ref{chapter:hybrid}, we study the two-field DBI inflation model with a potential that has the essential feature of the potential obtained with other corrections in addition to the leading term in string theory. In this model, inflation is driven by the motion of a D3 brane along the radial direction and at later times instabilities develop in the angular directions. It is shown that it is actually possible to satisfy the microphysical constraint with a turn in the trajectory in the field space. However, this particular choice of potential is excluded with the constraint on the local type non-Gaussianity by the latest CMB observations of the PLANCK satellite. 

We discuss the future perspective of DBI inflation models in the last chapter. 

\vskip10mm
\begin{center}
\large{\textbf{Units}}
\end{center}
In this thesis, we set the speed of light $c$, the gravitational constant $G$, Boltzmann's constant $k_{\rm{B}}$, the reduced Planck constant $\hbar$ and the reduced Planck mass $M_{\rm{P}} = \sqrt{\hbar c / 8 \pi G}$ to be unity in some equations following the conventional notation in cosmology. Those notations allow us to express any mechanical quantity in terms of only one unit. Therefore, we can use some units to express a physical quantity interchangeably. For example, the energy density and the mass density are interchangeable because the energy density equals the mass density multiplied by $c^2$ and because we set $c = 1$. The signs of the metric tensor are $-,+,+,+,...$ with a minus sign only for the time component while all the spatial components have positive signs. 


\afterabstract



\prefacesection{Preface}

The work of this thesis was carried out at the Institute of
Cosmology and Gravitation, University of Portsmouth, United
Kingdom. The author was supported by Leverhulme trust.
\\

The following chapter is based on published work:
\begin{itemize}
\item Chapter \ref{chapter:hybrid} - T. Kidani, K. Koyama and S. Mizuno,
``Non-Gaussianities in multi-field DBI inflation with a waterfall phase transition'', Phys. Rev. D {\bf 66} (2012) 083503
[arXiv:1207.4410 [astro-ph.CO]]

\end{itemize}

\afterpreface

\prefacesection{Acknowledgements}

To start the acknowledgment section, I would like to thank Kazuya Koyama for teaching me cosmology and for giving me the opportunity to research in such a wonderful place. It has been my great pleasure to be able to learn how to tackle a difficult problem working with excellent researchers at the Institute of Cosmology and Gravitation of the University of Portsmouth. I also would like to thank Shuntaro Mizuno for our collaboration and for his useful advice. I thank Jon Emery, Bridget Falck, Tim Higgs, Claire Le Cras, Ollie Steele, Harry Wilcox and David Wilkinson for proofreading this thesis. Many thanks to
\begin{itemize}
\item all the members of the ICG badminton club for letting me enjoy smashing the shuttlecocks. It was really fun to play badminton with you every week. 
\item all the members of the ICG volleyball club. Thanks especially to Robert Crittenden for organising the sessions. 
\item all the members of the ICG football club. Thanks especially to Jon Emery for organising the sessions. 
\item all the members of the Portsmouth Kendo club. It is my great pleasure to have been training with you. I learnt many things from you, English samurai! Thanks especially to Clive McNaught for founding such a good dojo. 
\item Jon Emery, Tim Higgs and Ollie Steele for letting me have fun times with you both inside the office and outside the office. The Pikachu costume that you gave me is my treasure! 
\end{itemize}
I could not write all the names of people whom I would like to thank. However, I need to stop thanking before the acknowledgement section becomes longer than all the other parts of this thesis. Finally, I would like to thank my parents for letting me have such great experiences. 


\afterpreface

\chapter{Introduction to inflation}\label{chapter:inflationintro}

In 17th century, thanks to Newtonian physics, it was found out that the motion of 
the solar planets can be explained with considerable accuracy by the physical laws which were discovered in the experiments on the earth. Even though it made it possible to study the motions of extraterrestrial matter by physics, people did not know the origin of planets and stars yet. 

In the 1940s, George Gamow and his colleagues Ralph A. Alpher and Robert Herman first 
explored the Big Bang model \cite{Weinberg:1988}. 
This model successfully explains how the Universe which we currently observe has evolved 
from a hot and dense state at the very early stage. 

In this chapter, we first introduce the standard Big Bang model. Then, some problems of 
the model are explained. Finally, we explain about the inflationary models which solve those problems and give the seeds for the large scale structure of the Universe. 

\section{\label{sec:sbbs}Standard Big Bang scenario}
\begin{figure}[!htb]
\centering
\includegraphics[width=12cm]{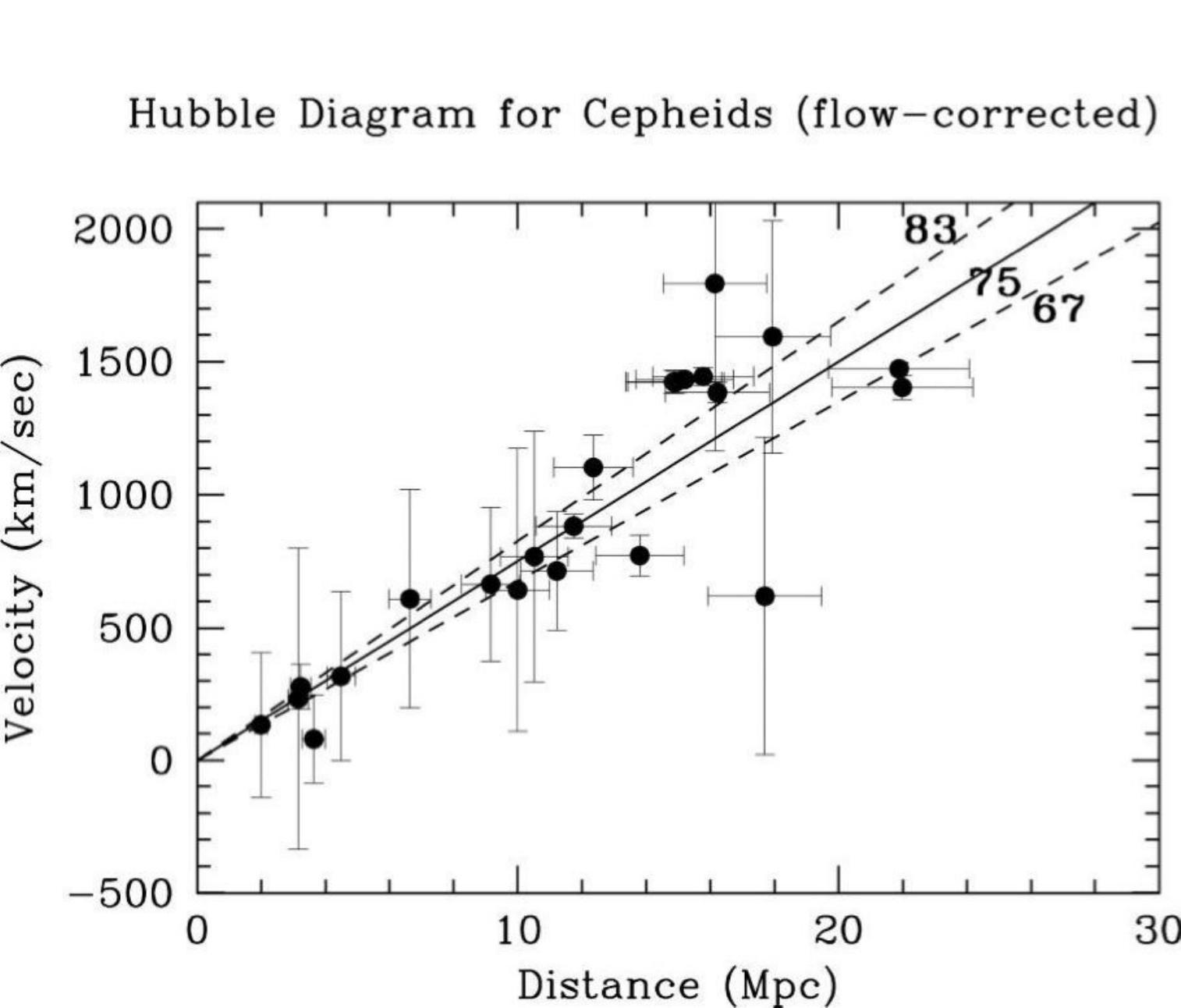}
\caption[Hubble's law]{The relation between the distances of Cepheids which are one of the `standard candles' from the earth and their recessional velocities relative to the earth (from Freedmann W. L. et al. 2000 \cite{Freedman:2001}). We can see that the best fit to these data yields a slope of $H = 75 \pm 10$ km/sec/Mpc, within the uncertainties. }{\label{fig11} }
\end{figure}

Some types of astronomical objects such as the type \Rmnum{1}a supernovae (SNe Ia) \cite{Liddle:2003} have characteristic absorption and emission lines in their spectra. If one of such objects is moving away from the earth, the wavelengths of these spectral features become longer because the light waves which propagate towards us from the object are stretched. This is called the \textbf{redshift}. The faster such objects move away from us, the redder the light waves emitted from them become. On the other hand, light emitted from an object which moves towards us becomes bluer. This is called the blueshift. Therefore, the velocities of objects can be measured by observing the redshifts or blueshifts of them. We can also measure the distances to galaxies from the careful observation and calibration of such astronomical objects called `standard candles' which are known to have the typical properties which can be used to measure the distances to those objects. For example, the SNe Ia are know to be standardisable. After applying an empirical correction to their observed light curve shapes and the peak magnitudes, they are known to be one of the best distance indicators \cite{Kattner:2012}. Basically, the further a standard candle is, the dimmer it looks when observed from the earth. From those observations, it was found that the recessional velocities of almost all the astronomical objects are roughly proportional to their distances from the earth as can be seen in Fig. \ref{fig11}. This is called \textbf{Hubble's law}. 
What does this mean? It seems like the earth is at a special place in the Universe from which almost all other objects move away at the velocities proportional to their distances. In this section, we introduce the current cosmologists' answer to this question. It is very simple and beautiful, but it also has some problems. We see what those problems are in this section as well. 

\subsection{The idea of the Big Band scenario}\label{ideaofbigbang}
\begin{figure}
\centering
\includegraphics[width=12cm]{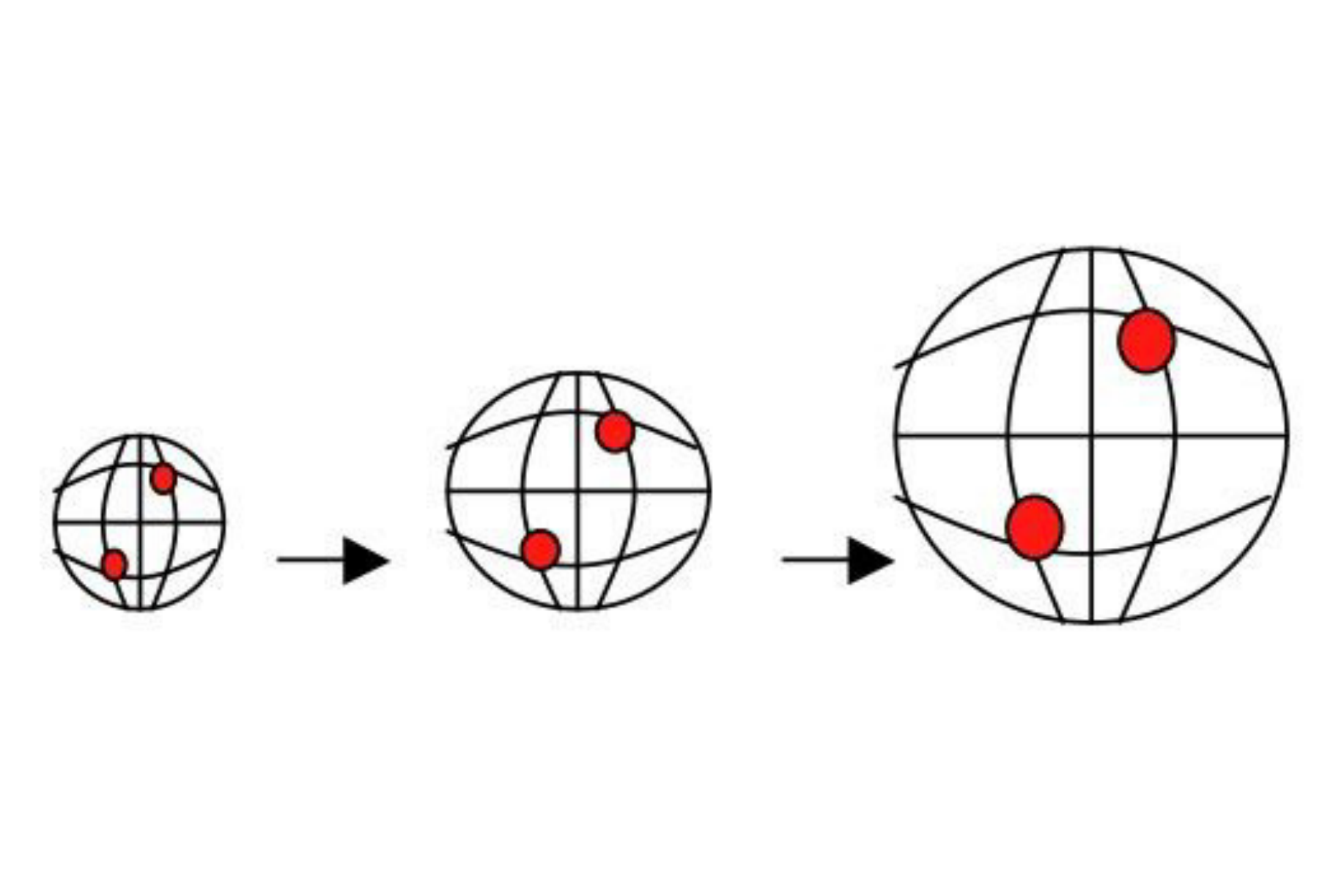}
\caption[Expanding balloon]{Expanding balloon as an analogy of the expanding Universe. The distance between two red stickers gets larger as the balloon expands. The larger the distance is, the faster two red stickers move away from each other regardless of where those stickers are on the balloon. This corresponds to Hubble's law.}{\label{fig12} }
\end{figure}

In the history of physics, people made a mistake by thinking that our planet is at a special place in the Universe around which all other astronomical objects are rotating. In 15th century, Copernicus corrected the mistake by stating that the earth is just one of the planets rotating around the Sun. Cosmologists remembered the lesson and guessed that our planet is not at a special place. Then, how can we explain Hubble's law? Imagine an expanding balloon with two stickers on it as in Fig. \ref{fig12}. 

If the balloon inflates uniformly, how fast those stickers go away from each other is proportional to how far they are from each other. Then, let us put many stickers on the balloon and assume that a randomly chosen sticker is the earth and other stickers are the standard candles. We can see that Hubble's law holds regardless of which sticker we choose to be the earth. Therefore, if we assume that our Universe is expanding uniformly like the balloon, we can actually explain the law without assuming we are at a special place in the Universe. 

\begin{figure}
\centering
\includegraphics[width=12cm]{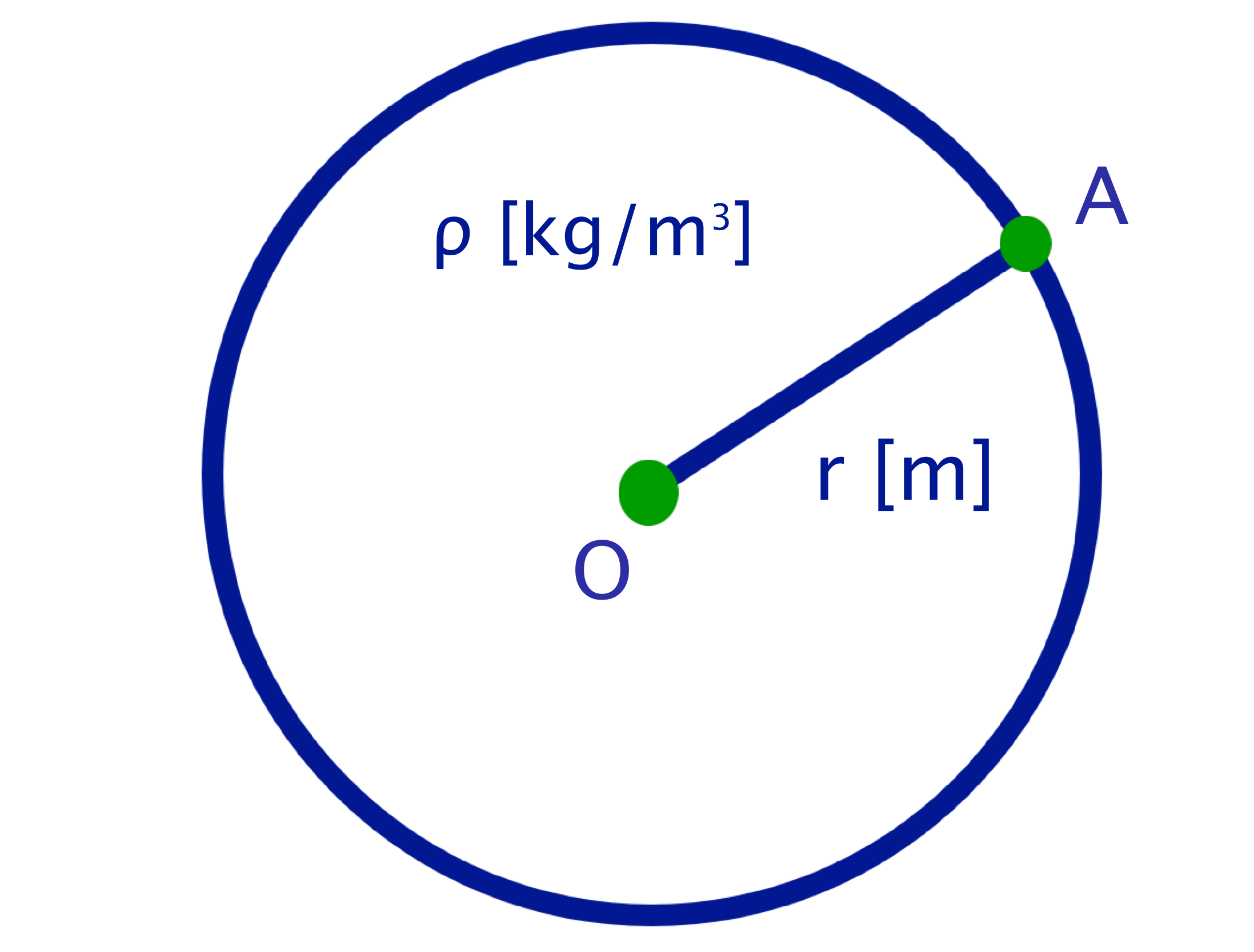}
\caption[Uniformly distributed matter]{Matter is distributed uniformly with a density $\rho$ [$\rm{kg}/\rm{m}^{3}$]. Distance from the point ``O" to the point ``A" at the `initial time' is defined to be $r$ [m].}{\label{fig13} }
\end{figure}

Recent redshift surveys show that our Universe is \textbf{homogeneous} and \textbf{isotropic} when we look at much larger scales than 100 Megaparsecs (Mpc). If the matter distribution is homogeneous, the density is the same everywhere. Isotropic distribution around the earth means that matter is distributed in the same way in all the directions if we observe the sky from the earth. For example, if matter is distributed in a spherically symmetric way, it can be isotropic and not homogeneous. On smaller scales, there are structures like galaxies, clusters and superclusters as explained in \cite{Mukhanov:2005}. 
Therefore, let us assume that matter distribution is homogeneous and isotropic when we study the global dynamics of the Universe. Let us pick up a random point ``O". Then, again, we pick another point ``A" randomly as in Fig. \ref{fig13}. The distance between those two points at the initial time is defined to be $r$ [m] while the energy density of the matter is $\rho$ [$\rm{kg}/\rm{m}^{3}$]. As we see below, we can study the dynamics of the Universe which is in analogy with that of the balloon explained above using the \textbf{General Relativity} (GR) under those assumptions. 

GR states that gravity is equivalent to the curvature of the space-time. How the space-time is curved is described by the metric tensor $g_{\mu\nu}$ which tells how the space-time interval $dx^{\mu}$ contributes to the line element $ds$ as
\begin{equation}
ds^{2} = g_{\mu\nu}dx^{\mu}dx^{\nu}.
\end{equation}
Note that we use the Einstein summation convention with which we sum terms over all values of indices if those indices appear both as the covariant and contravariant scripts in any expression. For example, the expression above means
\begin{equation}
g_{\mu\nu}dx^{\mu}dx^{\nu}=\sum\limits_{\mu,\nu=0}^3 g_{\mu\nu}dx^{\mu}dx^{\nu},
\end{equation}
in the four dimensional space-time which is described with the indices $\mu=0,1,2,3$. In the case that there is no gravity, the metric tensor reads
\begin{equation}
g_{\mu\nu}=\left( 
\begin{array}{@{\,}cccc@{\,}}
-1&0&0&0\\
0&1&0&0 \\
0&0&1&0\\
0&0&0&1
\end{array}
\right).
\end{equation}
This is the metric of the \textbf{Minkowski space-time} in which Special Relativity holds. As stated above, the Universe is homogeneous and isotropic on much larger scales than 100 Mpc. If we assume homogeneity and isotropy, the metric is determined uniquely as \cite{Mukhanov:2005}
\begin{equation}\label{walker}
ds^{2} = -dt^{2} + a^{2}(t) \left[\frac{dr^{2}}{1-Kr^{2}} + r^{2}\left(d\theta^{2} + \sin^{2}\theta d\phi^{2} \right) \right],
\end{equation}
where ($t$, $r$, $\theta$, $\phi$) are space-time spherical coordinates and $K$ is constant. This metric is called the \textbf{Friedmann-Robertson-Walker (FRW) metric}. $r$ is the radial coordinate at $t_{0}$ at which $a\left(t\right)$ is unity. We can express the position of the point A with respect to the point O in Fig. \ref{fig13} at a given time $t$ as
\begin{equation}\label{physicalcomoving}
\vec{l} = a\left(t\right)\vec{r},
\end{equation}
where $r=\lvert\vec{r}\rvert$ and $\vec{l}$ is the position of the object at a given time $t$. $a\left(t\right)$ has the meaning of the scale of $\vec{l}$, therefore it is called the \textbf{scale factor}. 

The space-time is ``curved" by matter. How the curvature of the space-time is generated is described in the \textbf{Einstein equation} as
\begin{equation}\label{einstein}
G_{\mu\nu} + \Lambda g_{\mu\nu} = 8\pi GT_{\mu\nu},
\end{equation}
where $\Lambda$ is a constant called the \textbf{cosmological constant}, $T_{\mu\nu}$ is the \textbf{energy momentum tensor} and $G_{\mu\nu}$ is the \textbf{Einstein tensor}. Those tensors are symmetric as follows
\begin{equation}
\left\{\begin{array}{@{\,}c@{\,}}
T_{\mu\nu} = T_{\nu\mu}\\
G_{\mu\nu} = G_{\nu\mu}\\
g_{\mu\nu} = g_{\nu\mu}
\end{array}\right. .
\end{equation}
Though we have used the word ``matter" without defining so far, let us define matter to be everything incorporated in the energy momentum tensor which generates gravity through the Einstein equation. For perfect fluid, the energy momentum tensor is given by
\begin{equation}\label{introductionperfectenergymomentum}
T^{\mu}_{\nu} = \left(\rho + P \right)u^{\mu}u_{\nu} + P\delta^{\mu}_{\nu},
\end{equation}
where $u^{\mu}$ is the fluid four-velocity, $\rho$ and $P$ are the energy density and the pressure of the fluid respectively. In a local rest frame where $u^{\mu}=\left(c,0,0,0\right)$, we have
\begin{equation}\label{energymomentumfirstdefined}
\left\{\begin{array}{@{\,}c@{\,}}
T^{00} = \rho \\
T^{i j} = P\delta_{i j}
\end{array}
\right. .
\end{equation}
The Einstein tensor is defined as
\begin{equation}
G_{\mu\nu} = R_{\mu\nu} - \frac{1}{2}R g_{\mu\nu},
\end{equation}
where the \textbf{Ricci scalar} R is defined by using the \textbf{Ricci tensor} $R_{\mu\nu}$ as
\begin{equation}
R = R^{\mu}_{\hspace{3pt}\mu} = g^{\mu\nu}R_{\mu\nu}. \label{Ricciscalar}
\end{equation}
The Ricci scalar is defined by using the \textbf{Riemann tensor} as follows
\begin{equation}
R_{\mu\nu} = R^{\lambda}_{\hspace{3pt}\mu \lambda \nu}.
\end{equation}
The Riemann tensor is written as
\begin{equation}
R^{\sigma}_{\hspace{3pt}\rho \nu \mu}=\Gamma^{\sigma}_{\hspace{3pt}\mu \rho, \nu}-\Gamma^{\sigma}_{\hspace{3pt}\nu \rho, \mu}+\Gamma^{\alpha}_{\hspace{3pt}\mu \rho}\Gamma^{\sigma}_{\hspace{3pt}\alpha \nu}-\Gamma^{\alpha}_{\hspace{3pt}\nu \rho}\Gamma^{\sigma}_{\hspace{3pt}\alpha \mu},
\end{equation}
where the Christoffel symbol is
\begin{equation}
\Gamma^{\rho}_{\hspace{3pt}\mu \nu}=\frac{1}{2}g^{\rho \sigma}\{g_{\nu \sigma,\mu}+g_{\mu \sigma,\nu}-g_{\mu \nu,\sigma}\}.
\end{equation}
Though we do not study General Relativity in detail in this thesis (see \cite{Wald:1984} for detailed explanation of those tensors), the Christoffel symbol defines the ``parallel transport" of a vector $v^{\mu}$ in the curved space-time as
\begin{equation}\label{parallel}
v^{\mu}\left(x+\delta x\right)=v^{\mu}\left(x\right)-\Gamma^{\mu}_{\hspace{3pt}\nu \lambda}\left(x\right)v^{\nu}\left(x\right)\delta x^{\lambda},
\end{equation}
where $v^{\mu}\left(x+\delta x\right)$ is the parallel transported $v^{\mu}\left(x\right)$ from $x$ to $x+\delta x$. It is worth noting that all components of the Christoffel symbol vanishes if there is no gravity and we can see from equation (\ref{parallel}) that the ``parallel transport" does not change any component of the vector as in the Euclidean space. 
Because the Riemann tensor can be defined by using only the metric tensors, the Einstein tensor can be rewritten by using only the metric tensors. Therefore, we can see that Einstein equation shows how matter curves the space-time through the metric tensor. 
If we substitute the FRW metric (\ref{walker}) into the Einstein equation (\ref{einstein}), we can obtain the gravitational equations for the homogeneous and isotropic space. The $\left(0,0\right)$ component of the Einstein equation is given by
\begin{equation}\label{friedmann}
H^{2}=\frac{\dot{a}\left(t\right)^{2}}{a\left(t\right)^{2}}=\frac{1}{3 M^{2}_{\rm{P}}} \rho - \frac{K\,c^{2}}{a\left(t\right)^{2}}+\frac{\Lambda c^{2}}{3},
\end{equation}
which is called the \textbf{Friedmann equation} where the constant $K$ is related to the spatial curvature and $\Lambda$ is the cosmological constant. The trace of the Einstein equation gives
\begin{equation}\label{acceleration}
\frac{\ddot{a}}{a}=-\frac{1}{6 M^{2}_{\rm{P}}}\left(\rho c^{2} + 3p\right) + \frac{\Lambda c^{2}}{3},
\end{equation}
which tells us about the acceleration of the expansion. Those equations describe the dynamics of the homogeneous Universe. 

By substituting equation (\ref{acceleration}) into equation (\ref{friedmann}) after differentiating with respect to the cosmic time, we obtain
\begin{equation}\label{continuity}
\dot{\rho}=-3H\left(\rho+\frac{P}{c^{2}}\right),
\end{equation}
which is called the continuity equation. The cosmological constant term could play an important role in explaining the accelerating expansion of the current Universe which is discovered with the observations of the type Ia supernovae. If we consider ordinary matter such as radiation whose equation of state is $p=\rho c^{2}/3$ or non-relativistic matter whose equation of state is $p=0$, the first two terms in the left-hand side of equation (\ref{acceleration}) are always negative. Therefore, we cannot explain why the acceleration $\ddot{a}$ can be positive without the cosmological constant term in General Relativity. 

\subsection{Success of the Big Bang scenario}
The Big Bang model explains Hubble's law as in subsection \ref{ideaofbigbang}. However, the biggest success of the Big Bang model is the prediction of the abundance of light elements in the early Universe. 

The most dominant chemical element in the Universe is hydrogen which constitutes about 75 $\%$ of all baryonic matter. The second most dominant chemical element is helium which makes up about 25 $\%$ of all baryonic matter. All the other chemical elements have only small abundances. If we assume that the large amount of $^{4}\rm{He}$ in the Universe had been produced in stars, we have a problem with observation as follows. The binding energy of $^{4}\rm{He}$ is 28.3 MeV. Therefore, when one nucleus of $^{4}\rm{He}$ is formed, the energy released per one baryon is $28.3/4 \sim 7.1$ Mev $\sim 1.1 \times 10^{-5}$ erg. If we assume that all the helium nuclei in the Universe were formed in the last 10 billion years, which is $3.2 \times 10^{17}$ s, the luminosity to mass ratio can be estimated roughly as \cite{Mukhanov:2005}
\begin{equation}
\frac{L}{M_{bar}} \sim \frac{1}{4}\frac{1.1 \times 10^{-5} \rm{erg}}{\left(1.7 \times 10^{-24} \rm{g}\right) \times \left(3.2 \times 10^{17} \rm{s}\right)} \sim 5 \frac{\rm{erg}}{\rm{g} \cdot \rm{s}} \sim 2.5 \frac{L_{\odot}}{M_{\odot}},
\end{equation}
where $M_{\odot}$ is the solar mass, $L_{\odot}$ is the solar luminosity and $M_{bar}$ is the average baryon mass in the nucleus which is about one quarter of the mass of the helium nucleus. However the observed value is $L / M_{bar} \leq 0.05 L_{\odot} / M_{\odot}$. This means that less than 2 $\%$ of $^{4}\rm{He}$ could be fused in stars if the luminosity of baryons observed on the earth was not much larger in the past than at present. 

The plausible explanation of the helium abundance which the Big Bang scenario provides is that it was produced in the very early stage of the Universe. According to the Big Bang scenario, as we go back in time, the temperature of the Universe goes up. In the very early stage of the Universe, the kinetic energy of the elementary particles was much higher than the binding energy of any kind of nucleus. Then, the helium nuclei were formed rapidly when the temperature drops well below the binding energy of helium which is about 28 MeV. Actually, primordial \textbf{nucleosynthesis} occurred at the temperature $\sim 0.1$ MeV which is a few minutes after the Big Bang. 

The abundances of the light elements produced in the nucleosynthesis are determined by the \textbf{Boltzmann equations} which govern the evolution of the distribution of the relativistic particles in phase space \cite{Lyth:2009}. To obtain accurate results, we need to use the numerical integration. However, analytic estimates exist \cite{Mukhanov:2005} and it agrees well with the numerical results for the $^{4}\rm{He}$ abundance. We show the analytic estimates briefly below. \\

As the temperature drops below a few hundred MeV, the quarks and gluons are confined and form baryons and mesons. Baryons are made of three quarks while the mesons are made of one quark and one anti-quark. Below 100 MeV, the Universe is filled with the primordial radiation ($\gamma$), neutrons ($n$), protons ($p$), electrons ($e^{-}$), positrons ($e^{+}$) and three neutrino species. Mesons, heavy baryons, $\mu^{-}$ and $\tau^{-}$ leptons are also present, but they become negligible compared to those abundant light particles as they get non-relativistic. 

When the temperature falls below 1 MeV, neutrinos decouple because weak interactions which keep neutrinos in thermal contact with each other and with the other particles become inefficient. Also, weak interactions maintain the chemical equilibrium between protons and neutrons as

\begin{equation}
n + \nu \rightleftarrows p + e^{-}, \,\,\,\,\,\,\,\,\,\, n + e^{+} \rightleftarrows p + \bar{\nu},
\end{equation}
where $\nu$ refers to the electron neutrino. Because weak interactions become inefficient, the neutron to proton ratio freezes out except for the neutron decay. Detailed calculation using the cross-section derived by Fermi theory can be found in \cite{Mukhanov:2005}. The relative concentration of neutrons is
\begin{equation}
X_{n} = \frac{n_{n}}{n_{n}+n_{p}},
\end{equation}
where $n_{n}$ and $n_{p}$ are the number densities of neutrons and protons respectively. The freeze-out concentration is derived as
\begin{equation}
X_{n}^{*} \simeq 0.158 + 0.005\left(N_{\nu} - 3 \right), \label{xnnucleo}
\end{equation}
where $N_{\nu}$ is the number of light neutrino species. Therefore, the neutron concentration after freeze-out reads
\begin{equation}
X_{n} = X_{n}^{*} \exp{\left(-t/\tau_{n} \right)},
\end{equation}
where the lifetime of a free neutron before the neutron decay
\begin{equation}
 n \rightarrow p + e^{-} + \bar{\nu},
\end{equation}
is $\tau_{n} \approx 886\,\left[\rm{s} \right]$. Helium-4 could be built directly in a four-body collision 
\begin{equation}
p + p + n + n \rightarrow ^{4}\rm{He}.
\end{equation}
However, the number densities of protons and neutrons during the period are too low to have such collisions sufficiently. Therefore, light complex nuclei are produced through a sequence of two-body reactions. The first step is the deuterium production as
\begin{equation}
p + n \rightleftarrows D + \gamma,
\end{equation}
where D is the deuterium. In order to produce heavier elements from the deuterium such as $^{4}$He, there needs to be sufficiently dense deuterium concentration. Actually, $^{4}$He is not produced until the temperature becomes around 0.08 MeV even though the binding energy of $^{4}$He is 28.3 MeV. This is because of a``\textit{deuterium bottleneck}", whereby the low abundance of deuterium suppresses the production of heavier elements. After the deuterium abundance rises, two-body reactions
\begin{equation}
D + D \rightarrow ^{3}\rm{He} + n, \,\,\,\,\,\,\,\,\,\, D + D \rightarrow T + n,
\end{equation}
become efficient where T is tritium. Then, tritium combines with deuterium to produce $^{4}$He as
\begin{equation}
T \, D \rightarrow ^{4}\rm{He} \, n.
\end{equation}
Even though $^{3}$He can have both the reactions
\begin{equation}
^{3}\rm{He} \, n \rightarrow T \, p, \label{heliumneutron}
\end{equation}
and
\begin{equation}
^{3}\rm{He} \, D \rightarrow ^{4}\rm{He} \, p, \label{heliumdeuterium}
\end{equation}
the reaction (\ref{heliumneutron}) is more efficient than the reaction (\ref{heliumdeuterium}). Also, the binding energy of $^{4}$He (28.3 MeV) is four times as large as the binding energies of the intermediate elements $^{3}$He (7.72 MeV) and T (6.92 MeV). Therefore, almost all the neutrons are fused into $^{4}$He through the reaction chain $n\,p \rightarrow D \rightarrow T \rightarrow ^{4} \rm{He}$ and $n\,p \rightarrow D \rightarrow ^{3} \rm{He} \rightarrow T \rightarrow ^{4} \rm{He}$. Therefore, the abundance of $^{4}$He is determined by the abundance of the available free neutrons at this time. The temperature at this time is
\begin{equation}
T_{\rm{MeV}} \simeq 0.07\left(1+0.03 \ln{\eta_{10}} \right), \label{temperaturenucleo}
\end{equation}
where $\eta_{10}$ is the baryon to photon ratio which is defined as
\begin{equation}
\eta_{10} \equiv 10^{10} \times \frac{n_{N}}{n_{\gamma}},
\end{equation}
where $n_{N}$ and $n_{\gamma}$ denote the number densities of the neutrons and photons respectively. The \textbf{Cosmic Microwave Background} (CMB) observations give us the value of the baryon to photon ratio. For example, we have $6.2<\eta_{10}<6.9$ at 68$\%$ confidence level in \cite{Spergel:2003}. In the early Universe, the main contribution to the energy density comes from relativistic particles. If we neglect the chemical potentials of the particles, the energy density $\tilde{\epsilon}_{r}$ is
\begin{equation}\label{temperatureenergyfirstchapfinal}
\tilde{\epsilon}_{r}=\kappa T^{4},
\end{equation}
where
\begin{equation}
\kappa \simeq 1.11 + 0.15 \left(N_{\nu} - 3 \right), \label{temperaturetime}
\end{equation}
after the electron-positron annihilation. From equations (\ref{friedmann}) and (\ref{temperatureenergyfirstchapfinal}), it is shown that the temperature is given by equation (\ref{temperaturenucleo}) at the time
\begin{equation}
t_{\rm{sec}} \simeq 269 \left(1 - 0.07\left(N_{\nu} - 3\right) - 0.06\ln{\eta_{10}} \right). \label{timenucleo}
\end{equation}
Note that we used $H=1/2t$ in the radiation domination. Because half of the total mass of $^{4}$He is due to protons, its final abundance by mass is
\begin{equation}
X_{^{4}\rm{He}} = 2 X_{n} = 2 X_{n}^{*} \exp{\left(-t_{\rm{sec}}/\tau_{n} \right)}. \label{abundancefinalnucleo}
\end{equation}
Substituting equations (\ref{xnnucleo}) and (\ref{timenucleo}) into equation (\ref{abundancefinalnucleo}), we obtain
\begin{equation}
X_{^{4}\rm{He}} \simeq 0.23 + 0.012\left(N_{\nu} - 3 \right) + 0.005 \ln{\eta_{10}}. \label{finalresultnuclehe}
\end{equation}
If we take into account the presence of an additional massless neutrino, the final abundance increases by about 1.2 $\%$. Also, if we substitute $\eta_{10}=6.5$ into equation (\ref{finalresultnuclehe}) for example, we obtain $X_{^{4}\rm{He}} \simeq 0.25$. We can see that this actually agrees well with what we observe. This also agrees well with the numerical results. The abundances of other light elements can be also theoretically predicted and they are in very good agreement with the observed element abundances. 

\subsection{Horizon Problem}
Even though the Big Bang scenario is successful explaining the expansion of the Universe and the abundance of the light elements, it has some problems as well. The first problem is called the \textbf{horizon problem}. Observations of the CMB give us the information about the radiation energy density distribution of the early Universe at the redshift $z \sim 1100$. It is well known that it is almost homogeneous with small fluctuations in all areas of the sky. This fact cannot be explained by the standard Big Bang scenario as shown below. With a change of variables, the FRW metric (\ref{walker}) can be rewritten as
\begin{equation}
ds^2 = - dt^2 + a(t)^2 \left[d\chi^2 + f\left(\chi\right)^{2}\left(d\theta^2 + \sin^2\theta d\phi^2 \right) \right],\label{RW}
\end{equation}
where
\begin{equation}
f(\chi) = \left\{\begin{array}{@{\,}cc@{\,}}
K^{-1/2}\sin(K^{1/2}\chi) & (K>0)\\
\chi & (K=0)\\
\lvert K \rvert \sinh(\lvert K \rvert \chi) & (K<0)
\end{array}
\right.
\end{equation}
We can obtain the distance which light has travelled in the radial direction since the decoupling as
\begin{equation}
\chi_{\rm{o}}\left(t \right) = \int^{t}_{t_{d}} \frac{dt'}{a(t')},\label{opticalh}
\end{equation}
where $t$ is the age of the Universe and $t_{d}$ is the age of the Universe at the decoupling. Decoupling is the event at which the Universe became electrically neutral and light could travel freely without being compton scattered with the electrically charged particles since then. $l_{\rm{o}} \left(t \right)  = a\left(t \right) \chi_{\rm{o}} \left(t \right) $ is called the \textbf{optical horizon} because we cannot observe anything at a distance greater than that. The distance that a photon could have traveled at $t$ since the Universe began at $a=0$ is called the \textbf{particle horizon}, which can be defined as
\begin{equation}
l_{\rm{p}} \left(t \right)  = a(t)\chi_{\rm{p}}(t),
\end{equation}
where the comoving distance $\chi_{\rm{p}}$ can be written
\begin{equation}
\chi_{\rm{p}} \left(t \right)  = \int^{t}_{0} \frac{dt'}{a(t')}.\label{particleh}
\end{equation}
As stated above, when the decoupling happened at the temperature 3000 K, photons start to travel freely. Because we can derive the relation between the redshift and the temperature of radiation $T$ as
\begin{equation}
\frac{a_{0}}{a} = 1+z \propto T,\label{scalered}
\end{equation}
where $a_{0}$ is the scale factor at present and the redshift $z$ is defined by
\begin{equation}
z = \frac{\lambda_{\rm{r}}-\lambda_{\rm{e}}}{\lambda_{\rm{e}}},
\end{equation}
where $\lambda_{\rm{r}}$ and $\lambda_{\rm{e}}$ are the wavelengths of light at the points of observation and emission respectively, we can obtain the redshift at the decoupling from equation (\ref{scalered}) as
\begin{equation}
z_{d} \simeq 1050,\label{zd}
\end{equation}
with the temperature of radiation $T \simeq 2.73K$ in the present Universe whose redshift is $z=1$. If we assume the radiation domination, the scale factor grows as
\begin{equation}
a\left(t \right) = \left(\frac{t}{t_{e}} \right)^{1/2}, \label{scalefactorradiation}
\end{equation}
where we define $t_{e}$ to be the time at the matter radiation equality. In the matter domination, the scale factor reads
\begin{equation}
a\left(t \right) = \left(\frac{t}{t_{e}} \right)^{2/3}. \label{scalefactormatter}
\end{equation}
Using equations (\ref{particleh}), (\ref{scalefactorradiation}) and (\ref{scalefactormatter}), we obtain the particle horizon at the decoupling as
\begin{equation}
\begin{split}
\chi_{\rm{p}}(t_{d}) &= \int^{t_{e}}_{0} \frac{dt'}{a(t')} + \int^{t_{d}}_{t_{e}} \frac{dt'}{a(t')}\\
 &= 2t_{e}\left[\left(\frac{t}{t_{e}}\right)^{1/2} \right]^{t_{e}}_{0} + 3t_{e}\left[\left(\frac{t}{t_{e}}\right)^{1/3} \right]^{t_{d}}_{t_{e}}\\
 &= 2t_{e} + 3t_{e}\left\{\left(\frac{a_{d}}{a_{e}} \right)^{1/2} -1 \right\}\\
 &= t_{e}\left(3 \sqrt{\frac{1+z_{e}}{1+z_{d}}} - 1 \right), \label{phdecoupilng}
\end{split}
\end{equation}
where we used the relation (\ref{scalered}). Note that we assumed that the transition from the radiation domination to the matter domination occurs instantaneously at $t_{e}$ (at the redshift $z_{e}$) which is earlier than the decoupling time $t_{d}$. Then, the optical horizon at the present Universe can be derived using equations (\ref{opticalh}) and (\ref{scalefactormatter}) as
\begin{equation}
\begin{split}
\chi_{\rm{o}}\left(t_{0} \right) &= \int^{t_{0}}_{t_{d}} \frac{dt'}{a(t')}\\
 &= 3 t_{e} \left\{\left(\frac{t_{0}}{t_{e}} \right)^{1/3} - \left(\frac{t_{d}}{t_{e}} \right)^{1/3} \right\}\\
 &= 3 t_{e} \left\{\left(\frac{a_{0}}{a_{e}} \right)^{1/2} - \left(\frac{a_{d}}{a_{e}} \right)^{1/2} \right\}\\
 &= 3 t_{e} \left(\sqrt{1+z_{e}} - \sqrt{\frac{1+z_{e}}{1+z_{d}}} \right).
\end{split}
\end{equation}
We used the relation (\ref{scalered}) again to convert the scale factors into the redshift parameters. Then, the ratio of the present optical horizon to the particle horizon at the decoupling is
\begin{equation}
\begin{split}
\frac{\chi_{\rm{o}}\left(t_{0} \right)}{\chi_{\rm{p}}\left(t_{d}\right)} &= \frac{3 t_{e} \left(\sqrt{1+z_{e}} - \sqrt{\frac{1+z_{e}}{1+z_{d}}} \right)}{t_{e}\left(3 \sqrt{\frac{1+z_{e}}{1+z_{d}}} - 1 \right)}\\
&= \frac{3 \left(\sqrt{\left(1+z_{e}\right)\left(1+z_{d}\right)} - \sqrt{1+z_{e}} \right)}{3 \sqrt{1+z_{e}} - \sqrt{1+z_{d}}}\\
&\sim 40,\label{horizonpro}
\end{split}
\end{equation}
where we used equation (\ref{zd}) and $z_{e} \simeq 3300$. This means that the observable area of the CMB today is much larger than the area of causality at the decoupling. Because nothing can propagate outside the particle horizon, it means that the Universe was homogeneous and isotropic in a large area which consisted of many areas that could not have interacted with each other at the time of decoupling. This cannot be explained only by the standard Big Band theory. 

\subsection{Flatness problem}\label{subsec:flatness}
We can rewrite the Friedmann equation (\ref{friedmann}) as
\begin{equation}
\left\lvert \frac{K}{a^2 H^2} \right\rvert = \lvert \Omega_{\rm{tot}} - 1 \rvert, \label{flatnesspro}
\end{equation}
where $\Omega_{\rm{tot}}$ is
\begin{equation}
\begin{split}
\Omega_{\rm{tot}} &= \Omega + \Omega_{\Lambda}\\
&= \frac{\rho}{3H^2} + \frac{\Lambda}{3H^2}.
\end{split}
\end{equation}
In equation (\ref{flatnesspro}), $\lvert K/a^2 H^2 \rvert$ can be rewritten as $\lvert K/\dot{a}^2 \rvert$. Therefore, in the matter domination, equation (\ref{flatnesspro}) can be rewritten as
\begin{equation}
\lvert \Omega_{\rm{tot}} - 1 \rvert \propto \lvert Kt^{2/3} \rvert,
\label{flatmatter}
\end{equation}
from equation (\ref{scalefactormatter}) and it can be rewritten in the radiation domination as
\begin{equation}
\lvert \Omega_{\rm{tot}} - 1 \rvert \propto \lvert Kt \rvert,
\label{flaradiation}
\end{equation}
from equation (\ref{scalefactorradiation}). Those equations mean that $\lvert \Omega_{\rm{tot}} - 1 \rvert$ is an increasing function of time in both the matter and radiation domination. Therefore, as we go backwards in time, the right hand side of equation (\ref{flatnesspro}) keeps decreasing. Because the current observations suggest that $\lvert \Omega_{\rm{tot}} - 1 \rvert$ is within a few percent of unity \cite{Ade:2013}, it must be even smaller in the past. We require $\lvert \Omega_{\rm{tot}} - 1 \rvert < \mathcal{O}\left(10^{-16}\right)$ at the nucleosynthesis and $\lvert \Omega_{\rm{tot}} - 1 \rvert < \mathcal{O}\left(10^{-64}\right)$ at the Planck epoch \cite{Bassett:2006}. This implies that the initial conditions must have been chosen accurately in order to have our current nearly flat Universe today. This also cannot be explained by the standard Big Bang model. This is the second problem of the standard Big Bang model called the \textbf{Flatness problem}. 
\subsection{Relic problem}
According to the standard particle physics, the physical laws were simpler in the early Universe before the gauge symmetries were broken. When such symmetries are broken, many unwanted relics such as monopoles, cosmic strings, and other topological defects are produced \cite{Bassett:2006}. Some of those particles behave as matter and hence their energy densities decrease as
\begin{equation}
\rho_{m} = \frac{\rho_{im}}{a^{3}},
\label{matterdensity}
\end{equation}
where the energy density of radiation decreases as
\begin{equation}
\rho_{r} = \frac{\rho_{ir}}{a^4},
\label{radiationdensity}
\end{equation}
with constants $\rho_{im}$ and $\rho_{ir}$ which are initial energy densities. Therefore, the energy densities of  such heavy particles decrease more slowly than those of radiation. It means that they are likely to be dominant in the present Universe. It would contradict a variety of observations such as those of the light element abundances. This problem is the third problem of the standard Big Bang scenario which is called the \textbf{relic problem}.

\section{\label{sec:inf}Inflation}
As explained in section \ref{sec:sbbs}, the standard Big Bang scenario has several problems which cannot be solved by itself. However, those problems can be evaded if we assume that the Universe expanded exponentially in the very early stage after the Big Bang. Such expansion is called ``inflation". In this section, the idea of inflation is first introduced explaining how such an expansion solves the horizon problem. Then, as an example, single field slow-roll inflation is introduced and we see how inflation solves other problems as well. 
\subsection{Idea of inflation}\label{subsec:ideainflation}
Among the three problems of the standard Big Bang scenario, let us think about the horizon problem first. As equation (\ref{horizonpro}) shows, the particle horizon at the decoupling is much smaller than the present optical horizon if we assume that the Universe has been dominated only by the ordinary matter such as radiation and matter whose scale factors behave as in equations (\ref{scalefactorradiation}) and (\ref{scalefactormatter}). However, if the Universe expanded exponentially as
\begin{equation}
a\left(t \right)=C \exp\left({D^{-1} t}\right),
\label{scalefatorinflation}
\end{equation}
in the first $t_{i}$ seconds with the constants $C$ and $D$, the particle horizon at the decoupling is
\begin{equation}
\begin{split}
\chi_{\rm{p}}(t_{d}) &= \int^{t_{i}}_{0} \frac{dt'}{a(t')} + \int^{t_{e}}_{t_{i}} \frac{dt'}{a(t')} + \int^{t_{d}}_{t_{e}} \frac{dt'}{a(t')}\\
 &= - C^{-1} D \left[\exp{\left(- D^{-1} t\right)} \right]^{t_{i}}_{0}  + 2t_{e}\left[\left(\frac{t}{t_{e}}\right)^{1/2} \right]^{t_{e}}_{t_{i}} + 3t_{e}\left[\left(\frac{t}{t_{e}}\right)^{1/3} \right]^{t_{d}}_{t_{e}}\\
 &= C^{-1} D \left\{1-\exp{\left(- D^{-1} t_{i}\right)} \right\} + 2t_{e}\left\{1-\frac{a_{i}}{a_{e}} \right\} + 3t_{e}\left\{\left(\frac{a_{d}}{a_{e}} \right)^{1/2} -1 \right\}\\
 &= C^{-1} D \left\{1-\exp{\left(- D^{-1} t_{i}\right)} \right\} + t_{e}\left(3 \sqrt{\frac{1+z_{e}}{1+z_{d}}} - 1 - 2 \frac{T_{e}}{T_{i}} \right),
 \label{orangejuice}
\end{split}
\end{equation}
where $t_{i}$ is the time at the end of inflation in which the Universe expands with the scale factor (\ref{scalefatorinflation}), $z_{i}$ is the redshift at $t_{i}$ and $T_{i}$ and $T_{e}$ are the temperatures of the Universe at $t_{i}$ and $t_{e}$ respectively. We obtain
\begin{equation}
C=\sqrt{\frac{t_{i}}{t_{e}}}\exp{(-D^{-1}t_{i})},
\end{equation}
by assuming the scale factor (\ref{scalefactorradiation}) is equal to the scale factor (\ref{scalefatorinflation}) at $t=t_{i}$. Then, we have
\begin{equation}
\begin{split}
C^{-1} D \left\{1-\exp\left({- D^{-1} t_{i}}\right) \right\}&=\sqrt{\frac{t_{e}}{t_{i}}}\exp{(D^{-1}t_{i})}D \left\{1-\exp\left({- D^{-1} t_{i}}\right) \right\}\\
&=\frac{t_{e}}{D^{-1}t_{i}}\sqrt{\frac{t_{i}}{t_{e}}}\left\{\exp{\left(D^{-1} t_{i}\right)}-1 \right\}.
\label{applejuice}
\end{split}
\end{equation}
If we assume $\exp{(D^{-1}t_{i})}=\exp{\left(60\right)}$ and use
\begin{equation}
\sqrt{\frac{t_{i}}{t_{e}}}=\frac{a_{i}}{a_{e}}=\frac{T_{e}}{T_{i}},
\end{equation}
during the radiation domination, with equations (\ref{orangejuice}) and (\ref{applejuice}), we have
\begin{equation}
\begin{split}
\chi_{\rm{p}}(t_{d}) &=\frac{t_{e}}{60}\frac{T_{e}}{T_{i}}\left\{\exp{\left(D^{-1} t_{i}\right)}-1 \right\} + t_{e}\left(3 \sqrt{\frac{1+z_{e}}{1+z_{d}}} - 1 - 2 \frac{T_{e}}{T_{i}} \right)\\
&\approx\frac{t_{e}}{60}\frac{1 \left[\rm{eV}\right]}{10^{21} \left[\rm{eV}\right]}\left\{\exp{60}-1 \right\} + t_{e}\left(3 \sqrt{\frac{1+3300}{1+1050}} - 1 - 2 \frac{1 \left[\rm{eV}\right]}{10^{21} \left[\rm{eV}\right]} \right)\\
&\approx 1900 t_{e},
\end{split}
\end{equation}
where we assumed $T_{e}=1 \left[\rm{eV}\right]$ and $T_{i}=10^{12} \left[\rm{GeV}\right]$. Using this result, equation (\ref{horizonpro}) is no longer correct and we have
\begin{equation}
\begin{split}
\frac{\chi_{\rm{o}}\left(t_{0} \right)}{\chi_{\rm{p}}\left(t_{d}\right)} &\approx \frac{3 t_{e} \left(\sqrt{1+z_{e}} - \sqrt{\frac{1+z_{e}}{1+z_{d}}} \right)}{1900t_{e}}\\
&\approx \frac{3 t_{e} \left(\sqrt{1+ 3300} - \sqrt{\frac{1+3300}{1+1050}} \right)}{1900t_{e}}\\
&\approx8.79\times10^{-2}.
\end{split}
\end{equation}
We can see that we no longer have the horizon problem because the radius of the optical horizon today is only around 8 percent of the radius of the particle horizon at the decoupling. Though this percentage changes depending on the parameters such as the temperature of the Universe at the end of inflation $t_{i}$ as we can see above, this means that inflation can solve the horizon problem. In subsection \ref{subsec:slow-roll}, we see that inflation also solves the other two problems of the standard Big Bang scenario. 

\subsection{Slow-roll inflation}\label{subsec:slow-roll}

How can we have such an expansion? The standard way is to assume that the energy density of the Universe was dominated by the scalar field called inflaton. Though we do not know what inflaton is, the experiments at the Large Hadron collider (LHC) in Switzerland strongly suggest the existence of the Higgs boson which constitutes a scalar field (see \cite{Consonni:2013} for details). Also, there are possibilities that the super symmetric (SUSY) scalar particles will be found in future experiments. Below, we see the most standard and simple example of inflation. 

From equation (\ref{acceleration}), we need
\begin{equation}
\rho + 3P < 0,\label{acceleratedexpansion}
\end{equation}
in order to make the expansion accelerating ($\ddot{a}>0$) if we set $\Lambda=0$. The reason for ignoring the cosmological constant is because the Universe would be dominated completely by the cosmological constant and different from our Universe which we observe today if it had dominated the Universe at the beginning because the energy density of the cosmological constant is constant while the energy density of radiation and that of matter decrease as in equations (\ref{matterdensity}) and (\ref{radiationdensity}). 

As a field which satisfies the condition (\ref{acceleratedexpansion}), let us consider a minimally-coupled scalar field for inflaton whose Lagrangian is given by
\begin{equation}
\begin{split}
L &= -\frac{1}{2}\partial_{\mu}\phi\partial^{\mu}\phi - V(\phi)\\
&= \frac{1}{2}\dot{\phi}^2 - V(\phi),\label{lagrangianinflaton}
\end{split}
\end{equation}
where we assume the spatial homogeneity of the Universe with the metric (\ref{walker}) where $K=0$. The energy-momentum tensor for inflaton is given by
\begin{equation}
T^{\mu}_{\hspace{5pt}\nu} = g^{\mu\rho}\partial_{\rho}\phi\partial_{\nu}\phi + L\delta^{\mu}_{\hspace{5pt}\nu}.\label{energymomentum}
\end{equation}
If we assume that the Universe is homogeneous and isotropic, the inflaton field can be regarded as \textbf{perfect fluid} whose energy-momentum tensor can be described as: $\rho = -T^{0}_{\hspace{5pt}0}$ and $T^{i}_{\hspace{5pt}i} = P\delta^{i}_{\hspace{5pt}i}$ where $i=1,2,3$. Because we are considering the Robertson-Walker metric, we can obtain $\rho$ and $P$ from equations (\ref{lagrangianinflaton}) and (\ref{energymomentum}) as
\begin{equation}
\rho_{\phi} = \frac{1}{2}\dot{\phi}^2 + V(\phi),\label{phirho}
\end{equation}
\begin{equation}
P_{\phi} = \frac{1}{2}\dot{\phi}^2 - V(\phi).\label{phip}
\end{equation}
By substituting $\rho$ and $P$ obtained above into equation ({\ref{continuity}}), we can obtain
\begin{equation}
\ddot{\phi} + 3H\dot{\phi} + V'(\phi) = 0,\label{phiequation}
\end{equation}
where $'\equiv d/d\phi$. Because $P_{\phi}$ can be rewritten as $P_{\phi} = - \rho_{\phi} + \dot{\phi}^2$, $\dot{\phi}^2$ must be small compared to $V(\phi)$ in order to satisfy the condition (\ref{acceleratedexpansion}) for the accelerating expansion. If we assume that
\begin{equation}
\dot{\phi}^2 \ll V(\phi),
\end{equation}
and
\begin{equation}
\lvert \ddot{\phi} \rvert \ll \lvert V'(\phi) \rvert,
\end{equation}
we can obtain $\rho \sim -P$ from equation (\ref{phirho}) and equation (\ref{phip}). Therefore, inflaton satisfies the condition (\ref{acceleratedexpansion}) and the expansion of the Universe becomes accelerating. Those assumptions are called the \textbf{slow-roll conditions}. If we introduce \cite{Lyth:2000}
\begin{equation}
\tilde{\epsilon} = \frac{M^{2}_{\rm{P}}}{2}\left(\frac{V'(\phi)}{V(\phi)} \right)^2,\label{sepsilon}
\end{equation}
\begin{equation}
\tilde{\eta} = M^{2}_{\rm{P}}\frac{V''(\phi)}{V(\phi)},\label{seta}
\end{equation}
which are called the \textbf{slow-roll parameters}, where $M^2_{\rm{P}} = (\hbar c/G)$ is the four dimensional Planck mass, slow-roll conditions can be written as
\begin{equation}
\tilde{\epsilon} \ll 1,
\end{equation}
\begin{equation}
\lvert \tilde{\eta} \rvert \ll 1.
\end{equation}
From equation (\ref{continuity}), the energy density is almost constant because we have $\rho \sim -P$. Then, from the Friedmann equation (\ref{friedmann}), the Hubble parameter $H$ is almost constant. Because $H\equiv\dot{a}/a$, we have
\begin{equation}
a\left(t \right)=C \exp\left({H t}\right),
\label{obtainedscale}
\end{equation}
where C is constant. We can see that the scale factor (\ref{obtainedscale}) is exactly the scale factor (\ref{scalefatorinflation}) which we need to solve the horizon problem as we saw above. Actually, with inflation, we can also solve other problems which are discussed in section \ref{sec:sbbs}. During inflation, equation (\ref{flatnesspro}) can be rewritten as
\begin{equation}
\left\lvert \Omega_{\rm{tot}} - 1 \right\rvert \propto \left\lvert \frac{K}{\exp{(2Ht)}} \right\rvert,
\end{equation}
from equation (\ref{obtainedscale}). Because this means that $\lvert \Omega_{\rm{tot}} - 1 \rvert$ decreases exponentially during inflation, it naturally explains why this quantity is extremely small at the beginning of the radiation domination as shown in subsection \ref{subsec:flatness}. Also, such exponential expansion of the Universe dilutes the density of the relic particles quickly. Therefore, the theory no longer predicts that the present Universe is dominated by such undesirable particles. In this way, inflation solves all three problems of the standard Big Bang scenario. 

After inflation ends, the energy of inflaton is converted into the energy of other elementary particles which our Universe is made of finally. This process is called the \textbf{reheating}. Reheating is essential because our Universe is obviously not made of inflaton currently. The regime of parametric resonance in which such elementary particles are produced from inflaton is called the \textbf{preheating}. I do not explain about the reheating in detail in this thesis because it is not directly related to the main topic. 

\subsection{CMB observables}\label{subsec:cmbobservables}
Observations have shown that the CMB is almost uniform, with small perturbations, $\Delta T$, present across the whole sky. The average temperature is 2.725 K and the amplitude of the fluctuations is given by \cite{Liddle:2003}
\begin{equation}
\frac{\Delta T}{T} \sim 10^{-5}.
\end{equation}
Those fluctuations are observed as the CMB temperature anisotropies in the observations. Inflation can naturally explain such fluctuations. Firstly, the comoving curvature perturbation $\mathcal{R}$ which will be introduced in subsection \ref{subsec:flatcomoving} is related to the quantum fluctuation of inflaton $\delta \phi$. In the canonical slow-roll inflation case, the relation is given by
\begin{equation}\label{examplecurvature}
\mathcal{R} = H \frac{\delta \phi}{\dot{\bar{\phi}}},
\end{equation}
where $H$ is the Hubble parameter and $\bar{\phi}$ is the homogeneous part of the scalar field. Secondly, the curvature perturbation is related to the temperature fluctuations of the CMB by the relation given in \cite{Lyth:2000}
\begin{equation}\label{sachswolfe}
\frac{\Delta T}{T} = \frac{1}{5} \mathcal{R}_{ls}, 
\end{equation}
where the subscript $ls$ denotes quantities at the last scattering when photons started to travel freely without Thompson scattered because almost all the charged particles have combined into atoms. Note that the \textbf{Sachs-Wolfe effect} (\ref{sachswolfe}) holds as long as we assume that the Universe has been matter-dominated from the last scattering to the present. This effect comes from the redshift at the last scattering surface and the fluctuations of the photon energy density on the last scattering surface causes the fluctuations of the CMB temperature through this effect \cite{Lyth:2009}. If we take into account the fact that the Universe has not been always matter-dominated from the last scattering to the present, we have an additional integrated effect which is acquired on the journey of photons from the last scattering surface to us, which is called the \textbf{Integrated Sachs-Wolfe effect}. 

With equations (\ref{examplecurvature}) and (\ref{sachswolfe}), we see that the quantum fluctuations of inflaton naturally produce the CMB temperature anisotropies. Therefore, the CMB observations strongly support inflationary models. At the same time, we can put constraints on the inflationary models with the statistical properties of the CMB temperature anisotropies. Below, let us introduce the correlation functions which are used to quantify the statistical properties of them following \cite{Lyth:2009}. Let us introduce a random field $g\left(\textbf{x}\right)$. This is a set of functions $g_{n}\left(\textbf{x}\right)$ each coming with a probability $P_{n}$. The set of functions is called the \textbf{ensemble} while each individual function is called a \textbf{realisation} of the ensemble. The two-point function is defined as
\begin{equation}
\left<g\left(\textbf{x}\right) g\left(\textbf{x}'\right) \right> = \sum_{n} P_{n} g_{n}\left(\textbf{x}\right) g_{n}\left(\textbf{x}'\right),
\end{equation}
and the N-point functions are defined similarly. The random field is usually statistically homogeneous and isotropic. This means that the probabilities attached to the realisations are invariant under translations and rotations. The translational invariance which is called the \textbf{ergodic} property implies that the ensemble average $\left<g\left(\textbf{x}\right) g\left(\textbf{x}'\right) \right>$ is equivalent to a spatial average at fixed $\textbf{x}' - \textbf{x}$ for a single realisation. The rotational invariance allows us to replace the ensemble average by an average with respect to the direction of the patch of the sky that we observe for a single realisation. With those statistical properties, we can observe the correlation functions of the CMB temperature anisotropies which are translated into the correlation functions of the curvature perturbation with equation (\ref{sachswolfe}). The correlation functions of the curvature perturbation are obtained with equation (\ref{examplecurvature}) because we obtain the correlation functions of the scalar fields as the vacuum expectation values of them as we will study in detail in section \ref{sec:equilateral}. If we assume that $g\left(\textbf{x}\right)$ is a Gaussian random field, its Fourier coefficients have no correlation except for the reality condition that is given by
\begin{equation}
\left<g\left(\textbf{k}\right) g\left(\textbf{k}'\right) \right> = \left(2\pi \right)^{3} \delta^{\left(3 \right)} \left(\textbf{k} + \textbf{k}' \right) \frac{2 \pi^{2}}{k^{3}} \mathcal{P}_{g},
\end{equation}
where $\mathcal{P}_{g}$ is the \textbf{power spectrum} of the random field $g\left(\textbf{x}\right)$. For a Gaussian random field, N-point functions vanish when N is an odd number as
\begin{equation}\label{oddpointfunctions}
\begin{split}
\left<g\left(\textbf{k}\right)\right> &= 0,\\
\left<g\left(\textbf{k}_{\textbf{1}}\right) g\left(\textbf{k}_{\textbf{2}}\right) g\left(\textbf{k}_{\textbf{3}}\right)\right> &= 0,\\
&\vdots
\end{split}
\end{equation}
When N is an even number, N-point functions can be expressed in terms of the power spectra as
\begin{equation}
\begin{split}
&\left<g\left(\textbf{k}_{\textbf{1}}\right) g\left(\textbf{k}_{\textbf{2}}\right) g\left(\textbf{k}_{\textbf{3}}\right) g\left(\textbf{k}_{\textbf{4}}\right) \right>\\
&= \left(2 \pi \right)^{6} \delta^{\left(3 \right)}\left(\textbf{k}_{\textbf{1}} + \textbf{k}_{\textbf{2}}\right) \delta^{\left(3 \right)}\left(\textbf{k}_{\textbf{3}} + \textbf{k}_{\textbf{4}}\right) \mathcal{P}_{g} \left(\textbf{k}_{\textbf{1}}\right) \mathcal{P}_{g} \left(\textbf{k}_{\textbf{3}}\right) + \rm{two\,\,cyclic\,\,terms},
\end{split}
\end{equation}
and similar expressions hold for higher correlation functions. This is called \textbf{Wick's theorem}. 

In the canonical slow-roll inflation models introduced in this section, by solving equation of motion for the linear perturbation, the power spectrum of the curvature perturbation is obtained as
\begin{equation}\label{slowrollcurvaturespectrum}
\mathcal{P}_{\mathcal{R}} = \left. \left(\frac{H}{\dot{\phi}} \right)^{2} \left(\frac{H}{2 \pi} \right)^{2} \right\rvert_{*} = \left. \frac{H^{2}}{8\pi^{2} M^{2}_{\rm{P}} \tilde{\epsilon}} \right\rvert_{*},
\end{equation}
where we have used the equations $3H^{2} M^{2}_{\rm{P}} = V$ and $3 H \dot{\phi} = - V'$ that are obtained from the Friedmann equation (\ref{friedmann}) and equation (\ref{phiequation}) respectively in the slow-roll limit. Note that the subscript $*$ indicates that the corresponding quantity is evaluated at horizon crossing $k = a H$. The spectral index for the curvature perturbation is defined as
\begin{equation}\label{spectralfirstdefinition}
n_{\rm{s}} - 1 = \frac{d\,\ln{\mathcal{P}_{\mathcal{R}}}}{d\,\ln{k}},
\end{equation}
which is equivalent to $\mathcal{P}_{\mathcal{R}} \propto k^{n_{\rm{s}}-1}$. The spectral index quantifies how the power spectrum depends on the wave number $k$ as we see in equation (\ref{spectralfirstdefinition}). The curvature perturbation is constant on super-horizon scales \cite{Langlois:2008} in a single field case and its power spectrum is evaluated when $k=aH$. We have $d\,\ln{k} = H^{*} \,dt^{*}$ taking into account the slow-roll approximation under which the rate of change of $H$ is negligible. From equation (\ref{phiequation}) in the slow-roll limit, we have $dt = - \left(3H/V\right)d\phi$, and we find
\begin{equation}\label{kderivativefirst}
\frac{d}{d\,\ln{k}} = - M^{2}_{\rm{P}} \frac{V'}{V} \frac{d}{d \phi}.
\end{equation}
The derivative of $\tilde{\epsilon}$ is given by
\begin{equation}\label{derivativeofepsilonfirst}
\frac{d\tilde{\epsilon}}{d\,\ln{k}} = 2 \tilde{\epsilon}^{*} \tilde{\eta}^{*} - 4 \tilde{\epsilon}^{*\,2},
\end{equation}
from equations (\ref{sepsilon}) and (\ref{kderivativefirst}). Using equations (\ref{slowrollcurvaturespectrum}), (\ref{spectralfirstdefinition}) and (\ref{derivativeofepsilonfirst}), we find \cite{Lyth:2000}
\begin{equation}\label{predictionspectralcurvature}
n_{\rm{s}} - 1 = - 6 \tilde{\epsilon}^{*} + 2 \tilde{\eta}^{*}.
\end{equation}
As we will see in section \ref{sec:linear}, there is also a tensorial metric perturbation. This represents gravitational waves. The power spectrum of the tensor perturbation is given by
\begin{equation}\label{tensorperturbationspectrum}
\mathcal{P}_{\rm{T}} = \left. \frac{8}{M^{2}_{\rm{P}}}\left(\frac{H}{2\pi} \right)^{2} \right\rvert_{*},
\end{equation}
and the spectral index for the tensor perturbation is given by
\begin{equation}
n_{\rm{T}} = \frac{d\, \ln{\mathcal{P}_{\rm{T}}}}{d\, \ln{k}} = - 2 \tilde{\epsilon}^{*}. 
\end{equation}
The tensor-to-scalar ratio is defined as
\begin{equation}\label{predictiontensortoscalarfirst}
r = \frac{\mathcal{P}_{\rm{T}}}{\mathcal{P}_{\mathcal{R}}} = 16 \tilde{\epsilon}^{*} = -8 n_{\rm{T}}.
\end{equation}
Finally, let us introduce the \textbf{non-Gaussianities} of the curvature perturbation. As in equation (\ref{oddpointfunctions}), the three-point function vanishes if the curvature perturbation obeys the Gaussian statistics. However, in the non-Gaussian case, it becomes
\begin{equation}\label{bispectrumdefinitionfirst}
 \left<R (k_1) R (k_2) R(k_3) \right> = \left(2 \pi \right)^{3} \delta^{(3)} (\textbf{k}_{\textbf{1}}+\textbf{k}_{\textbf{2}}+\textbf{k}_{\textbf{3}}) B(k_1, k_2, k_3),
\end{equation}
where $B(k_1, k_2, k_3)$ is called the \textbf{bispectrum}. The delta function corresponds to the invariance under translations while the fact that the bispectrum depends only on the lengths of the three sides of the triangles formed by the wave vectors corresponds to the invariance under rotations \cite{Lyth:2009}. As we see in chapter \ref{chapter:generalperturbations}, we can define different kinds of non-Gaussianities depending on what shape of the triangle formed by the wave vectors the bispectrum has the peak. $f_{\rm{NL}}^{equil}$ is the amplitude of the bispectrum that has its peak when the wave vectors form an equilateral triangle ($k_{1} = k_{2} = k_{3}$) while $f_{\rm{NL}}^{local}$ is the amplitude of the bispectrum that has its peak when the wave vectors form a squeezed triangle ($k_{i} \rightarrow 0$ and the other two k's have the same amplitude and the opposite directions because of the momentum conservation which comes from the delta function). In the canonical single field slow-roll inflation models, both $f_{\rm{NL}}^{equil}$ and $f_{\rm{NL}}^{local}$ are of the same order as the slow-roll parameters. According to the Planck satellite observations, the power spectrum of the curvature perturbation is given by \cite{Ade:2013}
\begin{equation}\label{plancksatteliteconstraintcurvaturespectrum}
\mathcal{P}_{\mathcal{R}} = \left(2.23 \pm 0.16 \right) \times 10^{-9},
\end{equation}
at the 68 $\%$ confidence level (CL). Also, the spectral index for the curvature perturbation and the tensor-to-scalar ratio are given by \cite{Ade:2013b}
\begin{equation}\label{observedspectralindexandratio}
\begin{split}
n_{\rm{s}} &= 0.9603 \pm 0.0073,\\
r &< 0.12,
\end{split}
\end{equation}
at the 68 $\%$ CL. Because this means $n_{\rm{s}} - 1 \sim 0.04$, we see that the prediction of the canonical slow-roll inflation (\ref{predictionspectralcurvature}) is compatible with the observation because the slow-roll parameters are much smaller than unity. For the same reason, the observed value for $r$ supports the prediction (\ref{predictiontensortoscalarfirst}). 

\begin{figure}
\centering
\includegraphics[width=15cm]{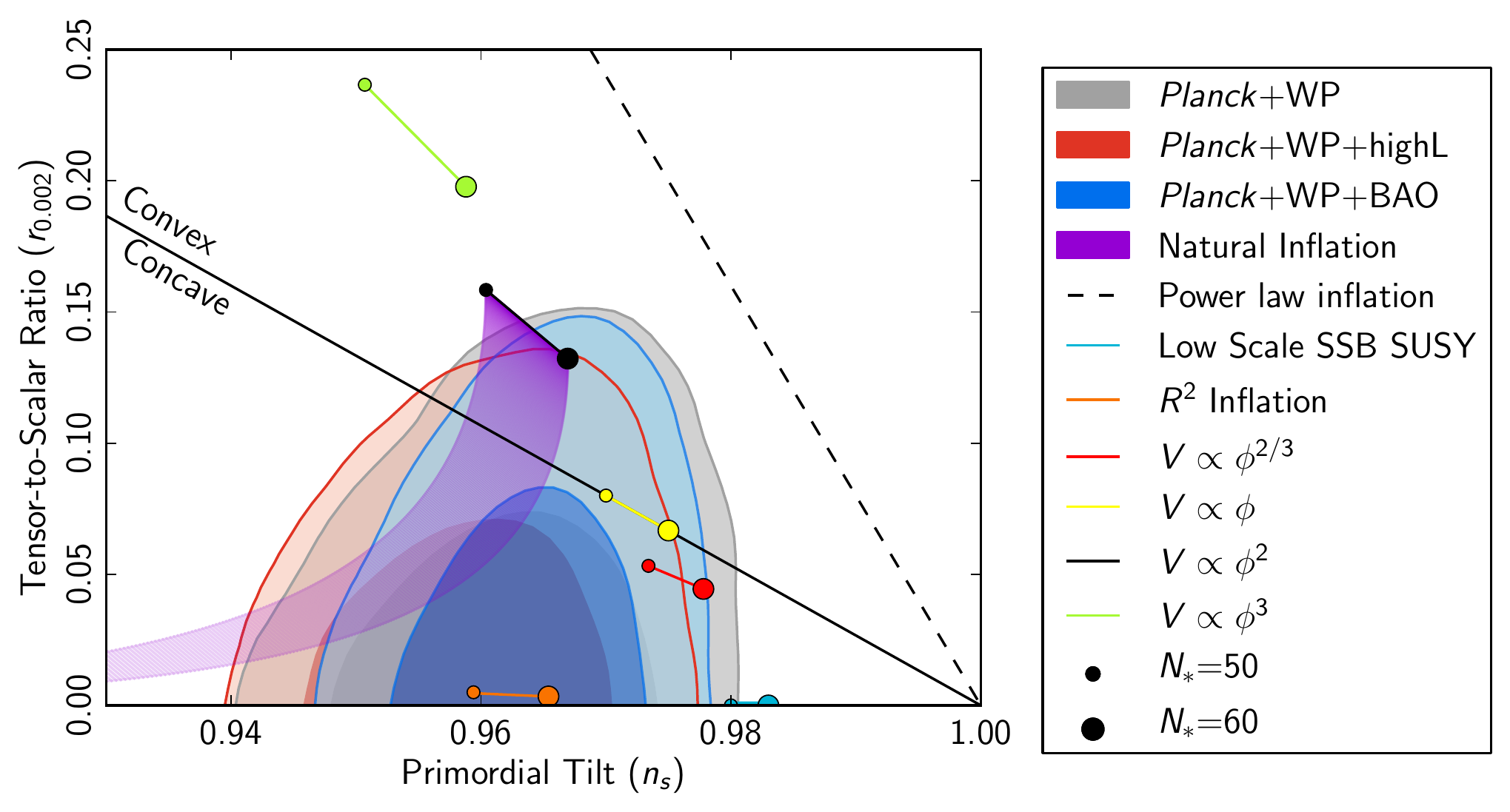}
\caption[Planck satellite constraints]{Marginalised joint 68 $\%$ and 95 $\%$ CL regions for $n_{\rm{s}}$ and $r$ from the Planck satellite observations in combination with other data sets compared to the theoretical predictions of selected inflationary models. $r_{0.002}$ means that the constraints are given at the pivot scale $k = 0.002$ Mpc$^{-1}$. The figure is taken from \cite{Ade:2013b} }{\label{fig14} }
\end{figure}

With the observational results (\ref{observedspectralindexandratio}), we can distinguish different inflation models as in figure \ref{fig14}. Though we don't go into detail about the models in figure \ref{fig14}, we see that the $R^{2}$ inflation model \cite{Starobinsky:1980} is compatible with the observed values of $n_{\rm{s}}$ and $r$ within the 68 $\%$ CL region while the canonical single field slow-roll inflation with a potential $V \propto \phi^{3}$ is excluded at 95 $\%$ CL if we require the number of e-folds to be from 50 to 60 because we need a sufficient number of e-folds to solve the horizon problem as we saw in subsection \ref{subsec:ideainflation}. The non-Gaussianity parameters observed by the Planck satellite are given by \cite{Ade:2013c}
\begin{equation}\label{fnlplanckobservationvaluesixtyeight}
f_{\rm{NL}}^{equil} = -42 \pm 75,
\end{equation}
and
\begin{equation}\label{fnllocalchapteronefinalvalue}
f_{\rm{NL}}^{local} = 2.7 \pm 5.8,
\end{equation}
at the 68 $\%$ CL. For both of the non-Gaussianity parameters, $f_{\rm{NL}} = 0$ is compatible with the observation. As stated above, the canonical single field slow-roll inflation models predict that both of the non-Gaussianity parameters are of the order of the slow-roll parameters. Therefore, the slow-roll inflation models satisfy the observational constraints on the non-Gaussianity parameters. If we consider a canonical slow-roll inflation model with a potential $V \propto \phi$, it seems that the model satisfies all the observational constraints shown above (see figure \ref{fig14}). However, we do not know what the inflaton is. We need to assume that there is a scalar field with the Lagrangian (\ref{lagrangianinflaton}) without knowing the physical meaning of the scalar field. In chapter \ref{chapter:stringinflation}, we introduce the \textbf{Dirac-Born-Infeld inflation} model that is motivated by string theory. In this model, the scalar fields are the spatial coordinates in the extra dimensions. Because $f_{\rm{NL}}^{equil}$ can take large values uniquely in this model, we can distinguish the DBI inflation models from the slow-roll inflation models if we observe a large value of $f_{\rm{NL}}^{equil}$ in the future experiments. The potentials of the DBI inflation models can be obtained by string theory. In chapter \ref{chapter:spinflation}, we study about the model with a potential which is obtained by string theoretical analysis. 
\chapter{Perturbations in general multi-field inflation}\label{chapter:generalperturbations}
\chaptermark{Perturbation theory}

As we saw in the previous chapter, the Universe can be approximated well with the homogeneous and isotropic metric (\ref{walker}) on large scales. However, as we know, the Universe is not homogeneous and isotropic at all on small scales. Therefore, we need to have small perturbations from the FRW metric. Inflation introduced in the previous chapter naturally generates small fluctuations that become the seeds of the structure of the Universe. In this chapter, we introduce linear perturbation theory for general inflation models. Then the ``in-in" formalism is used to calculate the equilateral non-Gaussianity for specific types of single field inflation models. We also study the local type non-Gaussianity in specific single field inflation models. Finally, the delta-N formalism is introduced and used in the calculation of the non-Gaussianities in multi-field inflation models. 

\section{\label{sec:linear}Linear perturbation}
Even though we have only studied the canonical single field slow-roll inflation in subsection \ref{subsec:slow-roll}, let us consider more general multi-field inflation whose action is given by
\begin{equation}
 S = \frac{1}{2} \int d^{4}x \sqrt{-g} \left[^{(4)}R + 2 P(X^{IJ},\phi^{I}) \right],
\label{actiongeneral}
\end{equation}
where we set $8 \pi G = 1$, $^{(4)}R$ is the four dimensional Ricci curvature, $\phi^{I}$ are the scalar fields with $I=1,2,...,N$ and
\begin{equation}
 X^{IJ} \equiv - \frac{1}{2} \partial _{\mu} \phi^{I} \partial ^{\mu} \phi^{J}. 
\end{equation}
As in \cite{Langlois:2008b}, we use the ADM approach as
\begin{equation}\label{admfirstintroduction}
ds^{2}=-N^{2}dt^{2}+h_{ij}\left(dx^{i}+N^{i}dt \right)\left(dx^{j}+N^{j}dt\right),
\end{equation}
where $N$ is the lapse and $N^{i}$ is the shift vector. Note that the field space metric $G_{IJ}$ is used to raise or lower the field space indices $I,J,K,...$. Note also that $G_{IJ}$ evolves with time and hence it is dynamic throughout this thesis. Then, the action (\ref{actiongeneral}) is rewritten as (see \cite{Wald:1984} for the derivation)
\begin{equation}\label{actiontransformed}
\begin{split}
 S &= \frac{1}{2} \int d^{4}x \sqrt{-g} \left[^{(4)}R + 2 P(X^{IJ},\phi^{I}) \right] \\
 &= \frac{1}{2}\int dt\,d^{3}x\sqrt{h}\left[N ^{(3)}R + \frac{1}{N}\left(-E^{2} + E_{ij}E^{ij} \right) + 2NP\right],
 \end{split}
\end{equation}
where $^{(3)}R$ is the Ricci curvature of the spatial metric $h_{ij}$ and h is the determinant of $h_{ij}$ where
\begin{equation}
E_{ij} \equiv \frac{1}{2}\dot{h}_{ij}-D_{(i}N_{j)},
\end{equation}
is proportional to the extrinsic curvature of the spatial slices with the spatial covariant derivatives $D_{i}$ associated with $h_{ij}$. If we consider the FRW metric with linear perturbations, the metric perturbations are given by
\begin{equation}\label{mostgeneralperturbation}
N=1+\alpha,\,\, N_{i}=\partial_{i} \psi + \bar{N}_{i},\,\, h_{ij}= a^{2}\left(t\right)\left[(1-2A)\delta_{ij} + \bar{h}_{\vert ij} + \partial_{(i} v_{j)} + t_{ij}\right],
\end{equation}
where $\alpha$, $\psi$, A and $\bar{h}$ are scalar perturbations, $\bar{N}_{i}$ and $v_{i}$ are vector perturbations and $t_{ij}$ is a tensor perturbation with the Kronecker delta $\delta_{ij}$. Note that $_{\vert i}$ denotes the spatial covariant derivative with $a^{2}\delta_{ij}$. Because the scalar modes of the equations only contain the scalar modes of the original metric perturbations as long as the metric perturbations are contracted with the quantities which come from the background metric or the derivatives (see \cite{Kodama:1984} for details), we can consider the scalar perturbation separately from the vector and the tensor perturbations. When we consider the metric perturbations in general, we have the freedom of changing the coordinate system by an infinitesimal coordinate transformation. A choice of a particular coordinate system is called a ``\textbf{gauge}''. We will study about the gauge issue in section \ref{sec:equilateral} in detail.  Also, we will work in the flat gauge where we set $A=0$ and $\bar{h}=0$. Then, we have
\begin{equation}\label{allperturbations}
N=1+\alpha,\,\, N_{i}=\partial_{i} \psi,\,\, h_{ij}= a^{2}\left(t\right) \delta_{ij},\,\, \phi^{I}\left(t, \textbf{x}\right)=\bar{\phi}^{I}\left(t\right) + Q^{I}\left(t, \textbf{x}\right),
\end{equation}
where $Q^{I}$ are the scalar field perturbations. The momentum constraint which we obtain by varying the action (\ref{actiontransformed}) with respect to $N_{i}$ gives
\begin{equation}\label{momentumconst}
\alpha = \frac{1}{2H} P_{<IJ>} \dot{\phi}^{I} Q^{J},
\end{equation}
where
\begin{equation}
P_{<IJ>} \equiv \frac{1}{2}\left(\frac{\partial P}{\partial X^{IJ}} + \frac{\partial P}{\partial X^{JI}}\right) = P_{JI}. 
\end{equation}
Also, the Hamiltonian constraint which we obtain by varying the action (\ref{actiontransformed}) with respect to $\alpha$ gives
\begin{equation}\label{hamiltonianconst}
-2H\left(\frac{\partial^{2} \psi}{a^{2}}\right) = 2 J \alpha + B_{IJ} \dot{\phi}^{J} Q^{I} + C_{I} Q^{I}, 
\end{equation}
with
\begin{equation}
\begin{split}
J&=P_{<IJ>}X^{IJ}-P-2X^{IJ}X^{KL}P_{<IJ>,<KL>},\\
B_{IJ}&=P_{<IJ>}+2X^{KL}P_{<IJ>,<KL>},\\
C_{I}&=-P_{,I}+2P_{<KL>,I}X^{KL}.
\end{split}
\end{equation}
If we expand the action (\ref{actiontransformed}) up to the second order in the linear perturbations (\ref{allperturbations}) and eliminate $\alpha$ using the constraint (\ref{momentumconst}), we obtain \cite{Langlois:2008c}
\begin{equation}\label{secondaction}
\begin{split}
S_{(2)}&=\frac{1}{2}\int dt d^{3}x a^{3} \left[\left(P_{<IJ>} + 2P_{<MJ>,<IK>}X^{MK}\right)\dot{Q}^{I}\dot{Q}^{J}- P_{<IJ>}h^{ij}\partial_{i}Q^{I}\partial_{j}Q^{J}\right.\\
&\left. - \mathcal{M}_{KL}Q^{K}Q^{L} + 2 \Omega_{KI}Q^{K}\dot{Q}^{I} \right],
\end{split}
\end{equation}
with the mass matrix
\begin{equation}
\begin{split}
\mathcal{M}_{KL}&=-P_{,KL}+3X^{MN}P_{<NK>}P_{<ML>}+\frac{1}{H}P_{<NL>}\dot{\phi}^{N}\left[2P_{<IJ>,K}X^{IJ}-P_{,K}\right]\\
&-\frac{1}{H^{2}}X^{MN}P_{<NK>}P_{<ML>}\left[X^{IJ}P_{<IJ>}+2P_{<IJ>,<AB>}X^{IJ}X^{AB}\right]\\
&-\frac{1}{a^{3}}\frac{d}{dt}\left(\frac{a^{3}}{H}P_{<AK>}P_{<LJ>}X^{AJ}\right),
\end{split}
\end{equation}
and
\begin{equation}
\Omega_{KI}=\dot{\phi}^{J}P_{<IJ>,K}-\frac{2}{H}P_{<LK>}P_{<MJ>,<NI>}X^{LN}X^{MJ}. 
\end{equation}
Note that terms with $\psi$ cancel each other and we do not need to use the Hamiltonian constraint (\ref{hamiltonianconst}). The equations of motion are obtained by varying the second order action (\ref{secondaction}) with respect to the scalar fields as
\begin{equation}\label{eqofmotiongeneral}
\begin{split}
K_{IJ}\ddot{Q}^{J}+\frac{k^{2}}{a^{2}}P_{<IJ>}Q^{J} &+\left(\dot{K}_{IJ}+3HK_{IJ}+\Omega_{JI} -\Omega_{IJ} \right)\dot{Q}^{J}\\
&+\left(\dot{\Omega}_{KI}+\mathcal{M}_{IK}+3H\Omega_{KI}\right)Q^{K}=0,
\end{split}
\end{equation}
with the coefficient
\begin{equation}
K_{IJ} \equiv P_{<IJ>} + 2 P_{<MJ>,<IK>}X^{MK}, 
\end{equation}
where $k$ in equation (\ref{eqofmotiongeneral}) is the angular wave number which is the magnitude of the wave vector $\textbf{k}$ in the Fourier decomposition given by \cite{Langlois:2008b}
\begin{equation}
Q^{I} \left(t, \textbf{x} \right) = \frac{1}{\left(2\pi \right)^{3/2}} \int d^{3}k \left\{\hat{a}^{I}_{\textbf{k}} Q^{I}_{k}\left(t \right) e^{i \textbf{k} \cdot \textbf{x}} + \hat{a}^{I \dagger}_{\textbf{k}} Q^{I *}_{k}\left(t \right) e^{- i \textbf{k} \cdot \textbf{x}} \right\},
\end{equation}
with the creation operator $\hat{a}^{I \dagger}_{\textbf{k}}$ and the annihilation operator $\hat{a}^{I}_{\textbf{k}}$ for the field $Q^{I}$ which satisfy the usual commutation relations
\begin{equation}
\begin{split}
\left[\hat{a}^{I}_{\textbf{k}}, \hat{a}^{I \dagger}_{\textbf{k}'} \right] = \left(2\pi\right)^{3}\delta^{\left(3\right)}\left(\textbf{k} - \textbf{k}' \right),\\
\left[\hat{a}^{I}_{\textbf{k}}, \hat{a}^{I}_{\textbf{k}'} \right] = \left[\hat{a}^{I \dagger}_{\textbf{k}}, \hat{a}^{I \dagger}_{\textbf{k}'} \right] = 0. 
\end{split}
\end{equation}
Note that $Q^{I *}_{k}$ is the complex conjugate of $Q^{I }_{k}$. The Klein-Gordon equation (\ref{eqofmotiongeneral}) is derived in several ways \cite{Wands:2008, Nibbelink:2008, vanTent:2001, vanTent:2004, Renaux-Petel:2009b} in order to study the variety of scalar field models \cite{Koyama:2011, Koyama:2011b, Boubekeur:2002, Burgess:2003, Nibbelink:2002, Wands:1994, Garcia-Bellido:1995, Malik:1999, Malik:2002}. 

If the matrix $K_{IJ}$ is invertible, the eigenvalues of the matrix $\left(K^{-1}\right)^{IL}P_{<LJ>}$ correspond to the sound speeds as we can see from equation (\ref{eqofmotiongeneral}). 
In the flat gauge, the gauge invariant quantity which is called the comoving curvature perturbation $\mathcal{R}$ is given by
\begin{equation}\label{comovingrelation}
\mathcal{R}=\left(\frac{H}{2P_{<IJ>}X^{IJ}} \right)P_{<KL>}\dot{\phi}^{K}Q^{L}. 
\end{equation}
We study about this quantity in section \ref{sec:equilateral}. 

\subsection{k-inflation}\label{kflationlinear}
\begin{figure}[!htb]
\centering
\includegraphics[width=12cm]{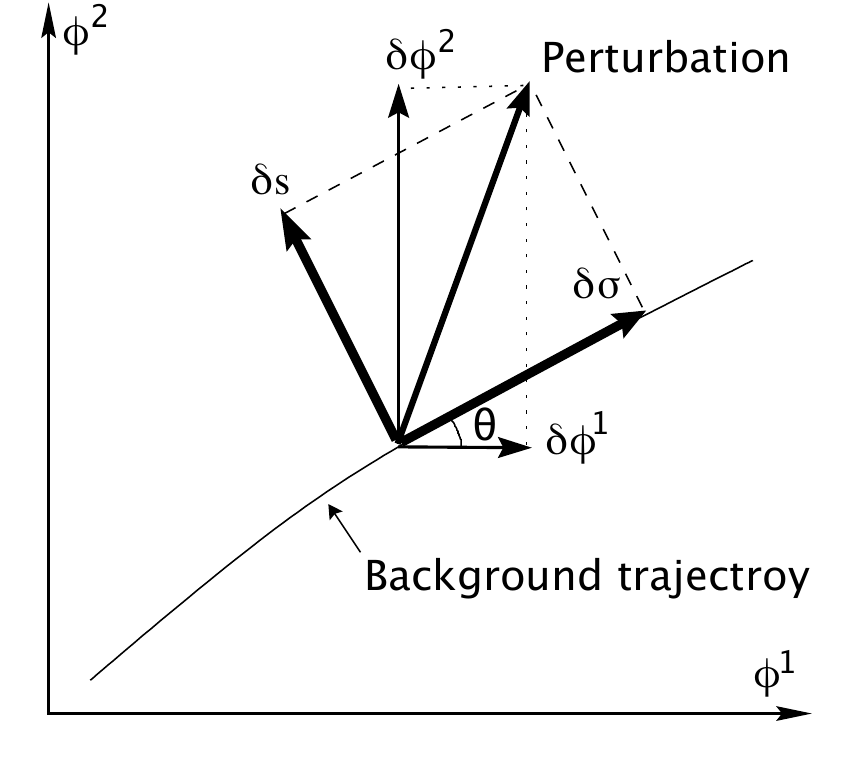}
\caption[Field decomposition]{An illustration of the decomposition of an arbitrary perturbation
into an adiabatic ($\delta \sigma$) and entropy ($\delta s$)
component. The angle of the tangent to the background trajectory
is denoted by $\theta$. The usual perturbation decomposition,
along the $\phi^{1}$ and $\phi^{2}$ axes, is also shown. The figure is adapted from \cite{Gordon:2001}}{\label{fig:decomposition} }
\end{figure}

In the case of k-inflation, the action is given by
\begin{equation}\label{kinflationaction}
 S = \frac{1}{2} \int d^{4}x \sqrt{-g} \left[^{(4)}R + 2 P(X,\phi^{I}) \right],
\end{equation}
where $X$ is defined as $X=G^{IJ}X_{IJ}$. Let us introduce the field space decomposition. We define the adiabatic unit vector as
\begin{equation}\label{basiskflation}
e^{I}_{\sigma}=\frac{\dot{\phi}^{I}}{\sqrt{2X}}.
\end{equation}
Below, let us consider two-field cases for simplicity. Then, the entropy unit vector $e^{I}_{s}$ which is orthonormal to the adiabatic vector is derived uniquely because of the orthonormality condition $G_{IJ}e^{I}_{\sigma}e^{J}_{s}=0$ and $G_{IJ}e^{I}_{s}e^{J}_{s}=1$. Let us introduce the canonical perturbation variables
\begin{equation}\label{relationofvandqinkflation}
v_{\sigma}=\frac{a\sqrt{P_{,X}}}{c_{s}}Q_{\sigma},\,\,\,\,\,v_{s}=a\sqrt{P_{,X}}Q_{s},
\end{equation}
where $P_{,X}$ denotes the derivative of $P$ with respect to $X$ and
\begin{equation}\label{definitionofqsigma}
Q_{\sigma} \equiv G_{IJ}Q^{I}e^{J}_{\sigma},\,\,\,Q_{s} \equiv G_{IJ}Q^{I}e^{J}_{s}.
\end{equation}
The reason why we introduce such a field decomposition is that the comoving curvature perturbation which is introduced in section \ref{sec:equilateral} is given by
\begin{equation}
\mathcal{R} = \frac{H}{\dot{\sigma}} Q_{\sigma}, 
\label{curvatureperturbationsimplerelation}
\end{equation}
where $\dot{\sigma} \equiv \sqrt{2 X}$. Therefore, such a decomposition makes numerical calculations easier because we obtain numerical values of the curvature perturbation simply by obtaining numerical values of the adiabatic scalar field perturbation. As we see in equation (\ref{basiskflation}), the adiabatic direction is the direction of the velocity in the field space as shown in figure \ref{fig:decomposition}. 

Then, the equations of motion are given by \cite{Langlois:2008b}
\begin{equation}
 v_{\sigma}'' - \xi v_{s}' + \left(c_{s}^{2} k^{2} - \frac{z''}{z} \right)v_{\sigma} - \frac{\left(z \xi \right)'}{z} v_{s} = 0,\label{equationofmotiononekflation}
\end{equation}
\begin{equation}
 v_{s}'' + \xi v_{\sigma}' + \left(k^{2} - \frac{\alpha''}{\alpha} + a^{2} \mu_{s}^{2} \right)v_{s} - \frac{z'}{z} \xi v_{\sigma} = 0,\label{equationofmotiontwokflation}
\end{equation}
where the prime denotes the derivative with respect to the conformal time $\tau = \int dt/a(t)$ and
\begin{equation}
 \xi \equiv \frac{a}{\dot{\sigma}P_{,X}c_{s}} \left[\left(1+c_{s}^{2}\right) P_{,s} - c_{s}^{2} \dot{\sigma}^{2} P_{,X s} \right],
\end{equation}
\begin{equation}
 \mu_{s}^{2} \equiv - \frac{{P}_{,ss}}{P_{,X}} +\frac{1}{2}\dot{\sigma}^2\tilde{R}- \frac{1}{2c^{2}_{s}X} \frac{P_{,s}^{2}}{P_{,X}^{2}} + 2  \frac{P_{,X s} P_{,s}}{P_{,X}^{2}},
\end{equation}
\begin{equation}
 z \equiv \frac{a \dot{\sigma}\sqrt{P_{,X}}}{c_{s} H}, \: \: \: \: \: \alpha \equiv a \sqrt{P_{,X}},
\end{equation}
\begin{equation}
c_{s} \equiv \sqrt{\frac{P_{,X}}{P_{,X} + 2 X P_{,X X}}} 
\end{equation}
with
\begin{equation}\label{carrotjuice}
P_{s} \equiv P_{,I} e_{s}^{I}, \:\:\: P_{,X s} \equiv P_{,X I} e_{s}^{I}, \:\:\: P_{,ss} \equiv \left(\mathcal{D}_{I} \mathcal{D}_{J} P \right) e_{s}^{I} e_{s}^{J},
\end{equation}
where $\mathcal{D}_{I}$ denotes the covariant derivative with respect to the field space metric $G_{IJ}$ and $\tilde{R}$ denotes the Riemann scalar curvature of the field space. From equations (\ref{equationofmotiononekflation}) and (\ref{equationofmotiontwokflation}), on small scales ($k \gg a H/c_{s}$), we can see that the adiabatic mode $v_{\sigma}$ propagates with the sound speed $c_{s}$ while the entropy mode $v_{s}$ propagates with the speed of light $c$. We briefly introduce the sound speed $c_{s}$ below. If the pressure $P$ depends only on the entropy $S$ and the energy density $\rho$, the pressure perturbation is given by
\begin{equation}
\delta P = \frac{\partial P}{\partial S} \delta S + \frac{\partial P}{\partial \rho} \delta \rho,
\end{equation}
and the adiabatic sound speed is defined as
\begin{equation}
c_{as} \equiv \left. \frac{\partial P}{\partial \rho} \right\rvert_{S}. 
\end{equation}
The adiabatic sound speed $c_{as}$ is the response of the pressure to a change in the energy density with constant entropy \cite{Christopherson:2009}. $c_{s}$ is the phase speed with which perturbations propagate. Even though the adiabatic sound speed $c_{as}$ is generally different from the phase speed $c_{s}$, the term ``sound speed" is confusingly used to denote the phase speed in the literature. In this thesis, the sound speed means the phase speed following the confusing convention in the literature. 
The parameter $\xi$ vanishes when the trajectory in the field space is straight. It is known that $\xi$ quantifies the coupling between the adiabatic and entropy modes which is directly related to the bending of the background trajectory in the field space \cite{Langlois:2008b}. 

In the k-inflation models, we can define the slow-roll parameters as \cite{Langlois:2008c}
\begin{equation}
 \epsilon = - \frac{\dot{H}}{H^{2}}, \:\:\:\:\: \eta = \frac{\dot{\epsilon}}{H \epsilon}, \:\:\:\:\: s = \frac{\dot{c_{s}}}{H c_{s}}.
\label{slowrollparametersgeneralcases}
\end{equation}
If the trajectory is not curved significantly, the coupling $\xi/a H$ becomes much smaller than one. When the slow-roll parameters are much smaller than unity, the approximations $z''/z \simeq 2 / \tau^{2}$ and $\alpha'' / \alpha \simeq 2 / \tau^{2}$ hold. With those conditions, we can approximate equations (\ref{equationofmotiononekflation}) and (\ref{equationofmotiontwokflation}) as the Bessel differential equations. Then, the solutions with the Bunch-Davis vacuum initial conditions are given by
\begin{equation}
 v_{\sigma k} \simeq \frac{1}{\sqrt{2 k c_{s}}} e^{-i k c_{s} \tau} \left(1 - \frac{i}{k c_{s} \tau} \right),\label{solonekinflation}
\end{equation}
\begin{equation}
 v_{s k} \simeq \frac{1}{\sqrt{2 k}} e^{-i k \tau} \left(1 - \frac{i}{k \tau} \right),\label{soltwokinflation}
\end{equation}
when $\mu^{2}_{s} / H^{2}$ is negligible for the entropy mode \cite{Langlois:2008b}. In this case, the curvature power spectrum on super-horizon scales reads
\begin{equation}
 \mathcal{P}_{\mathcal{R}_{*}} = \frac{k^{3}}{2 \pi^{2}} \lvert \mathcal{R} \rvert^{2} = \frac{k^{3}}{2 \pi^{2}} \frac{\lvert v_{\sigma k} \rvert^{2}}{z^{2}} \simeq \left. \frac{H^{4}}{8 \pi^{2} c_{s} X P_{,X}} \right\rvert_{*} \simeq \left. \frac{H^{2}}{8 \pi^{2} \epsilon c_{s}} \right\rvert_{*},
\label{curvaturepowerspectrumhorizonkflation}
\end{equation}
where the subscript $*$ indicates that the corresponding quantity is evaluated at sound horizon crossing $k c_{s} = a H$. Even though the power spectrum of the adiabatic fluctuations has the relation with the power spectrum of the entropy fluctuations as
\begin{equation}\label{relationadiabaticentropykinlfation}
 \mathcal{P}_{Q_{\sigma} *}= \left. \frac{k^{3}}{2 \pi^{2}} \lvert v_{\sigma k} \rvert^{2} \frac{c_{s}^{2}}{a^{2}P_{,X}}\right\rvert_{*} \simeq \left. \frac{H^{2}}{4\pi^{2}c_{s}P_{,X}}\right\rvert_{*} \simeq \left. \frac{\mathcal{P}_{Q_{s}}}{c_{s}}\right\rvert_{*},
\end{equation}
we will see that this totally depends on the definition of the basis vectors in subsection \ref{dbilinear} and we can define the basis vectors so that we have the same amplitude for the power spectra of the adiabatic fluctuations and of the entropy fluctuations. Note that the power spectrum of the tensor perturbations is given by equation (\ref{tensorperturbationspectrum}) also in the k-inflation models. 

\subsection{DBI inflation}\label{dbilinear}
As shown in section \ref{dbiintroduction}, the Lagrangian in equation (\ref{actiongeneral}) of the Dirac-Born-Infeld (DBI) inflation is given by
\begin{equation}
 P(X^{IJ},\phi^{I}) = \tilde{P}(\tilde{X},\phi^{I}) = - \frac{1}{f(\phi^{I})} \left(\sqrt{1-2 f(\phi^{I}) \tilde{X}} - 1\right) - V\left(\phi^{I}\right),
\label{multifieldlagrangian}
\end{equation}
where $V(\phi^I)$ is a function of the scalar fields and $\tilde{X}$ is defined in terms of the determinant
\begin{equation}
\begin{split}
\mathcal{D}&=\mbox{det} (\delta^{I}_{J} - 2 f X^{I}_{J} )\nonumber \\
&=1 - 2 f G_{IJ} X^{IJ} + 4 f^{2} X^{[I}_{I} X^{J]}_{J} - 8 f^{3} X^{[I}_{I} X^{J}_{J} X^{K]}_{K} + 16 f^{4} X^{[I}_{I} X^{J}_{J} X^{K}_{K} X^{L]}_{L},
\label{determinant}
\end{split}
\end{equation}
as
\begin{equation}
 \tilde{X} = \frac{(1-\mathcal{D})}{2 f}.
\end{equation}
Note that $f(\phi^{I})$ is defined
by the warp factor  $h(\phi^{I})$ and the brane tension $T_3$ as
\begin{equation}
 f(\phi^{I}) \equiv \frac{h(\phi^{I})}{T_{3}}.
\end{equation}
The sound speed is defined as
\begin{equation}
 c_{s} \equiv \sqrt{\frac{\tilde{P}_{,\tilde{X}}}{\tilde{P}_{,\tilde{X}} + 2 \tilde{X} \tilde{P}_{,\tilde{X} \tilde{X}}}} = \sqrt{1- 2 f \tilde{X}},
\end{equation}
where $,_{\tilde{X}}$ means the partial derivative with respect to $\tilde{X}$. Note that $\tilde{X}$ coincides with $X \equiv G_{IJ} X^{IJ}$ in the homogeneous background because all the spatial derivatives vanish. From the action (\ref{multifieldlagrangian}),
we can show that
\begin{equation}\label{pxtildeissoundspeedinverse}
 \tilde{P}_{,\tilde{X}} = \frac{1}{c_{s}}.
\end{equation}
As in section \ref{kflationlinear}, let us consider two-field cases for simplicity. In this section, we use the adiabatic basis vector given by
\begin{equation}\label{newadiabaticbasis}
 \tilde{e}^{I}_{\sigma}=\frac{\sqrt{c_{s}}\dot{\phi}^{I}}{\sqrt{2X}},
\end{equation}
and define the entropy basis vector with the conditions
\begin{equation}\label{orthoconditions}
\begin{split}
 G_{IJ} \tilde{e}_{s}^{I} \tilde{e}_{s}^{J} &= \frac{1}{c_{s}},\\
 G_{IJ} \tilde{e}_{\sigma}^{I} \tilde{e}_{s}^{J} &=0.
\end{split}
\end{equation}
If we assume the relation
\begin{equation}
Q^{I} = \tilde{Q}_{\sigma} \tilde{e}^{I}_{\sigma} + \tilde{Q}_{s} \tilde{e}^{I}_{s},
\end{equation}
we obtain
\begin{equation}\label{newdefinitionofqsigma}
\tilde{Q}_{\sigma} \equiv \frac{G_{IJ} Q^{I} \tilde{e}^{J}_{\sigma}}{c_{s}},\,\,\,\tilde{Q}_{s} \equiv G_{IJ} Q^{I} \tilde{e}^{J}_{s} c_{s}. 
\end{equation}
Let us define the canonically normalised fields as
\begin{equation}
 v_{\sigma}=\frac{a}{c_{s}} \tilde{Q}_{\sigma}, \: \: \: \: \: v_{s}=\frac{a}{c_{s}} \tilde{Q}_{s}.
\label{relationofvandq}
\end{equation}
Comparing equation (\ref{relationofvandqinkflation}) with equation (\ref{relationofvandq}) taking into account equation (\ref{pxtildeissoundspeedinverse}), we obtain the relations between the field perturbations with different basis vectors as
\begin{equation}\label{chaptertworelationbetweenqandqtilde}
Q_{\sigma} = \sqrt{c_{s}} \tilde{Q}_{\sigma}, \: \: \: \: \: Q_{s} = \frac{\tilde{Q}_{\sigma}}{\sqrt{c_{s}}}. 
\end{equation}
Then, with this field decomposition, the curvature perturbation is written as
\begin{equation}\label{chaptertwocurvaturewithnewbaseslinear}
 \mathcal{R} = \frac{H \sqrt{c_{s}}}{\dot{\sigma}} \tilde{Q}_{\sigma}. 
\end{equation}
From equations (\ref{newadiabaticbasis}), (\ref{orthoconditions}) and (\ref{newdefinitionofqsigma}), we can see that the canonical variables defined in equation (\ref{relationofvandqinkflation}) are exactly the same as those defined in equation (\ref{relationofvandq}) when we express them as functions of $Q^{1}$ and $Q^{2}$. 
Then, the equations of motion for $v_{\sigma}$ and $v_{s}$ are obtained as \cite{Langlois:2008c}
\begin{equation}
 v_{\sigma}'' - \xi v_{s}' + \left(c_{s}^{2} k^{2} - \frac{z''}{z} \right)v_{\sigma} - \frac{\left(z \xi \right)'}{z} v_{s} = 0,\label{equationofmotionone}
\end{equation}
\begin{equation}
 v_{s}'' + \xi v_{\sigma}' + \left(c_{s}^{2} k^{2} - \frac{\alpha''}{\alpha} + a^{2} \mu_{s}^{2} \right)v_{s} - \frac{z'}{z} \xi v_{\sigma} = 0,\label{equationofmotiontwo}
\end{equation}
where the prime denotes the derivative with respect to the conformal time $\tau$ and
\begin{equation}\label{xidefined}
 \xi \equiv \frac{a}{\dot{\sigma}} \left[\left(1+c_{s}^{2}\right) \tilde{P}_{,s} - c_{s}^{2} \dot{\sigma}^{2} \tilde{P}_{,\tilde{X} s} \right],
\end{equation}
\begin{equation}
 \mu_{s}^{2} \equiv -c_{s} \tilde{P}_{,ss} - \frac{1}{\dot{\sigma}^{2}} \tilde{P}_{,s}^{2} + 2 c_{2}^{2} \tilde{P}_{,\tilde{X} s} \tilde{P}_{,s},
\end{equation}
\begin{equation}
 z \equiv \frac{a \dot{\sigma}}{\sqrt{c_{s}} H}, \: \: \: \: \: \alpha \equiv a \frac{1}{\sqrt{c_{s}}},
\end{equation}
with
\begin{equation}\label{definitionsincludingsigmadotinchaptertwo}
 \dot{\sigma} \equiv \sqrt{2 X}, \:\:\: \tilde{P}_{s} \equiv \tilde{P}_{,I} e_{s}^{I} \sqrt{c_{s}}, \:\:\: \tilde{P}_{,\tilde{X} s} \equiv \tilde{P}_{,\tilde{X} I} e_{s}^{I} \sqrt{c_{s}}, \:\:\: \tilde{P}_{,ss} \equiv \left(\mathcal{D}_{I} \mathcal{D}_{J} \tilde{P} \right) e_{s}^{I} e_{s}^{J} c_{s},
\end{equation}
where $\mathcal{D}_{I}$ denotes the covariant derivative with respect to the field space metric $G_{IJ}$. 
From equations (\ref{equationofmotionone}) and (\ref{equationofmotiontwo}), on small scales ($k \gg a H/c_{s}$), we can see that both the adiabatic mode $v_{\sigma}$ and the entropy mode $v_{s}$ propagate with the sound speed $c_{s}$ in the case of DBI inflation. The slow-roll parameters in the DBI inflation models are defined in equation (\ref{slowrollparametersgeneralcases}). In a similar way to subsection \ref{kflationlinear}, when the coupling $\xi/a H$ and the slow-roll parameters are much smaller than unity, equations (\ref{equationofmotionone}) and (\ref{equationofmotiontwo}) are approximated as the Bessel differential equations and the solutions with the Bunch-Davis vacuum initial conditions are given by
\begin{equation}
 v_{\sigma k} \simeq \frac{1}{\sqrt{2 k c_{s}}} e^{-i k c_{s} \tau} \left(1 - \frac{i}{k c_{s} \tau} \right),\label{dbisolone}
\end{equation}
\begin{equation}
 v_{s k} \simeq \frac{1}{\sqrt{2 k c_{s}}} e^{-i k c_{s} \tau} \left(1 - \frac{i}{k c_{s} \tau} \right),\label{dbisoltwo}
\end{equation}
when $\mu^{2}_{s} / H^{2}$ is negligible for the entropy mode \cite{Langlois:2008c}. Note that the difference between equation (\ref{soltwokinflation}) and equation (\ref{dbisoltwo}) comes from the difference in the propagation speed of the entropy modes. With the solution (\ref{dbisolone}), the curvature perturbation on super-horizon scales reads
\begin{equation}
 \mathcal{P}_{\mathcal{R}_{*}} = \frac{k^{3}}{2 \pi^{2}} \lvert \mathcal{R} \rvert^{2} = \frac{k^{3}}{2 \pi^{2}} \frac{\lvert v_{\sigma k} \rvert^{2}}{z^{2}} \simeq \left. \frac{H^{4}}{4 \pi^{2} \dot{\sigma}^{2}} \right\rvert_{*} \simeq \left. \frac{H^{2}}{8 \pi^{2} \epsilon c_{s}} \right\rvert_{*},
\label{curvaturepowerspectrumhorizon}
\end{equation}
where the subscript $*$ indicates that the corresponding quantity is evaluated at sound horizon crossing $k c_{s} = a H$. We also have
\begin{equation}\label{powerspectrumoffieldsdbiwithkoyamasanbases}
 \mathcal{P}_{\tilde{Q}_{\sigma *}}=\left. \frac{k^{3}}{2 \pi^{2}} \lvert v_{\sigma k} \rvert^{2} \frac{c_{s}^{2}}{a^{2}}\right\rvert_{*} \simeq \left. \frac{H^{2}}{4\pi^{2}c_{s}}\right\rvert_{*} \simeq \mathcal{P}_{\tilde{Q}_{s} *}.
\end{equation}
We can see that the power spectra have the same value at horizon crossing for the adiabatic perturbation and the entropy perturbation unlike in subsection \ref{kflationlinear}. This is just because we used the different basis vectors. With the basis vectors in equation (\ref{definitionofqsigma}), we instead have
\begin{equation}\label{relationadiabaticentropydbi}
 \mathcal{P}_{Q_{\sigma *}}=\left. \frac{k^{3}}{2 \pi^{2}} \lvert v_{\sigma k} \rvert^{2} \frac{c_{s}^{3}}{a^{2}}\right\rvert_{*} \simeq \left. \frac{H^{2}}{4\pi^{2}}\right\rvert_{*} \simeq c^{2}_{s} \mathcal{P}_{Q_{s} *}.
\end{equation}
Note that the power spectrum of the tensor perturbations in the DBI inflation models is given by equation (\ref{tensorperturbationspectrum}). 

\section{\label{sec:equilateral}Calculation of the equilateral non-Gaussianity}
In this section, we first study the gauge transformation and then review the calculation of the three-point function of the curvature perturbation in single field k-inflation cases following \cite{Koyama:2010} using the in-in formalism introduced in appendix \ref{app:inin}. Then, we see the result for single field DBI inflation models as particular cases. The parameter $f_{\rm{NL}}^{equil}$ is introduced in this section. \\

\subsection{Flat gauge and comoving gauge}\label{subsec:flatcomoving}
The general linear perturbations (\ref{mostgeneralperturbation}) can be extended to higher order perturbations. We will work on higher order perturbations in this section because we need to calculate higher order quantities. Before working on them, let us review the gauge transformation briefly. A \textbf{gauge} is a coordinate choice which recovers the FRW metric in the limit of zero perturbations (see \cite{Lyth:2009, Malik:2013}). Because we always have degrees of freedom which we can eliminate by the infinitesimal coordinate transformation as
\begin{equation}\label{gaugetransformationcorrection}
 \left(\widetilde{\eta},\widetilde{\textbf{x}}\right) = \left(\eta+\delta \eta, \textbf{x} + \delta \textbf{x}\right),
\end{equation}
and the vector perturbation $\delta \textbf{x}$ can be decomposed into the scalar part $\bar{S}_{,i}$ and the vector part $V_{i}$, we can eliminate two scalar modes and one vector mode from the general perturbation (\ref{mostgeneralperturbation}). Note that $\eta$ is the conformal time while the components of $\textbf{x}$ denote the comoving spatial coordinates. The gauge fixing is merely a choice of the coordinate system that we use theoretically, thus any observable quantity should not depend on the gauge. The infinitesimal coordinate transformation (\ref{gaugetransformationcorrection}) is called the \textbf{gauge transformation}. We introduce the gauge transformations of the metric and matter variables up to first order below. We define that a tensor $\textbf{T}$ transforms into $\widetilde{\textbf{T}}$ due to a gauge transformation. Splitting a tensor up to first order as $\textbf{T} = \textbf{T}_{0} + \delta\textbf{T}_{1}$, the tensorial quantity transforms due to a gauge transformation at zeroth and first order, respectively, as \cite{Bruni:1997, Mukhanov:1997, Malik:2009}
\begin{equation}
\widetilde{\textbf{T}}_{0} = \textbf{T}_{0},
\end{equation}
\begin{equation}\label{firstordergaugetransformcorrection}
\widetilde{\delta\textbf{T}}_{1} = \delta\textbf{T}_{1} - \pounds_{\xi} \textbf{T}_{0},
\end{equation}
where the \textbf{Lie derivative} is defined for a scalar $\varphi$, a covariant vector $v_{\mu}$ and a covariant tensor $t_{\mu \nu}$ as \cite{Wald:1984, Malik:2009}
\begin{equation}\label{liederivativecorrectionscalar}
\pounds_{\xi} \varphi = \xi^{\lambda} \varphi_{,\lambda},
\end{equation}
\begin{equation}\label{liederivativecorrectionvector}
\pounds_{\xi} v_{\mu} = v_{\mu, \alpha} \xi^{\alpha} + v_{\alpha} \xi^{\alpha}_{, \mu},
\end{equation}
\begin{equation}\label{liederivativecorrection}
\pounds_{\xi} t_{\mu \nu} = t_{\mu \nu, \lambda} \xi^{\lambda} + t_{\mu \lambda} \xi^{\lambda}_{,\nu} + t_{\lambda \nu} \xi^{\lambda}_{,\mu},
\end{equation}
with the vector $\xi^{\mu} = \left(\delta \eta, \delta \textbf{x}\right)$. 
From equations (\ref{firstordergaugetransformcorrection}) and (\ref{liederivativecorrectionscalar}), the first order perturbation of the energy density $\rho$, which is a four scalar, transforms as
\begin{equation}\label{gaugetransformationrhocorrect}
\widetilde{\delta \rho} = \delta \rho - \rho'_{0} \delta \eta,
\end{equation}
where $'$ denotes the derivative with respect to the conformal time and $\rho_{0}$ is the background part of the energy density. We define the perturbation of the spatial part of the four velocity as
\begin{equation}\label{contravariantfourvelospatial}
u^{i} = a^{-1} v^{i},
\end{equation}
where $v^{i}$ can be split into a scalar part $v$ and a vector part $v_{vec}$ as
\begin{equation}
v^{i} = \delta^{ij} v_{,j} + v_{vec}^{i}. 
\end{equation}
Using the constraint
\begin{equation}
u_{\mu} u^{\mu} = -1,
\end{equation}
and the metric $g_{\mu\nu}$ that is obtained with equations (\ref{admfirstintroduction}) and (\ref{mostgeneralperturbation}) up to first order, we obtain
\begin{equation}\label{contrafourvelotime}
u^{0} = a^{-1} \left(1- \alpha \right),
\end{equation}
\begin{equation}\label{covariantfourcorrectionzero}
u_{0} = -a \left(1 + \alpha \right),
\end{equation}
\begin{equation}\label{covariantfourcorrectionspatial}
u_{i} = a \left(v_{i} + N_{i} \right).
\end{equation}
From equations (\ref{firstordergaugetransformcorrection}) and (\ref{liederivativecorrectionvector}), the gauge transformation of the first order perturbation of the covariant four vector is given by
\begin{equation}\label{gaugetranscorrectionfourvelo}
\widetilde{\delta u_{\mu}} = \delta u_{\mu} - u_{\left(0 \right) \mu} \delta \eta - u_{\left(0 \right) \lambda} \xi^{\lambda}_{,\mu},
\end{equation}
where the background part of the covariant four vector is given by
\begin{equation}
u_{\left(0 \right) \mu} = a \left(-1, 0, 0, 0\right),
\end{equation}
from equations (\ref{covariantfourcorrectionzero}) and (\ref{covariantfourcorrectionspatial}). The scalar part of the spatial components of equation (\ref{gaugetranscorrectionfourvelo}) gives
\begin{equation}\label{gaugetransformationofscalarv}
\widetilde{v_{, i}} + \widetilde{\psi_{, i}} = v_{, i} + \psi_{, i} + \delta \eta_{, i}, 
\end{equation}
note that $\psi$ is defined in equation (\ref{mostgeneralperturbation}). From equations (\ref{firstordergaugetransformcorrection}) and (\ref{liederivativecorrection}), the gauge transformations of the scalar perturbations of the metric, that can be obtained with equations (\ref{admfirstintroduction}) and (\ref{mostgeneralperturbation}), up to first order are given by \cite{Bassett:2006, Malik:2009, Langlois:2004, Malik:1998}
\begin{equation}
\widetilde{\alpha} = \alpha - \mathcal{H} \delta \eta - \delta \eta', 
\end{equation}
\begin{equation}\label{gaugetransformationofacorrection}
\widetilde{A} = A + \mathcal{H} \delta \eta, 
\end{equation}
\begin{equation}\label{gaugetransformationofpsicorrection}
\widetilde{\psi} = \psi + \frac{\delta \eta}{a} - \frac{S'}{a}, 
\end{equation}
\begin{equation}\label{gaugetransformationofhbarcorrect}
\widetilde{\bar{h}} = \bar{h} - 2S, 
\end{equation}
with
\begin{equation}
\mathcal{H} \equiv \frac{a'}{a} = a H, 
\end{equation}
where $S$ is the scalar part of $\delta \textbf{x}$ and the quantities $\alpha$, $A$, $\psi$ and $\bar{h}$ are defined in equation (\ref{mostgeneralperturbation}). From equations (\ref{gaugetransformationofscalarv}) and (\ref{gaugetransformationofpsicorrection}), we obtain the gauge transformation of the scalar part of the velocity perturbation as
\begin{equation}\label{gaugetransformationofvinsimpleform}
\widetilde{v} = v + S'. 
\end{equation}
If we define the comoving curvature perturbation as
\begin{equation}\label{correctionproperintroofcomocurv}
\mathcal{R} \equiv A - \mathcal{H} \left(v + \psi \right),
\end{equation}
it is obvious that this quantity is invariant under any gauge transformation as 
\begin{equation}
\widetilde{\mathcal{R}} = \mathcal{R}
\end{equation}
from equations (\ref{gaugetransformationofacorrection}), (\ref{gaugetransformationofpsicorrection}) and (\ref{gaugetransformationofvinsimpleform}). Substituting equations (\ref{contravariantfourvelospatial}), (\ref{contrafourvelotime}), (\ref{covariantfourcorrectionzero}) and (\ref{covariantfourcorrectionspatial}) into equation (\ref{introductionperfectenergymomentum}), we obtain the first order perturbations of the energy momentum tensor for perfect fluid as
\begin{equation}
\delta T^{0}_{0} = - \delta \rho,
\end{equation}
\begin{equation}
\delta T^{i}_{0} = - \left(\rho + P \right) v^{i},
\end{equation}
\begin{equation}\label{firstorderpertenergymomentumzeroi}
\delta T^{0}_{i} = \left(\rho + P \right)\left(v_{i} + N_{i} \right), 
\end{equation}
\begin{equation}
\delta T^{i}_{j} = - \delta P \delta^{i}_{j}. 
\end{equation}
The flat gauge (\ref{allperturbations}) which we used in section \ref{sec:linear} can be extended to higher orders 
\begin{equation}\label{flatgaugeseconorder}
\begin{split}
&\phi \left(\textbf{x}, t\right) = \bar{\phi} + \delta \phi \left(\textbf{x}, t\right),\\
&h_{ij} = a^{2} \hat{h}_{ij},\:\:\:\hat{h}_{ij} = \left(\delta_{ij} + \hat{t}_{ij} + \frac{1}{2}\hat{t}_{ik}\hat{t}^{k}_{j} + \cdots \right),
\end{split}
\end{equation}
where $\det{\hat{h}} = 1$ and $\hat{t}_{ij}$ is a tensor perturbation which we assume to be a second order quantity $\hat{t}_{ij} = \mathcal{O}\left(\delta \phi^{2} \right)$. It obeys the traceless and transverse conditions $\hat{t}^{i}_{i} = \partial^{i}\hat{t}_{ij}=0$ (indices are raised or lowered with $\delta_{ij}$ or $\delta^{ij}$ respectively). In the comoving gauge, the scalar degree of freedom is called the curvature perturbation $\mathcal{R}$ and the perturbations are given by
\begin{equation}\label{comovinggaugeseconorder}
\begin{split}
&\phi \left(\textbf{x}, t\right) = 0,\\
&h_{ij} = a^{2} e^{- 2 \mathcal{R}} \hat{h}_{ij},\:\:\:\hat{h}_{ij} = \left(\delta_{ij} + t_{ij} + \frac{1}{2}t_{ik}t^{k}_{j} + \cdots \right),
\end{split}
\end{equation}
where $\det{\hat{h}} = 1$ and $t_{ij}$ is a tensor perturbation which we assume to be a second order quantity $t_{ij} = \mathcal{O}\left(\delta \phi^{2} \right)$. It obeys the traceless and transverse conditions $t^{i}_{i} = \partial^{i}t_{ij}=0$. From equations (\ref{correctionproperintroofcomocurv}) and (\ref{firstorderpertenergymomentumzeroi}), the gauge invariant variable $\mathcal{R}$ is rewritten as
\begin{equation}\label{comovingcurvatureproperdefinition}
\mathcal{R} = A - \frac{\mathcal{H}}{\rho + P} \delta q,
\end{equation}
where $\delta q$ is the gauge-dependent 3-momentum perturbation given by
\begin{equation}\label{threemomentumcorrectiondefinition}
\partial_{i} \delta q \equiv \delta_{\left(S\right)} T^{0}_{i},
\end{equation}
where the subscript $\left(S\right)$ denotes the scalar part of the perturbation (\ref{firstorderpertenergymomentumzeroi}) \cite{Langlois:2010}. Note that $A$ is defined in equation (\ref{mostgeneralperturbation}). Because $\delta q = 0$ in the comoving gauge, $\mathcal{R}$ coincides with $A$. If we only consider the change of variables to the linear order between the flat gauge (\ref{flatgaugeseconorder}) and the comoving gauge (\ref{comovinggaugeseconorder}), equation (\ref{comovingrelation}) holds. Let us define $\mathcal{R}$ given in equation (\ref{comovingrelation}) as $\mathcal{R}_{n}$. If we take into account the second order terms, the comoving curvature perturbation is given by \cite{Koyama:2010, Arroja:2008}
\begin{equation}\label{gaugetransformationcurvature}
\mathcal{R} = \mathcal{R}_{n} - f \left(\mathcal{R}_{n} \right),
\end{equation}
where
\begin{equation}
\begin{split}
f(\mathcal{R}) = \frac{\eta}{4c_{s}^{2}}\mathcal{R}^{2} &+ \frac{1}{c^{s}_{s}H} \mathcal{R} \dot{\mathcal{R}} + \frac{1}{4a^{2}H^{2}}\left[-\left(\partial \mathcal{R}\right)\left(\partial \mathcal{R}\right) + \partial^{-2}\left(\partial_{i}\partial_{j}\left(\partial_{i}\mathcal{R} \partial_{j}\mathcal{R} \right) \right) \right]\\
& + \frac{1}{2a^{2}H^{2}}\left[-\left(\partial \mathcal{R}\right)\left(\partial \chi \right) + \partial^{-2}\left(\partial_{i}\partial_{j}\left(\partial_{i}\mathcal{R} \partial_{j}\chi \right) \right) \right],
\end{split}
\end{equation}
with a linear perturbation $\chi$ which is defined by
\begin{equation}
\partial^{2} \chi = - a^{2} \frac{\epsilon}{c^{2}_{s}} \mathcal{R}.
\end{equation}
Note that another gauge invariant variable $\zeta$ which is the curvature perturbation on uniform energy hypersurfaces is often used in the literature (see \cite{Koyama:2010, Maldacena:2005} for example). $\zeta$ is defined as
\begin{equation}\label{uniformnergycurvaturepert}
- \zeta = A + \frac{\mathcal{H}}{\rho'} \delta\rho,
\end{equation}
where
\begin{equation}
\delta \rho \equiv \delta T^{00},
\end{equation}
as we see in equation (\ref{energymomentumfirstdefined}). Note that $\zeta$ is gauge invariant from equations (\ref{gaugetransformationrhocorrect}) and (\ref{gaugetransformationofacorrection}). With equations (\ref{continuity}), (\ref{comovingcurvatureproperdefinition}) and (\ref{uniformnergycurvaturepert}), we obtain \cite{Malik:2009}
\begin{equation}\label{relationbetweencurvaturepertschaptwo}
\mathcal{R} = - \zeta - \frac{\mathcal{H}}{\rho'} \delta \rho_{m},
\end{equation}
where the gauge invariant comoving density perturbation $\delta \rho_{m}$ is defined by
\begin{equation}
\delta \rho_{m} = \delta\rho - 3 \mathcal{H} \delta q.
\end{equation}
Note that we can derive the gauge invariance of $\delta \rho_{m}$ from equations (\ref{gaugetransformationrhocorrect}), (\ref{gaugetransformationofpsicorrection}), (\ref{gaugetransformationofvinsimpleform}), (\ref{firstorderpertenergymomentumzeroi}) and (\ref{threemomentumcorrectiondefinition}). By combining the perturbed Einstein equations $\delta G = 8 \pi G \delta T_{\mu \nu}$, we obtain the gauge invariant generalisation of the Poisson equation as
\begin{equation}\label{poissongeneralised}
\frac{k^{2}}{a^{2}} \Psi = - 4 \pi G \delta \rho_{m},
\end{equation}
where
\begin{equation}\label{afterdefinitionofbardeenpot}
\Psi = A + a H \left(\frac{\bar{h}'}{2} - a \psi \right),
\end{equation}
where $\Psi$ is gauge invariant from equations (\ref{gaugetransformationofacorrection}), (\ref{gaugetransformationofpsicorrection}) and (\ref{gaugetransformationofhbarcorrect}). From equations (\ref{friedmann}), (\ref{continuity}) (\ref{relationbetweencurvaturepertschaptwo}) and (\ref{poissongeneralised}), we obtain the relation between $\mathcal{R}$ and $\zeta$ as \cite{Bassett:2006}
\begin{equation}\label{bassettcurvaturerelation}
\mathcal{R} = - \zeta - \frac{2 \rho}{9 c^{2}_{s} \left(\rho + P \right)}\left(\frac{kc_{s}}{aH}\right)^{2} \Psi,
\end{equation}
note that $\bar{h}$ and $\psi$ are defined in equation (\ref{mostgeneralperturbation}). Therefore, on super-horizon scales ($kc_{s}/aH\ll1$), $\mathcal{R}$ coincides with $-\zeta$. We use $\mathcal{R}$ consistently for the correlation functions of the curvature perturbation. 

\subsection{Single field k-inflation models}
Let us consider only single field k-inflation models with the leading order terms in slow-roll expansion in the action (\ref{kinflationaction}) in the flat gauge which is introduced in section \ref{sec:linear}. Then, the second and the third order actions are obtained as
\begin{equation}
S_{2} = \int dt\,d^{3}x\frac{a^{3}P_{,X}}{2}\left[\frac{1}{c_{s}^{2}}\delta\dot{\phi}^2-\frac{1}{a^{2}}\left(\partial\delta\phi\right)^2\right],
\end{equation}
\begin{equation}
S_{3} = \int dt\,dx^{3} \left[\left(P_{,XX}\frac{\dot{\phi}}{2} + P_{,XXX} \frac{\dot{\phi}^3}{6} \right) a^{3}\delta\dot{\phi}^3 - P_{,XX}\frac{\dot{\phi}}{2}a\delta\phi\left(\partial \delta \phi \right)^{2} \right],
\end{equation}
respectively. The perturbation in the interaction picture can be given by
\begin{equation}\label{redwine}
\begin{split}
\delta\phi\left(\tau, \textbf{x} \right) = \frac{1}{\left(2\pi\right)^{3}}\int d^{3}\textbf{k}\,\delta\phi\left(\tau, \textbf{k}\right)e^{i\textbf{k} \cdot \textbf{x}},\\
\delta\phi\left(\tau, \textbf{k} \right) = u\left(\tau, \textbf{k} \right)a\left(\textbf{k} \right) + u^{\ast}\left(\tau, -\textbf{k} \right)a^{\dagger}\left(-\textbf{k} \right),
\end{split}
\end{equation}
where $a\left(\textbf{k} \right)$ and $a^{\dagger}\left(-\textbf{k} \right)$ are the annihilation and creation operators respectively, whose commutation relations are given by
\begin{equation}
\begin{split}
\left[a\left(\textbf{k}_{\textbf{1}} \right), a^{\dagger}\left(\textbf{k}_{\textbf{2}} \right) \right] = \left(2\pi\right)^{3}\delta^{\left(3\right)}\left(\textbf{k}_{\textbf{1}} - \textbf{k}_{\textbf{2}}  \right),\\
\left[a\left(\textbf{k}_{\textbf{1}} \right), a\left(\textbf{k}_{\textbf{2}} \right) \right] = \left[a^{\dagger}\left(\textbf{k}_{\textbf{1}} \right), a^{\dagger}\left(\textbf{k}_{\textbf{2}} \right) \right] = 0. 
\end{split}
\end{equation}
At the leading order, the solution for the mode function reads
\begin{equation}
u\left(\tau, \textbf{k} \right) = \frac{H}{\sqrt{2c_{s}P_{,X}}}\frac{1}{k^{3/2}} \left(1 + i k c_{s}\tau \right) e^{-i k c_{s} \tau}. 
\end{equation}
The leading order term which corresponds to $N=1$ term in the in-in formalism (\ref{ininresulttwo}) gives the vacuum expectation value of the three point function of the perturbations in the interaction picture as \cite{Maldacena:2005, Chen:2007}
\begin{equation}
\begin{split}
&\left<\Omega \left\lvert \delta\phi\left(t, \textbf{k}_{\textbf{1}} \right) \delta\phi\left(t, \textbf{k}_{\textbf{2}} \right) \delta\phi \left(t, \textbf{k}_{\textbf{3}} \right) \right\lvert \Omega\right>\\
&\approx -i\int^{t}_{t_{0}} d\tilde{t} \left<0 \left\lvert \left[\delta\phi\left(t, \textbf{k}_{\textbf{1}} \right) \delta\phi\left(t, \textbf{k}_{\textbf{2}} \right) \delta\phi \left(t, \textbf{k}_{\textbf{3}} \right), H_{I}\left(\tilde{t} \right) \right] \right\lvert 0\right>,
\end{split}
\end{equation}
where $t_{0}$ is the initial time when the scale of the field fluctuation is deep inside the horizon, t is some time after the horizon exit and $H_{I}$ denotes the interaction Hamiltonian which is given by $H_{I}=-L_{3}$. $\left\lvert \Omega \right>$ is the interacting vacuum which is different from the free field vacuum $\left\lvert 0 \right>$. Note that we specify the vacuum only in this section to make the calculations explicit. Note that $N=0$ term in equation (\ref{ininresulttwo}) vanishes because the number of the annihilation operators must be different from that of the creation operators in the $N=0$ term because of equations (\ref{redwine}). If we use the conformal time, $\tau_{0}$ and $\tau$ which correspond to $t_{0}$ and $t$ respectively can be approximated with $-\infty$ and $0$ respectively because $\tau \approx -\left(aH \right)^{-1}$. After similar calculations to the calculations in \cite{Chen:2007, Arroja:2008b}, we obtain
\begin{equation}\label{kinflationthreepoint}
\begin{split}
&\left<\Omega \left\lvert \delta\phi\left(0, \textbf{k}_{\textbf{1}} \right) \delta\phi\left(0, \textbf{k}_{\textbf{2}} \right) \delta\phi \left(0, \textbf{k}_{\textbf{3}} \right) \right\lvert \Omega\right>\\
&=-\left(2\pi\right)^{3} \delta^{\left(3\right)}\left(\textbf{k}_{\textbf{1}}+\textbf{k}_{\textbf{2}}+\textbf{k}_{\textbf{3}}\right)\frac{H^{4}}{\sqrt{2\epsilon}c_{s}^{2}\left(P_{,X}\right)^{3/2}}\frac{1}{\Pi^{3}_{i=1}k^{3}_{i}}\mathcal{A}^{k-inf}_{\phi}\left(k_{1},k_{2},k_{3}\right),
\end{split}
\end{equation}
where
\begin{equation}\label{kinflationainthreepoint}
\begin{split}
\mathcal{A}^{k-inf}_{\phi} &= - \frac{3\lambda}{\Sigma}\frac{k_{1}^{2}k_{2}^{2}k_{3}^{2}}{K^{3}} + \left(\frac{1}{c_{s}^{2}} - 1\right)\frac{k^{2}_{1}\textbf{k}_{\textbf{2}}\cdot \textbf{k}_{\textbf{3}}}{4K}\left(1+\frac{k_{2}+k_{3}}{K} + 2\frac{k_{2}k_{3}}{K^{2}}\right) + 2 \: \rm{cyclic \: terms}\\
&= \left(\frac{1}{c_{s}^{2}} -1 - \frac{2\lambda}{\Sigma}\right)\frac{3k^{2}_{1}k^{2}_{2}k^{2}_{3}}{2K^{3}} + \frac{1-c_{s}^{2}}{c_{s}^{2}} \left(-\frac{1}{K}\sum_{i>j} k^{2}_{i}k^{2}_{j} + \frac{1}{2K^{2}}\sum_{i \neq j} k^{2}_{i}k^{3}_{j} + \frac{1}{8} \sum_{i} k^{3}_{i} \right),
\end{split}
\end{equation}
with
\begin{equation}
\Sigma = X  P_{,X} + 2 X^{2} P_{,XX} = \frac{H^{2}\epsilon}{c_{s}^{2}},
\end{equation}
\begin{equation}
\lambda = X^{2}  P_{,XX} + \frac{2}{3} X^{3} P_{,XXX}.
\end{equation}
Note that we explicitly show that the correlation functions are evaluated at $\tau=0$ only in this section. If all the slow-roll parameters (\ref{slowrollparametersgeneralcases}) are much smaller than unity, the relation between the curvature perturbation and the field perturbation is simply given by
\begin{equation}\label{simplerelationkflation}
\mathcal{R}\left(\textbf{k}\right) = \frac{H}{\dot{\sigma}} \delta\phi\left(\textbf{k}\right),
\end{equation}
and the three-point function of the curvature perturbation coming from the three-point function of the scalar field perturbation in the k-inflation models at the leading order in slow-roll expansion is simply given by
\begin{equation}\label{curvaturethreepointdefinition}
\begin{split}
&\left<\Omega \left\lvert \mathcal{R}\left(0, \textbf{k}_{\textbf{1}} \right) \mathcal{R}\left(0, \textbf{k}_{\textbf{2}} \right) \mathcal{R}\left(0, \textbf{k}_{\textbf{3}} \right) \right\lvert \Omega\right>^{\left(3\right)}\\
& = \left(\frac{H}{\dot{\sigma}}\right)^{3} \left<\Omega \left\lvert \delta\phi\left(0, \textbf{k}_{\textbf{1}} \right) \delta\phi\left(0, \textbf{k}_{\textbf{2}} \right) \delta\phi \left(0, \textbf{k}_{\textbf{3}} \right) \right\lvert \Omega\right>\\
& = - \left(2\pi\right)^{7} \delta^{\left(3\right)}\left(\textbf{k}_{\textbf{1}}+\textbf{k}_{\textbf{2}}+\textbf{k}_{\textbf{3}}\right) \left(\mathcal{P}_{\mathcal{R}}\right)^{2} \frac{1}{\Pi_{i}k^{3}_{i}} \mathcal{A}^{k-inf}_{\phi} \left(k_{1},k_{2},k_{3}\right), 
\end{split}
\end{equation}
where $\mathcal{A}^{k-inf}_{\phi}$ is given in equation (\ref{kinflationainthreepoint}) and $\mathcal{P}_{\mathcal{R}}$ is given in equation (\ref{curvaturepowerspectrumhorizon}). Note that there are other contributions such as the part coming from the four-point function of the scalar field perturbation. These contributions are negligible in the single field DBI inflation models as we see in section \ref{sec:local}, thus we ignore them here. If we take into account the higher order terms in slow-roll expansion, equation (\ref{simplerelationkflation}) does not hold as we see in equation (\ref{gaugetransformationcurvature}). To take such terms into account, it is possible to calculate the bispectrum of the curvature perturbation in the comoving gauge. The bispectrum of the curvature perturbation with $\mathcal{O}\left(\epsilon\right)$ terms is given by \cite{Chen:2007}
\begin{equation}\label{bispectrumcurvatureslowroll}
\begin{split}
&\left<\Omega \left\lvert \mathcal{R}\left(0, \textbf{k}_{\textbf{1}} \right) \mathcal{R}\left(0, \textbf{k}_{\textbf{2}} \right) \mathcal{R}\left(0, \textbf{k}_{\textbf{3}} \right) \right\lvert \Omega\right>^{\left(3\right)}\\
& = - \left(2\pi\right)^{7} \delta^{\left(3\right)}\left(\textbf{k}_{\textbf{1}}+\textbf{k}_{\textbf{2}}+\textbf{k}_{\textbf{3}}\right) \left(\mathcal{P}_{\mathcal{R}}\right)^{2} \frac{1}{\Pi_{i}k^{3}_{i}} \mathcal{A}^{k-inf} \left(k_{1},k_{2},k_{3}\right), 
\end{split}
\end{equation}
where
\begin{equation}\label{uptofirstorderkinflationbispectrum}
\begin{split}
\mathcal{A}^{k-inf} &=  \left(\frac{1}{c_{s}^{2}} -1 - \frac{2\lambda}{\Sigma}\right)\frac{3k^{2}_{1}k^{2}_{2}k^{2}_{3}}{2K^{3}} + \frac{1-c_{s}^{2}}{c_{s}^{2}} \left(-\frac{1}{K}\sum_{i>j} k^{2}_{i}k^{2}_{j} + \frac{1}{2K^{2}}\sum_{i \neq j} k^{2}_{i}k^{3}_{j} + \frac{1}{8} \sum_{i} k^{3}_{i} \right)\\
& + \frac{\epsilon}{c^{2}_{s}}\left(- \frac{1}{8} \sum_{i} k^{3}_{i} + \frac{1}{8}\sum_{i \neq j} k^{2}_{i}k^{2}_{j} + \frac{1}{K}\sum_{i>j} k^{2}_{i}k^{2}_{j} \right) + \frac{\eta}{c^{2}_{s}}\left(\frac{1}{8} \sum_{i} k^{3}_{i} \right)\\
& + \frac{s}{c^{2}_{s}} \left(- \frac{1}{4} \sum_{i} k^{3}_{i} - \frac{1}{K}\sum_{i>j} k^{2}_{i}k^{2}_{j} + \frac{1}{2K^{2}}\sum_{i \neq j} k^{2}_{i}k^{3}_{j} \right).
\end{split}
\end{equation}
In the standard slow-roll inflation, the first two terms and the last term vanish because $c_{s} = 1$ and $\lambda = 0$. We see that the bispectrum obtained by the calculation in the comoving gauge (\ref{bispectrumcurvatureslowroll}) is consistent with the bispectrum (\ref{curvaturethreepointdefinition}) at the leading order in slow-roll expansion. 

\subsection{Single field DBI inflation models}
In the single field DBI inflation models, the action (\ref{multifieldlagrangian}) takes the form of
\begin{equation}
P\left(\phi,X \right) = -\frac{1}{f\left(\phi \right)} \left(\sqrt{1 - 2 X f\left(\phi \right)} - 1\right) + V\left(\phi \right).
\end{equation}
Therefore, single field DBI inflation models are included in the k-inflation models whose actions are written as in equation (\ref{kinflationaction}). Note that multi-field DBI inflation models are not included in the k-inflation models as you can see from the action (\ref{multifieldlagrangian}) and hence their perturbations have different dynamics from the k-inflation models \cite{Langlois:2008c}. This means that we can apply the result (\ref{kinflationthreepoint}) only to single field DBI inflation models. In this case, $\lambda$ is given by
\begin{equation}\label{dbithreepointintro}
\lambda = \frac{\Sigma}{2} \left(\frac{1}{c^{2}_{s}} - 1 \right).
\end{equation}
From equations (\ref{kinflationthreepoint}) and (\ref{dbithreepointintro}), the vacuum expectation value of the three point function of the scalar field perturbation in the single field DBI inflation at the leading order is given by
\begin{equation}\label{dbifieldthreepoint}
\begin{split}
&\left<\Omega \left\lvert \delta\phi\left(0, \textbf{k}_{\textbf{1}} \right) \delta\phi\left(0, \textbf{k}_{\textbf{2}} \right) \delta\phi \left(0, \textbf{k}_{\textbf{3}} \right) \right\lvert \Omega\right>\\
&=-\left(2\pi\right)^{3} \delta^{\left(3\right)}\left(\textbf{k}_{\textbf{1}}+\textbf{k}_{\textbf{2}}+\textbf{k}_{\textbf{3}}\right)\frac{H^{4}}{\sqrt{2\epsilon}c_{s}^{2}\left(P_{,X}\right)^{3/2}}\frac{1}{\Pi^{3}_{i=1}k^{3}_{i}}\mathcal{A}^{DBI}_{\phi}\left(k_{1},k_{2},k_{3}\right),
\end{split}
\end{equation}
where
\begin{equation}\label{dbifieldathreepoint}
\begin{split}
\mathcal{A}^{DBI}_{\phi} &= \left(\frac{1}{c_{s}^{2}} - 1\right) \left[ - \frac{3}{2} \frac{k_{1}^{2}k_{2}^{2}k_{3}^{2}}{K^{3}} + \frac{k^{2}_{1}\textbf{k}_{\textbf{2}}\cdot \textbf{k}_{\textbf{3}}}{4K}\left(1+\frac{k_{2}+k_{3}}{K} + 2\frac{k_{2}k_{3}}{K^{2}}\right) + 2 \: \rm{cyclic \: terms} \right] \\
&= \frac{1-c_{s}^{2}}{c_{s}^{2}} \left(-\frac{1}{K}\sum_{i>j} k^{2}_{i}k^{2}_{j} + \frac{1}{2K^{2}}\sum_{i \neq j} k^{2}_{i}k^{3}_{j} + \frac{1}{8} \sum_{i} k^{3}_{i} \right).
\end{split}
\end{equation}
From equations (\ref{bispectrumcurvatureslowroll}) and (\ref{dbithreepointintro}), the vacuum expectation value of the three point function of the curvature perturbation in the single field DBI inflation up to the $\mathcal{O}\left(\epsilon\right)$ terms is given by
\begin{equation}\label{singledbithreepoint}
\begin{split}
&\left<\Omega \left\lvert \mathcal{R}\left(0, \textbf{k}_{\textbf{1}} \right) \mathcal{R}\left(0, \textbf{k}_{\textbf{2}} \right) \mathcal{R}\left(0, \textbf{k}_{\textbf{3}} \right) \right\lvert \Omega\right>^{\left(3\right)}\\
& = - \left(2\pi\right)^{7} \delta^{\left(3\right)}\left(\textbf{k}_{\textbf{1}}+\textbf{k}_{\textbf{2}}+\textbf{k}_{\textbf{3}}\right) \left(\mathcal{P}_{\mathcal{R}}\right)^{2} \frac{1}{\Pi_{i}k^{3}_{i}} \mathcal{A}^{DBI} \left(k_{1},k_{2},k_{3}\right), 
\end{split}
\end{equation}
where
\begin{equation}\label{singledbia}
\begin{split}
\mathcal{A}^{DBI} &= \frac{1-c_{s}^{2}}{c_{s}^{2}} \left(-\frac{1}{K}\sum_{i>j} k^{2}_{i}k^{2}_{j} + \frac{1}{2K^{2}}\sum_{i \neq j} k^{2}_{i}k^{3}_{j} + \frac{1}{8} \sum_{i} k^{3}_{i} \right)\\
& + \frac{\epsilon}{c^{2}_{s}}\left(- \frac{1}{8} \sum_{i} k^{3}_{i} + \frac{1}{8}\sum_{i \neq j} k^{2}_{i}k^{2}_{j} + \frac{1}{K}\sum_{i>j} k^{2}_{i}k^{2}_{j} \right) + \frac{\eta}{c^{2}_{s}}\left(\frac{1}{8} \sum_{i} k^{3}_{i} \right)\\
& + \frac{s}{c^{2}_{s}} \left(- \frac{1}{4} \sum_{i} k^{3}_{i} - \frac{1}{K}\sum_{i>j} k^{2}_{i}k^{2}_{j} + \frac{1}{2K^{2}}\sum_{i \neq j} k^{2}_{i}k^{3}_{j} \right),
\end{split}
\end{equation}
because the first term in equation (\ref{uptofirstorderkinflationbispectrum}) vanishes. This function has a peak at the equilateral configuration $k_{1} \sim k_{2} \sim k_{3}$ as shown in \cite{Creminelli:2006}. The parameter $f_{\rm{NL}}^{equil}$ is defined by
\begin{equation}\label{fnlequilateraldefinition}
\begin{split}
&\left<\Omega \left\lvert \mathcal{R}\left(0, \textbf{k}_{\textbf{1}} \right) \mathcal{R}\left(0, \textbf{k}_{\textbf{2}} \right) \mathcal{R}\left(0, \textbf{k}_{\textbf{3}} \right) \right\lvert \Omega\right>^{\left(3\right)}\\
&= - (2\pi)^{7} \delta^{\left(3\right)}\left(\textbf{k}_{\textbf{1}}+\textbf{k}_{\textbf{2}}+\textbf{k}_{\textbf{3}}\right) \left(\frac{3}{10} f_{\rm{NL}}^{equil} \left(\mathcal{P}_{\mathcal{R}_{*}} \right)^{2} \right) \frac{\sum_{i}k^{3}_{i}}{\Pi_{i}k^{3}_{i}}.
\end{split}
\end{equation}
Obviously, the actual three-point function of the curvature perturbation which is given by equation (\ref{singledbithreepoint}) is different as a function of $k_{i}$ from the ``three-point function of the curvature perturbation" in equation (\ref{fnlequilateraldefinition}). $f_{\rm{NL}}^{equil}$ is defined so that the actual three-point function of the curvature perturbation has the same value with the function in equation (\ref{fnlequilateraldefinition}) at the equilateral configuration $k_{1} \sim k_{2} \sim k_{3}$ where it has its maximum. By setting $k_{1} = k_{2} = k_{3} = \tilde{k}$,  we have
\begin{equation}\label{limitequilateral}
\mathcal{A}^{DBI} = -\frac{7}{24} \left(\frac{1}{c^{2}_{s}} - 1 \right) \tilde{k}^{3},
\end{equation}
at the leading order. From equations (\ref{singledbithreepoint}) and (\ref{limitequilateral}), in this limit, we have
\begin{equation}\label{equilateralfnlderivation}
\begin{split}
&\left<\Omega \left\lvert \mathcal{R}\left(0, \textbf{k}_{\textbf{1}} \right) \mathcal{R}\left(0, \textbf{k}_{\textbf{2}} \right) \mathcal{R}\left(0, \textbf{k}_{\textbf{3}} \right) \right\lvert \Omega\right>^{\left(3\right)}\\
&= - (2\pi)^{7} \delta^{\left(3\right)}\left(\textbf{k}_{\textbf{1}}+\textbf{k}_{\textbf{2}}+\textbf{k}_{\textbf{3}}\right) \left(-\frac{7}{24} \left(\frac{1}{c^{2}_{s}} - 1 \right) \frac{\tilde{k}^{3}}{\Pi_{i}k^{3}_{i}} \right) \left(\frac{H^{2}}{8 \pi^2 \epsilon c_{s}} \right)^{2}\\
&= - (2\pi)^{7} \delta^{\left(3\right)}\left(\textbf{k}_{\textbf{1}}+\textbf{k}_{\textbf{2}}+\textbf{k}_{\textbf{3}}\right) \left(-\frac{7}{72} \left(\frac{1}{c^{2}_{s}} - 1 \right) \frac{\sum_{i}k^{3}_{i}}{\Pi_{i}k^{3}_{i}} \right) \left(\mathcal{P}_{\mathcal{R}_{*}} \right)^{2}\\
&= - (2\pi)^{7} \delta^{\left(3\right)}\left(\textbf{k}_{\textbf{1}}+\textbf{k}_{\textbf{2}}+\textbf{k}_{\textbf{3}}\right) \frac{3}{10} \left(-\frac{35}{108} \left(\frac{1}{c^{2}_{s}} - 1 \right) \frac{\sum_{i}k^{3}_{i}}{\Pi_{i}k^{3}_{i}} \right) \left(\mathcal{P}_{\mathcal{R}_{*}} \right)^{2}.
\end{split}
\end{equation}
Comparing equation (\ref{fnlequilateraldefinition}) with equation (\ref{equilateralfnlderivation}), we obtain
\begin{equation}\label{fnlequilateralfinal}
f_{\rm{NL}}^{equil} = -\frac{35}{108} \left(\frac{1}{c^{2}_{s}} - 1 \right). 
\end{equation}

\section{\label{sec:deltaN}$\delta$ N-formalism}
\begin{figure}[!htb]
\centering
\includegraphics[width=8cm]{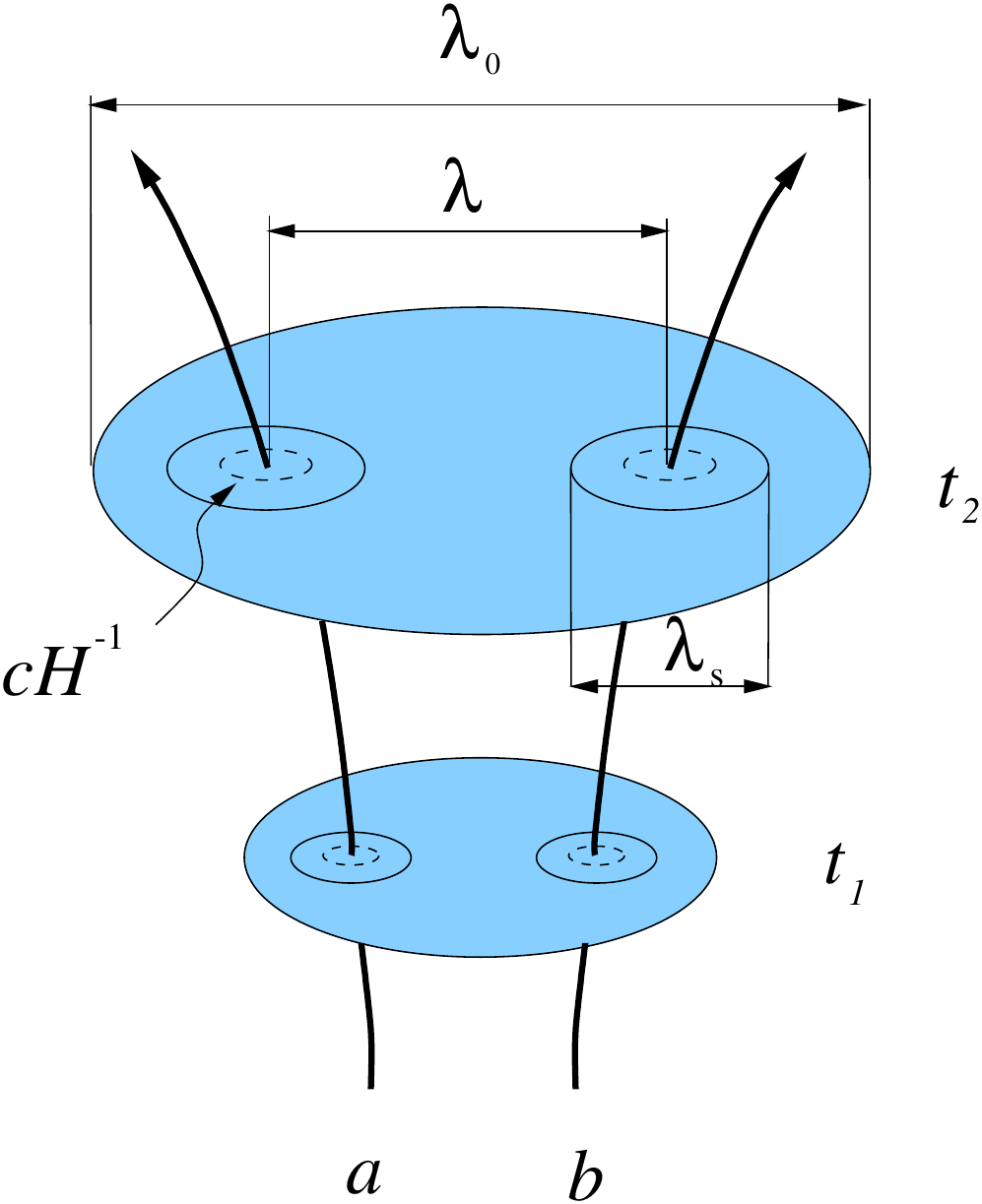}
\caption[Separate Universe assumption]{Schematic figure about the separate Universe assumption with the symbols explained in the text. The figure is taken from \cite{Wands:2010}}{\label{fig:separateuniverse}}
\end{figure}

In this section, we review the $\delta$N-formalism \cite{Sasaki:1996, Lyth:2005b, Naruko:2012}. In this formalism, we make the ``separate Universe assumption" \cite{Lyth:2009, Wands:2010}. In this non-trivial assumption, each super-horizon sized region of the Universe evolves like a separate FRW Universe where density and pressure may take different values, but are locally homogeneous. There has to be some scale of such a locally homogeneous region $\lambda_{\rm{s}}$ for which this assumption is valid to useful accuracy. Those locally homogeneous regions with different energy densities are in a much larger region whose scale is $\lambda_{0}$ which should be much bigger than our present horizon size. When locally homogeneous regions are pieced together over a large scale $\lambda$, they represent the long wavelength perturbations under consideration. Therefore, we require a hierarchy of scales as
\begin{equation}
\lambda_{0} \gg \lambda \gg \lambda_{\rm{s}} \gtrsim cH^{-1},
\end{equation}
which is described in figure \ref{fig:separateuniverse}. Note that the $\delta$N-formalism is valid on the basis of the separate Universe assumption. 

Let us consider the standard $\left(3 + 1\right)$ decomposition of the metric (\ref{admfirstintroduction}) (ADM formalism). Then, the unit time-like vector orthonormal to the constant time (spatial) hypersurface $n^{\mu}$ has the components
\begin{equation}
n_{\mu} = \left[-N_{l}, 0 \right],\:\:\:\:\:n^{\mu} = \left[\frac{1}{N_{l}},-\frac{N^{i}}{N_{l}} \right].
\end{equation}
Note that the lapse function $N$ in the metric (\ref{admfirstintroduction}) is redefined as $N_{l}$ only in this section to avoid confusion with the number of e-folds. Then, $\theta_{n} = \nabla_{\mu}n^{\mu}$ is the volume expansion rate of the hypersurfaces along the integral curve of $n^{\mu}$ where $\nabla_{\mu}$ is the covariant derivative \cite{Sasaki:1996}. The local Hubble parameter is defined as
\begin{equation}\label{deltanexpansionparameter}
\tilde{H} = \frac{1}{3} \theta_{n} = \frac{1}{N_{l}} \left(\frac{\dot{a}}{a} - \dot{A}\right) + \mathcal{O}\left(\mathcal{E}^{2} \right),
\end{equation}
where $\mathcal{E}=k/aH$. It is shown that the $\left(0,0\right)$-component of the Einstein equation with the perturbations gives \cite{Lyth:2005b}
\begin{equation}
\tilde{H}^{2} = \frac{1}{3 M^{2}_{\rm{P}}} \rho + \mathcal{O}\left(\mathcal{E}^{2} \right). 
\end{equation}
Therefore, on super-horizon scales, the number of e-folds of the expansion along the integral curve of the 4 velocity is given by
\begin{equation}\label{generalresultdeltan}
\begin{split}
N \left(t_{2},t_{1};\textbf{x}\right) &= \int^{\tau_{2}}_{\tau_{1}} \tilde{H} d\tau\\
&= \frac{1}{3} \int^{t_{2}}_{t_{1}} \theta_{n} N_{l} dt\\
&= \ln{\left[\frac{a\left(t_{2}\right)}{a\left(t_{1}\right)}\right]} - A\left(t_{2};\textbf{x}\right) + A\left(t_{1};\textbf{x}\right),
\end{split}
\end{equation}
where $\tau$ is the proper time along the curve and we used equation (\ref{deltanexpansionparameter}). This is the general result of the $\delta$N-formalism. It means that the change in $A$, going from one slice to another, is equal to the difference between the actual number of e-folds $N \left(t_{2},t_{1};\textbf{x}\right)$ and the background value $N_{0}\left(t_{2},t_{1}\right) = \ln{\left[a\left(t_{2}\right)/a\left(t_{1}\right)\right]}$. Let us consider two different ways of slicing. In the slicing $\mathfrak{A}$, it starts on a flat slice at $t=t_{1}$ and ends on a uniform density slice at $t=t_{2}$ while it starts on a flat slice at $t=t_{1}$ and ends on a flat slice at $t=t_{2}$ in the slicing $\mathfrak{B}$. From equation (\ref{generalresultdeltan}), the difference in $A$ between the slicing $\mathfrak{A}$ and $\mathfrak{B}$ at $t=t_{2}$ gives
\begin{equation}\label{deltanfinalresult}
\begin{split}
- \mathcal{R} = \zeta &= - A_{\mathfrak{A}}\left(t_{2};\textbf{x}\right)\\
& = N_{\mathfrak{A}}\left(t_{2},t_{1};\textbf{x}\right) - N_{0}\left(t_{2},t_{1}\right) = \delta N_{F}\left(t_{2},t_{1};\textbf{x}\right), 
\end{split}
\end{equation}
where $\delta N_{F}\left(t_{2},t_{1};\textbf{x}\right)$ is the difference in the number of e-folds between the final uniform density slice and the initial flat slice and we have used the fact that $A=0$ on a flat slice. The first equality in equation (\ref{deltanfinalresult}) holds only on super-horizon scales as we see in equation (\ref{bassettcurvaturerelation}) while the second equality comes from equation (\ref{uniformnergycurvaturepert}) because $\delta \rho = 0$ on a uniform density slice. During inflation, the number of e-folds depends on the dynamics of the scalar fields. In the slow-roll cases, the equations of motion for the scalar fields are approximated with the first order differential equations in the cosmic time $t$. Then, all the dynamics is determined only by the initial values of the fields $\phi^{I}$ although we also need the initial values of the time derivatives of the fields $\dot{\phi^{I}}$ to solve second order differential equations in $t$. Therefore, $ \delta N_{F}\left(t_{2},t_{1};\textbf{x}\right)$ is expanded as
\begin{equation}\label{usedlaterdeltan}
\begin{split}
-\mathcal{R} &= \delta N_{F}\left(t_{2}, \phi^{I}\left(t_{1},\textbf{x}\right)\right)\\
&= \sum_{I} N_{,I} \left(t_{1}, t_{2} \right) \phi^{I} \left(t_{1},\textbf{x}\right)\\ 
& + \sum_{I,J} \frac{N_{,IJ} \left(t_{1}, t_{2} \right)}{2} \left( \phi^{I} \left(t_{1},\textbf{x}\right) \phi^{J} \left(t_{1},\textbf{x}\right) - \left< \phi^{I} \left(t_{1},\textbf{x}\right) \phi^{J} \left(t_{1},\textbf{x}\right) \right> \right) + \cdots,
\end{split}
\end{equation}
where the subscript $_{,I}$ denotes the derivative with respect to the scalar fields, $t_{1}$ needs to be after the horizon exit because $\delta$N-formalism is valid only on super-horizon scales and
\begin{equation}
N_{I} = \frac{\partial N}{\partial \phi^{I}},\:\:\:\:\:N_{IJ} = \frac{\partial^{2} N}{\partial \phi^{I} \partial \phi^{J}}.
\end{equation}
Note that $\left< \phi^{I} \left(t_{1},\textbf{x}\right) \phi^{J} \left(t_{1},\textbf{x}\right) \right>$ is subtracted so that the ensemble average of the random field $ \left<\mathcal{R}  \right>$ vanishes. 

\begin{figure}[!htb]
\centering
\includegraphics[width=12cm]{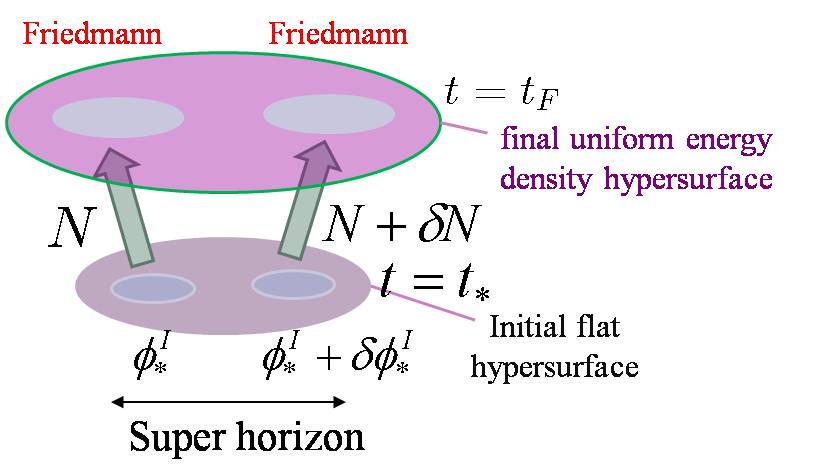}
\caption[$\delta$N-formalism]{Schematic figure about the $\delta$N-formalism. Different patches of the Universe expand differently because of the fluctuations of the scalar fields on the initial slice. The figure is taken from \cite{Tanaka:2010}}{\label{fig:deltan}}
\end{figure}

Figure \ref{fig:deltan} shows how the fluctuations of the scalar fields result in the fluctuations in the number of e-folds. A patch of the Universe with a higher energy density than the background because of the scalar field fluctuation expands more on the basis of the separate Universe assumption. 

If we consider two-field inflation models for simplicity, we can expand $\delta N$ in terms of the values of the fields (\ref{definitionofqsigma}) at sound horizon crossing up to the second order. In this case, equation (\ref{usedlaterdeltan}) reads
\begin{equation}
\begin{split}
 -\mathcal{R} &= \delta N \left(\phi_{\sigma}^{*}, \dot{\phi}_{\sigma}^{*}, \phi_{s}^{*}, \dot{\phi}_{s}^{*}, t_{2} \right) \simeq \delta N \left(\phi_{\sigma}^{*}, \phi_{s}^{*}, t_{2} \right) \\
 &\simeq N_{,\sigma} \left(t_{1}, t_{2} \right) Q_{\sigma}^{*} + N_{,s}  \left(t_{1}, t_{2} \right) Q_{s}^{*} \\
 & + \frac{1}{2} N_{,\sigma \sigma}  \left(t_{1}, t_{2} \right) \left (Q_{\sigma}^{*2} - \left<Q_{\sigma}^{*2} \right> \right) + \frac{1}{2} N_{,ss}  \left(t_{1}, t_{2} \right) \left( Q_{s}^{*2} - \left<Q_{s}^{*2} \right> \right),\label{ndecomposition}
 \end{split}
\end{equation}
where $\phi_{\sigma}$ denotes the scalar field in the instantaneous adiabatic direction which is defined by the basis vector $e^{I}_{\sigma}$ in equation (\ref{basiskflation}) as $\phi_{\sigma} \equiv G_{IJ} \phi^{I} e^{I}_{\sigma}$ and $\phi_{s}$ denotes the scalar field in the instantaneous entropic direction which is defined by the basis vector $e^{I}_{s}$ in subsection \ref{kflationlinear} as $\phi_{s} \equiv G_{IJ} \phi^{I} e^{I}_{s}$. Note that the subscripts $, \sigma$ and $,s$ denote the derivatives with respect to $\phi_{\sigma}$ and $\phi_{s}$ respectively and the subscript $*$ means that the quantity is evaluated at the sound horizon exit at $t = t_{1}$. Note also that we ignored the cross term $Q_{\sigma} Q_{s}$ because we assume that the two fields are independent quantum fields around the sound horizon exit and hence this term does not contribute to the vacuum expectation value of the correlation functions of the curvature perturbation. With the basis vectors $\tilde{e}^{I}_{\sigma}$ and $\tilde{e}^{I}_{s}$ defined in section \ref{dbilinear}, equation (\ref{usedlaterdeltan}) reads
\begin{equation}
\begin{split}
 -\mathcal{R} &= \delta N \left(\tilde{\phi}_{\sigma}^{*}, \dot{\tilde{\phi}}_{\sigma}^{*}, \tilde{\phi}_{s}^{*}, \dot{\tilde{\phi}}_{s}^{*}, t_{2} \right) \simeq \delta N \left(\tilde{\phi}_{\sigma}^{*}, \tilde{\phi}_{s}^{*}, t_{2} \right) \\
 &\simeq N_{,\tilde{\sigma}} \left(t_{1}, t_{2} \right) \tilde{Q}_{\sigma}^{*} + N_{,\tilde{s}}  \left(t_{1}, t_{2} \right) \tilde{Q}_{s}^{*} \\
 & + \frac{1}{2} N_{,\tilde{\sigma} \tilde{\sigma}}  \left(t_{1}, t_{2} \right) \left (\tilde{Q}_{\sigma}^{*2} - \left<\tilde{Q}_{\sigma}^{*2} \right> \right) + \frac{1}{2} N_{,\tilde{s}\tilde{s}}  \left(t_{1}, t_{2} \right) \left( \tilde{Q}_{s}^{*2} - \left<\tilde{Q}_{s}^{*2} \right> \right),\label{ndecompositionwiththenewbases}
 \end{split}
\end{equation}
where $\tilde{\phi}_{\sigma}$ denotes the scalar field in the instantaneous adiabatic direction which is defined by the basis vector $\tilde{e}^{I}_{\sigma}$ in equation (\ref{newadiabaticbasis}) as $\tilde{\phi}_{\sigma} \equiv G_{IJ} \phi^{I} \tilde{e}^{I}_{\sigma} / c_{s}$ and $\tilde{\phi}_{s}$ denotes the scalar field in the instantaneous entropic direction which is defined by the basis vector $\tilde{e}^{I}_{s}$ in equation (\ref{orthoconditions}) as $\tilde{\phi}_{s} \equiv G_{IJ} \phi^{I} \tilde{e}^{I}_{s} c_{s}$ because we have the relation $\phi^{I} = \tilde{\phi}_{\sigma} \tilde{e}^{I}_{\sigma} + \tilde{\phi}_{s} \tilde{e}^{I}_{s}$. Note that the subscripts $, \tilde{\sigma}$ and $,\tilde{s}$ denote the derivatives with respect to $\tilde{\phi}_{\sigma}$ and $\tilde{\phi}_{s}$ respectively and the subscript $*$ means that the quantity is evaluated at the sound horizon exit at $t = t_{1}$. Note also that we ignored the cross term $\tilde{Q}_{\sigma} \tilde{Q}_{s}$ because we assume that the two fields are independent quantum fields around the sound horizon exit and hence this term does not contribute to the vacuum expectation value of the correlation functions of the curvature perturbation. 

\section{\label{sec:local}Calculation of the local type non-Gaussianity}
Let us introduce the local type non-Gaussianity in this section following the references \cite{Koyama:2010, Creminelli:2006, Babich:2004}. The non-linearity in the curvature perturbation which produces the local type non-Gaussianity is often given by
\begin{equation}\label{firstdefinitionlocalcurvature}
\mathcal{R} = \mathcal{R}_{n} - \frac{3}{5} f_{\rm{NL}}^{local} \left(\mathcal{R}^{2}_{n} - \left<\mathcal{R}^{2}_{n}\right> \right),
\end{equation}
where $\mathcal{R}_{n}$ obeys Gaussian statistics. The Fourier transformation of the quadratic part is written as
\begin{equation}\label{localcalculationsquare}
\mathcal{R}^{2}_{n} = \int \frac{d^{3}k}{\left(2 \pi \right)^{3}} \left[\mathcal{R}_{n} \star \mathcal{R}_{n} \right] \left(\textbf{k} \right) e^{i \textbf{k} \cdot \textbf{x}},
\end{equation}
where we can easily derive the relation
\begin{equation}\label{minorcorrectionone}
\left[\mathcal{R}_{n} \star \mathcal{R}_{n} \right] \left(\textbf{k} \right) = \int \frac{d^{3}q}{\left(2 \pi \right)^{3}} \mathcal{R}_{n} \left(\textbf{q} \right) \mathcal{R}_{n} \left(\textbf{k} - \textbf{q} \right),
\end{equation}
by using the inverse Fourier transformation. Note that $\star$ denotes the convolution operation. 

Then, the three point function of the curvature perturbation reads
\begin{equation}\label{localshapederivation}
\begin{split}
&\left<\mathcal{R} \left(\textbf{k}_{\textbf{1}}\right) \mathcal{R} \left(\textbf{k}_{\textbf{2}}\right) \mathcal{R} \left(\textbf{k}_{\textbf{3}} \right) \right>\\
&= - \frac{3}{5} f_{\rm{NL}}^{local} \left<\mathcal{R}_{n} \left(\textbf{k}_{\textbf{1}}\right) \mathcal{R}_{n} \left(\textbf{k}_{\textbf{2}}\right) \left( \left[\mathcal{R}_{n} \star \mathcal{R}_{n} \right] \left(\textbf{k}_{\textbf{3}} \right) - \left< \left[\mathcal{R}_{n} \star \mathcal{R}_{n} \right] \left(\textbf{k}_{\textbf{3}}\right) \right> \right)  \right> + 2\:\rm{cyclic \: terms}\\
&= - \frac{3}{5} f_{\rm{NL}}^{local} \int \frac{dq^{3}}{\left(2 \pi \right)^{3}} \left\{\left<\mathcal{R}_{n} \left(\textbf{k}_{\textbf{1}}\right) \mathcal{R}_{n} \left(\textbf{k}_{\textbf{2}}\right) \right>\left<\mathcal{R}_{n} \left(\textbf{q} \right) \mathcal{R}_{n} \left(\textbf{k}_{\textbf{3}} - \textbf{q}\right)\right> \right.\\
&\left. -  \left<\mathcal{R}_{n} \left(\textbf{k}_{\textbf{1}}\right) \mathcal{R}_{n} \left(\textbf{k}_{\textbf{2}}\right) \right> \left<\mathcal{R}_{n} \left(\textbf{q} \right) \mathcal{R}_{n} \left(\textbf{k}_{\textbf{3}} - \textbf{q}\right)\right> + \left<\mathcal{R}_{n} \left(\textbf{k}_{\textbf{1}}\right) \mathcal{R}_{n} \left(\textbf{q} \right) \right>\left<\mathcal{R}_{n} \left(\textbf{k}_{\textbf{2}} \right) \mathcal{R}_{n} \left(\textbf{k}_{\textbf{3}} - \textbf{q}\right)\right>\right. \\
&\left. + \left<\mathcal{R}_{n} \left(\textbf{k}_{\textbf{1}}\right) \mathcal{R}_{n} \left(\textbf{k}_{\textbf{3}} - \textbf{q} \right)\right>\left<\mathcal{R}_{n} \left(\textbf{k}_{\textbf{2}}\right) \mathcal{R}_{n} \left(\textbf{q} \right)\right>\right\} + \cdots\\
&= - \frac{3}{5} f_{\rm{NL}}^{local} \left(2\pi\right)^{3} \int dq^{3} \left\{\delta^{\left(3 \right)}\left(\textbf{k}_{\textbf{1}} + \textbf{q} \right) \delta^{\left(3 \right)}\left(\textbf{k}_{\textbf{2}} + \textbf{k}_{\textbf{3}} - \textbf{q} \right) \left<\mathcal{R}^{2}_{n}\right> \left(k_{1} \right) \left<\mathcal{R}^{2}_{n}\right> \left(k_{2} \right) \right. \\
&\left. + \delta^{\left(3 \right)}\left( \textbf{k}_{\textbf{1}} + \textbf{k}_{\textbf{3}} - \textbf{q} \right) \delta^{\left(3 \right)}\left(\textbf{k}_{\textbf{2}} + \textbf{q}\right) \right\} \left<\mathcal{R}^{2}_{n}\right> \left(k_{1} \right) \left<\mathcal{R}^{2}_{n}\right> \left(k_{2} \right) + \cdots\\
&= - \frac{3}{10} f_{\rm{NL}}^{local} \left(2\pi\right)^{7} \frac{\left(\mathcal{P}_{\mathcal{R}}\right)^{2}}{k_{1}^{3}k_{2}^{3}}  \delta^{\left(3 \right)}\left( \textbf{k}_{\textbf{1}} + \textbf{k}_{\textbf{2}} + \textbf{k}_{\textbf{3}} \right) + \cdots,
\end{split}
\end{equation}
where we have used Wick's theorem (see \cite{Lyth:2009}) to decompose the four point function into the combinations of the two point functions, the vacuum expectation value of equation (\ref{localcalculationsquare}) to obtain the Fourier mode of $\left<\mathcal{R}^{2}_{n}\right>$ and the following relations
\begin{equation}
\left<\mathcal{R}_{n} \left(\textbf{k} \right) \mathcal{R}_{n} \left(\textbf{k}'\right) \right> = \left(2 \pi \right)^{3} \delta^{\left(3 \right)} (\textbf{k} + \textbf{k}') \left<\mathcal{R}^{2}_{n}\right> \left(k \right),
\end{equation}
\begin{equation}
\mathcal{P}_{\mathcal{R}} = \frac{k^{3}}{2 \pi^{2}} \left<\mathcal{R}^{2}_{n}\right> \left(k \right). 
\end{equation}
Note that the three point functions of quantities which obey Gaussian statistics vanish and $\cdots$ denotes the terms which come from the 2 cyclic terms. From equation (\ref{localshapederivation}), the bispectrum of the curvature perturbation is given by
\begin{equation}\label{localthreespecialdefinition}
\left<\mathcal{R} \left(\textbf{k}_{\textbf{1}}\right) \mathcal{R} \left(\textbf{k}_{\textbf{2}}\right) \mathcal{R} \left(\textbf{k}_{\textbf{3}} \right) \right> = - (2 \pi)^{7} \delta^{\left(3\right)} \left(\textbf{k}_{\textbf{1}} +\textbf{k}_{\textbf{2}} +\textbf{k}_{\textbf{3}} \right) \left(\mathcal{P}_{\mathcal{R}}\right)^{2} F\left(k_{1},k_{2},k_{3} \right),
\end{equation}
where
\begin{equation}\label{localsinglegeneraldefinition}
F\left(k_{1},k_{2},k_{3} \right) = \left(\frac{3}{10} f_{\rm{NL}}^{local}\right) \left(\frac{1}{k_{1}^{3}k_{2}^{3}} + \frac{1}{k_{2}^{3}k_{3}^{3}} + \frac{1}{k_{1}^{3}k_{3}^{3}} \right).
\end{equation}
In single field DBI inflation models, $f_{\rm{NL}}^{local}$ is obtained in terms of the slow-roll parameters. If we take a squeezed limit in which $k_{3} \rightarrow 0$ and $k_{1} = k_{2} = \tilde{k}$, the function (\ref{singledbia}) reads
\begin{equation}\label{squeezedsingledbilocal}
\lim_{k_{3} \rightarrow 0} \mathcal{A}^{DBI} = \frac{\tilde{k}^{3}}{4 c^{2}_{s}} \left(2\epsilon + \eta - 3s \right).
\end{equation}
Note that the first term in the function (\ref{singledbia}) vanishes in this limit. Comparing the three point function (\ref{singledbithreepoint}) defined with the function (\ref{squeezedsingledbilocal}) with the local type three point function (\ref{localthreespecialdefinition}) defined with the function (\ref{localsinglegeneraldefinition}) in the squeezed limit, we obtain
\begin{equation}
f_{\rm{NL}}^{local} = \frac{5 c^{2}_{s}}{12} \left(2\epsilon + \eta - 3s \right).
\end{equation}
Therefore, $f_{\rm{NL}}^{local}$ in single field DBI inflation models is of the order of the slow-roll parameters. 

\section{Observables in multi-field inflation}\label{sec:observable sinmultifield}
In this section, we review the dynamics of multi-field inflation models and introduce the observables in multi-field DBI inflation models following \cite{Langlois:2008b}. 

Let us define an entropy perturbation as
\begin{equation}\label{definedentropypert}
\mathcal{S} = c_{s} \frac{H}{\dot{\sigma}} Q_{s} = \frac{H \sqrt{c_{s}}}{\dot{\sigma}} \tilde{Q}_{s},
\end{equation}
where $\tilde{Q}_{s}$ is defined in equation (\ref{newdefinitionofqsigma}). The relation between the curvature perturbation and the entropy perturbation is given by
\begin{equation}
\dot{\mathcal{R}} = \frac{\xi}{a}\mathcal{S} + \frac{H}{\dot{H}} \frac{c^{2}_{s} k^{2}}{a^{2}}\Psi,
\end{equation}
where $\Psi$ is defined in equation (\ref{afterdefinitionofbardeenpot}) and $\xi$ is defined in equation (\ref{xidefined}). We see that the entropy perturbation is the only source of the curvature perturbation on super-horizon scales. As shown in \cite{Langlois:2008b}, the interaction parameter $\xi$ is directly related to the bending of the background trajectory. When there is a curve in the trajectory in the field space resulting in the rotation of the adiabatic/entropy basis, $\xi$ takes a non-zero value. On super-horizon scales, the evolution of those perturbations are given by
\begin{equation}\label{interactionevolutionequationentropy}
\dot{\mathcal{R}} \approx \alpha H \mathcal{S},\:\:\:\:\:\dot{\mathcal{S}} \approx \beta H \mathcal{S},
\end{equation}
where
\begin{equation}
 \alpha = \frac{\Xi}{c_{s} H},\label{alphamouse}
\end{equation}
\begin{equation}
 \beta = \frac{s}{2} - \frac{\eta}{2} - \frac{1}{3H^{2}}\left(\mu^{2}_{\rm{s}} + \frac{\Xi^{2}}{c^{2}_{\rm{s}}} \right),\label{betamouse}
\end{equation}
and
\begin{equation}\label{interactionredefinition}
 \Xi = \frac{c_{s}}{a}\xi.
\end{equation}
When such a conversion from the entropy perturbation to the curvature perturbation exists, from equations (\ref{interactionevolutionequationentropy}) we quantify the conversion as
\begin{equation}\label{transfermatrix}
\left( 
\begin{array}{@{\,}c@{\,}}
\mathcal{R}\\
\mathcal{S}
\end{array}
\right) = \left( 
\begin{array}{@{\,}cc@{\,}}
1&T_{\mathcal{R} \mathcal{S}}\\
0&T_{\mathcal{S} \mathcal{S}}\\
\end{array}
\right) \left( 
\begin{array}{@{\,}c@{\,}}
\mathcal{R}\\
\mathcal{S}
\end{array}
\right)_{*}
\end{equation}
where the subscript $*$ indicates that the corresponding quantity is evaluated at sound horizon crossing $kc_{s} = aH$ with
\begin{equation}
 T_{\mathcal{R} \mathcal{S}}(t_{*},t) = \int ^{t}_{t_{*}} \alpha(t')T_{\mathcal{S} \mathcal{S}}(t_{*},t')H(t')dt',\label{trs}
\end{equation}
\begin{equation}
 T_{\mathcal{S} \mathcal{S}}(t_{*},t) = \exp{\left(\int ^{t}_{t_{*}} \beta(t')H(t')dt'\right)}. \label{tss}
\end{equation}
Hence, the power spectrum of the curvature perturbation is given by
\begin{equation}\label{chaptertwotransferdefined}
\mathcal{P}_{\mathcal{R}} = \left(1 + T^{2}_{\mathcal{R} \mathcal{S}}\right) \mathcal{P}_{\mathcal{R}_{*}},
\end{equation}
where $\mathcal{P}_{\mathcal{R}_{*}}$ is given by equation (\ref{curvaturepowerspectrumhorizon}) if the slow-roll parameters and $\xi/a H$ are much smaller than unity around the horizon crossing. If we define
\begin{equation}
 \sin{\Theta} \equiv \frac{T_{\mathcal{R} \mathcal{S}}}{\sqrt{1 + T^{2}_{\mathcal{R} \mathcal{S}}}},\,\,\,\cos{\Theta} \equiv \frac{1}{\sqrt{1 + T^{2}_{\mathcal{R} \mathcal{S}}}},\label{sintheta}
\end{equation}
the final value of the power spectrum of the curvature perturbation is given by
\begin{equation}
\mathcal{P}_{\mathcal{R}} = \frac{\mathcal{P}_{\mathcal{R}_{*}}}{\cos^{2}{\Theta}} = \left(\frac{H^{2}}{8\pi^{2} \epsilon c_{s}} \right)\frac{1}{\cos^{2}{\Theta}}.\label{prmouse}
\end{equation}
From equations (\ref{tensorperturbationspectrum}) and (\ref{prmouse}), the tensor-to-scalar ratio is given by
\begin{equation}\label{chaptertwofinaltensortoscalar}
 r = 16 \epsilon c_{s} \cos^{2}{\Theta}.
\end{equation}
The spectral index for the curvature perturbation is obtained from equations (\ref{spectralfirstdefinition}) and (\ref{prmouse}) if we assume all the slow-roll parameters (\ref{slowrollparametersgeneralcases}) are much smaller than unity as 
\begin{equation}\label{chaptertwospectralindexfinal}
\begin{split}
n_{\mathcal{R}} &= \frac{d\,\ln{\mathcal{P}_{\mathcal{R}}}}{d\,\ln{k}} + 1\\
 & = n_{\mathcal{R}_{*}} + H^{-1}_{*} \sin{\left( 2 \Theta \right)} \frac{\partial T_{\mathcal{R} \mathcal{S}}}{\partial t^{*}}\\
 & = n_{\mathcal{R}_{*}} - \alpha_{*} \sin{\left( 2 \Theta \right)} - 2 \beta_{*} \sin^{2}{\Theta},
\end{split}
\end{equation}
with
\begin{equation}\label{spectraldbihorizon}
n_{\mathcal{R}_{*}} - 1 = - 2 \epsilon_{*} - \eta_{*} - s_{*},
\end{equation}
where we have used
\begin{equation}
H^{-1}_{*} \frac{\partial T_{\mathcal{S} \mathcal{S}}}{\partial t^{*}} = - T_{\mathcal{S} \mathcal{S}} \beta_{*} ,\:\:\:\:\:H^{-1}_{*} \frac{\partial T_{\mathcal{R} \mathcal{S}}}{\partial t^{*}} = - \alpha_{*} - T_{\mathcal{R} \mathcal{S}} \beta_{*},
\end{equation}
which are obtained from equations (\ref{trs}) and (\ref{tss}). Note that we have also used $d\,\ln{k} = H^{*} dt^{*}$ which is valid with the slow-roll approximation. Because the slow-roll parameters (\ref{slowrollparametersgeneralcases}) are expressed in terms of the slow-roll parameters (\ref{sepsilon}) and (\ref{seta}) as
\begin{equation}
\epsilon = \tilde{\epsilon},\:\:\: \eta = - 2 \tilde{\eta} + 4 \epsilon, \:\:\: s = 0,
\end{equation}
in the standard slow-roll inflation models, we see that the spectral index (\ref{spectraldbihorizon}) is consistent with the spectral index (\ref{predictionspectralcurvature}) in the canonical slow-roll inflation models. 
The calculation of the bispectrum of the curvature perturbation in two-field DBI inflation models is given in \cite{Langlois:2008c}. The third order action with the decomposition (\ref{definitionofqsigma}) at the leading order with the slow-roll approximation and the small sound speed limit $c_{s}\ll 1$ is given by
\begin{equation}\label{thirdordertwofielddbi}
\begin{split}
S_{\left(3 \right)} &= \int \, dt \, d^{3}x \, \left\{\frac{a^{3}}{2 c^{5}_{s} \dot{\sigma}} \left[ \dot{Q}^{3}_{\sigma} + c^{2}_{s} \dot{Q}_{\sigma} \dot{Q}^{2}_{s} \right] \right.\\
& \left. - \frac{a}{2 c^{3}_{s}\dot{\sigma}} \left[\dot{Q}_{\sigma}\left(\nabla Q_{\sigma} \right)^{2} - c^{2}_{s} \dot{Q}_{\sigma} \left(\nabla Q_{s} \right)^{2} + 2 c^{2}_{s} \dot{Q}_{s} \nabla Q_{\sigma} \nabla Q_{s} \right] \right\},
\end{split}
\end{equation}
where we have used the fact that $f \sim 1/\dot{\sigma}^{2}$ is valid in the small sound speed limit. With the third order action (\ref{thirdordertwofielddbi}), the three-point functions of the scalar field perturbations are obtained using the in-in formalism as introduced in section \ref{sec:equilateral} as
\begin{equation}\label{adiabaticfieldthreepoint}
\begin{split}
&\left< Q_{\sigma}\left(\textbf{k}_{\textbf{1}} \right) Q_{\sigma}\left(\textbf{k}_{\textbf{2}} \right) Q_{\sigma} \left(\textbf{k}_{\textbf{3}} \right) \right>\\
&=-\left(2\pi\right)^{3} \delta^{\left(3\right)}\left(\textbf{k}_{\textbf{1}}+\textbf{k}_{\textbf{2}}+\textbf{k}_{\textbf{3}}\right)\frac{H^{4}}{\sqrt{2\epsilon}c_{s}^{2}\left(P_{,X}\right)^{3/2}}\frac{1}{\Pi^{3}_{i=1}k^{3}_{i}}\mathcal{A}^{DBI}_{Q_{\sigma}Q_{\sigma}Q_{\sigma}}\left(k_{1},k_{2},k_{3}\right),
\end{split}
\end{equation}
where
\begin{equation}\label{adiabaticfieldathreepoint}
\mathcal{A}^{DBI}_{Q_{\sigma}Q_{\sigma}Q_{\sigma}} = \frac{1}{c_{s}^{2}} \left[ - \frac{3}{2} \frac{k_{1}^{2}k_{2}^{2}k_{3}^{2}}{K^{3}} + \frac{k^{2}_{1}\textbf{k}_{\textbf{2}}\cdot \textbf{k}_{\textbf{3}}}{4K}\left(1+\frac{k_{2}+k_{3}}{K} + 2\frac{k_{2}k_{3}}{K^{2}}\right) + 2 \: \rm{cyclic \: terms} \right], 
\end{equation}
and
\begin{equation}\label{entropyfieldthreepoint}
\begin{split}
&\left< Q_{\sigma}\left(\textbf{k}_{\textbf{1}} \right) Q_{s}\left(\textbf{k}_{\textbf{2}} \right) Q_{s} \left(\textbf{k}_{\textbf{3}} \right) \right>\\
&=-\left(2\pi\right)^{3} \delta^{\left(3\right)}\left(\textbf{k}_{\textbf{1}}+\textbf{k}_{\textbf{2}}+\textbf{k}_{\textbf{3}}\right)\frac{H^{4}}{\sqrt{2\epsilon}c_{s}^{2}\left(P_{,X}\right)^{3/2}}\frac{1}{\Pi^{3}_{i=1}k^{3}_{i}}\mathcal{A}^{DBI}_{Q_{\sigma}Q_{s}Q_{s}}\left(k_{1},k_{2},k_{3}\right),
\end{split}
\end{equation}
where
\begin{equation}\label{entropyfieldathreepoint}
\begin{split}
\mathcal{A}^{DBI}_{Q_{\sigma}Q_{s}Q_{s}} &= \frac{1}{c_{s}^{4}} \left[ - \frac{1}{2} \frac{k_{1}^{2}k_{2}^{2}k_{3}^{2}}{K^{3}} - \frac{k^{2}_{1}\textbf{k}_{\textbf{2}}\cdot \textbf{k}_{\textbf{3}}}{4K}\left(1+\frac{k_{2}+k_{3}}{K} + 2\frac{k_{2}k_{3}}{K^{2}}\right) \right.\\
& \left. + \frac{k^{2}_{3}\textbf{k}_{\textbf{1}}\cdot \textbf{k}_{\textbf{2}}}{4K}\left(1+\frac{k_{1}+k_{2}}{K} + 2\frac{k_{1}k_{2}}{K^{2}}\right) + \frac{k^{2}_{2}\textbf{k}_{\textbf{1}}\cdot \textbf{k}_{\textbf{3}}}{4K}\left(1+\frac{k_{1}+k_{3}}{K} + 2\frac{k_{1}k_{3}}{K^{2}}\right) \right], 
\end{split}
\end{equation}
and all other three-point functions of the field perturbations vanish. It is obvious that the three-point function of the adiabatic field perturbation (\ref{adiabaticfieldthreepoint}) coincides with the three-point function of the field perturbation in the single field case (\ref{dbifieldthreepoint}) at the leading order in the small sound speed limit $c_{s} \ll 1$. Using the $\delta$N formalism (\ref{usedlaterdeltan}) with equation (\ref{ndecomposition}), the three point function of the curvature perturbation is given by
\begin{equation}\label{generaluptosecondorderbispectrumdbicurvature}
\begin{split}
&\left<\mathcal{R} \left(\textbf{k}_{\textbf{1}}\right) \mathcal{R} \left(\textbf{k}_{\textbf{2}}\right) \mathcal{R} \left(\textbf{k}_{\textbf{3}} \right) \right> = \sum_{I,J,K = \sigma,s} N_{,I}N_{,J}N_{,K} \left< Q_{I}^{*}\left(\textbf{k}_{\textbf{1}} \right) Q_{J}^{*}\left(\textbf{k}_{\textbf{2}} \right) Q_{K}^{*} \left(\textbf{k}_{\textbf{3}} \right) \right>\\
& + \sum_{I,J,K = \sigma,s} \frac{N_{,II}N_{,J}N_{,K}}{2} \left< \left( \left[Q_{I}^{*} \star Q_{I}^{*} \right] \left(\textbf{k}_{\textbf{1}} \right)  - \left< \left[Q_{I}^{*} \star Q_{I}^{*} \right] \left(\textbf{k}_{\textbf{1}}\right) \right> \right) Q_{J}^{*} \left(\textbf{k}_{\textbf{2}} \right) Q_{K}^{*} \left(\textbf{k}_{\textbf{3}}\right) \right>\\
& + \sum_{I,J,K = \sigma,s} \frac{N_{,I}N_{,JJ}N_{,K}}{2} \left< Q_{I}^{*} \left(\textbf{k}_{\textbf{1}} \right) \left( \left[Q_{J}^{*} \star Q_{J}^{*} \right] \left(\textbf{k}_{\textbf{2}} \right) - \left< \left[Q_{J}^{*} \star Q_{J}^{*} \right] \left(\textbf{k}_{\textbf{2}}\right) \right> \right) Q_{K}^{*} \left(\textbf{k}_{\textbf{3}}\right) \right>\\
& + \sum_{I,J,K = \sigma,s} \frac{N_{,I}N_{,J}N_{,KK}}{2} \left< Q_{I}^{*} \left(\textbf{k}_{\textbf{1}} \right) Q_{J}^{*} \left(\textbf{k}_{\textbf{2}}\right) \left( \left[Q_{K}^{*} \star Q_{K}^{*} \right] \left(\textbf{k}_{\textbf{3}} \right) - \left< \left[Q_{K}^{*} \star Q_{K}^{*} \right] \left(\textbf{k}_{\textbf{3}}\right) \right> \right) \right>,
\end{split}
\end{equation}
where the subscript $*$ denote a quantity evaluated at the horizon exit. If we assume that the slow-roll parameters (\ref{slowrollparametersgeneralcases}) and the sound speed $c_{s}$ are much smaller than unity at the horizon exit, from equations (\ref{curvatureperturbationsimplerelation}), (\ref{definedentropypert}) and (\ref{transfermatrix}) compared with equation (\ref{ndecomposition}), we obtain
\begin{equation}\label{lineardeltancoefficients}
N_{,\sigma} = \frac{H}{\dot{\sigma}},\:\:\:\:\:N_{,s} = T_{\mathcal{R}\mathcal{S}}\left(\frac{c_{s}H}{\dot{\sigma}}\right), 
\end{equation}
while we have the relations
\begin{equation}\label{lineardeltancoefficientswiththenewbases}
N_{,\tilde{\sigma}} = \frac{H \sqrt{c_{s}}}{\dot{\sigma}},\:\:\:\:\:N_{,\tilde{s}} = T_{\mathcal{R}\mathcal{S}}\left(\frac{H \sqrt{c_{s}}}{\dot{\sigma}}\right), 
\end{equation}
from equations (\ref{chaptertwocurvaturewithnewbaseslinear}), (\ref{definedentropypert}) and (\ref{transfermatrix}) compared with equation (\ref{ndecompositionwiththenewbases}). With equations (\ref{lineardeltancoefficients}) and (\ref{lineardeltancoefficientswiththenewbases}), the first line in the right hand side of equation (\ref{generaluptosecondorderbispectrumdbicurvature}) is given by
\begin{equation}\label{orderthreecurvaturedefined}
\begin{split}
&\left<\mathcal{R} \left(\textbf{k}_{\textbf{1}}\right) \mathcal{R} \left(\textbf{k}_{\textbf{2}}\right) \mathcal{R} \left(\textbf{k}_{\textbf{3}} \right) \right>^{\left(3\right)} = \sum_{I,J,K = \sigma,s} N_{,I}N_{,J}N_{,K} \left< Q_{I}^{*}\left(\textbf{k}_{\textbf{1}} \right) Q_{J}^{*}\left(\textbf{k}_{\textbf{2}} \right) Q_{K}^{*} \left(\textbf{k}_{\textbf{3}} \right) \right>\\
& = N_{,\sigma}^{3} \left< Q_{\sigma}^{*}\left(\textbf{k}_{\textbf{1}} \right) Q_{\sigma}^{*}\left(\textbf{k}_{\textbf{2}} \right) Q_{\sigma}^{*} \left(\textbf{k}_{\textbf{3}} \right) \right> + N_{,\sigma} N_{,s}^{2} \left( \left< Q_{\sigma}^{*}\left(\textbf{k}_{\textbf{1}} \right) Q_{s}^{*}\left(\textbf{k}_{\textbf{2}} \right) Q_{s}^{*} \left(\textbf{k}_{\textbf{3}} \right) \right> + 2\:\rm{cyclic \: terms} \right)\\
& = N_{,\tilde{\sigma}}^{3} \left< \tilde{Q}_{\sigma}^{*}\left(\textbf{k}_{\textbf{1}} \right) \tilde{Q}_{\sigma}^{*}\left(\textbf{k}_{\textbf{2}} \right) \tilde{Q}_{\sigma}^{*} \left(\textbf{k}_{\textbf{3}} \right) \right> + N_{,\tilde{\sigma}} N_{,\tilde{s}}^{2} \left( \left< \tilde{Q}_{\sigma}^{*}\left(\textbf{k}_{\textbf{1}} \right) \tilde{Q}_{s}^{*}\left(\textbf{k}_{\textbf{2}} \right) \tilde{Q}_{s}^{*} \left(\textbf{k}_{\textbf{3}} \right) \right> + 2\:\rm{cyclic \: terms} \right)\\
&= \left(\frac{H}{\dot{\sigma}} \right)^{3} \left< Q_{\sigma}^{*}\left(\textbf{k}_{\textbf{1}} \right) Q_{\sigma}^{*}\left(\textbf{k}_{\textbf{2}} \right) Q_{\sigma}^{*} \left(\textbf{k}_{\textbf{3}} \right) \right> \left(1 + T_{\mathcal{R}\mathcal{S}}^{2} \right)\\
&= \left(\frac{H \sqrt{c_{s}}}{\dot{\sigma}} \right)^{3} \left< \tilde{Q}_{\sigma}^{*}\left(\textbf{k}_{\textbf{1}} \right) \tilde{Q}_{\sigma}^{*}\left(\textbf{k}_{\textbf{2}} \right) \tilde{Q}_{\sigma}^{*} \left(\textbf{k}_{\textbf{3}} \right) \right> \left(1 + T_{\mathcal{R}\mathcal{S}}^{2} \right),
\end{split}
\end{equation}
from equations (\ref{chaptertworelationbetweenqandqtilde}), (\ref{adiabaticfieldathreepoint}) and (\ref{entropyfieldathreepoint}). From equations (\ref{fnlequilateraldefinition}), (\ref{fnlequilateralfinal}), (\ref{prmouse}) and (\ref{orderthreecurvaturedefined}), we obtain
\begin{equation}\label{dbimultiequilateralfinalresult}
f_{\rm{NL}}^{equil} = - \frac{35}{108} \left(\frac{1}{c^{2}_{s}} - 1\right) \frac{1}{1 + T_{\mathcal{R}\mathcal{S}}^{2}} = - \frac{35}{108} \left(\frac{1}{c^{2}_{s}} - 1\right) \cos^{2}{\Theta}.
\end{equation}
All the four-point functions in equation (\ref{generaluptosecondorderbispectrumdbicurvature}) come from the quadratic terms in equation (\ref{ndecomposition}). Actually, the quadratic terms in equation (\ref{ndecomposition}) correspond to the quadratic term in equation (\ref{firstdefinitionlocalcurvature}) and we obtain the local type non-Gaussianity as in equation (\ref{localshapederivation}). From equation (\ref{localshapederivation}), it is obvious that only the four-point functions in equation (\ref{generaluptosecondorderbispectrumdbicurvature}) with the subscripts $I = J = K = \sigma$ or $I = J = K = s$ have non-zero values. Therefore, from equations (\ref{ndecomposition}) and (\ref{firstdefinitionlocalcurvature}), we obtain the relation
\begin{equation}\label{localgeneralresultfordbi}
\frac{3}{5} f_{\rm{NL}}^{local} = \frac{1}{2} N_{,\sigma \sigma} N_{,\sigma}^{2} \times \frac{c^{4}_{s}}{\left\{\left(c_{s} N_{,\sigma}\right)^{2} + N_{,s}^{2} \right\}^{2}} + \frac{1}{2} N_{,ss} N_{,s}^{2} \times \frac{1}{\left\{\left(c_{s} N_{,\sigma}\right)^{2} + N_{,s}^{2} \right\}^{2}},
\end{equation}
where we used the relation
\begin{equation}
\mathcal{P}_{\mathcal{R}} = \frac{\left(c_{s} N_{,\sigma}\right)^{2} + N_{,s}^{2}}{c^{2}_{s}} \mathcal{P}^{*}_{Q_{\sigma}} = \left\{\left(c_{s} N_{,\sigma}\right)^{2} + N_{,s}^{2}\right\} \mathcal{P}^{*}_{Q_{s}}, 
\end{equation}
which is obtained from equations (\ref{relationadiabaticentropydbi}), (\ref{transfermatrix}) and (\ref{lineardeltancoefficients}) assuming the slow-roll parameters and the sound speed are much smaller than unity at the horizon exit. Considering the definitions of the derivatives of $N$ with respect to $\tilde{\phi}_{\sigma}$ and $\tilde{\phi}_{s}$ in equation (\ref{ndecompositionwiththenewbases}), we obtain 
\begin{equation}\label{relationbetweennderivativesandntildederivatives}
N_{,\sigma} = \frac{N_{,\tilde{\sigma}}}{\sqrt{c_{s}}},\,\,\,N_{,\sigma \sigma} = \frac{N_{,\tilde{\sigma} \tilde{\sigma}}}{c_{s}},\,\,\,N_{,s} = \sqrt{c_{s}} N_{,\tilde{s}},\,\,\,N_{,ss} = c_{s} N_{,\tilde{s} \tilde{s}},
\end{equation}
from the relations $\phi_{\sigma} = \tilde{\phi}_{\sigma} \sqrt{c_{s}}$ and $\phi_{s} = \tilde{\phi}_{s} / \sqrt{c_{s}}$ which come from the definitions of $\tilde{\phi}_{\sigma}$ and $\tilde{\phi}_{s}$ in equation (\ref{ndecompositionwiththenewbases}). From equations (\ref{localgeneralresultfordbi}) and (\ref{relationbetweennderivativesandntildederivatives}), we have

\begin{equation}
\begin{split}
f_{\rm{NL}}^{local} &= \frac{5}{6} \frac{c^{4}_{s} N_{,\sigma \sigma} N_{,\sigma}^{2} + N_{,ss} N_{,s}^{2}}{\left\{(c_{s} N_{,\sigma})^{2} + N_{,s}^{2} \right\}^{2}}\\
&= \frac{5}{6} \frac{N_{, \tilde{\sigma} \tilde{\sigma}} N_{, \tilde{\sigma}}^{2} + N_{, \tilde{s} \tilde{s}} N_{, \tilde{s}}^{2}}{\left\{N_{, \tilde{\sigma}}^{2} + N_{, \tilde{s}}^{2} \right\}^{2}}, 
\end{split}
\end{equation}

in the two-field DBI inflation models. When $N_{,\sigma \sigma}$ is smaller or of the same order as $N_{,ss}$ and $N_{,\sigma}$ is smaller or of the same order as $N_{,s}$, we have
\begin{equation}\label{chaptertwofinalfnllocalformula}
f_{\rm{NL}}^{local} = \frac{5}{6} \frac{N_{,ss}}{N_{,s}^{2}} = \frac{5}{6} \frac{N_{,\tilde{s} \tilde{s}}}{N_{, \tilde{s}}^{2}},
\end{equation}
because of the small sound speed assumption. 
\chapter{String inflation}\label{chapter:stringinflation}
In particle physics, the \textbf{standard model} has been carefully confirmed experimentally. In this model, gravity is mediated by the spin-2 graviton and electromagnetic, weak and strong nuclear forces are mediated by the spin-1 $SU(3)\times SU(2)\times SU(1)$ gauge bosons. Though this model is very successful, it also has several problems \cite{Polchinski:1998}. First, we need roughly twenty free parameters in the Lagrangian and we do not know what determines these parameters. Secondly, the unification of gravity with the quantum theory yields the non-renormalisable quantum field theory in which we cannot absorb the divergence with a finite number of counter terms. Thirdly, the theory breaks down at the singularities of general relativity. In this chapter, we first introduce the Dirac-Born-Infeld inflation motivated by string theory, which is thought to be the best candidate so far to able to solve those problems of the standard model. We also see the stringent microphysical constraint combined with the observations that disfavours the single field DBI inflation models. Then, how multi-field DBI inflation can satisfy this constraint is shown. Finally, we introduce some specific models of DBI inflation. 
\chaptermark{Inflation motivated by string theory}
\section{String theory and Dirac-Born-Infeld (DBI) inflation}\label{dbiintroduction}
The problems of the standard model introduced above can be solved by several ideas. One is the grand unification. This successfully unifies the three gauge fields and predicts one of the free parameters. The second idea is that the space-time has more than 4 dimensions. This opens the possibility of unifying the gauge interactions and gravity. The third one is the supersymmetry which helps with the divergence and naturalness problems. \textbf{String theory} is the only known model which can contain all those three ideas. In this theory, the graviton and all other elementary particles are one dimensional objects, strings \cite{Polchinski:1998}. Supersymmetric string theory can be constructed consistently only in 10 dimensions. It was also discovered that the \textbf{p-branes} which are extended higher dimensional objects than strings (1-branes) play a fundamental role in the theory. Especially, p-branes on which open strings can end are called the \textbf{D-branes}. Roughly speaking, open strings describe the non-gravitational sector. The open strings are attached to the D-branes while the closed strings of the gravitational sector can move freely in the \textbf{bulk} which is the ten dimensional space-time. Classically, this means that matter and radiation field are localised on the brane while gravity propagates in the bulk \cite{Maartens:2010}. In the Dirac-Born-Infeld (DBI) inflation \cite{Silverstein:2004, Alishahiha:2004, Underwood:2008}, we consider a D3-brane in which our Universe is confined to in the bulk as shown in figure \ref{braneinthebulk}. As we see below, the inflaton field is a spatial coordinate of the extra dimension in this model. Therefore, we know what inflaton is unlike many other inflation models where the origin of inflaton is not specified. 

\begin{figure}[!htb]
\centering
\includegraphics[width=12cm]{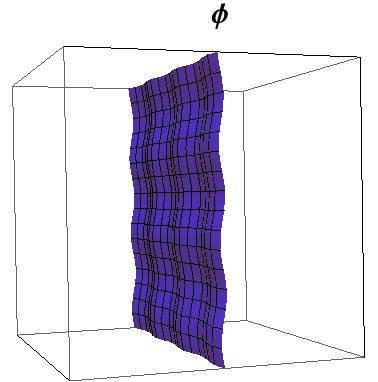}
\caption[Hubble's law]{Schematic figure of the D3-brane in the bulk which is the ten dimensional space-time. The $\phi$ axis represents an extra dimension. We have six extra dimensions in total. The other two axes represent two of the dimensions of our four dimensional Universe. The brane fluctuates in the $\phi$ direction and this corresponds to the scalar field perturbation.}{\label{braneinthebulk} }
\end{figure}

We consider the bulk whose warped geometry is given by \cite{Langlois:2008c}
\begin{equation}\label{dbigeneralmetric}
ds^{2} = h^{-1/2} \left(y^{K}\right) g_{\mu \nu} dx^{\mu} dx^{\nu} + h^{1/2} \left(y^{K} \right) G_{IJ} \left(y^{K} \right) dy^{I} dy^{J} \equiv H_{AB} dY^{A} dY^{B}, 
\end{equation}
where $Y^{A}=\left\{x^{\mu}, y^{I} \right\}$ with the indices $\mu=0,1,2,3$ and $I=1,...,6$. The kinetic part of the DBI Lagrangian reads
\begin{equation}\label{kineticdbilagrangian}
P_{\rm{kinetic}}=-T_{3}\sqrt{- \det{\gamma_{\mu \nu}}},
\end{equation}
where $T_{3}$ is the brane tension of a D3-brane and the induced metric on the D3-brane is defined as
\begin{equation}\label{inducedmetric}
\gamma_{\mu \nu} = H_{AB} \partial_{\mu} Y_{(b)}^{A} \partial_{\nu} Y_{(b)}^{B},
\end{equation}
with $Y_{(b)}^{A}\left(x^{\mu}\right) = \left(x^{\mu},  \eta^{I}\left(x^{\mu}\right) \right)$ which are the ten dimensional coordinates which specify the brane position in the bulk where $x^{\mu}$ is the four dimensional space-time coordinates on the brane. From equations (\ref{dbigeneralmetric}) and (\ref{inducedmetric}), the kinetic part of the Lagrangian (\ref{kineticdbilagrangian}) is rewritten as
\begin{equation}
P_{\rm{kinetic}}=-T_{3} h^{-1} \sqrt{-g} \sqrt{\det \left(\delta^{\mu}_{\nu} + h G_{IJ}  \partial^{\mu} \eta^{I} \partial_{\nu} \eta^{J}\right)}. 
\end{equation}
Let us rescale the variables as
\begin{equation}\label{rescalingthecoordinates}
f=\frac{h}{T_{3}}, \,\,\, \phi^{I} = \sqrt{T_{3}} \eta^{I}. 
\end{equation}
Then, by adding the potential term which comes from the brane's interactions with the bulk fields and other branes, we obtain the DBI Lagrangian as
\begin{equation}\label{dbigenerallagrangian}
P\left(X^{IJ}, \phi^{I} \right) = - \frac{1}{f\left(\phi^{I} \right)}\left(\sqrt{\mathcal{D}} - 1 \right) - V \left(\phi^{I} \right),
\end{equation}
where
\begin{equation}\label{fourbyfourmatrix}
\mathcal{D}=\det \left(\delta^{\mu}_{\nu} + f G_{IJ}  \partial^{\mu} \phi^{I} \partial_{\nu} \phi^{J} \right). 
\end{equation}
We can express the determinant of $4 \times 4$ matrix (\ref{fourbyfourmatrix}) as the determinant of $N \times N$ matrix in terms of $X^{I}_{J}=G_{IK}X^{KJ}$ as 
\begin{equation}
\mathcal{D}=\det \left(\delta^{J}_{I} - 2 f X^{J}_{I} \right).
\end{equation}
Then, as shown in \cite{Langlois:2008b}, the $N \times N$ matrix can be expressed by
\begin{equation}\label{kineticmatrixdeterminant}
\mathcal{D}=1 - 2 f \tilde{X},
\end{equation}
with
\begin{equation}\label{tildexdefinition}
\tilde{X} \equiv X + \mathcal{F} \left(X^{IJ}, \phi^{K} \right),
\end{equation}
where
\begin{equation}\label{dbibackgroundf}
\mathcal{F}\left(X^{IJ}, \phi^{K} \right) = - 2 f X^{[I}_{I} X^{J]}_{J} + 4 f^{2} X^{[I}_{I} X^{J}_{J} X^{K]}_{K} - 8 f^{3} X^{[I}_{I} X^{J}_{J} X^{K}_{K} X^{L]}_{L}, 
\end{equation}
note that $[\,\,]$ denotes the antisymmetric bracket. From equations (\ref{dbigenerallagrangian}), (\ref{kineticmatrixdeterminant}) and (\ref{tildexdefinition}), we can see that the DBI action coincides with the k-inflation action (\ref{kinflationaction}) if $\mathcal{F}=0$. In single field cases ($I = 1$), it is clear that $\mathcal{F}$ vanishes and we have
\begin{equation}\label{singledbid}
\mathcal{D} = 1+f(\phi) \partial_{\rm{\mu}}\phi \partial^{\rm{\mu}}\phi. 
\end{equation}
Even in multi-field cases, if we consider homogeneous scalar fields whose spatial derivatives vanish, $\mathcal{F}=0$ in the background because of the anti-symmetrisation on field indices in equation (\ref{dbibackgroundf}). However, even in these cases, the scalar field perturbations are inhomogeneous and $\mathcal{F}$ does not vanish if we consider the perturbations. Therefore, the dynamics of the perturbations in multi-field DBI inflation models is different from that in the multi-field k-inflation models as we saw in chapter \ref{chapter:generalperturbations}. 
We can obtain the equations of motion for the scalar fields by varying the action (\ref{actiongeneral}) with the DBI Lagrangian (\ref{dbigenerallagrangian}) with respect to the scalar fields as
\begin{equation}\label{dbieqofmotiongeneralfirst}
G_{IJ}\left(\ddot{\phi}^{I}+3H\dot{\phi}^{I}-\frac{\dot{c_{\rm{s}}}}{c_{\rm{s}}}\dot{\phi}^{I}\right)+\left(G_{IJ,K}-\frac{1}{2}G_{IK,J}\right)\dot{\phi}^{I}\dot{\phi}^{K}+c_{\rm{s}}V_{,J}-\frac{\left(1-c_{\rm{s}}\right)^{2}}{2}\frac{f_{,J}}{f^{2}}=0,
\end{equation}
where $_{,J}$ denotes the partial derivative with respect to $\phi^{J}$. If $G_{IJ}\left(\phi^{K}\right)$ takes a simple diagonal form, $G_{IJ}\left(\phi^{K}\right)=A_{I}\left(\phi^{K}\right)\delta_{IJ}$, the field equations for the scalar fields are given by
\begin{equation}\label{eqofmotionai}
A_{I}\left(\ddot{\phi}^{I}+3H\dot{\phi}^{I}-\frac{\dot{c_{\rm{s}}}}{c_{\rm{s}}}\dot{\phi}^{I}\right)+\sum_{J}\left[A_{I,J}\dot{\phi}^{I}\dot{\phi}^{J}-\frac{1}{2}A_{J,I}\left(\dot{\phi}^{J}\right)^{2}\right]+c_{\rm{s}}V_{,J}-\frac{\left(1-c_{\rm{s}}\right)^{2}}{2}\frac{f_{,J}}{f^{2}}=0, 
\end{equation}
and the sound speed is given by
\begin{equation}\label{csai}
c_{\rm{s}}=\sqrt{1-f\sum_{I}A_{I}\left(\phi^{I}\right)^{2}},
\end{equation}
note that the Einstein summation convention is not used in equations (\ref{eqofmotionai}) and (\ref{csai}).
The Friedmann equation for the general inflation is derived as the time-time component of the Einstein equations which are obtained by varying the action (\ref{actiongeneral}) with respect to the metric $g_{\mu \nu}$ as
\begin{equation}\label{generaldbifriedmann}
H^{2}=\frac{1}{3M_{P}^2}\left(2 P_{<IJ>}X^{IJ} - P \right),
\end{equation}
we also obtain
\begin{equation}\label{hdotdbi}
\dot{H} M_{P}^2 = - X^{IJ} P_{<IJ>},
\end{equation}
by combining the derivative of the Friedmann equation (\ref{friedmann}) with respect to time and the continuity equation (\ref{continuity}) with the Planck mass $M_{P}$. In the cases of the DBI inflation, the following equations hold \cite{Langlois:2008b} as
\begin{equation}\label{pijderived}
P_{<IJ>} = c_{s} \tilde{G}_{IJ},
\end{equation}
\begin{equation}\label{gtildederived}
 \tilde{G}^{I}_{J} = \perp^{I}_{J} + \frac{1}{1-2fX}e^{I}e_{J},
\end{equation}
and
\begin{equation}
\perp^{I}_{J} = \delta^{I}_{J} - e^{I}e_{J},
\end{equation}
where $e^{I}$ is defined in equation (\ref{basiskflation}) and $X^{IJ} = \dot{\phi}^{I} \dot{\phi}^{J}/2$ in the homogeneous background. Substituting equation (\ref{pijderived}) into the Friedmann equation (\ref{generaldbifriedmann}), we obtain
\begin{equation}\label{friedmanngeneralderived}
3 H^{2} M_{P}^2 = \frac{1}{f\left(\phi^{I} \right)}\left(\frac{1}{c_{s}} - 1 \right) + V\left(\phi^{I} \right),
\end{equation}
and in a similar way, equation (\ref{hdotdbi}) in the DBI inflation reads
\begin{equation}\label{hubbledotdbi}
\dot{H} M_{P}^2 = -\frac{1}{2f\left(\phi^{I} \right)} \left(c_{s} - \frac{1}{c_{s}} \right) = -\frac{X}{c_{s}}. 
\end{equation}
where $X=G_{IJ}\dot{\phi}^{I}\dot{\phi}^{J}/2$ in the homogeneous background. 

\section{Constraints on DBI inflation from the CMB observation}
In this section, we introduce single field DBI inflation models and some cosmological parameters whose values can be constrained by the observations of the \textbf{Cosmic Microwave Background} (CMB). Then, the observational constraints on a simple model are studied. Finally, we see the general observational constraints together with the theoretical constraints from string theory that disfavour the single field models. 

\subsection{Single field DBI inflation}\label{subsec:singledbiinflationintroduction}
In the single field DBI model, the Friedmann equation (\ref{friedmanngeneralderived}) reads
\begin{equation}
\begin{split}
 H^{2} &= \frac{1}{3 M_{P}^2} \left(\frac{\gamma^{2}}{\gamma + 1}\dot{\phi}^{2} + V\left(\phi \right) \right)\\
 &= \frac{1}{3 M_{P}^2} \left(\frac{\gamma - 1}{f\left(\phi \right)} + V(\phi) \right),\label{friedmanntwo}
 \end{split}
\end{equation}
where $M_{P}$ is the reduced Planck mass with
\begin{equation}
\begin{split}
c_{s} &= \frac{1}{\gamma}\\
 &= \sqrt{1-2fX}\\
 &= \sqrt{1-2 T^{-1}X},\label{gammatwo}
\end{split}
\end{equation}
where 
\begin{equation}
T \equiv f^{-1} = T_{3}/h,
\end{equation}
and we have $X = \dot{\phi}^{2}/2$ because of the homogeneity. It is useful to know that the \textbf{sound speed} $c_{\rm{s}}$ can be expressed as follows
\begin{equation}
c_{s} \equiv \frac{P_{,\rm{X}}}{P_{,\rm{X}}+P_{,\rm{XX}}} = \frac{1}{P_{,\rm{X}}},\label{cstwo}
\end{equation}
in the DBI inflation because of the special form of the kinetic term. The equation of motion for the scalar field (\ref{dbieqofmotiongeneralfirst}) in the single field cases reads
\begin{equation}
\ddot{\phi} + 3H \dot{\phi} - \frac{\dot{c}_{\rm{s}}}{c_{\rm{s}}} \dot{\phi} + c_{\rm{s}} \partial_{\rm{\phi}} \left(V + \frac{c_{\rm{s}}-1}{f} \right)=0.\label{klein}
\end{equation}
Slow roll parameters (\ref{slowrollparametersgeneralcases}) in single field DBI inflation models are given as follows
\begin{equation}
\epsilon = - \frac{\dot{H}}{H^{2}} = \frac{X P_{,\rm{X}}}{M^{2}_{P}H^{2}} = \frac{2M^{2}_{P}}{\gamma}\left(\frac{H'}{H}\right)^{2}, \label{epsilontwo}
\end{equation}
\begin{equation}
 \eta = 2 \epsilon - \frac{2M^{2}_{P}}{\gamma}\frac{H''}{H} + s, 
\end{equation}
\begin{equation}
 s = \frac{\dot{c_{\rm{s}}}}{c_{\rm{s}}H} = \frac{2M^{2}_{P}}{\gamma}\frac{H'}{H}\frac{\gamma'}{\gamma}.\label{s}
\end{equation}
Let us use a new slow-roll parameter
\begin{equation}\label{etatwo}
\bar{\eta} = \frac{2M^{2}_{P}}{\gamma}\frac{H''}{H},
\end{equation}
instead of $\eta$ in this section. By varying the action with respect to the metric, we obtain the energy momentum tensor. In the FRW metric, we can obtain the energy density and the pressure
\begin{equation}
 \rho = \frac{\gamma}{f} + (V - f^{-1}), \label{rhotwo}
\end{equation}
\begin{equation}
 \rho = -\frac{1}{f \gamma} - (V - f^{-1}). \label{ptwo}
\end{equation}
By differentiating the Friedmann equation and substituting equation (\ref{continuity}), we obtain
\begin{equation}
 \dot{\phi} = - \frac{2 M^{2}_{P}H'}{\gamma} = - \frac{2 M^{2}_{P}H'}{\sqrt{1 + 4 M^{4}_{\rm{P}}T^{-1}H'^{2}}}. \label{twentynine}
\end{equation}
Note that we used equations (\ref{rhotwo}) and (\ref{ptwo}) to derive equation (\ref{twentynine}). 
The amplitudes and spectral indices of the two point functions of the scalar and tensor perturbations in single field DBI inflation models are given by
\begin{equation}
 \mathcal{P}_{\mathcal{R}} = \frac{1}{8 \pi^{2} M^{2}_{P}}\frac{H^{2}}{c_{\rm{s}}\epsilon} = \frac{H^{4}}{4\pi^{2}\dot{\phi}^{2}},\label{scalarspectrum}
\end{equation}
\begin{equation}
\mathcal{P}_{\rm{T}} = \frac{2}{\pi^{2}}\frac{H^{2}}{M^{2}_{P}}, 
\end{equation}
\begin{equation}
 1-n_{\rm{s}} = 4\epsilon - 2 \bar{\eta} + 2 s, \label{stwo}
\end{equation}
\begin{equation}
 n_{\rm{t}}=-2\epsilon,
\end{equation}
where $\mathcal{P}_{\mathcal{R}}$ and $\mathcal{P}_{\rm{T}}$ are amplitudes of the two-point functions of the scalar and tensor perturbations, and $n_{\rm{s}}$ and $n_{\rm{t}}$ are the spectral indices for the scalar and tensor perturbations respectively. The tensor-to-scalar ratio is defined as
\begin{equation}
 r \equiv \frac{\mathcal{P}_{\rm{T}}}{\mathcal{P}_{\mathcal{R}}} = -8c_{\rm{s}}n_{\rm{t}} = 16c_{\rm{s}}\epsilon. \label{rtwo}
\end{equation}
Hence, a sound speed different from unity leads to a violation of the standard inflationary consistency equation.
A more important consequence of a small sound speed is that departures from pure Gaussian statistics may be large. In DBI inflation, the bispectrum of the curvature perturbation defined by equation (\ref{bispectrumdefinitionfirst}) has a peak at the equilateral triangle made of $k_1$, $k_2$ and $k_3$ as we saw in section \ref{sec:equilateral}. The amplitude of the bispectrum at the equilateral limit is defined as $f_{\rm{NL}}^{equil}$ as given in equation (\ref{fnlequilateralfinal}). Finally, combining equations (\ref{fnlequilateralfinal}), (\ref{hubbledotdbi}), (\ref{gammatwo}), (\ref{epsilontwo}), (\ref{scalarspectrum}) and (\ref{rtwo}), we can obtain
\begin{equation}
 \frac{T_{*}}{M^{4}_{\rm{P}}} = \frac{\pi^{2}}{16} r^{2} \mathcal{P}_{\mathcal{R}} \left(1 + \frac{1}{3f_{\rm{NL}}^{equil}} \right),\label{tensionsingle}
\end{equation}
where `$*$' means quantities evaluated during the observable inflation.

\subsection{Analysis of a simple model}\label{subsec:analysisofsimplemodeldbi}
We study the observational constraints in a simple model with a quadratic potential following \cite{Silverstein:2004, Alishahiha:2004} in this section. Let us consider a single field DBI inflation model with a warp factor
\begin{equation}\label{dbisingleexamplewarp}
f \left(\phi \right) = \frac{\lambda}{\phi^{4}},
\end{equation}
with a constant $\lambda$, and a potential
\begin{equation}\label{dbisingleexamplepotential}
V \left(\phi \right) = m^{2} \phi^{2},
\end{equation}
where $m$ is the mass of inflaton. From equations (\ref{friedmanntwo}) and (\ref{twentynine}), we obtain
\begin{equation}\label{potentialexpressionseconddbisingleexample}
\begin{split}
V &= 3 M^{2}_{P} H^{2} - \frac{\sqrt{1 + 4 M^{4}_{\rm{P}}T^{-1}H'^{2}}}{f} + \frac{1}{f}\\
&= 3 M^{2}_{P} H^{2} - \sqrt{1 + 4 M^{4}_{\rm{P}} \lambda H'^{2}/\phi^{4}} \frac{\phi^{4}}{\lambda} + \frac{\phi^{4}}{\lambda},
\end{split}
\end{equation}
where we used equation (\ref{dbisingleexamplewarp}). In order to investigate the late time dynamics when $\phi$ is small, let us use the simple anzatz,
\begin{equation}\label{anzatzsingledbiexample}
H = h_{1}\phi + \cdots.
\end{equation}
By substituting equation (\ref{anzatzsingledbiexample}) into equation (\ref{potentialexpressionseconddbisingleexample}), we obtain the contributions to the coefficient of the quadratic term from both the $H^{2}$ term and the square root term. We find the potential is given by
\begin{equation}\label{examplesingledbipotentialderived}
V = \left(3h_{1}^{2} - \frac{2 h_{1}}{\sqrt{\lambda}} \right) M^{2}_{P} \phi^{2} + \mathcal{O} \left(\phi^{4} \right) = m^{2} \phi^{2},
\end{equation}
which means that the $\mathcal{O} \left(\phi^{4} \right)$ terms vanish if we take into account all the terms in equation (\ref{anzatzsingledbiexample}). By substituting the ansatz (\ref{anzatzsingledbiexample}) into equation (\ref{twentynine}), we obtain
\begin{equation}
\dot{\phi} = \frac{- 2 h_{1} \phi^{2}}{\sqrt{\phi^{4}/M^{4}_{P} + 4 \lambda h^{2}_{1}}},
\end{equation}
where we used equation (\ref{dbisingleexamplewarp}). Because we assume $\phi / M_{P} \ll 1$, we simply ignore the $\phi^{4}$ term in the denominator and obtain the solution of the differential equation as
\begin{equation}\label{latetimedynamicsphi}
\phi \approx \frac{\sqrt{\lambda}}{t}.
\end{equation}
Combining equation (\ref{latetimedynamicsphi}) with the ansatz (\ref{anzatzsingledbiexample}), we find the late time behaviour of the scale factor as
\begin{equation}
a \left(t \right) = a_{0} t^{h_{1} \sqrt{\lambda}},
\end{equation}
where $a_{0}$ is constant. When the sound speed is small $c_{s} \ll 1$, the Friedmann equation (\ref{friedmanntwo}) can be solved for $\gamma$ with equations (\ref{dbisingleexamplewarp}), (\ref{dbisingleexamplepotential}), (\ref{anzatzsingledbiexample}) and (\ref{latetimedynamicsphi}) as
\begin{equation}\label{dbisingleexamplegammalatetime}
\begin{split}
\gamma &\approx f \left(3 M^{2}_{P} H^{2} - V \right)\\
& = t^{2} \left(3 M^{2}_{P} h^{2}_{1} - m^{2} \right)\\
& = \frac{2 M^{2}_{P} h_{1}}{\sqrt{\lambda}} t^{2},
\end{split}
\end{equation}
where equation (\ref{examplesingledbipotentialderived}) was used to derive the last equality. From equations (\ref{anzatzsingledbiexample}), (\ref{latetimedynamicsphi}) and (\ref{dbisingleexamplegammalatetime}), the slow-roll parameter $\epsilon$ (\ref{epsilontwo}) reads
\begin{equation}\label{examplesingledbiepsilonexpression}
\epsilon = \frac{1}{h_{1}\sqrt{\lambda}}. 
\end{equation}
When $\epsilon \ll 1$, from equations (\ref{examplesingledbipotentialderived}) and (\ref{examplesingledbiepsilonexpression}), we obtain the relation
\begin{equation}\label{relationmandhinexampleofdbi}
m^{2} = \left(3h_{1}^{2} - \frac{2h_{1}^{2}}{h_{1} \sqrt{\lambda}} \right) M^{2}_{P} = 3 h_{1}^{2} \left(1 - \frac{2 \epsilon}{3} \right) M^{2}_{P} \approx 3 h_{1}^{2} M^{2}_{P}. 
\end{equation}
From equations (\ref{anzatzsingledbiexample}), (\ref{latetimedynamicsphi}), (\ref{dbisingleexamplegammalatetime}), (\ref{examplesingledbiepsilonexpression}) and (\ref{relationmandhinexampleofdbi}), we obtain
\begin{equation}\label{alltheresultsindbisingleexample}
\phi = \frac{\sqrt{\lambda}}{t}, \:\:\:\:\: \gamma = c^{-1}_{s} = \sqrt{\frac{4}{3 \lambda}} M_{P} m t^{2}, \:\:\:\:\: H = \frac{1}{\epsilon t}, \:\:\:\:\: \epsilon = \sqrt{\frac{3}{\lambda}}\frac{M_{P}}{m},
\end{equation}
when the field $\phi / M_{P}$ and the sound speed $c_{s}$ are much smaller than unity. With equation (\ref{alltheresultsindbisingleexample}), the power spectrum of the curvature perturbation (\ref{scalarspectrum}) reads
\begin{equation}\label{curvaturespectrumexampledbisingle}
\mathcal{P}_{\mathcal{R}} = \frac{1}{8 \pi^{2} M^{2}_{P}}\frac{H^{2}}{c_{\rm{s}}\epsilon} = \frac{1}{4 \pi^{2} \epsilon^{4} \lambda},
\end{equation}
and the spectral index for the scalar perturbation (\ref{stwo}) is given by
\begin{equation}
1 - n_{\rm{s}} = \mathcal{O} \left(\epsilon^{2} \right),
\end{equation}
because the slow roll parameters (\ref{s}) and (\ref{etatwo}) are given by
\begin{equation}\label{relationchapthreesimple}
 \bar{\eta} = 0, \:\:\:\:\: s = -2 \epsilon,
\end{equation}
with equations (\ref{anzatzsingledbiexample}) and (\ref{alltheresultsindbisingleexample}). The tensor-to-scalar ratio and $f_{\rm{NL}}^{equil}$ are given by equations (\ref{rtwo}) and (\ref{fnlequilateralfinal}) respectively. If we choose a parameter set so that the observational constraint on the power spectrum of the curvature perturbation is satisfied with a fixed value of the slow-roll parameter $\epsilon$, the values of all the other observables introduced above are determined. For example, if we take $\epsilon = 1/20$, we obtain
\begin{equation}\label{dbisingleexamplelambdavalue}
\lambda \approx 10^{12},
\end{equation}
from equation (\ref{curvaturespectrumexampledbisingle}) with the observational constraint on the curvature perturbation (\ref{plancksatteliteconstraintcurvaturespectrum}). From equations (\ref{alltheresultsindbisingleexample}) and (\ref{dbisingleexamplelambdavalue}), we obtain the value of the mass of inflaton as
\begin{equation}\label{dbisingleexamplemassvalue}
\frac{m}{M_{P}} \approx 3.5 \times 10^{-5}.
\end{equation}
From equations (\ref{alltheresultsindbisingleexample}), (\ref{dbisingleexamplelambdavalue}) and (\ref{dbisingleexamplemassvalue}), the sound speed is given by
\begin{equation}\label{dbisingleexamplesoundspeedbyphi}
c_{s} \approx \frac{\phi^{2}}{40 M^{2}_{P}}. 
\end{equation}
By substituting equation (\ref{dbisingleexamplesoundspeedbyphi}) into equation (\ref{rtwo}) with $\epsilon = 1/20$, we obtain
\begin{equation}\label{dbisingleexampletensorscalarresult}
r \approx \frac{\phi^{2}}{50 M^{2}_{P}},
\end{equation}
where equation (\ref{fnlequilateralfinal}) combined with equation (\ref{dbisingleexamplesoundspeedbyphi}) gives the value of $f_{\rm{NL}}^{equil}$ as
\begin{equation}\label{dbisingleexamplefnlequilresult}
f_{\rm{NL}}^{equil} \approx -\frac{35}{108} \left(\frac{1.6 M^{4}_{P} \times 10^{3}}{\phi^{4}} -1 \right) \approx \frac{520 M^{4}_{P}}{\phi^{4}},
\end{equation}
where we used the assumption $\phi/M_{P} \ll 1$. The tensor-to-scalar ratio (\ref{dbisingleexampletensorscalarresult}) satisfies the observational constraint (\ref{observedspectralindexandratio}) with a sub-Planckian field value $\phi/M_{P} < 1$. However, the value of $f_{\rm{NL}}^{equil}$ in this model exhibits too much deviation from the Gaussian distribution to be compatible with the Planck satellite observations (\ref{fnlplanckobservationvaluesixtyeight}) because equation (\ref{dbisingleexamplefnlequilresult}) means $f_{\rm{NL}}^{equil} > 520$ if we consider a sub-Planckian field value. From equations (\ref{dbisingleexampletensorscalarresult}) and (\ref{dbisingleexamplefnlequilresult}), we see that the value of $f_{\rm{NL}}^{equil}$ becomes even larger if the value of $r$ is smaller. We will see that this relation generally holds in single field DBI inflation models in subsection \ref{subsec:observationalconstraintsdbisingle}. Finally, let us mention the meaning of the parameter $\lambda$. If we consider a compactification of string theory on $AdS_{5} \times X_{5}$ where $AdS_{5}$ is a five dimensional anti De Sitter space and $X_{5}$ is a five dimensional Einstein manifold, we have the relation \cite{Kachru:2003}
\begin{equation}\label{relationlambdaandninexampledbisingle}
\lambda \sim C N,
\end{equation}
where $C$ is a constant which depends on the form of $X_{5}$ and $N$ is the number of D3-branes. With a specific form of $X_{5}$, the relation (\ref{relationlambdaandninexampledbisingle}) is given by $\lambda \sim N$. In this case, we see that the number of the D3-branes $N$ is $10^{12}$ from equation (\ref{dbisingleexamplelambdavalue}). However, if we assume that the geometry of $X_{5}$ is a $Z^{l}_{n}$ orbifold \cite{Polchinski:1998}, the relation (\ref{relationlambdaandninexampledbisingle}) is given by $\lambda \sim n^{l} N$ for which no integer quantum number needs to be particularly large \cite{Alishahiha:2004}. 

\subsection{Observational constraints on single field cases}\label{subsec:observationalconstraintsdbisingle}
In this subsection, we will first derive the upper bound of the tensor-to-scalar ratio $r$ defined in subsection \ref{subsec:singledbiinflationintroduction} and introduce the general observational constraints on the single field DBI inflation following the derivation of Lidsey and Huston (2007) \cite{Lidsey:2007}. 

Baumann and McAllister (2006) derived an upper bound on the tensor-to-scalar ratio by analysing the higher dimensional structure of the manifold as follows \cite{Baumann:2006}. Let us consider a warped geometry whose metric is given by
\begin{equation}
ds^{2} = h^{-1/2} (y) g_{\mu \nu} dx^{\mu}dx^{\nu} + h^{1/2}(y)g_{ij} dy^{i}dy^{j}.
\end{equation}
Let us consider the internal space with a conical throat whose metric is locally of the form
\begin{equation}
g_{ij} dy^{i}dy^{j} = d\xi^{2} + \xi^{2} ds^{5}_{X_{5}}
\end{equation}
for some five manifold $X_{5}$. Note that $\xi$ is the radial coordinate in the throat. According to Baumann and McAllister (2006) \cite{Baumann:2006}, many such warped throats can be approximated by the the geometry $AdS_{5} \times X_{5}$ with the warp factor
\begin{equation}
h(\xi) = \left(\frac{R}{\xi} \right)^{4}, \label{warpedfactor}
\end{equation}
where $R$ is the radius of curvature of the AdS space for a small range of $\xi$. Then, in some models, we have the relation \cite{Baumann:2006, Gubser:1999}
\begin{equation}
\frac{R^{4}}{(\alpha')^{2}} = 4\pi g_{\rm{s}}N \frac{\pi^{3}}{\rm{Vol}(X_{5})}, \label{backgroundflux}
\end{equation}
where $\rm{Vol}(X_{5})$ is the dimensionless volume of the space $X_{5}$ with a unit radius, $\alpha'$ is the inverse string tension, $N$ is the background charge and $g_{\rm{s}}$ is the string coupling. Usually, we expect $\rm{Vol}(X_{5}) = \mathcal{O}(\pi^{3})$.
\\The following relation between the four dimensional Plank mass $M_{P}$, the warped volume of the compact space $V_{6}$, the inverse string tension $\alpha'$ and the string coupling $g_{\rm{s}}$ is obtained by the dimensional reduction from ten dimensions to four dimensions as
\begin{equation}
 M^{2}_{P} = \frac{V_{6}}{\kappa_{10}^{2}},\label{bulkvolume}
\end{equation}
where $\kappa_{10}^{2} = \frac{1}{2}(2\pi)^{7}g^{2}_{\rm{s}}(\alpha')^{4}$. The warped volume of the internal space is defined as
\begin{equation}
 V_{6} = \int d^{6}y \sqrt{g} h(\xi).
\end{equation}
In general, the compactified volume is a sum of bulk and throat contributions as
\begin{equation}
V_{6} = V_{6, bulk} + V_{6, throat}. \label{summation}
\end{equation}
The warped throat contribution is defined as
\begin{equation}
\begin{split}
V_{6, throat} &= \rm{Vol}(X_{5}) \int^{\xi_{\rm{UV}}}_{0} d\xi \: \xi^{5} h(\xi)\\
&= \frac{1}{2} \rm{Vol}(X_{5}) R^{4} \xi^{2}_{\rm{UV}}\\
&= 2 \pi^{4} g_{\rm{s}} N (\alpha')^{2} \xi^{2}_{\rm{UV}},\label{throatvolume}
\end{split}
\end{equation}
where $\xi = 0$ is the radial coordinate at the tip of the throat and $\xi_{\rm{UV}}$ is the radial coordinate where the ultraviolet end of the throat is glued into the bulk. Note that we used equations (\ref{warpedfactor}) and (\ref{backgroundflux}) to derive equation (\ref{throatvolume}). From equation (\ref{summation}), we obtain a conservative upper bound on the throat volume as
\begin{equation}
 V_{6} > V_{6, throat},
\end{equation}
and the following inequality from equation (\ref{bulkvolume}) as
\begin{equation}\label{relationchapthreempandvsix}
 M^{2}_{P} > \frac{V_{6, throat}}{\kappa_{10}^{2}}.
\end{equation}
Let us now consider the brane inflation in which the scalar fields describe the position of the D3-brane in the bulk. The canonically normalised inflaton field is written as
\begin{equation}
\phi^{2} = T_{3} \xi^{2},\:T_{3}=\frac{1}{(2\pi)^{3}}\frac{1}{g_{\rm{s}}(\alpha')^{2}}.\label{definephi}
\end{equation}
In this thesis, we will only consider the ultra-violet (UV) DBI inflation scenario where the brane is moving towards the tip of the throat while it is moving away from the tip of the throat in the infra-red (IR) DBI inflation scanario. Note that $\xi$ is the radial coordinate and hence has the dimension of length [L] while $\phi$ has the dimension of mass [M] because $\sqrt{T}_{3}$ has the dimension of squared mass $\left[M^{2}\right]$.
The maximal radial displacement of the brane in the throat is the length of the throat from the tip $\xi_{\rm{IR}} \simeq 0$ to the ultraviolet end $\xi_{\rm{UV}}$. One cannot increase the length of the throat arbitrarily because of the inequality
\begin{equation}
 \left(\frac{\phi_{\rm{UV}}}{M_{P}}\right)^{2} = \frac{T_{3}\xi^{2}_{\rm{UV}}}{M^{2}_{P}} < \frac{T_{3}\kappa^{2}_{10}\xi^{2}_{\rm{UV}}}{V_{6,throat}},\label{onewater}
\end{equation}
from equations (\ref{relationchapthreempandvsix}) and (\ref{definephi}). Substituting equations (\ref{throatvolume}), (\ref{definephi}) and the definition of $\kappa$ in terms of $g_{\rm{s}}$ and $\alpha'$ into the inequality (\ref{onewater}) gives the important constraint on the maximal field variation in four dimensional Plank units as follows
\begin{equation}
\left(\frac{\phi_{\rm{UV}}}{M_{P}}\right)^{2} < \frac{4}{N}.\label{conditionphi}
\end{equation}
We can see that the upper bound of the field variation depends only on the background charge N and is independent of the choice of $X_{5}$. $N=0$ corresponds to an unwarped throat, and we require $N>1$ at least and we need $N \gg 1$ in practice for backreaction effects to be negligible.
\\The condition (\ref{conditionphi}) can be converted into a corresponding constraint on the tensor-to-scalar ratio by using following expression derived by Boubekeur and Lyth (2005) \cite{Lyth:2005}
\begin{equation}
\begin{split}
 \frac{1}{M^{2}_{P}}\left(\frac{d\phi}{d\mathcal{N}}\right)^{2} &=  \frac{1}{M^{2}_{P}} \frac{\dot{\phi}^{2}}{H^{2}}\\
&= \frac{2}{P_{,\rm{X}}}\left(\frac{XP_{,\rm{X}}}{M^{2}_{P}H^{2}}\right)\\
&= 2\epsilon c_{\rm{s}}\\
&= \frac{r}{8},\label{lythbound}
\end{split}
\end{equation}
where we used equations (\ref{cstwo}), (\ref{epsilontwo}), (\ref{rtwo}) and $\mathcal{N} \equiv \int dt \: H$. If we integrate equation (\ref{lythbound}) from the tip of the throat ($\phi=0$) to the end of inflation ($\phi = \phi_{\rm{end}}$), we can obtain
\begin{equation}
\frac{\phi_{\rm{end}}}{M_{P}} = \sqrt{\frac{r_{*}}{8}}\mathcal{N}_{\rm{eff}}, \label{integratelyth}
\end{equation}
where $\mathcal{N}_{\rm{eff}} \equiv \int^{N_{\rm{end}}}_{0}d\mathcal{N}\:(r/r_{*})^{1/2}$ is a model-dependent parameter that quantifies how $r$ varies during the final stages of inflation. If we define observable region of inflation as $\phi = \phi_{*}$, we have $\phi_{\rm{UV}} > \phi_{*} > \phi_{\rm{end}} > 0$ because the brane moves towards the tip of the throat in UV DBI inflation. Therefore, combining equations (\ref{conditionphi}) and (\ref{integratelyth}), we obtain
\begin{equation}
 r_{*} < \frac{32}{N(\mathcal{N}_{\rm{eff}})^{2}}. \label{restrictiondbi}
\end{equation}
Typically, one expects $30 \lesssim \mathcal{N}_{\rm{eff}} \lesssim 60$, though it can be smaller if slow-roll inflation ends after the obsevable scales cross the horizon. Because we need $N \gg 1$ as stated above, the constraint (\ref{restrictiondbi}) imposes a strong restriction on DBI inflationary models. On the other hand, the value of $\mathcal{N}_{\rm{eff}}$ is uncertain.\\
Therefore, we can also try to derive the constraint on the range of values covered by the stages of observable inflation as follows. Firstly, we define $\Delta \phi_{*} = T_{3} \Delta \xi_{*}$ as the variation of the inflaton field during the observable inflation. We also have $V_{6,*} < V_{6,\rm{throat}}$ if we define $V_{6,*}$ as the volume of the part of the throat where inflaton moves during the observable inflation. Combining this inequality with the inequality (\ref{onewater}) and multiplying $\Delta \xi^{2}_{*} / \xi^{2}_{\rm{UV}}$, we have
\begin{equation}
\left(\frac{\Delta \phi_{*}}{M_{P}}\right)^{2} < \frac{T_{3}\kappa^{2}_{10} \Delta \xi^{2}_{*}}{V_{6,*}}. \label{constraintobserve}
\end{equation}
Note that `$*$` denotes quantities to be evaluated at the observable epoch. The observation of the CMB that directly constrain the promordial tensor perturbations only cover the multipole values in the range $2 \lesssim l \lesssim 100$. The corresponding scales to this range are leaving the horizon in the e-folds of $\Delta \mathcal{N} \simeq 4$ during inflation. Therefore, this corresponds to a narrow range of inflaton values. Therefore, the fraction of the throat volume that is observable is derived from equation (\ref{throatvolume}) approximately as
\begin{equation}
 \lvert V_{6, *} \rvert \simeq \rm{Vol}(X_{5}) \lvert\Delta \xi_{*} \rvert \xi^{5}_{*} h(\xi_{*}).
\end{equation}
Because of the inequality $\xi_{*} > \xi_{\rm{end}}>0$ which implies $\xi_{*} > \lvert \Delta \xi_{*} \rvert$, we have
\begin{equation}
 \lvert V_{6, *}\rvert > \rm{Vol}(X_{5}) \lvert\Delta \xi_{*} \rvert^{6} h(\xi_{*}).\label{xicondition}
\end{equation}
Substituting equation (\ref{tensionsingle}) and the condition (\ref{xicondition}) into equation (\ref{constraintobserve}) and using $\kappa^{2}_{10} = \pi T^{-2}_{3}$, we obtain
\begin{equation}
 \left(\frac{\Delta \phi_{*}}{M_{P}}\right)^{6} < \frac{\pi^{3}}{16 \rm{Vol}(X_{5})}r^{2}_{*}\mathcal{P}_{\mathcal{R}}\left(1+ \frac{1}{3f_{\rm{NL}}^{equil}} \right).
\end{equation}
Note that we used $\Delta \phi_{*} = \sqrt{T_{3}} \Delta \xi_{*}$ to derive this equation. Hence, using the Lyth's expression in the form $(\Delta \phi_{*}/M_{P})^{2} \simeq r(\Delta \mathcal{N}_{*})^{2}$ results in a very general upper limit on the tensor-to-scalar ratio:
\begin{equation}
 r_{*} < \frac{32 \pi^{3}}{(\Delta \mathcal{N}_{*})^{6} \rm{Vol}(X_{5})} \mathcal{P}_{\mathcal{R}}\left(1+ \frac{1}{3f_{\rm{NL}}^{equil}} \right).\label{onewatertwo}
\end{equation}
This condition weakly depends on the non-Gaussianity in the case that $f_{\rm{NL}}^{equil} > 5$ which is still compatible with the Planck satellite observations \cite{Ade:2013c} and we may neglect the factor with this parameter. Substituting the Planck normalisation $\mathcal{P}_{\mathcal{R}} = 2.23 \times 10^{-9}$ \cite{Ade:2013c}, the condition (\ref{onewatertwo}) reads
\begin{equation}
 r_{*} < \frac{2.23 \times 10^{-6}}{(\Delta \mathcal{N}_{*})^{6} \rm{Vol}(X_{5})}.
\end{equation}
$\Delta \mathcal{N}_{*} \sim 1$ is likely from the observation, whereas $\rm{Vol}(X_{5}) \simeq \mathcal{O}(\pi^{3})$ as stated above \cite{Baumann:2006}. 
Then, we obtain the upper bound on the tensor-to-scalar ratio for the standard UV DBI inflation
\begin{equation}
r_{*} < 10^{-7}. \label{rcondition}
\end{equation}
This is significantly below the sensitivity of future CMB polarization experiments $r \gtrsim 10^{-4}$. Actually, this upper limit results in difficulties for the single field DBI inflation models as follows.\\
Differentiating equation (\ref{gammatwo}) with respect to cosmic time and using equations (\ref{hubbledotdbi}) and (\ref{twentynine}), we can express $\bar{\eta}$ with respect to the other slow-roll parameters as
\begin{equation}
 \bar{\eta} = \pm \frac{M^{2}_{P} T' \lvert H'\rvert }{\gamma TH} + s + \frac{s}{\gamma^{2}-1}, \label{watch}
\end{equation}
where plus (minus) sign corresponds to a brane moving down (up) the warped throat. Then, substituting equation (\ref{stwo}) in to equation (\ref{watch}), we obtain
\begin{equation}
 1 - n_{\rm{s}} = 4 \epsilon + \frac{2s}{1 - \gamma^{2}} \mp \frac{2M^{2}_{P}}{\gamma}\frac{T'\lvert H'\rvert}{TH},\label{watchtwo}
\end{equation}
where minus (plus) sign corresponds to a brane moving down (up) the warped throat. In the UV DBI inflation, we take the minus sign, for example. The second term in the right hand side of equation (\ref{watchtwo}) can be converted into observable parameter by defining the `tilt` of the non-linearity parameter
\begin{equation}\label{fly}
 n_{\rm{NL}} \equiv \frac{d \: \ln{f_{\rm{NL}}^{equil}}}{d \: \ln{k}}.
\end{equation}
If we assume the slow-roll approximation, we have
\begin{equation}\label{aroundtheworld}
\frac{d \: \ln{k}}{dt} \sim H.
\end{equation}
With equations (\ref{fnlequilateralfinal}) and (\ref{aroundtheworld}), equation (\ref{fly}) implies
\begin{equation}
 s = \frac{-3f_{\rm{NL}}^{equil} n_{\rm{NL}}}{2(1+3f_{\rm{NL}}^{equil})}.\label{mario}
\end{equation}
Substituting equations (\ref{fnlequilateralfinal}), (\ref{stwo}) and (\ref{rtwo}) into equation (\ref{watchtwo}), we obtain
\begin{equation}
 1 - n_{\rm{s}} = \frac{r}{4}\sqrt{1 + 3f_{\rm{NL}}^{equil}} + \frac{n_{\rm{NL}}}{1 + 3f_{\rm{NL}}^{equil}} \mp \sqrt{\frac{r}{8}}\left(\frac{T'}{T}M_{P} \right).\label{mousetwo}
\end{equation}
From equation (\ref{warpedfactor}), we can see $dh/d\xi \leq 0$. Therefore, in the case of the UV DBI inflation, the third term in the right hand side of equation (\ref{mousetwo}) is negative definite, which implies that
\begin{equation}
\frac{r}{4}\sqrt{1 + 3f_{\rm{NL}}^{equil}} + \frac{n_{\rm{NL}}}{1 + 3f_{\rm{NL}}^{equil}} > 1 - n_{\rm{s}}.\label{yoshi}
\end{equation}
This is a consistency relation in UV DBI inflation. Firstly, let us assume the tensor-to-scalar ratio is negligible. Because the Planck data \cite{Ade:2013c} strongly favours a red spectral index with $n_{\rm{s}} \sim 0.96$, $\epsilon$ cannot be smaller than $\sim 10^{-2}$ from equation (\ref{stwo}). Therefore, from equation (\ref{rtwo}), the sound speed must be small as $c_{\rm{s}} \lesssim 10^{-6}$ in order to let $r$ to be negligible satisfying the condition (\ref{rcondition}). Then, from equation(\ref{fnlequilateralfinal}), $f_{\rm{NL}}^{equil}$ is necessarily very large as $f_{\rm{NL}}^{equil} \gtrsim 10^{12}$. This is incompatible with the Planck satellite observations \cite{Ade:2013c} which gives the limit $\lvert f_{\rm{NL}}^{equil} \rvert < 120$. In addition to that, we need
\begin{equation}
 n_{\rm{NL}} \simeq -2s > 3(1-n_{\rm{s}})f_{\rm{NL}}^{equil} > 0.1 f_{\rm{NL}}^{equil},
\end{equation}
to satifsy $1-n_{\rm{s}} > 0.033$. However, when $f_{\rm{NL}}^{equil}\gg1$, this would violate the slow-roll conditions because $\lvert s \rvert$ becomes much larger than unity.\\
Then, let us assume that $r$ is not negligible. Also in this case, we have a large non-Gaussianity as $f_{\rm{NL}}^{equil} \gtrsim 10^{12}$ because the argument about $f_{\rm{NL}}^{equil}$ above does not depend on the value of $r$. Again this is too large as stated above. In addition to that, because the second term in the left hand side of equation (\ref{yoshi}) can be neglected when $f_{\rm{NL}}^{equil} \gg 1$, we have
\begin{equation}
 r_{*} > \frac{4(1 - n_{\rm{s}})}{\sqrt{3f_{\rm{NL}}^{equil}}} > \frac{1 - n_{\rm{s}}}{5}, \label{mousefour}
\end{equation}
where we used $\lvert f_{\rm{NL}}^{equil} \rvert < 120$. $r > 0.002$ for $1 - n_{\rm{s}} \simeq 0.038$ that is favoured by the Planck satellite observations for non-negligible $r$ \cite{Ade:2013b}, this is incompatible with the upper limit (\ref{rcondition}). Therefore, we conclude that it is difficult for the single field models to satisfy the bounds on $r$ with the Planck data in a single field UV DBI inflation.

\section{Multi-field DBI inflation}\label{sec:multifielddbiinflationthree}
Not only because the single field DBI inflation is incompatible with the observation as we saw in the previous section, it is very natural for us to consider the multi-field DBI inflation in the string theory. As mentioned in subsection \ref{dbiintroduction}, we have ten dimensions in string theory with the supersymmetry. It means that we have six extra dimensions in addition to our usual four dimensional space-time. Therefore, it is quite natural to think that we have multi-field in the DBI inflation because the scalar fields describe the position of the brane in the internal space. For multi-field DBI inflation, as we see in section \ref{sec:observable sinmultifield}, the observables are different from the single field DBI inflation models. In equations (\ref{equationofmotionone}) and (\ref{equationofmotiontwo}), we see that $\xi$ quantifies the interaction between the adiabatic and entropic components. 
Therefore, from equations (\ref{alphamouse}), (\ref{interactionredefinition}) and (\ref{trs}), we see that $T_{RS}=0$ if there is no interaction ($\Xi = 0$). Then, let us assume that there is some interaction which is equivalent to $\Xi \neq 0$. In this case, we may have non-zero $T_{RS}$ as long as $\beta \neq 0$.
Furthermore, $T_{RS}$ may become even larger if the slow-roll conditions are broken because $\beta$ can be larger with $s>1$ and $\eta>1$ and it makes $T_{SS}$ larger as we can see from equations (\ref{betamouse}) and (\ref{tss}). Then, $T_{RS}$ can be large and hence $\sin{\Theta}$ can be non-zero as we see in equation ({\ref{sintheta}}). This makes $\cos^{2}{\Theta}$ to be smaller than unity and we can clearly see that $\mathcal{P}_{\mathcal{R}}$ is enhanced in this case from equation (\ref{prmouse}). As we introduced in section \ref{sec:observable sinmultifield}, the interaction parameter $\Xi$ takes non-zero value when the trajectory is bent. Therefore, a curve in the background trajectory makes $\lvert T_{RS} \rvert$ take a non-zero value. 
\\
In multi-field DBI inflation models, we have the relation \cite{Langlois:2009}
\begin{equation}
 1 - n_{\rm{s}} \simeq \frac{\sqrt{3 \lvert f_{\rm{NL}}^{equil} \rvert}r}{4 \cos^{3}{\Theta}} - \frac{\dot{f}}{Hf} + \alpha_{*} \sin{2 \Theta} + 2 \beta_{*} \sin^{2}{\Theta},\label{mousethree}
\end{equation}
where $f$ is the warp factor and $\alpha_{*}$ and $\beta_{*}$ are evaluated at the sound horizon exit using equations (\ref{alphamouse}) and (\ref{betamouse}).
Equation (\ref{mousethree}) is different from equation (\ref{mousetwo}) which is valid for the single field case. In the single field case, the last two terms in equation (\ref{mousethree}) can be neglected and we obtain equation (\ref{mousefour}) if we assume $\dot{f}>0$ because $\cos{\Theta}$ is unity in the single field cases. From equation (\ref{mousethree}), we see that the inequality $1-n_{\rm{s}} \ll 1$ is not necessary even if we have very small $r$ when the transfer function $T_{RS}$ is large because the last two terms can increase the value of the right hand side of equation (\ref{mousethree}). In this way, we can avoid the observational constraint in the multi-field models.
Also, from equation (\ref{dbimultiequilateralfinalresult}), we see that $f_{\rm{NL}}^{equil}$ can take a small value which satisfies the observational constraint even if the sound speed is small as $c_{\rm{s}} \lesssim 10^{-6}$ if $T_{RS}$ is sufficiently large. Therefore, multi-field DBI inflation models can satisfy the microphysical constraint combined with the observations which are shown in this section. As we saw above, the trajectory in the field space affects the value of $T_{RS}$. Therefore, we need concrete models to see if there are actual models which satisfy the microphysical constraint. 

\section{Specific models}\label{sec:specificmodelsofdbi}
As we saw in section \ref{sec:multifielddbiinflationthree}, we need concrete DBI inflation models with specific potentials in order to see if a curve in the background trajectory can make the values of the observables compatible with the observations by numerical calculations. In this section, we briefly explain how we derive a potential in the Lagrangian (\ref{dbigenerallagrangian}) in string theory. 

We first explain the reason why we need to consider a specific set-up of the DBI inflation models in string theory. Let us consider brane-anti-brane inflation models in which a brane and an anti-brane are initially separated by a distance $r$ on the compactification manifold M. If we assume $r \gg M^{-1}_{10,P}$ where $M_{10,P}$ is the ten dimensional Planck scale \cite{Dvali:1999}, the force which the brane feels is well approximated by the Coulomb attraction given by \cite{Kachru:2003}
\begin{equation}\label{potentialinstandardsetupinbraneworld}
V \left(\phi \right) = 2 T_{3} \left(1 - \frac{1}{2 \pi^{2}}\frac{T_{3}^{3}}{M^{8}_{10,P} \phi^{4}} \right),
\end{equation}
where $T_{3}$ is the tension of a D3-brane and the canonically normalised field $\phi = \sqrt{T_{3}} r$ is defined as in equation (\ref{rescalingthecoordinates}). From equations (\ref{seta}) and (\ref{potentialinstandardsetupinbraneworld}), the standard slow-roll parameter $\tilde{\eta}$ is given by
\begin{equation}
\tilde{\eta} = -\frac{10}{\pi^{3}} \left(\frac{L}{r}\right)^{6} = - 0.3 \left(\frac{L}{r}\right)^{6},
\end{equation}
where the four dimensional Planck mass is defined as $M^{2}_{P} = M^{8}_{10,P} L^{6}$ with the volume of M that is given by $L^{6}$. Therefore, $\tilde{\eta} \ll 1$ is possible only if $r > L$. However, two branes cannot be separated by a distance $r$ which is greater than L which is the size of the manifold M. This is called the \textbf{$\eta$ problem}. Also, we have a problem in stabilising the moduli fields which control the shape and size of the compactification manifold \cite{Kachru:2003}. If we assume that the main contribution to the inflationary energy is from the D3-brane tension, we have
\begin{equation}\label{modulistabilizationproblempotential}
V \left(\phi, L \right) \sim \frac{2 T_{3}}{L^{12}},
\end{equation}
where $r \gg M^{-1}_{10,P}$. As we see in equation (\ref{modulistabilizationproblempotential}), L is not stabilised and it rolls down the potential to the direction of large $L$. Therefore, in the absence of a stabilisation mechanism which fixes $L$ with sufficient mass so that the variation of $L$ is negligible, we have a steep potential which makes the standard slow-roll parameters (\ref{sepsilon}) and (\ref{seta}) larger than unity. This is the \textbf{moduli stabilisation problem}. 

\begin{figure}[!htb]
\centering
\includegraphics[width=14cm]{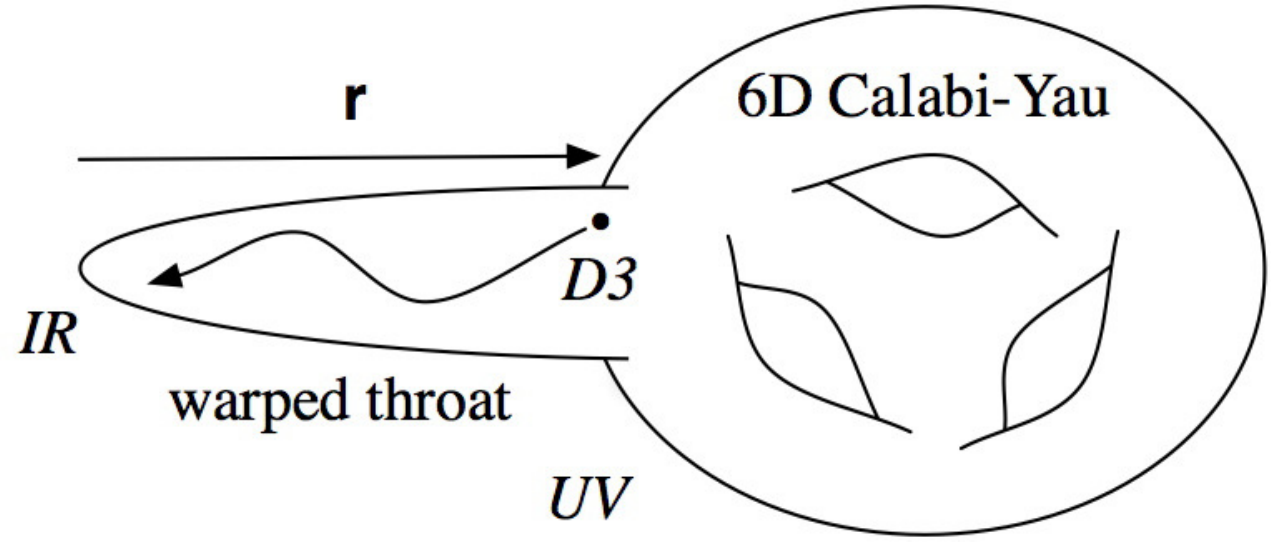}
\caption[Internal space]{Schematic figure about the internal space which is a space in the extra dimensions.The mobile D3-brane moves in the background geometry being attracted by an anti-brane located at the tip of the throat. The throat is smoothly glued to a Calabi-Yau manifold in the UV region. The figure is taken from \cite{Easson:2008}}{\label{fig:warpedthroatcalabiyau}}
\end{figure}

Even though inflation can occur with steep potentials in the DBI inflation models because the slow-roll parameters (\ref{epsilontwo}), (\ref{s}) and (\ref{etatwo}) are less than unity even with steep potentials if the sound speed is sufficiently small, new models have been provided to stabilise the undesired moduli fields in string theory. However, explicitly calculable models are scarce. Hence, the warped deformed conifold is an ideal geometry for string inflationary model building as a concrete calculable model \cite{Chen:2010}. The potential (\ref{potentialinstandardsetupinbraneworld}) is modified by considering branes and anti-branes in a warped geometry and the $\eta$ problem is evaded successfully \cite{Kachru:2003}. Therefore, we consider a D3-brane in a smooth warped deformed conifold throat which can easily develop in a bulk compact Calabi-Yau manifold near the conifold singularity as in figure \ref{fig:warpedthroatcalabiyau} \cite{Chen:2010}. 

As explained in section \ref{dbiintroduction}, the coordinates in the internal space play roles of the scalar fields which cause inflation. The warped geometry is given in equation (\ref{dbigeneralmetric}). The potential experienced by a D3-brane in such a throat is given by \cite{Chen:2010}
\begin{equation}\label{potentiallastsectionchapthree}
V = T_{3} \Phi_{-},
\end{equation}
where $T_{3}$ is the D3-brane tension and 
\begin{equation}\label{fiveformandwarpfactor}
\Phi_{-} = h^{-1} - \alpha
\end{equation}
is a combination of the throat warp factor in equation (\ref{dbigeneralmetric}) and the five form field strength $\alpha$ given in \cite{Baumann:2009}. The equation of motion for $\Phi_{-}$ is derived from the IIB supergravity action \cite{Baumann:2010} as
\begin{equation}\label{equationofmotionforpotentialdependingonembedding}
\nabla^{2} \Phi_{-} = \frac{g_{s}}{24 h^{2}} \left\lvert G_{-} \right\rvert + \mathcal{R}_{4} + h \left\lvert \nabla \Phi_{-} \right\rvert^{3} + \mathcal{S}_{\rm{local}},
\end{equation}
where $\nabla^{2}$ is the Laplacian with respect to the 6D metric $G_{IJ}$ in equation (\ref{dbigeneralmetric}), $h$ is the warp factor, $g_{s} = e^{\phi}$ is the string coupling with the dilation field $\phi$, $ \mathcal{R}_{4}$ is the four dimensional Ricci scalar, $\mathcal{S}_{\rm{local}}$ is the localised sources and 
\begin{equation}
G_{-} = \left(\star_{6} - i \right) G_{3},
\end{equation}
is the imaginary anti-self-dual (IASD) component of the complex three form flux $G_{3}$ where $\star_{6}$ is the six dimensional Hodge star operator. If the flux $G_{3}$ is imaginary self-dual (ISD),  i.e. $\star_{6} G_{3} = i G_{3}$, the background metric $G_{IJ}$ is a Calabi-Yau metric with the five form field strength given by $\alpha = h^{-1}$. Therefore, from equation (\ref{fiveformandwarpfactor}), we have $V = T_{3} \Phi_{-} = 0$ in this case. The mode $\Phi_{-}$ characterises the perturbation away from the ISD background. In principle, equation (\ref{equationofmotionforpotentialdependingonembedding}) can be solved and the potential (\ref{potentiallastsectionchapthree}) is obtained if we know all the information about the background geometry including the warp factor $h$, the metric of the internal space $G_{IJ}$ and how we embed the D3-branes. 
\chapter{Spinflation}\label{chapter:spinflation}
\chaptermark{Spinflation}

In this chapter, we study a multi-field DBI inflation model with a potential derived in string theory. As shown in chapter \ref{chapter:stringinflation}, the conversion of the entropy perturbation into the curvature perturbation gives multi-field DBI inflation models the possibility of satisfying the microphysical constraint combined with the CMB observations. In order to know how much conversion occurs, we need to specify the model. Because the DBI inflation models are motivated by string theory, it is important to investigate potentials derived in string theory. Here, we study the simplest two-field DBI inflation model with a potential that is derived in string theory. In \cite{Gregory:2012}, a potential for the two-field model was derived in string theory by embedding a warped throat into a compact Calabi-Yau space with all moduli fields stabilised and the background dynamics of DBI inflation was studied. We first introduce the model and reexamine the background dynamics in this model. Then, we see that the results of our numerical calculations can be explained with the analytic formulae introduced in subsection \ref{subsec:analysisofsimplemodeldbi} with some approximations. We also show how the amplitude of the curvature perturbation is enhanced with different initial conditions for the angular field. Finally, we show that this model is not compatible with the Planck satellite observations in the regime where those approximations hold and discuss the cases where those approximations cannot be used. 

\section{The model}\label{sec:spinflationfirst}
In this section, we introduce the phase space variables that we used in our numerical code. Then, we show the set-up of the multi-field DBI inflation model introduced in \cite{Gregory:2012} using those variables. 

\subsection{Phase space variables}\label{phasespacevarispin}
Let us consider cases where the field space metric is given by
\begin{equation}\label{simplediagonalform}
 G_{IJ} = A_{I} \delta_{IJ},
\end{equation}
with the Kronecker delta $\delta_{IJ}$. Note that the right hand side of equation (\ref{simplediagonalform}) is a diagonal matrix with the $\left(I,I\right)$-component $A_{I}$. For the numerical computation, it is useful to introduce the phase space variables. We consider general two-field DBI inflation models with the fields $\phi^{a}$ and $\phi^{b}$. In general, when the metric takes a simple diagonal form (\ref{simplediagonalform}), the sound speed is given by
\begin{equation}\label{cssimplediagonal}
c_{s}=\sqrt{1-f\sum_{I}A_{I}\left(\phi^{I}\right)^{2}}.
\end{equation}
If we define the phase space variables as
\begin{equation}\label{phasespacevari}
x_{I}=-\sqrt{fA_{I}}\dot{\phi}^{I},
\end{equation}
the sound speed in equation (\ref{cssimplediagonal}) is rewritten as
\begin{equation}\label{newsoundsimplediagonal}
c_{s}=\sqrt{1-x_{a}^{2}-x_{b}^{2}}. 
\end{equation}
As we can see, the expression in the square root of equation (\ref{newsoundsimplediagonal}) is simpler than that of equation (\ref{cssimplediagonal}). As such, the computation time is reduced with this method. Below, we show how to derive the equations of motion for the scalar fields with those new variables. If we assume
\begin{equation}\label{simpledependenceanddiagonal}
f\left(\phi^{K}\right)=f\left(\phi^{a}\right),\,\,\,\,\,A_{I}\left(\phi^{K}\right)=A_{I}\left(\phi^{a}\right),
\end{equation}
as in the model of \cite{Gregory:2012}, the equations (\ref{eqofmotionai}) are given by
\begin{equation}\label{eqofmotionaione}
\begin{split}
A_{a}\left(\ddot{\phi}^{a}+3H\dot{\phi}^{a}-\frac{\dot{c_{s}}}{c_{s}}\dot{\phi}^{a}\right)+\frac{1}{2}&\left(A_{a,\phi^{a}}\left(\dot{\phi}^{a}\right)^{2}-A_{b,\phi^{a}}\left(\dot{\phi}^{b}\right)^{2}\right)\\
&+c_{s}V_{,\phi^{a}}-\frac{\left(1-c_{s}\right)^{2}}{2}\frac{f_{,\phi^{a}}}{f^{2}}=0, 
\end{split}
\end{equation}
\begin{equation}\label{eqofmotionaitwo}
A_{b}\left(\ddot{\phi}^{b}+3H\dot{\phi}^{b}-\frac{\dot{c_{s}}}{c_{s}}\dot{\phi}^{b}\right)+A_{b,\phi^{a}}\dot{\phi}^{a}\dot{\phi}^{b}+c_{s}V_{,\phi^{b}}=0. 
\end{equation}
From the equations (\ref{phasespacevari}), (\ref{newsoundsimplediagonal}) and (\ref{simpledependenceanddiagonal}), we obtain
\begin{equation}
\begin{split}
\ddot{\phi^{I}}&=\left[\frac{\dot{x_{I}}}{x_{I}}-\frac{1}{2}\left(\frac{\dot{f}}{f}+\frac{\dot{A}_{I}}{A_{I}}\right)\right]\dot{\phi}^{I}\\
&=-\frac{\dot{x}_{I}}{\sqrt{fA_{I}}}-\frac{1}{2}\frac{x_{I}}{\sqrt{fA_{I}}}\frac{x_{a}}{\sqrt{fA_{I}}}\left(\frac{f_{\phi_{a}}}{f}+\frac{A_{I,\phi_{a}}}{A_{I}}\right),
\end{split}
\end{equation}
\begin{equation}
\dot{c_{s}}=\frac{1}{2}\frac{\left(1-x_{a}^{2}-x_{b}^{2}\right)^{.}}{\left(1-x_{a}^{2}-x_{b}^{2}\right)}c_{s}=\frac{-x_{a}\dot{x}_{a}-x_{b}\dot{x}_{b}}{c_{s}},
\end{equation}
where we have used
\begin{equation}
\dot{f}=f_{\phi^{a}}\dot{\phi}^{a}=-f_{\phi^{a}}\frac{x_{a}}{\sqrt{fA_{a}}},
\end{equation}
\begin{equation}
\dot{A}_{a}=A_{a,\phi^{a}}\dot{\phi}^{a}=-A_{a,\phi^{a}}\frac{x_{a}}{\sqrt{fA_{a}}}.
\end{equation}
Making use of the relations above, the equations (\ref{eqofmotionaione}) and (\ref{eqofmotionaitwo}) become
\begin{equation}
\begin{split}
\dot{x_{a}}=-\frac{x_{a}x_{b}}{1-x_{b}^{2}}\dot{x}_{b}+\frac{c_{s}^{2}}{1-x_{b}^{2}}&\left[-\frac{1}{2\sqrt{fA_{a}}}\frac{(A_{b})_{,\phi^{a}}}{A_{b}}x_{b}^{2}-3Hx_{a}+\frac{\sqrt{f}}{\sqrt{A_{a}}}c_{s}V_{,\phi^{a}}\right.\\
&\left.+\frac{f_{,\phi^{a}}}{f}\frac{1}{\sqrt{fA_{a}}}\left(-1+c_{s}+\frac{1}{2}x_{b}^{2}\right)\right], 
\end{split}
\end{equation}
\begin{equation}
\dot{x_{b}}=-\frac{x_{a}x_{b}}{1-x_{a}^{2}}\dot{x}_{a}+\frac{c_{s}^{2}}{1-x_{a}^{2}}\left[\frac{1}{2\sqrt{fA_{a}}}\left(\frac{(A_{b})_{,\phi^{a}}}{A_{b}}-\frac{f_{\phi^{a}}}{f}\right)x_{a}x_{b}-3Hx_{b}+\frac{\sqrt{f}}{\sqrt{A_{b}}}c_{s}V_{,\phi^{b}}\right],
\end{equation}
which lead to
\begin{equation}\label{phaseeqofmotiongeneone}
\begin{split}
\dot{x_{a}}=&\left(1-x_{a}^{2}\right)\left[-\frac{1}{2\sqrt{fA_{a}}}\frac{(A_{b})_{,\phi^{a}}}{A_{b}}x_{b}^{2}-3Hx_{a}+\frac{\sqrt{f}}{\sqrt{A_{a}}}c_{s}V_{,\phi^{a}}+\frac{f_{,\phi^{a}}}{f}\frac{1}{\sqrt{fA_{a}}}\left(-1+c_{s}+\frac{1}{2}x_{b}^{2}\right)\right]\\
&-x_{a}x_{b}\left[\frac{1}{2\sqrt{fA_{a}}}\left(\frac{(A_{b})_{,\phi^{a}}}{A_{b}}-\frac{f_{\phi^{a}}}{f}\right)x_{a}x_{b}-3Hx_{b}+\frac{\sqrt{f}}{\sqrt{A_{b}}}c_{s}V_{,\phi^{b}}\right],
\end{split}
\end{equation}
\begin{equation}\label{phaseeqofmotiongenetwo}
\begin{split}
\dot{x_{b}}=&-x_{a}x_{b}\left[-\frac{1}{2\sqrt{fA_{a}}}\frac{(A_{b})_{,\phi^{a}}}{A_{b}}x_{b}^{2}-3Hx_{a}+\frac{\sqrt{f}}{\sqrt{A_{a}}}c_{s}V_{,\phi^{a}}+\frac{f_{,\phi^{a}}}{f}\frac{1}{\sqrt{fA_{a}}}\left(-1+c_{s}+\frac{1}{2}x_{b}^{2}\right)\right]\\
&+\left(1-x_{b}^{2}\right)\left[\frac{1}{2\sqrt{fA_{a}}}\left(\frac{(A_{b})_{,\phi^{a}}}{A_{b}}-\frac{f_{\phi^{a}}}{f}\right)x_{a}x_{b}-3Hx_{b}+\frac{\sqrt{f}}{\sqrt{A_{b}}}c_{s}V_{,\phi^{b}}\right].
\end{split}
\end{equation}
These are the equations of motion that we solve in our numerical codes.

\subsection{Spinflation model}
Let us define $\chi$ and $\theta$ to be the radial and angular coordinates in the warped throat in the internal space respectively. The internal metric is described by
\begin{equation}
ds^{2}=\tilde{g}_{mn}dy^{m}dy^{n}=\kappa^{4/3}\left[\frac{d\chi^{2}}{6K\left(\chi\right)^{2}}+B\left(\chi\right)d\theta^{2}\right],
\end{equation}
with
\begin{equation}
K\left(\chi\right)=\frac{\left(\sinh{\chi}\cosh{\chi}-\chi\right)^{1/3}}{\sinh{\chi}},\,\,\,\,\,B\left(\chi\right)=\frac{1}{2}K\left(\chi\right)\cosh{\chi}, 
\end{equation}
and the deformation parameter $\kappa$. Note that $\eta^{I} = \left(\chi, \theta\right)$ in equation (\ref{inducedmetric}). The proper radial coordinate is defined as 
\begin{equation}\label{spincanonicalfielddefined}
r\left(\chi\right)=\frac{\kappa^{2/3}}{\sqrt{6}}\int^{\chi}_{0}\frac{dx}{K\left(x\right)},
\end{equation}
where the canonical scalar field is given by $\phi \equiv \sqrt{T_{3}} r$. The warp factor in this model is given by \cite{Klebanov:2000}
\begin{equation}\label{warpfactor}
h\left(\chi\right) \equiv e^{-4A} = 2\left(g_{\rm{s}}M\alpha '\right)^{2}\kappa^{-8/3}I\left(\chi\right),
\end{equation}
where
\begin{equation}\label{spinfirstdefinitionofi}
I\left(\chi\right)\equiv\int^{\infty}_{\chi}dx\frac{x\coth{x}-1}{\sinh^2{x}}\left(\sinh{x}\cosh{x}-x\right)^{1/3},
\end{equation}
with the parameter $M$, the string coupling $g_{\rm{s}}$ and the inverse string tension $\alpha'$. The equations of motion (\ref{eqofmotionai}) for the radial scalar field $\chi$ and angular scalar field $\theta$ are given by 
\begin{equation}\label{equationofmotionchibackground}
\begin{split}
\ddot{\chi}=&-\frac{3H}{\gamma^{2}}\dot{\chi}-4A_{,\chi}\left(\gamma^{-1}-1\right)\dot{\chi}^{2}-12\kappa^{-4/3}K^{2}A_{,\chi}e^{4A}\left(\gamma^{-1}-1\right)^{2}\\
&+\frac{K_{,\chi}}{K}\dot{\chi}^{2}+3K^{2}B_{,\chi}\dot{\theta}^{2}+e^{-4A}\dot{\theta}\dot{\chi}\frac{U_{,\theta}}{\gamma}-\left(6K^{2}\kappa^{-4/3}-e^{-4A}\dot{\chi}^{2}\right)\frac{U_{,\chi}}{\gamma},
\end{split}
\end{equation}
\begin{equation}\label{equationofmotionthetabackground}
\ddot{\theta}=-\frac{3H}{\gamma^{2}}\dot{\theta}-4A_{,\chi}\left(\gamma^{-1}-1\right)\dot{\chi}\dot{\theta}-\frac{B_{,\chi}}{B}\dot{\chi}\dot{\theta}+e^{-4A}\dot{\chi}\dot{\theta}\frac{U_{,\chi}}{\gamma}-\left(\frac{\kappa^{-4/3}}{B}-e^{-4A}\dot{\theta}^{2}\right)\frac{U_{,\theta}}{\gamma},
\end{equation}
where the subscripts denote derivatives with respect to the fields as $_{,\chi}\equiv\frac{\partial}{\partial\chi}$ and $_{,\theta}\equiv\frac{\partial}{\partial\theta}$. In our numerical code, we used the phase space variables introduced in subsection \ref{phasespacevarispin} for the numerical efficiency. The equation of motion for the phase space variables (\ref{phaseeqofmotiongeneone}) and (\ref{phaseeqofmotiongenetwo}) are given by
\begin{equation}\label{phaseeqofmotionspinflationone}
\begin{split}
\dot{x_{\chi}}=&\left(1-x_{\chi}^{2}\right)\left[-\frac{1}{2}\sqrt{\frac{6K\left(\chi\right)^{2}}{f\kappa^{4/3}}}\frac{B\left(\chi\right)_{,\chi}}{B\left(\chi\right)}x_{\theta}^{2}-3Hx_{\chi}+\sqrt{\frac{6fK\left(\chi\right)^{2}}{\kappa^{4/3}}}c_{\rm{s}}V_{,\chi}\right. \\
&\left.+\frac{I\left(\chi\right)_{,\chi}}{I\left(\chi\right)}\sqrt{\frac{6K\left(\chi\right)^{2}}{f\kappa^{4/3}}}\left(-1+c_{\rm{s}}+\frac{1}{2}x_{\theta}^{2}\right)\right]\\
&-x_{\chi}x_{\theta}\left[\frac{1}{2}\sqrt{\frac{6K\left(\chi\right)^{2}}{f\kappa^{4/3}}}\left(\frac{B\left(\chi\right)_{,\chi}}{B\left(\chi\right)}-\frac{I\left(\chi\right)_{,\chi}}{I\left(\chi\right)}\right)x_{\chi}x_{\theta}-3Hx_{\theta}+\frac{\sqrt{f}}{\sqrt{\kappa^{4/3}B\left(\chi\right)}}c_{\rm{s}}V_{,\theta}\right],
\end{split}
\end{equation}
\begin{equation}\label{phaseeqofmotionspinflationtwo}
\begin{split}
\dot{x_{\theta}}=&-x_{\chi}x_{\theta}\left[-\frac{1}{2}\sqrt{\frac{6K\left(\chi\right)^{2}}{f\kappa^{4/3}}}\frac{B\left(\chi\right)_{,\chi}}{B\left(\chi\right)}x_{\theta}^{2}-3Hx_{\chi}+\sqrt{\frac{6fK\left(\chi\right)^{2}}{\kappa^{4/3}}}c_{\rm{s}}V_{,\chi}\right. \\
&\left.+\frac{I\left(\chi\right)_{,\chi}}{I\left(\chi\right)}\sqrt{\frac{6K\left(\chi\right)^{2}}{f\kappa^{4/3}}}\left(-1+c_{\rm{s}}+\frac{1}{2}x_{\theta}^{2}\right)\right]\\
&+\left(1-x_{\theta}^{2}\right)\left[\frac{1}{2}\sqrt{\frac{6K\left(\chi\right)^{2}}{f\kappa^{4/3}}}\left(\frac{B\left(\chi\right)_{,\chi}}{B\left(\chi\right)}-\frac{I\left(\chi\right)_{,\chi}}{I\left(\chi\right)}\right)x_{\chi}x_{\theta}-3Hx_{\theta}+\frac{\sqrt{f}}{\sqrt{\kappa^{4/3}B\left(\chi\right)}}c_{\rm{s}}V_{,\theta}\right],
\end{split}
\end{equation}
with the phase space variables
\begin{equation}
x_{\chi} = - \sqrt{\frac{\kappa^{4/3}}{6 K^{2}}} \dot{\chi},\,\,\,\,\,x_{\theta} = - \sqrt{\kappa^{4/3} B} \dot{\theta}. 
\end{equation}
The sound speed of the scalar field perturbations is given by
\begin{equation}\label{spinflationsoundspeedandfactor}
\begin{split}
c_{\rm{s}}=\gamma^{-1}&=\sqrt{1-e^{-4A}\tilde{g}_{mn}\dot{y}^{m}\dot{y}^{n}}\\
 & = \sqrt{1 - h \kappa^{4/3} \left(\frac{\dot{\chi}^{2}}{6K^{2}}+B \dot{\theta}^{2} \right)}, 
\end{split}
\end{equation}
where $y^{\rm{m}}=\left(\chi,\,\theta\right)$. 

In \cite{Gregory:2012}, the equation of motion (\ref{equationofmotionforpotentialdependingonembedding}) is solved including linearised perturbations around the ISD solution with a warped throat embedded into a compact Calabi-Yau space with all moduli fields stabilised. Because the dominant source for $\Phi_{-}$ is $G_{-}$ and it sources only second order perturbations, the perturbations of $\Phi_{-}$ around the ISD condition $\left(\Phi_{-}=0\right)$ satisfy
\begin{equation}\label{firsteqodspinlaplace}
\nabla^{2} \Phi_{-} = 0,
\end{equation}
at linear level. For a general warped deformed conifold, the Laplacian of equation (\ref{firsteqodspinlaplace}) takes a simple form when we are only interested in the low lying states that are dependent on only one angular coordinate $\theta$. In this case, the leading order term of the eigenfunction with the lowest angular mode $\ell=1$ is given by
\begin{equation}\label{eigenfunctionspin}
\Phi_{-} \propto \left(\cosh{\chi} \sinh{\chi} - \chi \right)^{1/3} \cos{\theta}.
\end{equation}
Adding the mass term that arises from the effects of the bulk geometry, the potential is derived with the eigenfunction (\ref{eigenfunctionspin}) as
\begin{equation}\label{spinflationpotential}
V\left(\phi^{I} \right) = T_{3} \Phi_{-} + \frac{1}{2} m_{0}^{2} \phi^{2} = T_{3} U = T_{3} \left[\frac{1}{2}m_{0}^{2}\left\{r\left(\chi\right)^{2}+c_{2}K\left(\chi\right)\sinh{\chi}\cos{\theta}\right\}+U_{0} \right],
\end{equation}
with an arbitrary constant $c_{2}$ which is smaller, or of a similar magnitude, to the deformation parameter: $c_{2} \sim \kappa^{4/3}$. Note that the constant $U_{0}$ is chosen so that the global minimum of $V$ is $V=0$ . 

\begin{figure}[!htb]
\centering
\includegraphics[width=12cm]{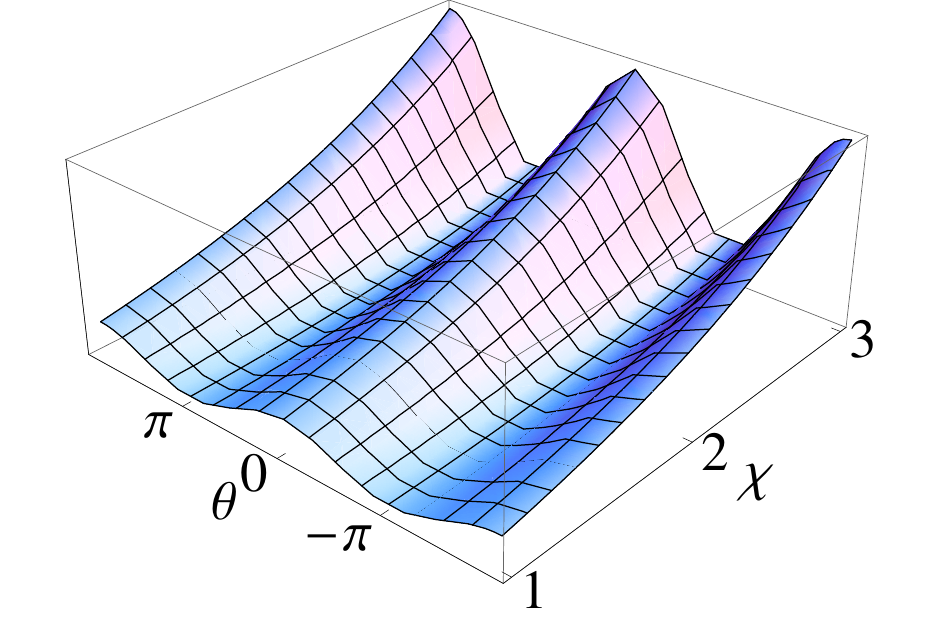}
\caption[Spinflation potential]{The potential (\ref{spinflationpotential}) with $g_{\rm{s}}=1/2\pi\,\left[\mathcal{M}^{0} \right]$, $\alpha'=1\,\left[\mathcal{M}^{-2} \right]$, $M=10^{6}\pi\,\left[\mathcal{M}^{0} \right]$, $\kappa=10^{-11}\,\left[\mathcal{M}^{-3/2} \right]$ and $m_{0}=4.5\times10^{-5}\,\left[\mathcal{M} \right]$. Note $\mathcal{M}$ is the mass unit defined in equation (\ref{massunitdefinitionspin}). The minima of the potential in the angular direction are at $\theta=\left(2N+1\right)\pi$ with an integer $N$ while the maxima are along the lines $\theta = 2N \pi$. The radial potential is quadratic in terms of $r\left(\chi\right)$.}{\label{spinflationpotentialshape}}
\end{figure}

Figure \ref{spinflationpotentialshape} shows the shape of the potential. From equations (\ref{rescalingthecoordinates}), (\ref{friedmanngeneralderived}) and (\ref{spinflationpotential}), the Friedmann equation reads
\begin{equation}\label{friedmannspinflation}
H^{2}=\frac{T_{3}}{3M^{2}_{P}}\left[\frac{1}{h}\left(\gamma-1\right)+U\right]. 
\end{equation}
Combining the time derivative of equation (\ref{friedmannspinflation}) with respect to the cosmic time $t$ and the continuity equation $\dot{\rho} = - 3 H \left(E + P \right)$, we obtain
\begin{equation}\label{spinflationhubblederivative}
\dot{H} = - \frac{T_{3}}{2 M^{2}_{P} h} \left(\gamma - \gamma^{-1} \right). 
\end{equation}
Note that $E$ denotes the energy density which is equivalent to $3M^{2}_{P} H^{2}$ while $P$ is the pressure which is equivalent to the Lagrangian given in equation (\ref{dbigenerallagrangian}). In \cite{Baumann:2007}, it is shown that the volume of the compactification is constrained by the Planck mass. Because the volume of the warped throat must be smaller than the total volume of the internal space, we have the relation
\begin{equation}\label{planckmassbound}
M^{2}_{P} \geq \frac{\kappa^{4/3}g_{\rm{s}}M^{2}T_{3}}{6 \pi} J\left(\chi_{\rm{UV}} \right),
\end{equation}
where $J\left(\chi \right) = \int d\chi I \left(\chi \right) \sinh^{2}{\chi}$ with the UV cut-off of the throat at $\chi = \chi_{\rm{UV}}$. Note that $T_{3}$ is the D3-brane tension given in equation (\ref{definephi}). Note that the mass dimension $\left[\mathcal{M} \right]$ is determined by the relation (\ref{planckmassbound}), which is rewritten as
\begin{equation}\label{massunitdefinitionspin}
M^{2}_{P} = \frac{\bar{\kappa}^{4/3}g_{\rm{s}}M^{2}\bar{T}_{3}}{6 \pi} N J\left(\chi_{\rm{UV}} \right) \mathcal{M}^{2},
\end{equation}
with the dimensionless parameters $N\geq1$, $\bar{T}_{3} = T_{3}/\mathcal{M}^{4}$ and $\bar{\kappa}^{4/3} = \kappa^{4/3} \mathcal{M}^{2}$. Note that we use only the dimensionless parameters in the numerical calculations and that $g_{\rm{s}}$ and $M$ are dimensionless. Below, all the values of the parameters are given in the mass unit $\left[\mathcal{M} \right]$ unless stated otherwise. The dimensions of the parameters are: $g_{\rm{s}}\,\left[\mathcal{M}^{0} \right]$, $M\,\left[\mathcal{M}^{0} \right]$, $T_{s}\,\left[\mathcal{M}^{4} \right]$, $\kappa^{4/3}\,\left[\mathcal{M}^{-2} \right]$ and $\alpha'\,\left[\mathcal{M}^{-2} \right]$. By saturating the Planck mass bound (\ref{planckmassbound}), equations (\ref{friedmannspinflation}) and (\ref{spinflationhubblederivative}) are rewritten as
\begin{equation}\label{spinflationsaturatedfriedmann}
H^{2}=\frac{T_{3}}{3M^{2}_{P}}\left[\frac{1}{h}\left(\gamma-1\right)+U\right] \rightarrow \frac{2 \pi}{\kappa^{4/3}g_{\rm{s}}M^{2}J \left(\chi_{\rm{UV}}\right)} \left[\frac{1}{h}\left(\gamma-1\right)+U\right], 
\end{equation}
\begin{equation}\label{spinflationsaturatedhubblederivative}
\dot{H} = - \frac{T_{3}}{2 M^{2}_{P} h} \left(\gamma - \gamma^{-1} \right) \rightarrow - \frac{3 \pi}{\kappa^{4/3}g_{\rm{s}}M^{2}J \left(\chi_{\rm{UV}}\right) h} \left(\gamma - \gamma^{-1} \right).
\end{equation}
The system of equations (\ref{phaseeqofmotionspinflationone}), (\ref{phaseeqofmotionspinflationtwo}), (\ref{spinflationsaturatedfriedmann}) and (\ref{spinflationsaturatedhubblederivative}) can be solved numerically if we set the values of the parameters. 

\section{Background dynamics}\label{sec:backgroundbehaviourspin}
\begin{figure}[htp]
  \centering
   \begin{tabular}{cc} 
    \includegraphics[width=70mm]{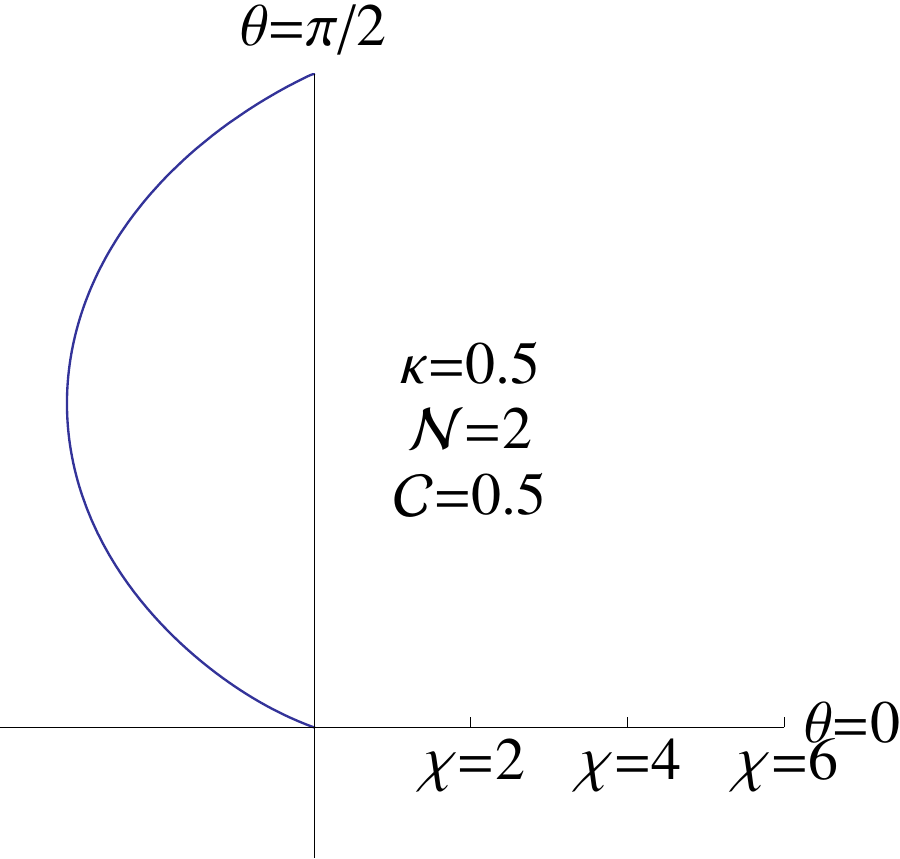}&
    \includegraphics[width=70mm]{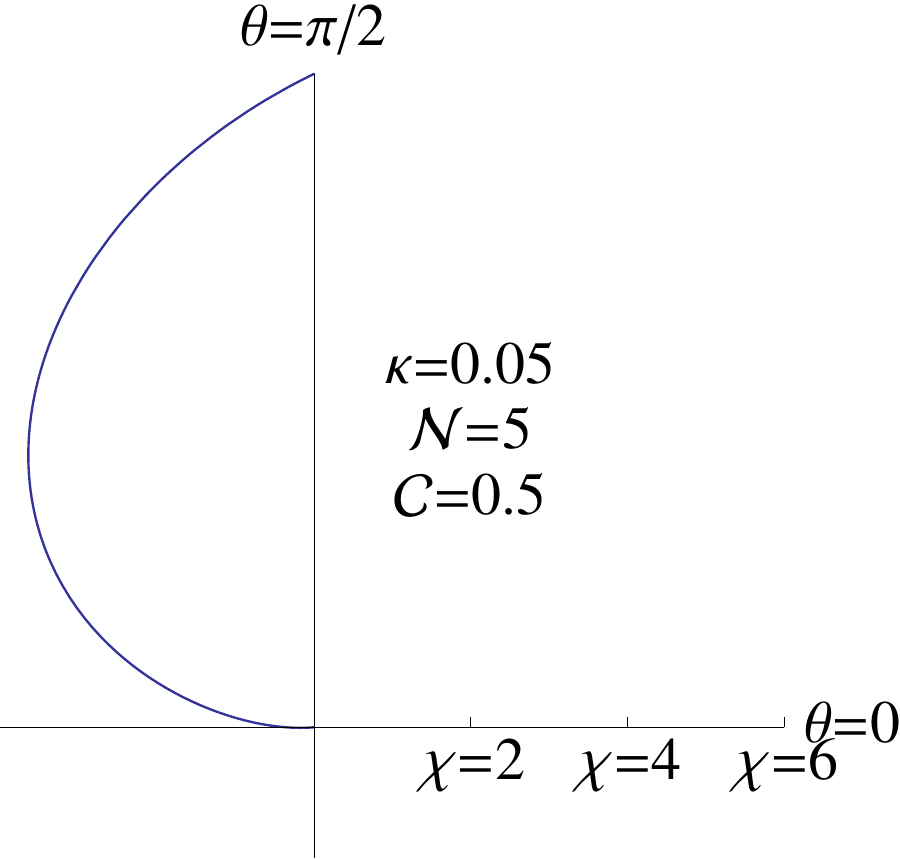}\\
    \includegraphics[width=70mm]{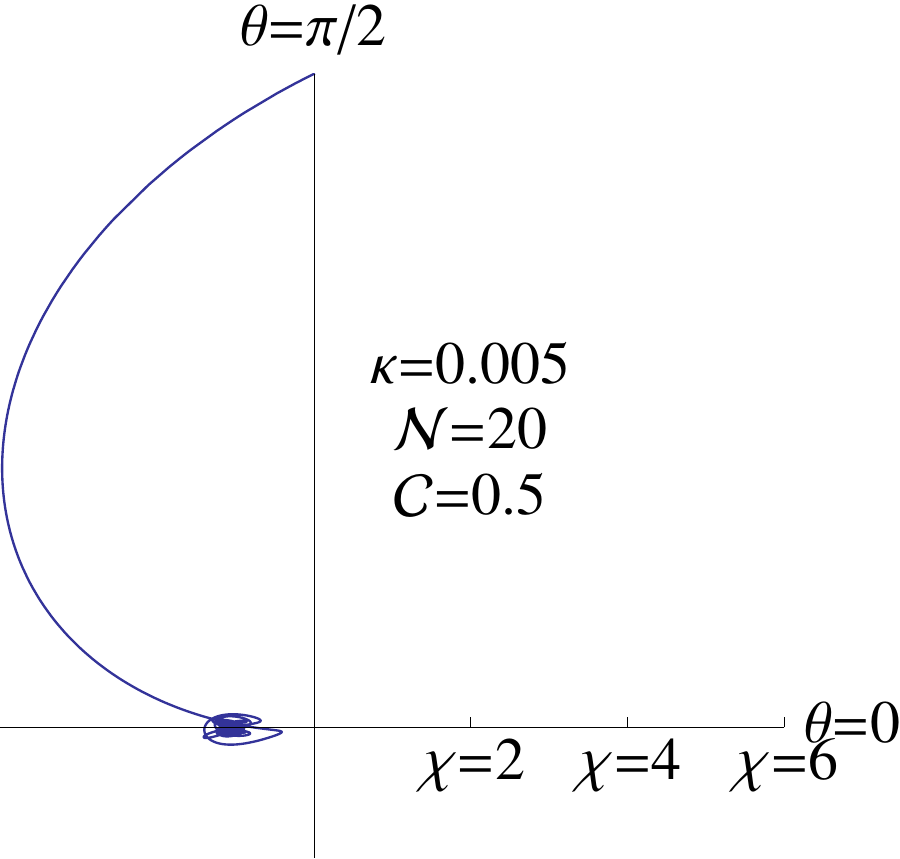}&
    \includegraphics[width=70mm]{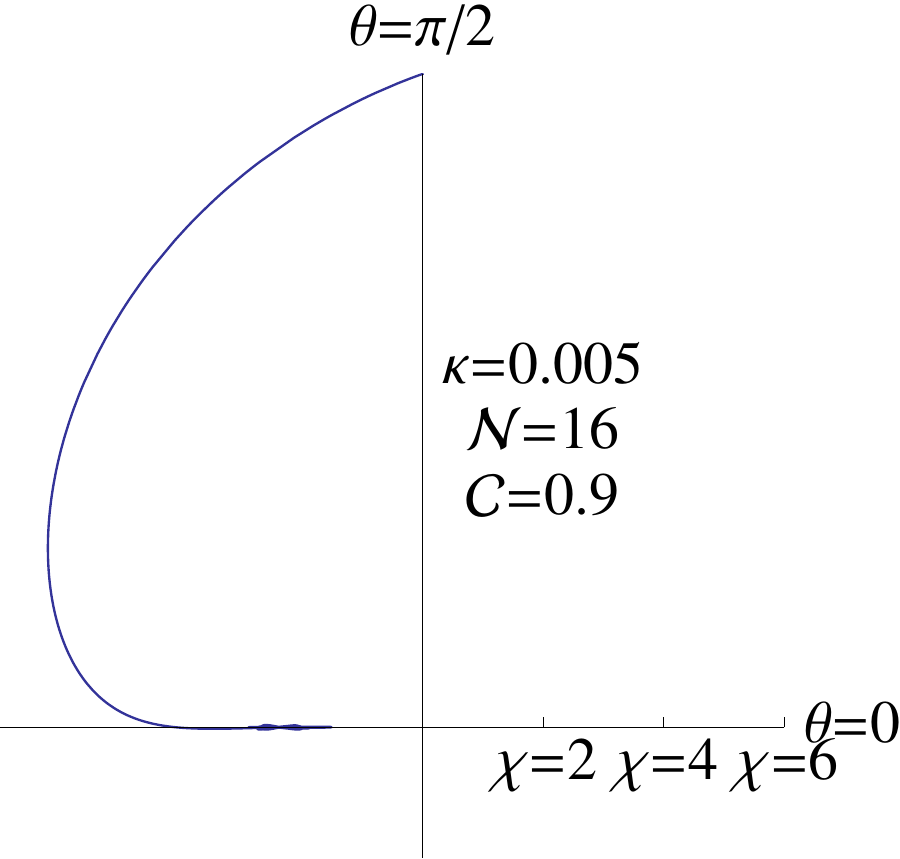}\\
    \includegraphics[width=70mm]{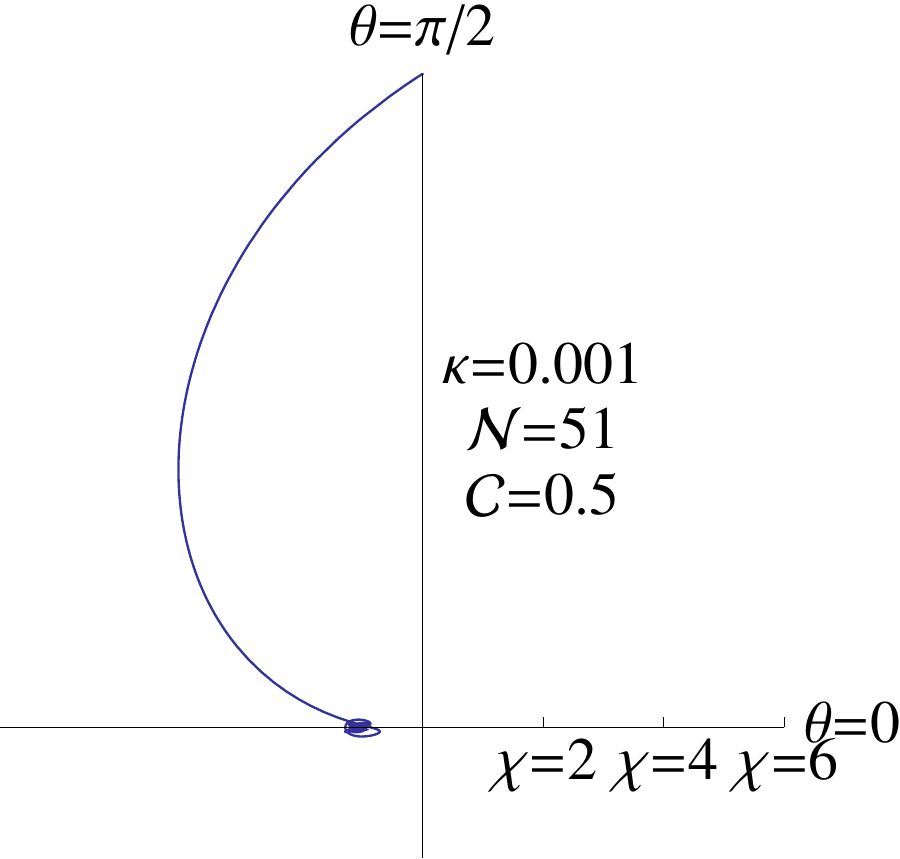}&
    \includegraphics[width=70mm]{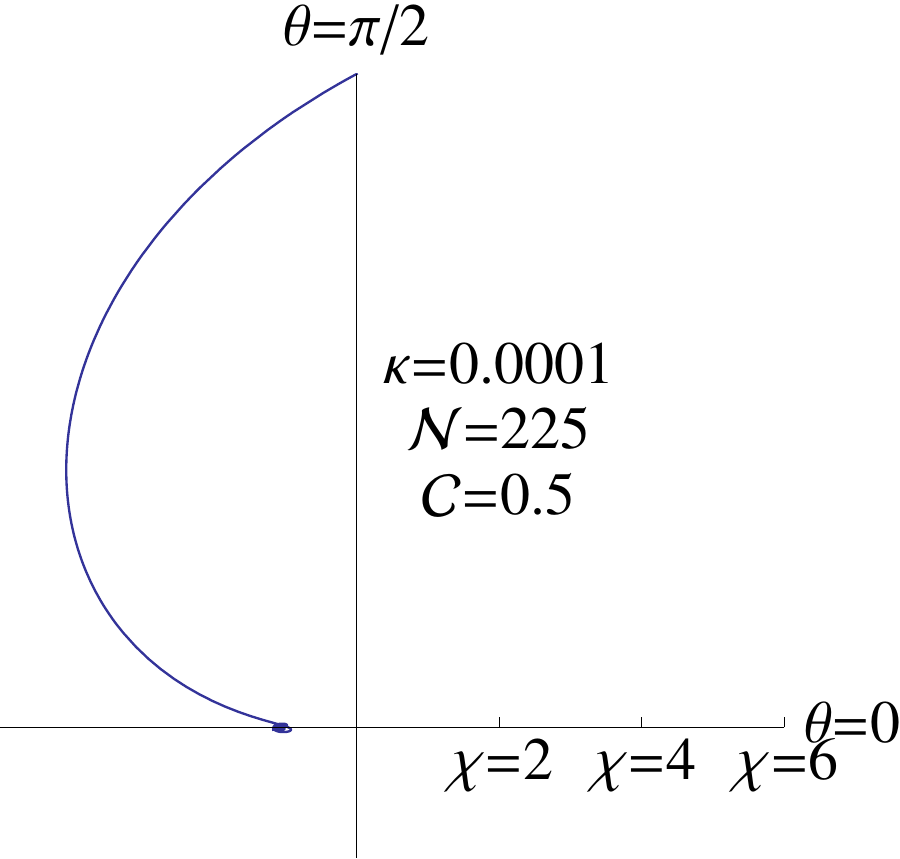}\\
  \end{tabular}
  \caption[Background trajectories]{Background inflationary trajectories with a flux parameter $g_{\rm{s}}M = 100\,\left[\mathcal{M}^{0} \right]$, inflaton mass $m_{0} = 5\,\left[\mathcal{M} \right]$ and the saturated Planck mass. Note $\mathcal{M}$ is the mass unit defined in equation (\ref{massunitdefinitionspin}). The horizontal axes denote $\chi \cos{\theta}$ while the vertical axes denote $\chi \sin{\theta}$. The value of $\kappa$ and $\mathcal{C} = c_{2} \kappa^{-4/3} [\mathcal{M}^{0}]$, are indicated. The same parameter sets are studied in \cite{Gregory:2012}. }\label{backgroundreproduction}
\end{figure}

We show the numerical results for the background trajectories with six different parameter sets following \cite{Gregory:2012} in figure \ref{backgroundreproduction}. The flux parameter $g_{\rm{s}}M$ is set to 100 and the inflaton mass $m_{0}$ is set to 5 while we changed the values of $\kappa$ and $c_{2}$. Note that we assume $\chi_{\rm{UV}} = 10$ in this section. The trajectory starts with $\chi=10$ and $\theta=\pi/2$. The initial radial brane velocity is taken to vanish while the angular brane velocity is highly relativistic. Although the results are different from those in the published version of \cite{Gregory:2012} due to a numerical problem, we confirmed some of their findings about the background dynamics \cite{Thanks:Ruth} as follows. 
\begin{itemize}
\item Decreasing the deformation parameter $\kappa$ slows down the brane and increase the number of e-folds. 
\item For all the parameter sets in figure \ref{backgroundreproduction}, regardless of the angular dependence or initial momenta, the brane rapidly becomes highly relativistic in the radial direction and makes its first sweep down the throat. 
\item Increasing the angular perturbation $c_{2}$ shifts the minimum of the potential. 
\end{itemize}

\begin{figure}[!htb]
\centering
\includegraphics[width=12cm]{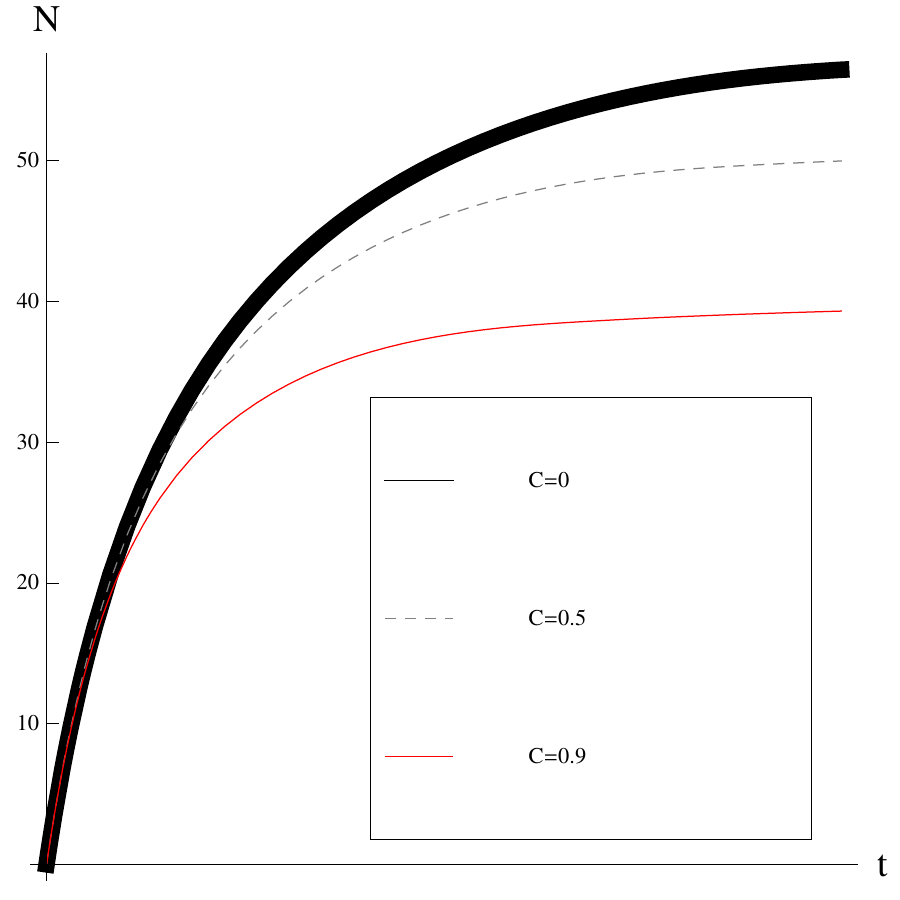}
\caption[Angular dependence]{The evolution of the number of e-folds along inflationary trajectories with $m_{0}=5\,\left[\mathcal{M} \right]$, $g_{\rm{s}}M=100\,\left[\mathcal{M}^{0} \right]$ and $\kappa=0.001\,\left[\mathcal{M}^{-3/2} \right]$.  Note $\mathcal{M}$ is the mass unit defined in equation (\ref{massunitdefinitionspin}). The thick black curve, the dashed grey curve and the red solid curve describe the trajectories with $c_{2}=0\,\left[\mathcal{M}^{-2} \right]$, $c_{2}=0.5 \kappa^{4/3}\,\left[\mathcal{M}^{-2} \right]$ and $c_{2}=0.9 \kappa^{4/3}\,\left[\mathcal{M}^{-2} \right]$ respectively. Note that $\mathcal{C} = c_{2} \kappa^{-4/3} [\mathcal{M}^{0}]$. The amount of inflation decreases as the angular dependence is increased.}{\label{angulardependence} }
\end{figure}

Below, we introduce the new findings in our numerical results. As we can see in the middle pair of the plots in figure \ref{backgroundreproduction}, increasing the angular dependence changes the trajectory and decreases the number of e-folds. This is because the values of the slow-roll parameters increase due to the angular motion. Figure \ref{angulardependence} shows the decrease in the number of e-folds due to the increase in the angular dependence in more detail with different parameter sets. This is opposite to the finding about the angular dependence in \cite{Gregory:2012}. The number of e-folds decreases by about 20 $\%$ when increasing $C=c_{2}\kappa^{-4/3}$ from 0.5 to 0.9 regardless of the value of $\kappa$ as shown in figure \ref{backgroundreproduction} and figure \ref{angulardependence}. We also checked that the effect on the number of e-folds stays the same even if we change the inflaton mass $m_{0}$. Therefore, the angular terms have some impacts on the background dynamics even though they are still subdominant.

In the top pair of the plots in figure \ref{backgroundreproduction}, we show the trajectories until the brane reaches the tip of the warped throat. With those parameter sets, the brane goes to the tip of the throat without reaching the minimum of the potential. This means that the radial coordinate $\chi$ overshoots the minimum in the radial direction before the angular coordinate $\theta$ reaches its minimum at $\theta=\pi$. This is because the velocity of the brane is high with a large deformation parameter as explained below. When the brane moves relativistically, the sound speed $c_{s}$ approaches unity and the speed limiting effect appears \cite{Underwood:2008}. Equation (\ref{spinflationsoundspeedandfactor}) shows that the maximum speed of the brane is higher with a large deformation parameter $\kappa$ because of the factor $\kappa^{-8/3}$ in equation (\ref{warpfactor}). In other plots in figure \ref{backgroundreproduction}, the brane velocity is suppressed by the small deformation parameters and the brane moves slowly enough to settle at the minimum of the potential after some oscillations. The oscillations are smaller with a smaller deformation parameter because of the speed limiting effect. We checked that the increase in the number of e-folds during the oscillations around the minimum of the potential is negligible regardless of the choice of the parameter set and inflation occurs mainly in the initial sweep down the throat. 

\section{Multi-field effects}\label{sec:multifieldeffectspin}
In this section, we first show that the spinflation model can be approximated with the simple model introduced in subsection \ref{subsec:analysisofsimplemodeldbi} using our numerical results when the coupling between the radial and angular fields is small around horizon crossing. Then, we study the cases where the couplings are too large to use those approximations. We show how the power spectrum of the curvature perturbation is enhanced with our numerical results. 

\subsection{Analytic formulae}\label{subsec:analyticconfirmspinlast}
The warp factor (\ref{warpfactor}) is approximated as \cite{Gregory:2012, Klebanov:2000}
\begin{equation}\label{largechiwarpthroat}
h\left(\chi\right) \sim \frac{27}{8} \frac{\left(g_{\rm{s}}M\alpha' \right)^2}{r\left(\chi \right)^4} \left(\ln{\frac{r\left(\chi \right)^{3}}{\kappa^2}} + \ln{\frac{4\sqrt{2}}{3\sqrt{3}}} - \frac{1}{4} \right),
\end{equation}
for large $\chi > 1$. The angular term in the potential (\ref{spinflationpotential}) is always smaller than the radial term because $c_{2}$ is smaller than $\kappa^{4/3}$ while $r\left(\chi \right)^{2}$ is of the order of $\kappa^{4/3}$. Also, the constant term $U_{0}$ is small by definition because the global minimum of the potential is at a point where $\chi \ll 1$ where both terms are negligible. Even though the angular term is not negligible in general, the radial term affects the dynamics dominantly when the motion of the brane is mainly in the radial direction. In such cases, the potential is approximated as
\begin{equation}\label{largechipotential}
V\left(\phi^{I} \right) \sim T_{3} U = T_{3} \left[ \frac{1}{2}m_{0}^{2} r\left(\chi\right)^{2} \right]. 
\end{equation}
When we compare the spinflation model in this chapter with the simple model in subsection \ref{subsec:analysisofsimplemodeldbi}, we can identify the fields $\chi$ and $\theta$ in this chapter with the dimensionless coordinates in equation (\ref{inducedmetric}) as mentioned in section \ref{sec:spinflationfirst}. Therefore, the canonical field $\phi$ with a mass dimension $\left[\mathcal{M} \right]$ is given by
\begin{equation}\label{spinexamplecanonicalfield}
\phi \left(\chi\right) = \sqrt{T_{3}} r \left(\chi\right) = \sqrt{T_{3}} \frac{\kappa^{2/3}}{\sqrt{6}}\int^{\chi}_{0}\frac{dx}{K\left(x\right)},
\end{equation}
where the dimensions of $T_{3}$ and $\kappa^{2/3}$ are $\left[\mathcal{M}^{4} \right]$ and $\left[\mathcal{M}^{-1} \right]$ respectively. 

Regarding the logarithmic dependence of $r$ as constant, the warp factor (\ref{largechiwarpthroat}) is approximated by equation (\ref{dbisingleexamplewarp}) with
\begin{equation}\label{largechilambda}
\begin{split}
\lambda &\equiv \frac{27\,T_{3}}{8} \left(g_{\rm{s}}M\alpha' \right)^{2} \left(\ln{\frac{r\left(\chi \right)^{3}}{\kappa^2}} + \ln{\frac{4\sqrt{2}}{3\sqrt{3}}} - \frac{1}{4} \right)\\ 
&= \frac{27}{64 \pi^3} g_{\rm{s}} M^{2} \left(\ln{\frac{r\left(\chi \right)^{3}}{\kappa^2}} + \ln{\frac{4\sqrt{2}}{3\sqrt{3}}} - \frac{1}{4} \right),
\end{split}
\end{equation}
where we used equation (\ref{definephi}). Note that $f$ in equation (\ref{dbisingleexamplewarp}) is the rescaled warp factor $f = h/T_{3}$ with $h$ in equation (\ref{largechiwarpthroat}) as defined in equation (\ref{rescalingthecoordinates}). The potential (\ref{largechipotential}) is approximated by equation (\ref{dbisingleexamplepotential}) with
\begin{equation}
m \equiv m_{0}.
\end{equation}
We now show that the analytic formulae in subsection \ref{subsec:analysisofsimplemodeldbi} predict the numerical results with considerable accuracy. We consider a model with $g_{\rm{s}} = 1/2\pi \left[\mathcal{M}^{0} \right]$, $M = 1.2 \times 10^{6} \pi \left[\mathcal{M}^{0} \right]$, $m_{0} = 3 \times 10^{-3} \left[\mathcal{M} \right]$, $\kappa = 10^{-8} \left[\mathcal{M}^{-3/2} \right]$, $N = 1 \left[\mathcal{M}^{0} \right]$, $\alpha' = 10 \left[\mathcal{M}^{-2} \right]$, $\chi_{\rm{UV}} = 10 \left[\mathcal{M}^{0} \right]$ and $C = c_{2} \kappa^{-4/3} = 0.5 \left[\mathcal{M}^{0} \right]$. Note that all the quantities have the units associated with the mass unit $\mathcal{M}$ defined in equation (\ref{massunitdefinitionspin}). With those parameters, the mass unit $\mathcal{M}$ is defined by equation (\ref{massunitdefinitionspin}) as
\begin{equation}\label{massunitspinexample}
\mathcal{M} \simeq 0.0138 M_{\rm{P}}. 
\end{equation}
Therefore, for example, the string scale $1/\alpha \, \left[\mathcal{M}^{2} \right]$ in the Planck units is given by
\begin{equation}
\frac{\mathcal{M}^{2}}{\alpha' M^{2}_{\rm{P}}} = 1.91 \times 10^{-5} \left[M^{2}_{\rm{P}} \right]. 
\end{equation}

\begin{figure}[htp]
  \centering
   \begin{tabular}{cc} 
    \includegraphics[width=70mm]{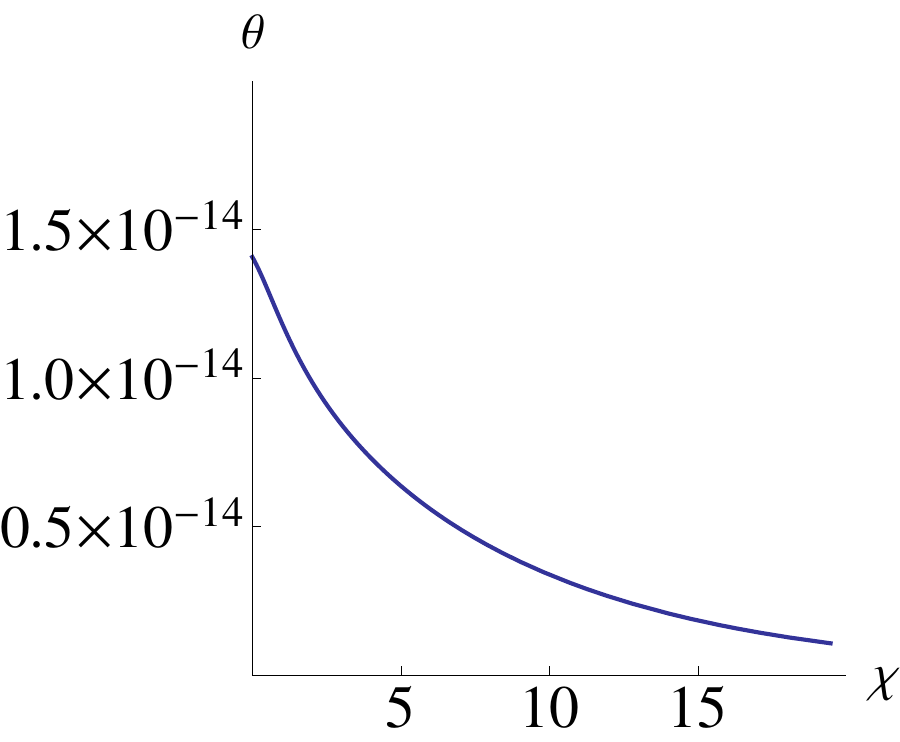}&
    \includegraphics[width=70mm]{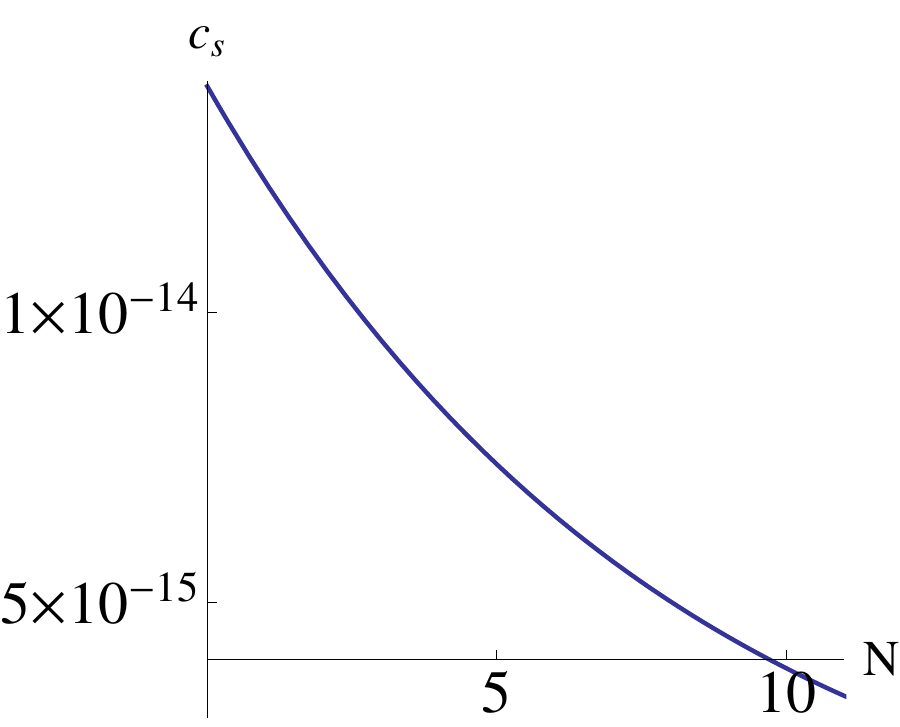}
  \end{tabular}
  \caption[Background trajectory and sound speed]{Left: The back ground trajectory in the $\chi$-$\theta$ plane of the brane moving down the throat along the maximum of the potential in the angular direction with a small displacement from the maximum. Right: The evolution of the sound speed in the early stage of inflation with respect to the number of e-folds.}\label{spinchithetaandsound}
\end{figure}

Figure \ref{spinchithetaandsound} shows the numerical results for the trajectory in the field space and the sound speed. The initial position of the brane is $\left(\chi,\,\theta \right) = \left(20, 10^{-15} \right)$. The initial velocity is only in the radial direction, even though we confirmed that the velocity becomes highly relativistic only in the radial direction, regardless of the initial velocity, when the trajectory is close to the maximum. In the left panel of figure \ref{spinchithetaandsound}, it is shown that the trajectory is bent towards the angular direction slowly and the deviation from the maximum of the potential in the angular direction becomes larger gradually. Below, we consider the perturbation that exits the horizon around $N \sim 2$. Using the numerical results, the value of the canonical field (\ref{spinexamplecanonicalfield}) around horizon crossing in the Planck units is
\begin{equation}
\frac{\phi \left(\chi \right)}{M_{\rm{P}}} = 4.50 \times 10^{-5} \frac{\mathcal{M}}{M_{\rm{P}}} = 6.22 \times 10^{-7} \left[M_{\rm{P}} \right], 
\end{equation}
where we used equation (\ref{massunitspinexample}). From equation (\ref{alltheresultsindbisingleexample}), the sound speed is given by
\begin{equation}\label{soundspeedpredictionspin}
c_{s} = \sqrt{\frac{3}{\lambda}} \frac{M_{\rm{P}}}{2 \bar{m}_{0} \mathcal{M}} \left(\frac{\phi}{M_{\rm{P}}}\right)^{2} \simeq 1.05 \times 10^{-14},
\end{equation}
with the dimensionless parameters $\bar{m}_{0} = m_{0}/\mathcal{M}$ and $\lambda = 5.78 \times 10^{11}$ that is given by equation (\ref{largechilambda}). In the right panel of figure \ref{spinchithetaandsound}, we see that the analytic formula (\ref{soundspeedpredictionspin}) predicts the sound speed around $N \sim 2$ with great accuracy.

\begin{figure}[htp]
  \centering
   \begin{tabular}{ccc} 
    \includegraphics[width=45mm]{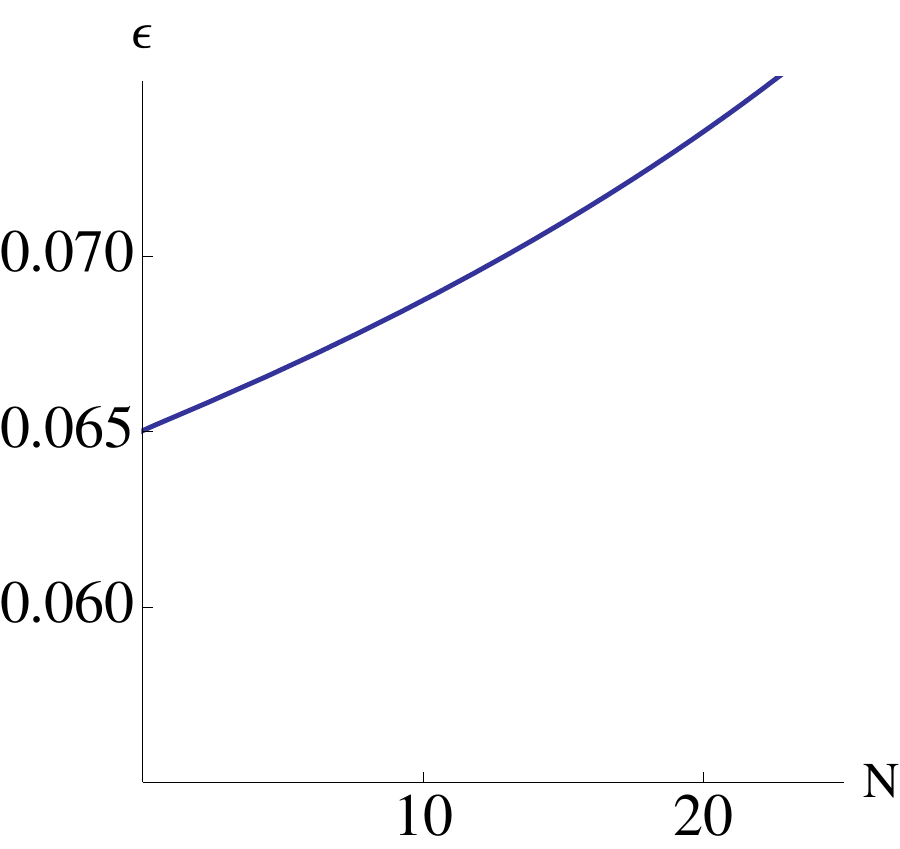}&
    \includegraphics[width=45mm]{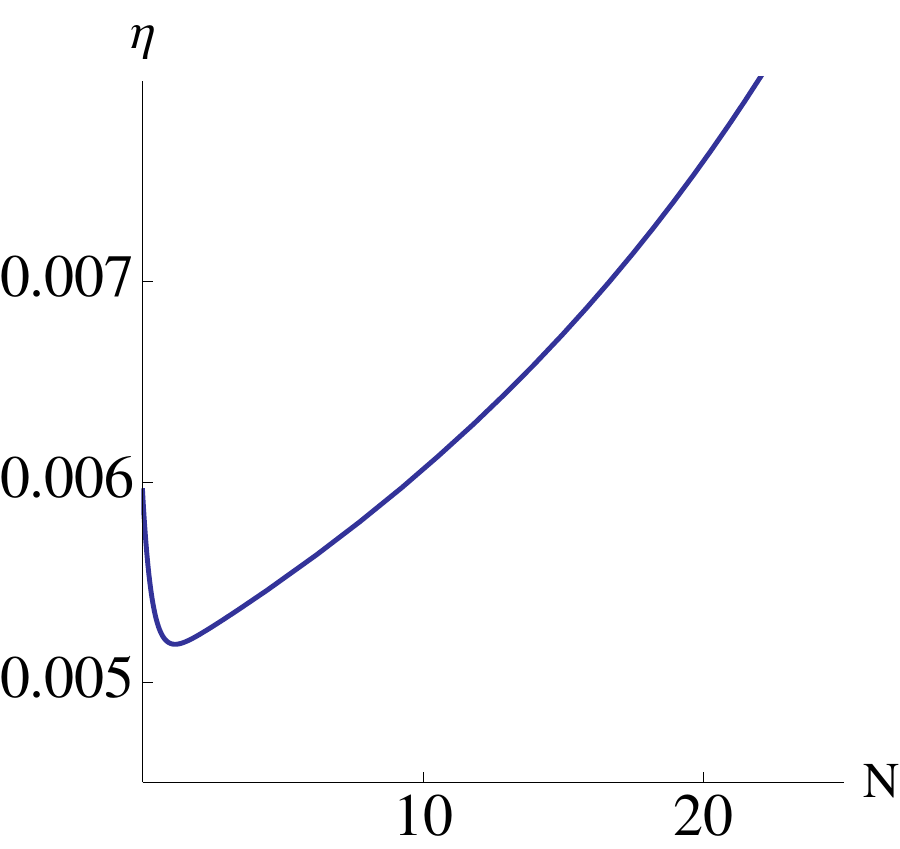}&
    \includegraphics[width=45mm]{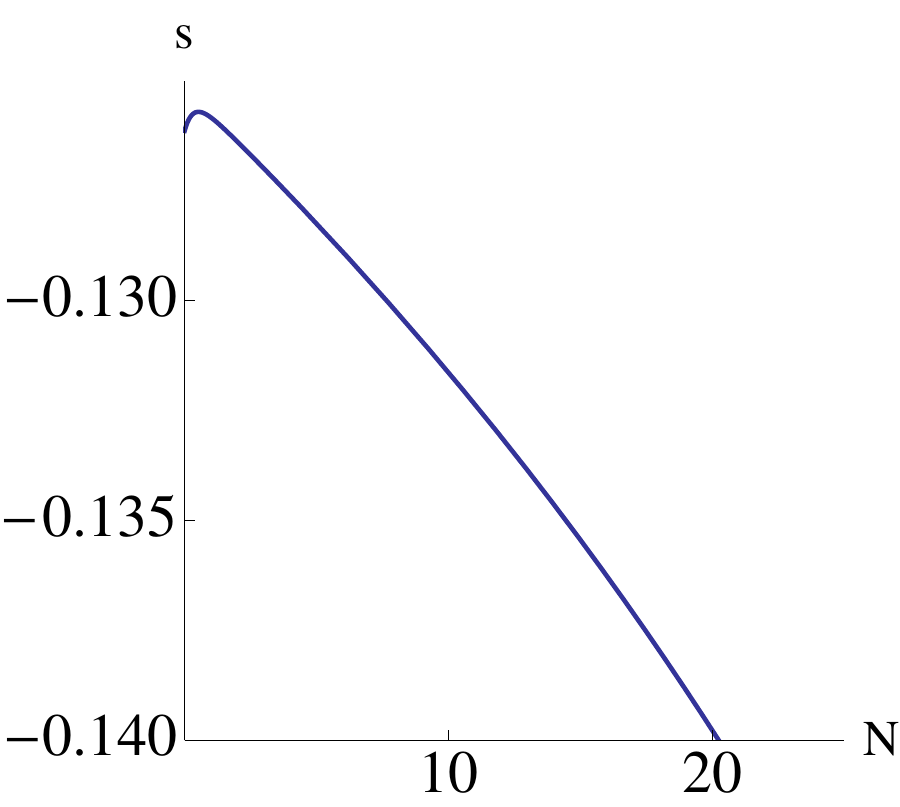}   
  \end{tabular}
  \caption[Slow-roll parameters]{Left: The slow-roll parameter $\epsilon$. Middle: The slow-roll parameter $\eta$. Right: The slow-roll parameter $s$. All the horizontal axes denote the number of e-folds.}\label{spinslowrollexample}
\end{figure}

Figure \ref{spinslowrollexample} shows the behaviour of the slow-roll parameters. From equation (\ref{alltheresultsindbisingleexample}), the analytic prediction of the slow-roll parameter $\epsilon$ is given by
\begin{equation}\label{spinepsilonpredictionexample}
\epsilon = \sqrt{\frac{3}{\lambda}} \frac{M_{\rm{P}}}{\bar{m}_{0} \mathcal{M}} = 0.0545. 
\end{equation}
In the left panel of figure \ref{spinslowrollexample}, we see that the value of $\epsilon$ is predicted with the analytic formula (\ref{spinepsilonpredictionexample}) with around 20 $\%$ error. In the middle panel, $\eta$ is much smaller than $\epsilon$, whereas it is expected to vanish in equation (\ref{relationchapthreesimple}). The right panel shows that the second relation in equation (\ref{relationchapthreesimple}) holds as $s \simeq -2 \epsilon$. 

\begin{figure}[htp]
  \centering
   \begin{tabular}{ccc} 
    \includegraphics[width=45mm]{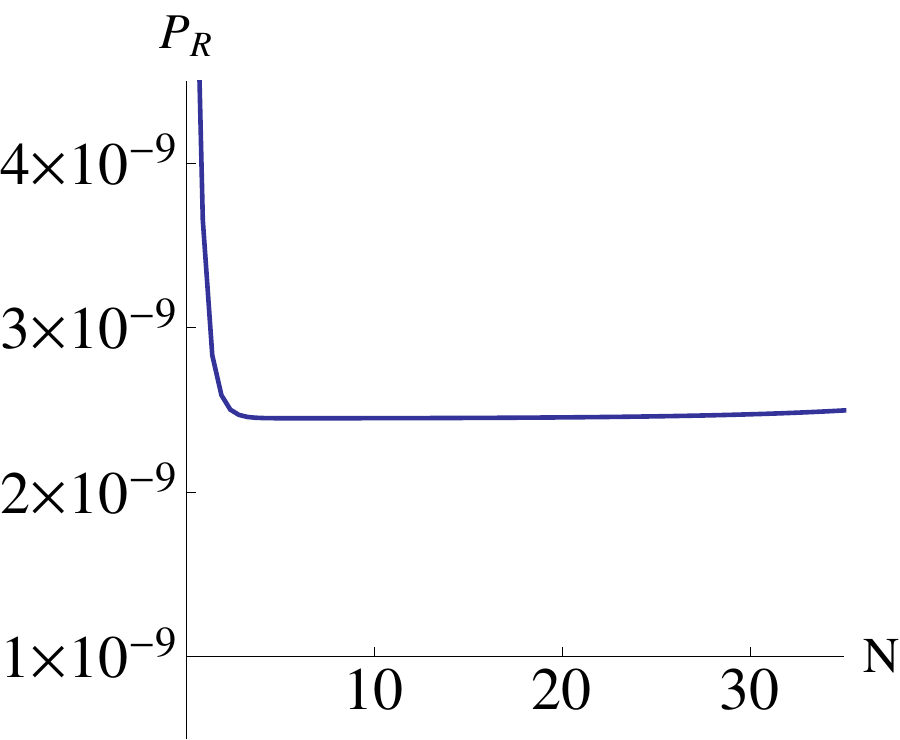}&
    \includegraphics[width=45mm]{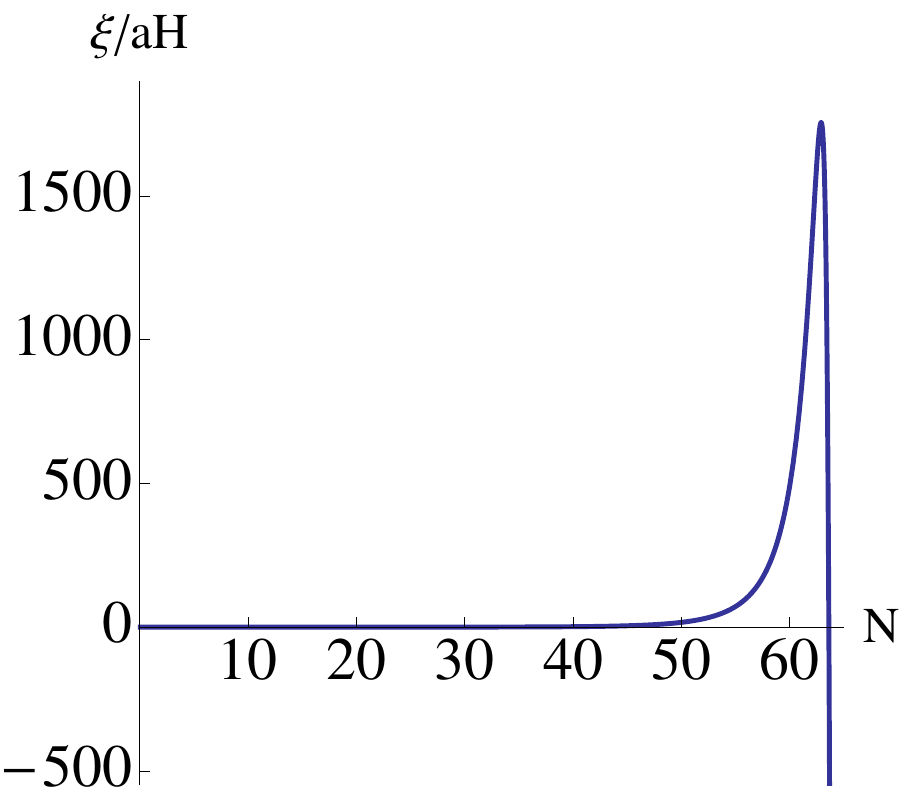}
    \includegraphics[width=45mm]{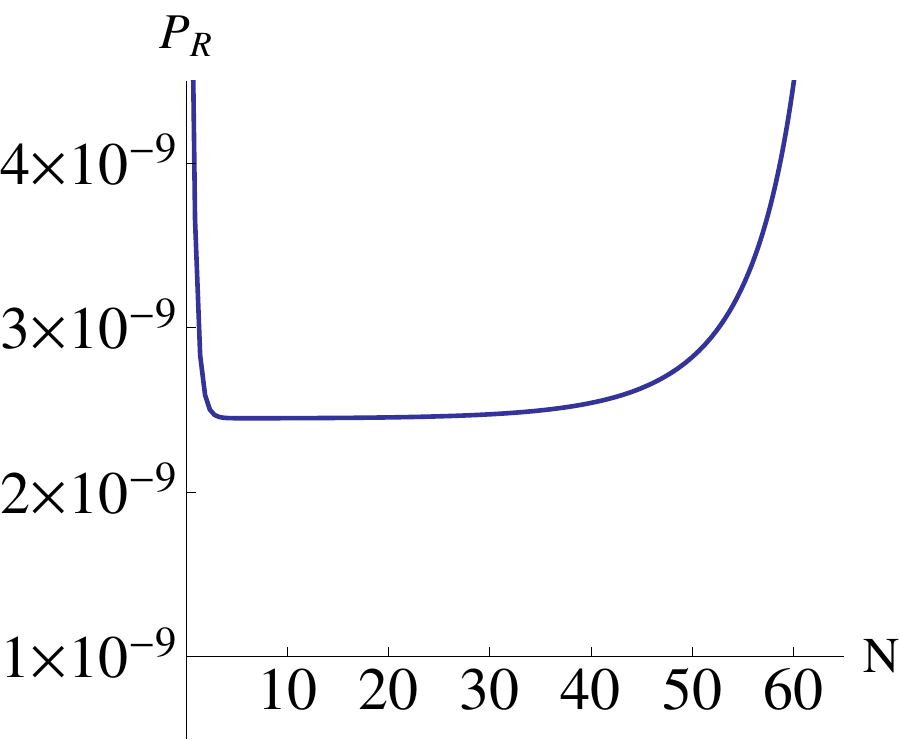}
  \end{tabular}
  \caption[Power spectrum of the curvature perturbation]{Left: The curvature power spectrum in the early stage of inflation. Middle: The coupling. It increases rapidly in the late stage of inflation. Right:The behaviour of the curvature power spectrum until the end of inflation. }\label{spincurvatureandcouplingexample}
\end{figure}

As stated above, the trajectory is bent towards the angular direction gradually. As shown in the middle panel of figure \ref{spincurvatureandcouplingexample}, the coupling is negligible in the early stage of inflation and becomes larger rapidly in the late stage. Therefore, the curvature power spectrum $\mathcal{P}_{\mathcal{R}}$ is almost constant in the early stage of inflation as shown in the left panel of figure \ref{spincurvatureandcouplingexample}. Because this is the effective single field phase, it shows the same behaviour as the power spectrum of the curvature perturbation in the single field inflation models \cite{Nalson:2013, Nalson:2013b}. The value of the power spectrum is predicted with equation (\ref{curvaturespectrumexampledbisingle}) as
\begin{equation}\label{curvatureexamplespin}
\mathcal{P}_{\mathcal{R}_{*}} = \frac{1}{4 \pi^{2} \epsilon^4 \lambda} \simeq 2.42 \times 10^{-9},
\end{equation}
where we used $\epsilon = 0.065$ and $\lambda = 5.78 \times 10^{11}$ that is obtained with equation (\ref{largechilambda}). In the left panel of figure \ref{spincurvatureandcouplingexample}, it is shown that the approximated analytic formula (\ref{curvatureexamplespin}) predicts the power spectrum of the curvature perturbation with great accuracy when the brane has the effective single field dynamics. In the right panel, we see that the curvature power spectrum is enhanced in the late stage of inflation because of the coupling between the adiabatic and entropy perturbations. Even though it is enhanced only by the factor of 2 with those parameters and the initial conditions, the conversion of the entropy perturbation into the curvature perturbation becomes larger if we make the initial displacement from the angular maximum larger. 

\subsection{Trajectories along the minimum}\label{minimumtrajectories}
\begin{figure}[htb]
  \centering
   \begin{tabular}{cc} 
    \includegraphics[width=70mm]{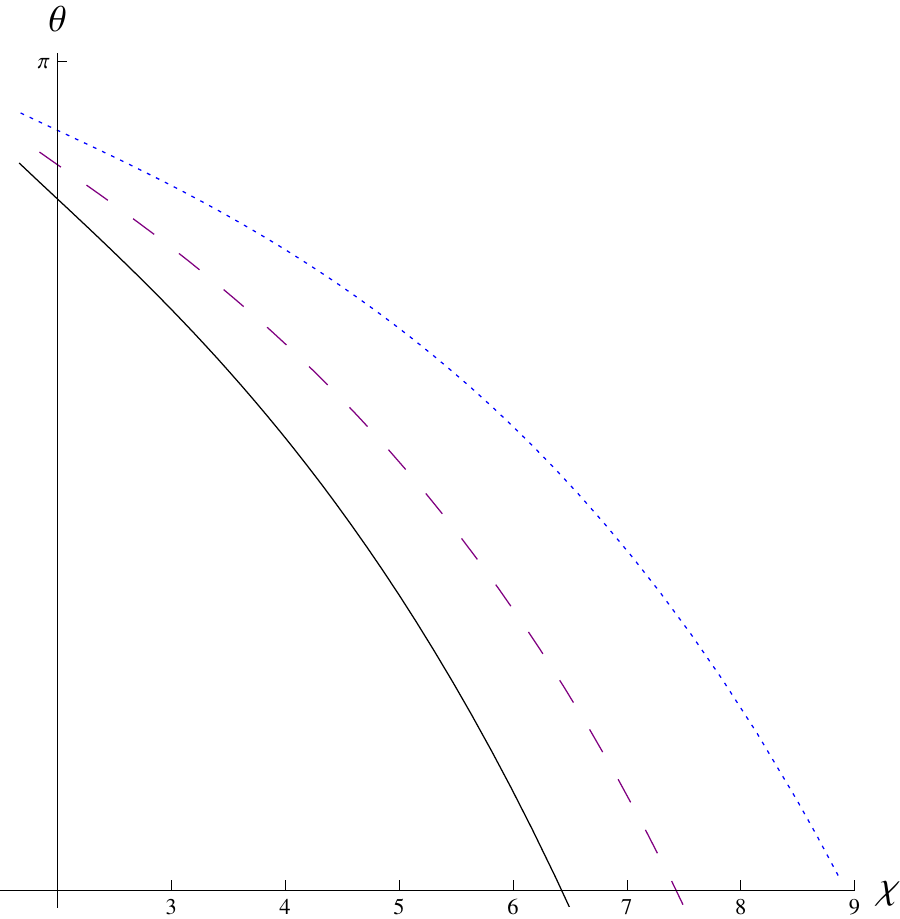}&
    \includegraphics[width=70mm]{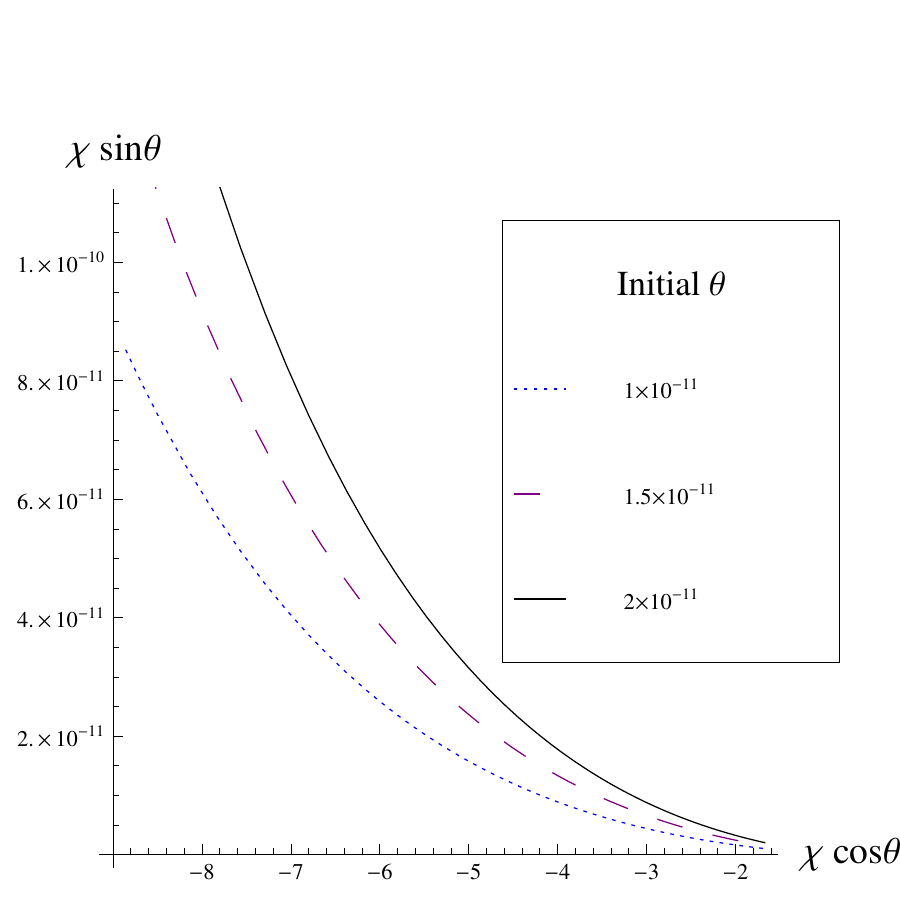}
  \end{tabular}
  \caption[Trajectories along the minimum]{Trajectories along the minimum of the potential at $\theta=\pi$. Left: The trajectories in the $\chi$-$\theta$ plane. Right: The trajectories in the phase space.}\label{trajectoriesalongminimumspinflation}
\end{figure}

\begin{figure}[!htb]
  \centering
   \begin{tabular}{ccc} 
    \includegraphics[width=45mm]{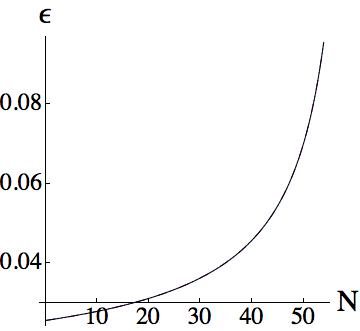}&
    \includegraphics[width=45mm]{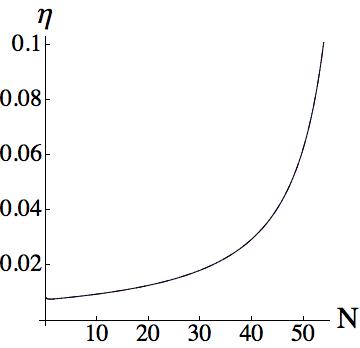}&
     \includegraphics[width=45mm]{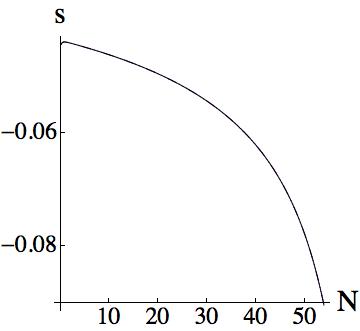}   
  \end{tabular}
  \caption[Slow-roll parameters]{Left: The slow-roll parameter $\epsilon$. Middle: The slow-roll parameter $\eta$. Left: The slow-roll parameter $s$. All the slow-roll parameters behave in the same way for all the trajectories with $\theta = 1 \times 10^{-11}$, $1.5 \times 10^{-11}$ and $2 \times 10^{-11}$ because the displacements are small. The slow-roll approximation holds until the end of inflation.}\label{spinflationslowrollparametersmini}
\end{figure}

\begin{figure}[!htb]
\centering
\includegraphics[width=10cm]{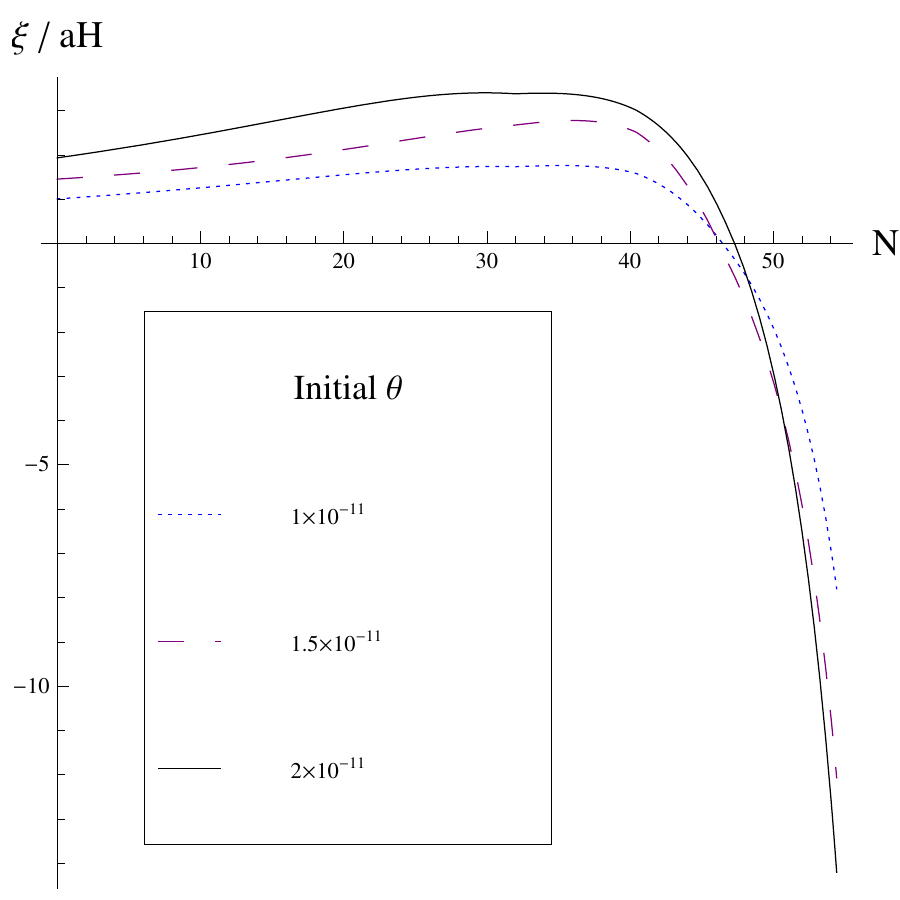}
\caption[Coupling along the minimum]{The evolution of the coupling $\xi/a H$ in terms of the number of e-folds $N$ along the minimum of the potential in the angular direction.}{\label{couplingspinflationmini}}
\end{figure}

\begin{figure}[!htb]
\centering
\includegraphics[width=12cm]{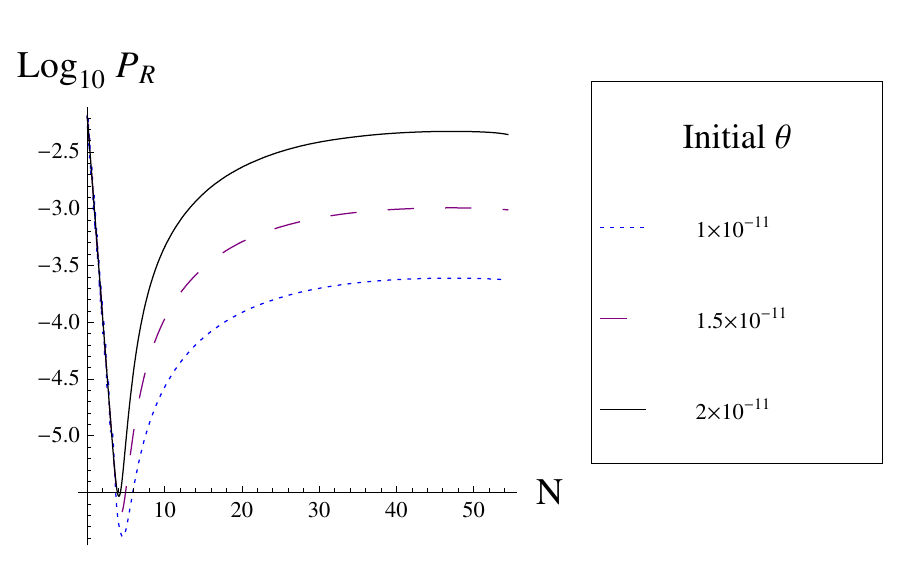}
\caption[Power spectrum of the curvature perturbation]{The evolution of the power spectrum of the curvature perturbation in terms of the number of e-folds $N$ along the minimum of the potential in the angular direction. The scale of interest exits the horizon around $N \sim 5$ for all the trajectories. }{\label{curvaturepowerspectrumminispinflation}}
\end{figure}

In this subsection, we show how the power spectrum of the curvature perturbation is enhanced when we consider trajectories that start with slight deviations from the minimum of the potential in the angular direction. As shown in figure \ref{spinflationpotentialshape}, the potential has its minima in the angular direction at $\theta=\left(2N+1 \right)\pi$ where $N$ is an integer number. We assume that the trajectories start at $\left(\chi,\theta\right)=\left(9,\pi - \delta\theta\right)$ with $\delta\theta \ll 1$. The reason that we consider such trajectories is that the coupling $\xi/a H$ is small when the trajectory is a gentle curve around the minimum. If the coupling is not too large around horizon crossing, we can use analytic expressions that are useful to study the model, such as the solutions of the equations of motion for the linear perturbations (\ref{dbisolone}) and (\ref{dbisoltwo}), by starting the numerical calculations when the scale of interest is well within the horizon $k \gg a H/c_{s}$. Our numerical calculations show that the brane quickly becomes highly relativistic in the radial direction bending slowly towards the angular direction even if we set the initial velocity highly relativistic only in the angular direction. On the other hand, if the trajectory starts in the middle of the hill of the potential, it is bent towards the angular direction even if the initial velocity is only in the radial direction producing large coupling terms with $\xi/a H \gg 1$. Therefore, even though we have more conversion of the entropy perturbation to the curvature perturbation with a larger coupling, we study those cases in which the coupling is small and see how much conversion we have in those cases. Below, we show the numerical results for three different trajectories with $\delta\theta = 1 \times 10^{-11}$, $\delta\theta = 1.5 \times 10^{-11}$ and $\delta\theta = 2 \times 10^{-11}$ which will be described with a blue dotted line, a purple dashed line and a black solid line respectively in figures \ref{trajectoriesalongminimumspinflation}, \ref{spinflationslowrollparametersmini}, \ref{couplingspinflationmini} and \ref{curvaturepowerspectrumminispinflation}. 

In figure \ref{trajectoriesalongminimumspinflation}, three trajectories with the different initial displacements from the minimum are shown. In the left panel, the trajectories are shown in the $\chi$-$\theta$ plane. As the brane goes down the throat, it approaches the minimum at $\theta=\pi$ in all the trajectories. The more the initial displacement from the minimum is, the more the trajectory is bent. In the right panel, the trajectories are shown in the phase space. We see that the angle $\theta$ gradually approaches $\pi$ with the radius $\chi$ decreasing. In figure \ref{spinflationslowrollparametersmini}, the slow-roll parameters are shown. Because the trajectories are close to each other, all the slow-roll parameters behave in almost the same way. All the slow-roll parameters are much smaller than unity until the end of inflation. 

The numerical results for the coupling are shown in figure \ref{couplingspinflationmini}. A trajectory with a larger initial displacement has a larger coupling until the late stage of inflation. The coupling becomes smaller as inflation proceeds because the trajectories approach the minimum of the potential. 
Finally, we show the evolution of the power spectrum of the curvature perturbation that has been obtained numerically in figure \ref{curvaturepowerspectrumminispinflation}. The values of the power spectrum of the curvature perturbation around horizon crossing are of the same order as $\sim 2.5 \times 10^{-6}$ in all the trajectories. However, the final values are $2.5 \times 10^{-4}$, $1 \times 10^{-3}$ and $4.8 \times 10^{-3}$ for the trajectories with the initial displacements $\delta\theta = 1 \times 10^{-11}$, $1.5 \times 10^{-11}$ and $2 \times 10^{-11}$ respectively. The trajectory with $\delta\theta =1 \times 10^{-11}$ has the amplitude of the curvature perturbation that is ten times smaller than that of the trajectory with $\delta\theta =1 \times 10^{-11}$. For the trajectory with $\delta\theta =2 \times 10^{-11}$, $\cos{\Theta}$ defined in equation (\ref{sintheta}) is given by $\cos^{2}{\Theta} \simeq 5 \times 10^{-4}$. From figure \ref{curvaturepowerspectrumminispinflation}, it is clear that the conversion is larger if the initial displacement is larger. 

\subsection{Trajectories along the maximum}
\begin{figure}[htb]
  \centering
   \begin{tabular}{cc} 
    \includegraphics[width=70mm]{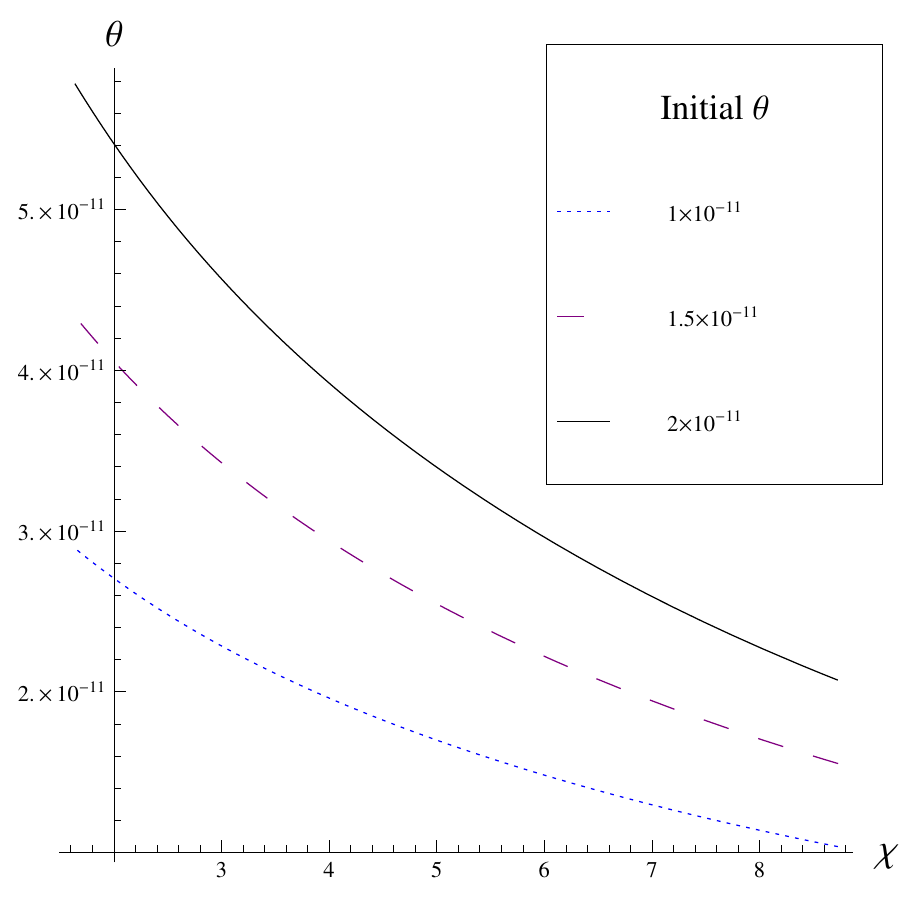}&
    \includegraphics[width=70mm]{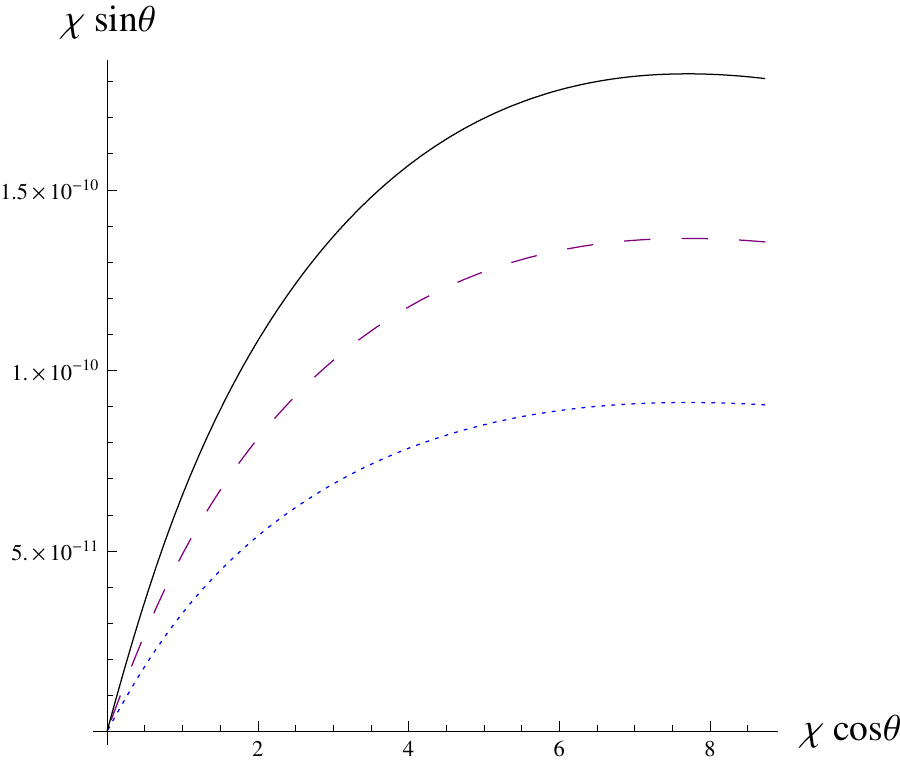}
  \end{tabular}
  \caption[Trajectories along the maximum]{Trajectories along the maximum of the potential at $\theta=0$. Left: The trajectories in the $\chi$-$\theta$ plane. Right: The trajectories in the phase space.}\label{trajectoriesalongmaximumspinflation}
\end{figure}

\begin{figure}[!htb]
  \centering
   \begin{tabular}{ccc} 
    \includegraphics[width=45mm]{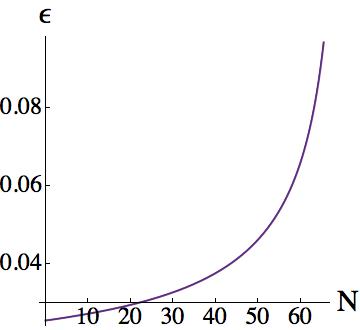}&
    \includegraphics[width=45mm]{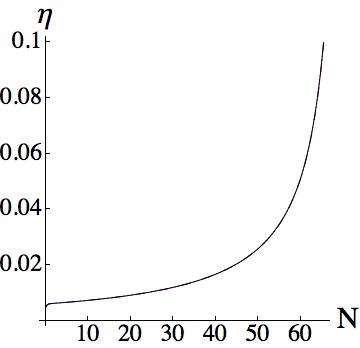}&
     \includegraphics[width=45mm]{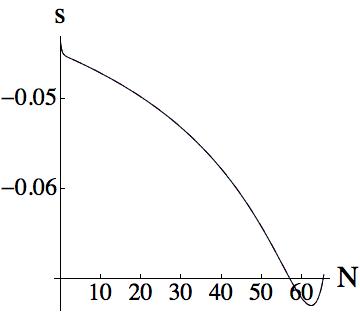}   
  \end{tabular}
  \caption[Slow-roll parameters]{Left: The slow-roll parameter $\epsilon$. Middle: The slow-roll parameter $\eta$. Left: The slow-roll parameter $s$. All the slow-roll parameters behave in the same way for all the trajectories with $\theta = 1 \times 10^{-11}$, $1.5 \times 10^{-11}$ and $2 \times 10^{-11}$ because the displacements are small. The slow-roll approximation holds until the end of inflation.}\label{spinflationslowrollparametersmax}
\end{figure}

\begin{figure}[htb]
\centering
\includegraphics[width=10cm]{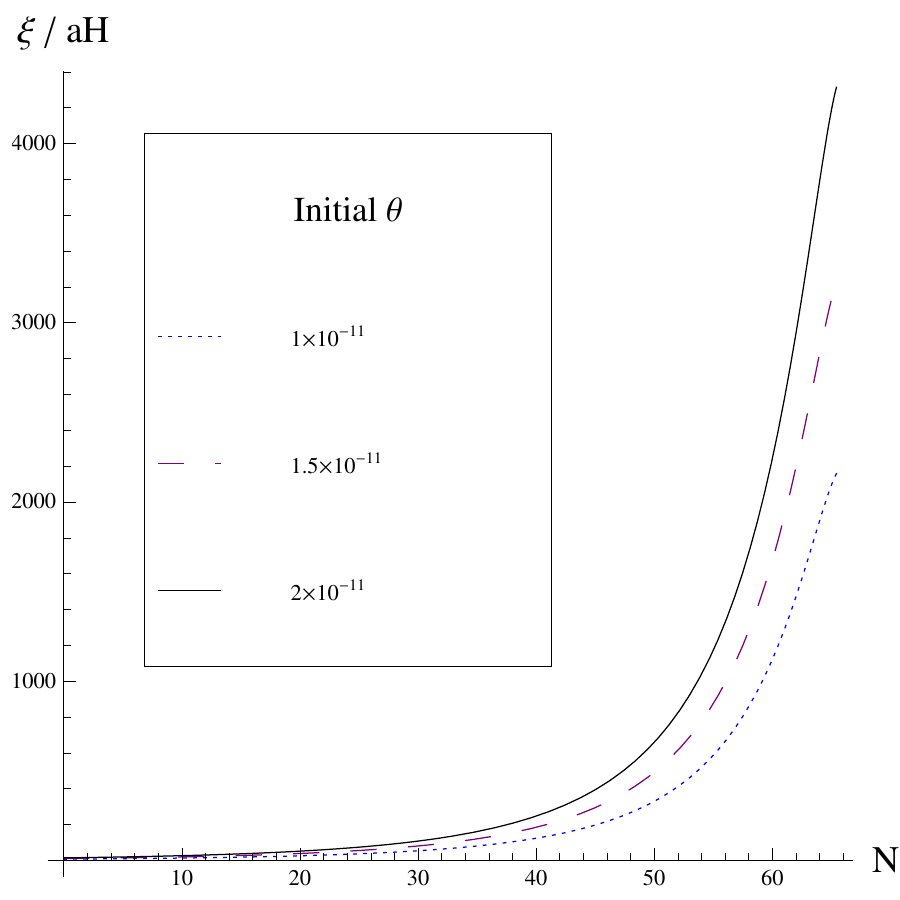}
\caption[Coupling along the maxmum]{The evolution of the coupling $\xi/a H$ in terms of the number of e-folds $N$ along the maximum of the potential in the angular direction.}{\label{couplingspinflationmax}}
\end{figure}

\begin{figure}
\centering
\includegraphics[width=10cm]{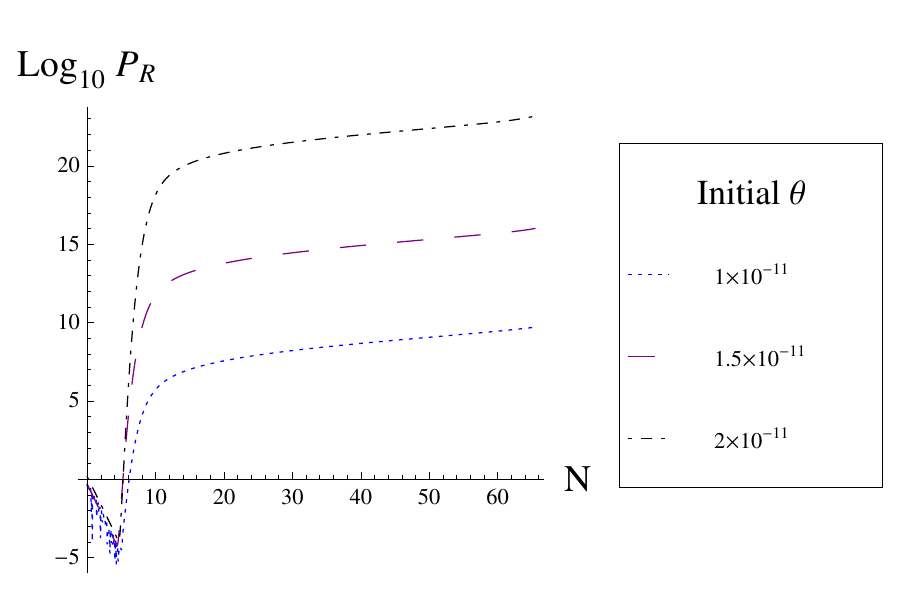}
\caption[Power spectrum of the curvature perturbation]{The evolution of the power spectrum of the curvature perturbation in terms of the number of e-folds $N$ along the maximum of the potential in the angular direction. The scale of interest exits the horizon around $N \sim 7$ for all the trajectories. }{\label{spinflationcurvaturemax}}
\end{figure}

In this subsection, we investigate the cases in which trajectories start with small deviations from the maximum of the potential in the angular direction. We set the initial conditions to $\left(\chi,\theta\right)=\left(9, \tilde{\delta}\theta\right)$ with $\tilde{\delta}\theta \ll 1$. We show the numerical results for three different trajectories with $\tilde{\delta}\theta = 1 \times 10^{-11}$, $\tilde{\delta}\theta = 1.5 \times 10^{-11}$ and $\tilde{\delta}\theta = 2 \times 10^{-11}$, which will be shown with a blue dotted line, a purple dashed line and a black solid line respectively in figures \ref{trajectoriesalongmaximumspinflation}, \ref{spinflationslowrollparametersmax}, \ref{couplingspinflationmax} and \ref{spinflationcurvaturemax}. 

In this case, the displacement from the maximum of the potential increases as inflation proceeds as shown in the left panel of figure \ref{trajectoriesalongmaximumspinflation}. The right panel shows that the brane goes to the tip of the throat $\chi = 0$ without reaching the minimum of the potential in the angular direction at $\theta = \pi$. The slow-roll parameters are shown in figure \ref{spinflationslowrollparametersmax}. Slow-roll approximation holds until the end of inflation. 

The coupling exhibits interesting behaviours in figure \ref{couplingspinflationmax}. Unlike the cases studied in subsection \ref{minimumtrajectories}, the coupling increases as the number of e-folds increases. In addition to that, the difference between the trajectories also increases. This means that the coupling at the end of inflation could be large even if it is almost negligible around horizon crossing. Figure \ref{spinflationcurvaturemax} shows the numerical result for the power spectrum of the curvature perturbation. 

Even though the amplitudes of $\tilde{\delta}\theta$ are the same as the amplitudes of $\delta \theta$ in the subsection \ref{minimumtrajectories} at the initial time when the scales of interest are well within the horizons, the coupling is larger with the trajectories along the maximum of the potential because $\tilde{\delta}\theta$ keeps growing. The values of $\xi/a H$ at horizon crossing are $10$, $15$, $22$ for the trajectories with $\tilde{\delta}\theta = 1 \times 10^{-11}$, $1.5 \times 10^{-11}$ and $2 \times 10^{-11}$ respectively. Therefore, we can no longer use the analytic solutions for the linear perturbations and the curvature power spectrum takes different values around horizon crossing with different trajectories. The values of the curvature power spectrum around horizon crossing are obtained numerically as $200$, $4\times10^{7}$ and $2 \times 10^{11}$ for the trajectories with $\tilde{\delta}\theta = 1 \times 10^{-11}$, $1.5 \times 10^{-11}$ and $2 \times 10^{-11}$ respectively while the values of the final curvature power spectrum are $\sim10^{9}$,$\sim10^{16}$ and $\sim10^{24}$. 
Therefore, the values of $\cos^{2}{\Theta}$ are $\sim10^{-7}$, $\sim10^{-9}$ and $\sim10^{-15}$ for the trajectories with $\tilde{\delta}\theta = 1 \times 10^{-11}$, $1.5 \times 10^{-11}$ and $2 \times 10^{-11}$. Those are much smaller than the values of $\cos^{2}{\Theta}$ in the trajectories along the minimum with the same amplitudes of the displacements. In general, it is safe to assume that $\cos^{2}{\Theta}$ keeps decreasing as the amplitude of the displacement from either the minimum or maximum increases. 

\section{Observational constraints}
As shown in chapter \ref{chapter:stringinflation}, the conversion of the entropy perturbation into the curvature perturbation due to curves in the trajectory in the field space plays an important role in suppressing $f_{\rm{NL}}^{equil}$. This mechanism gives multi-field DBI inflation models the possibility of satisfying the observational constraints by relaxing the stringent microphysical constraint that disfavours single field DBI inflation models. Here, the analytic formulae derived in subsection \ref{subsec:analysisofsimplemodeldbi} are used to show that the spinflation model studied in this chapter is excluded by the Planck satellite observations even with the conversion mechanism of the entropy perturbation when the the dynamics is effectively single field around horizon crossing. Finally, we show an example where those approximations do not hold.

\subsection{Cases with small couplings around the horizon exits}\label{subsec:constraintsspinflationgeneral}
\begin{figure}
  \centering
   \begin{tabular}{cc} 
    \includegraphics[width=70mm]{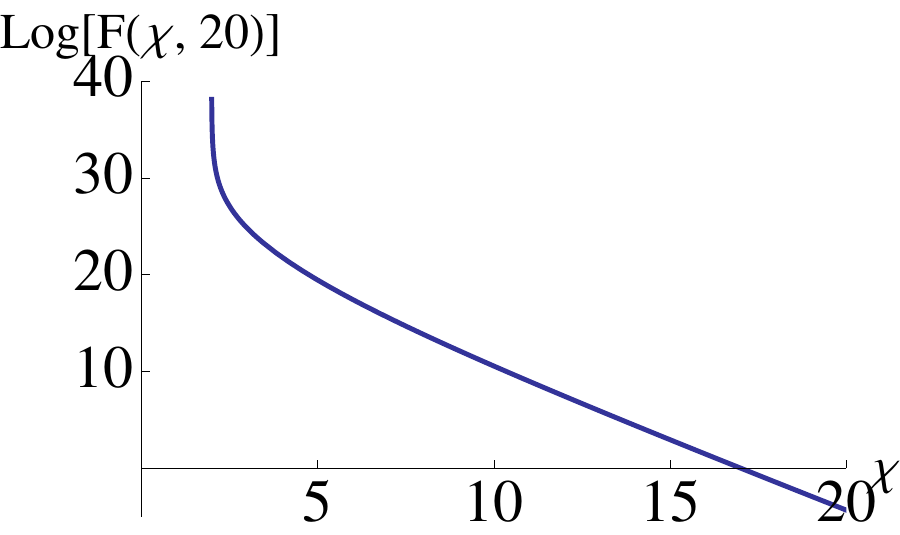}&
    \includegraphics[width=70mm]{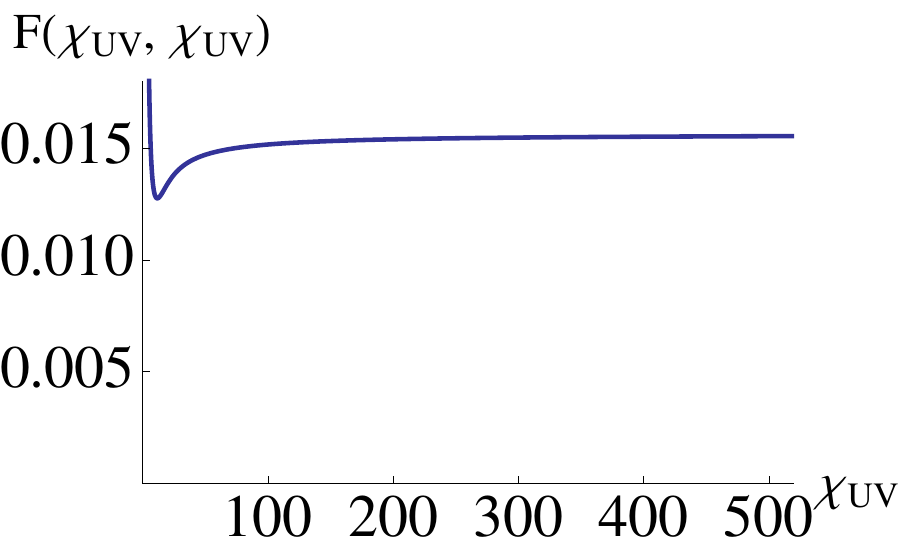}
  \end{tabular}
  \caption[Geometrical function]{Left: The semi-log plot of $F\left(\chi, \chi_{\rm{UV}}\right)$ with respect to $\chi$ for $\chi_{\rm{UV}} = 20$. It keeps decreasing exponentially. Right: The plot of $F\left(\chi_{\rm{UV}}, \chi_{\rm{UV}} \right)$ with respect to $\chi_{\rm{UV}}$. It takes a constant value asymptotically. }\label{geometricalfunctionplots}
\end{figure}

As shown in section \ref{sec:multifielddbiinflationthree}, the microphysical constraint that disfavours the single field DBI inflation models is possibly satisfied when the power spectrum of the curvature perturbation is enhanced after the horizon exit. In this section, we consider the cases where the dynamics is effectively single field until the perturbations considered are stretched to super-horizon scales as in the example in subsection \ref{subsec:analyticconfirmspinlast}. The enhancement is quantified by the transfer function as in equation (\ref{chaptertwotransferdefined}). Using equation (\ref{prmouse}), the ratio of the power spectrum of the curvature perturbation at the end of inflation $\mathcal{P}_{\mathcal{R}}$ to the power spectrum of the curvature perturbation around horizon crossing $\mathcal{P}_{\mathcal{R}_{*}}$ is given by
\begin{equation}\label{spincosthetadefinitionfinal}
\cos^{-2}{\Theta} = \frac{\mathcal{P}_{\mathcal{R}}}{\mathcal{P}_{\mathcal{R}_{*}}}.
\end{equation}
Because $\mathcal{P}_{\mathcal{R}}$ at the end of inflation needs to satisfy the constraint (\ref{plancksatteliteconstraintcurvaturespectrum}) by the Planck satellite observations, we have $\mathcal{P}_{\mathcal{R}} \sim 2.2 \times 10^{-9}$. Because we need $\cos^{-2}{\Theta} \gg 1$ to make the multi-field DBI inflation model compatible with the Planck satellite observations for the equilateral non-Gaussianity, we require
\begin{equation}\label{spinanaconpr}
\mathcal{P}_{\mathcal{R}_{*}} < 10^{-9}.
\end{equation}
Using the approximated analytic expression (\ref{curvaturespectrumexampledbisingle}), equation (\ref{spinanaconpr}) gives the lower bound of $\lambda$ as
\begin{equation}\label{lowerboundlambdaspin}
\lambda > \frac{10^{9}}{4 \pi^2 \epsilon^4},
\end{equation}
with the slow-roll parameter $\epsilon$. From equations (\ref{dbimultiequilateralfinalresult}) and (\ref{spincosthetadefinitionfinal}), we obtain
\begin{equation}\label{fnlequilspinfirst}
\begin{split}
f_{\rm{NL}}^{equil} &\approx -\frac{\cos^{2}{\Theta}}{3 c_{s}^{2}}\\
&= -\frac{1}{3 c_{s}^{2}}\frac{\mathcal{P}_{\mathcal{R}_{*}}}{\mathcal{P}_{\mathcal{R}}}. 
\end{split}
\end{equation}
Using the constraint on $f_{\rm{NL}}^{equil}$ by the Planck satellite observations $\left\lvert f_{\rm{NL}}^{equil} \right\rvert < 100$, equation (\ref{fnlequilspinfirst}) leads to
\begin{equation}\label{spinconstraintprandcs}
\frac{\mathcal{P}_{\mathcal{R}_{*}}}{c_{s}^{2}} < 6 \times 10^{-7},
\end{equation}
where we have used $\mathcal{P}_{\mathcal{R}} \sim 2.2 \times 10^{-9}$. The inequality (\ref{spinconstraintprandcs}) is rewritten as
\begin{equation}\label{constraintspinlastchapphi}
\frac{10^{7}}{6 \pi^{2} \epsilon^{6} \lambda} < \left(\frac{\phi}{M_{\rm{P}}} \right)^{4},
\end{equation}
using equation (\ref{curvaturespectrumexampledbisingle}) and the relation
\begin{equation}
c_{s} = \frac{\epsilon}{2} \left(\frac{\phi}{M_{\rm{P}}} \right)^{2},
\end{equation}
which is derived from equation (\ref{alltheresultsindbisingleexample}). From equations (\ref{massunitdefinitionspin}) and (\ref{spinexamplecanonicalfield}), the canonical field in the Planck units is given by
\begin{equation}\label{phioverplanckmassspinconstraint}
\begin{split}
\frac{\phi \left(\chi\right)}{M_{\rm{P}}} &= \frac{\sqrt{T_{3}} \kappa^{2/3}}{M_{\rm{P}}} \frac{1}{\sqrt{6}}\int^{\chi}_{0}\frac{dx}{K\left(x\right)}\\
&= \sqrt{\bar{T}_{3}} \bar{\kappa}^{2/3} \frac{\mathcal{M}}{M_{\rm{P}}} \frac{1}{\sqrt{6}}\int^{\chi}_{0}\frac{dx}{K\left(x\right)}\\
&= \sqrt{\frac{6 \pi}{\bar{\kappa}^{4/3}g_{\rm{s}}M^{2}\bar{T}_{3} N J \left(\chi_{\rm{UV}}\right)}} \sqrt{\bar{T}_{3}} \bar{\kappa}^{2/3} \frac{1}{\sqrt{6}}\int^{\chi}_{0}\frac{dx}{K\left(x\right)}\\
&= \sqrt{\frac{\pi}{g_{\rm{s}}M^{2} N J \left(\chi_{\rm{UV}}\right)}} \int^{\chi}_{0}\frac{dx}{K\left(x\right)}.
\end{split}
\end{equation}
From equations (\ref{constraintspinlastchapphi}) and (\ref{phioverplanckmassspinconstraint}), we obtain the inequality
\begin{equation}
\left(\sqrt{\frac{\pi}{g_{\rm{s}}M^{2} N J \left(\chi_{\rm{UV}}\right)}} \int^{\chi}_{0}\frac{dx}{K\left(x\right)} \right)^{4} > \frac{10^{7}}{6 \pi^{2} \epsilon^{6} \lambda},
\end{equation}
which leads to
\begin{equation}\label{preupperboundforlambdaspin}
\frac{\pi^2}{N^{2} J \left(\chi_{\rm{UV}}\right)^{2}} \left[ \frac{27}{64 \pi^3 \lambda} \left(\ln{\frac{r\left(\chi \right)^{3}}{\kappa^2}} + \ln{\frac{4\sqrt{2}}{3\sqrt{3}}} - \frac{1}{4} \right) \right]^{2} \left(\int^{\chi}_{0}\frac{dx}{K\left(x\right)} \right)^{4} > \frac{10^{7}}{6 \pi^{2} \epsilon^{6} \lambda}. 
\end{equation}
from equation (\ref{largechilambda}). Simplifying equation (\ref{preupperboundforlambdaspin}), we obtain the upper bound of $\lambda$ as
\begin{equation}\label{lambdaupperboundspin}
\lambda < \frac{3}{2 \pi^{2}} \left(\frac{27}{32}\right)^{2} 10^{-7} \epsilon^{6} \frac{\left(\ln{\frac{r\left(\chi \right)^{3}}{\kappa^2}} + \ln{\frac{4\sqrt{2}}{3\sqrt{3}}} - \frac{1}{4} \right)^{2}}{J \left(\chi_{\rm{UV}}\right)^{2}} \left(\int^{\chi}_{0}\frac{dx}{K\left(x\right)} \right)^{4}.
\end{equation}
Because we have both the lower bound (\ref{lowerboundlambdaspin}) and the upper bound (\ref{lambdaupperboundspin}) of $\lambda$, the lower bound must be smaller than the upper bound
\begin{equation}
\frac{10^{9}}{4 \pi^2 \epsilon^4} < \frac{3}{2 \pi^{2}} \left(\frac{27}{32}\right)^{2} 10^{-7} \epsilon^{6} \frac{\left(\ln{\frac{r\left(\chi \right)^{3}}{\kappa^2}} + \ln{\frac{4\sqrt{2}}{3\sqrt{3}}} - \frac{1}{4} \right)^{2}}{J \left(\chi_{\rm{UV}}\right)^{2}} \left(\int^{\chi}_{0}\frac{dx}{K\left(x\right)} \right)^{4},
\end{equation}
which is rewritten as
\begin{equation}\label{fchiconditionfinalspin}
F\left(\chi, \chi_{\rm{UV}} \right) < 4.27 \times 10^{-16} \epsilon^{10},
\end{equation}
with
\begin{equation}\label{explicitdefinitionoffchichiuvspin}
F\left(\chi, \chi_{\rm{UV}} \right) \equiv \frac{J \left(\chi_{\rm{UV}}\right)^{2}}{\left[\ln{ \left(\frac{1}{\sqrt{6}}\int^{\chi}_{0}\frac{dx}{K\left(x\right)} \right)^3} + \ln{\frac{4\sqrt{2}}{3\sqrt{3}}} - \frac{1}{4} \right]^{2} \left(\int^{\chi}_{0}\frac{dx}{K\left(x\right)} \right)^{4}},
\end{equation}
where we have used equation (\ref{spincanonicalfielddefined}). If the condition (\ref{fchiconditionfinalspin}) is not satisfied, $\lambda$ cannot take any value that is larger than the lower bound (\ref{lowerboundlambdaspin}) and smaller than the upper bound (\ref{lambdaupperboundspin}) at the same time. Because the function $F\left(\chi \right)$ is dependent only on $\chi$ and $\chi_{UV}$, this is a general condition that is independent of all other parameters. Numerically, we obtain
\begin{equation}
L \left(\chi_{N} \right) \equiv \ln{ \left(\frac{1}{\sqrt{6}}\int^{\chi_{N}}_{0}\frac{dx}{K\left(x\right)} \right)^3} + \ln{\frac{4\sqrt{2}}{3\sqrt{3}}} - \frac{1}{4} = 0,
\end{equation}
where $\chi_{N} = 1.9966$. As $\chi$ increases from $\chi_{N}$, $L \left(\chi \right)$ increases monotonically because we have
\begin{equation}
\frac{d}{d \chi} \left(\int^{\chi}_{0}\frac{dx}{K\left(x\right)} \right) = \frac{1}{K\left(x\right)} >0.
\end{equation}
Because we consider the case $\chi \gg 1$, we study the behaviour of $F\left(\chi, \chi_{\rm{UV}} \right)$ only in the region $\chi > 2$ below. Therefore, the denominator of $F\left(\chi, \chi_{\rm{UV}} \right)$ in equation (\ref{explicitdefinitionoffchichiuvspin}) is a monotonically increasing function with $\chi$. This means that $\chi=\chi_{\rm{UV}}$ minimises $F\left(\chi, \chi_{\rm{UV}} \right)$ and the condition (\ref{fchiconditionfinalspin}) is rewritten as
\begin{equation}\label{fchiuvfinalconditionspin}
F\left(\chi_{\rm{UV}}, \chi_{\rm{UV}} \right) < 4.27 \times 10^{-16} \epsilon^{10} < 4.27 \times 10^{-16},
\end{equation}
where we used $\epsilon < 1$ during inflation. Choosing $\chi = \chi_{\rm{UV}}$ means considering the perturbation that exits the horizon when the brane is at $\chi = \chi_{\rm{UV}}$. As shown in the left panel of figure \ref{geometricalfunctionplots}, $F\left(\chi, \chi_{\rm{UV}} \right)$ keeps decreasing exponentially. The plot is for $\chi_{\rm{UV}} = 20$ and $F\left(\chi_{\rm{UV}}, \chi_{\rm{UV}} \right) \approx 0.0134$ in this case. This does not satisfy the condition (\ref{fchiuvfinalconditionspin}). 

Let us show the behaviour of $F\left(\chi_{\rm{UV}}, \chi_{\rm{UV}} \right)$ with respect to $\chi_{\rm{UV}}$ below. For large $\chi$, we obtain
\begin{equation}
K\left(\chi \right) \approx 2^{1/3} \exp{\left(-\frac{1}{3} \chi\right)},
\end{equation}
which leads to
\begin{equation}\label{spincoordinateapproximation}
\int^{\chi}_{0}\frac{dx}{K\left(x\right)} \approx \frac{3}{2^{1/3}} \exp{\left(\frac{1}{3} \chi \right)}. 
\end{equation}
Using equation (\ref{spincoordinateapproximation}), we obtain
\begin{equation}\label{spinapproximationofthelnfunctioninf}
\ln{ \left(\frac{1}{\sqrt{6}}\int^{\chi}_{0}\frac{dx}{K\left(x\right)} \right)^3} + \ln{\frac{4\sqrt{2}}{3\sqrt{3}}} - \frac{1}{4} \approx \chi. 
\end{equation}
The function $I\left( \chi \right)$ in equation (\ref{spinfirstdefinitionofi}) is approximated as
\begin{equation}\label{chapfourapproxifuninlargechi}
I \left( \chi \right) \approx \frac{3}{4^{1/3}} \chi \exp{\left(- \frac{4}{3} \chi \right)},
\end{equation}
for large $\chi$ from the approximated expression in \cite{Gregory:2012}. For sufficiently large $\chi_{\rm{UV}}$, we obtain
\begin{equation}\label{spinfinalapproxforjuv}
\begin{split}
J \left(\chi_{\rm{UV}} \right) &= \int^{\chi_{\rm{UV}}}_{0} d\chi I \left(\chi \right) \sinh^{2}{\chi}\\
&\approx \int^{\chi_{\rm{UV}}}_{\chi_{\rm{t}}} d\chi I \left(\chi \right) \sinh^{2}{\chi}\\
&\approx \frac{3^{2}}{2^{11/3}} \chi_{\rm{UV}} \exp{\left(\frac{2}{3} \chi_{\rm{UV}} \right)},
\end{split}
\end{equation}
using equation (\ref{chapfourapproxifuninlargechi}) where $1 \ll \chi_{\rm{t}} \ll \chi_{\rm{UV}}$. From equations (\ref{explicitdefinitionoffchichiuvspin}), (\ref{spincoordinateapproximation}), (\ref{spinapproximationofthelnfunctioninf}) and (\ref{spinfinalapproxforjuv}), we obtain
\begin{equation}
F\left(\chi_{\rm{UV}}, \chi_{\rm{UV}} \right) \approx \frac{1}{2^{6}} \approx 0.0156,
\end{equation}
for large $\chi_{\rm{UV}}$. In the right panel of figure \ref{geometricalfunctionplots}, we see that $F\left(\chi_{\rm{UV}}, \chi_{\rm{UV}} \right)$ actually approaches $0.0156$. It also shows that $F\left(\chi_{\rm{UV}}, \chi_{\rm{UV}} \right)$ does not become smaller than $0.01$ in the region $2 <\chi_{\rm{UV}}$ before it becomes constant. Therefore, we conclude that the necessary condition (\ref{fchiuvfinalconditionspin}) is not satisfied regardless of the value of $\chi_{\rm{UV}}$. Because the condition (\ref{fchiuvfinalconditionspin}) is independent of any other parameter, this model is excluded by the observations in the regime where we can use those approximated formulae. Note that this strong constraint comes from the fact that the sound speed and the amplitude of the curvature perturbation is controlled essentially by one factor that consists of the model parameters $g_{\rm{s}} M^2$ as in equations (\ref{largechilambda}) and (\ref{phioverplanckmassspinconstraint}). Due to this relation, it is not possible to satisfy (\ref{spinanaconpr}) and (\ref{spinconstraintprandcs}) simultaneously. 

\subsection{Cases with large couplings around the horizon exits}
In the previous subsection \ref{subsec:constraintsspinflationgeneral}, we have shown that the spinflation model is excluded by the observations when the dynamics is effectively single field around horizon crossing. We assumed the effective single field dynamics around horizon crossing because all the formulae derived in section \ref{sec:observable sinmultifield} have been derived with this assumption. We can use all the formulae in section \ref{sec:observable sinmultifield} to estimate the amount of conversion of the entropy perturbation into the curvature perturbation only if the coupling is sufficiently small around horizon crossing. If the coupling is large around horizon crossing as $\xi/aH > 1$, the adiabatic perturbation is coupled to the entropy perturbation around horizon crossing and the power spectrum of the curvature perturbation around horizon crossing can no longer be estimated with equation (\ref{curvaturepowerspectrumhorizon}). The expression for $f_{\rm{NL}}^{equil}$ (\ref{dbimultiequilateralfinalresult}) is also not valid in such cases because we used the expression for the curvature power spectrum which is valid only with a small coupling around horizon crossing  in deriving this expression. Because the analytic formulae in subsection \ref{subsec:analyticconfirmspinlast} are derived assuming the single field dynamics, the conclusion in subsection \ref{subsec:constraintsspinflationgeneral} is no longer valid in cases with large couplings around horizon crossing. However, out numerical calculations show that it is difficult to maintain the almost scale-invariant curvature power spectrum which is compatible with the observations when the coupling is large around horizon crossing. 

If the coupling between adiabatic and entropy modes cannot be neglected at horizon crossing, we cannot use the result for non-Gaussianity obtained in section \ref{sec:observable sinmultifield} where we assumed that the conversion happens on super-horizon scales. A similar situation arises in a quasi-single field inflation proposed in \cite{Chen:2010b}. They studied a model where there is one slow-roll direction while all other isocurvature fields have masses at least of the order of $H$. In this case, it was shown that large bispectra with shapes between the local and equilateral shapes arise. To study such cases in our model, we need to perform the full calculations with the in-in formalism to obtain the non-Gausianity parameters. 
\chapter{Hybrid inflation model}\label{chapter:hybrid}
\chaptermark{Hybrid inflation model}

As we saw in chapter \ref{chapter:spinflation}, the DBI inflation model with the simplest two-field potential derived in string theory is excluded by the observations in the regime where we have the effective single field dynamics around horizon crossing. However, potentials derived in string theory are generally more complicated. The potential for the angular directions for a D3 brane in the deformed warped conifold was calculated in \cite{Baumann:2008} and the impact of angular motion on DBI inflation has been studied in \cite{Easson:2008, Gregory:2012}. It was shown that the angular directions can become unstable in a particular embedding of D7 brane in the warped conifold and the angular instability connects different extreme trajectories \cite{Chen:2010}. These potentials are calculated assuming that the backreaction of the moving brane is negligible and strictly speaking, it cannot be applied to DBI inflation directly. However, it is natural to consider that a similar transition due to the angular instability happens also in DBI inflation. The potential derived in \cite{Chen:2010} has a similar feature to the potential in hybrid inflation models. In this chapter, we analyse a two-field DBI model with a potential which has such a feature. Using this multi-field potential, we will explicitly study predictions for observables such as the spectral index, tensor-to-scalar ratio and non-Gaussainities. Finally, we show that we can avoid the stringent constraint that rules out the single field UV DBI inflation models as explained in section \ref{sec:multifielddbiinflationthree}. However, it is difficult to obtain a small value of $f_{\rm{NL}}^{local}$ that is compatible with the Planck satellite observations with our specific choice of the potential. 

\section{The model}\label{sec:modelofhybridinflation}
In this section, we first introduce a single field model with a constant sound speed which is used for the radial part of the two-field model studied in this chapter. Then, the two-field model is introduced and we analyse the model in order to derive the effective forms of the potential before and after the waterfall phase transition . 

\subsection{DBI inflation with a constant sound speed}\label{constantsoundspeed}
The equation of motion for the single field DBI model is given by equation (\ref{klein}). In \cite{Copeland:2010}, it was shown that when $V (\phi)$ and $f (\phi)$ are given by
\begin{equation}
 V (\phi) = V_{0} \phi^{-q},\label{potentialmizuno}
\end{equation}
\begin{equation}
 f(\phi) = f_{0} \phi^{q+2},\label{warpfactormizuno}
\end{equation}
with constants $V_{0}$, $f_{0}$ and $q$, equation (\ref{klein}) has a late time attractor inflationary solution with a constant sound speed that is given by 
\begin{equation}
 c_{s} = \sqrt{\frac{3}{16 f_{0} V_{0} + 3}}. \label{attractorsoundspeed}
\end{equation}
Throughout this chapter, as a concrete example, we consider the case with $q=4$. This attractor solution is potential dominated, which means that the potential term is much larger than the kinetic term along the attractor solution. The slow-roll parameters defined in equation (\ref{slowrollparametersgeneralcases}) are much smaller than unity for the attractor solution because $s$ vanishes because of the constant sound speed and $\epsilon$ and $\eta$ are much smaller than unity because it is a potential dominated attractor solution. 

\subsection{Two-field model}\label{twofield}
\begin{figure}[!htb]
\centering
\includegraphics[width=14cm]{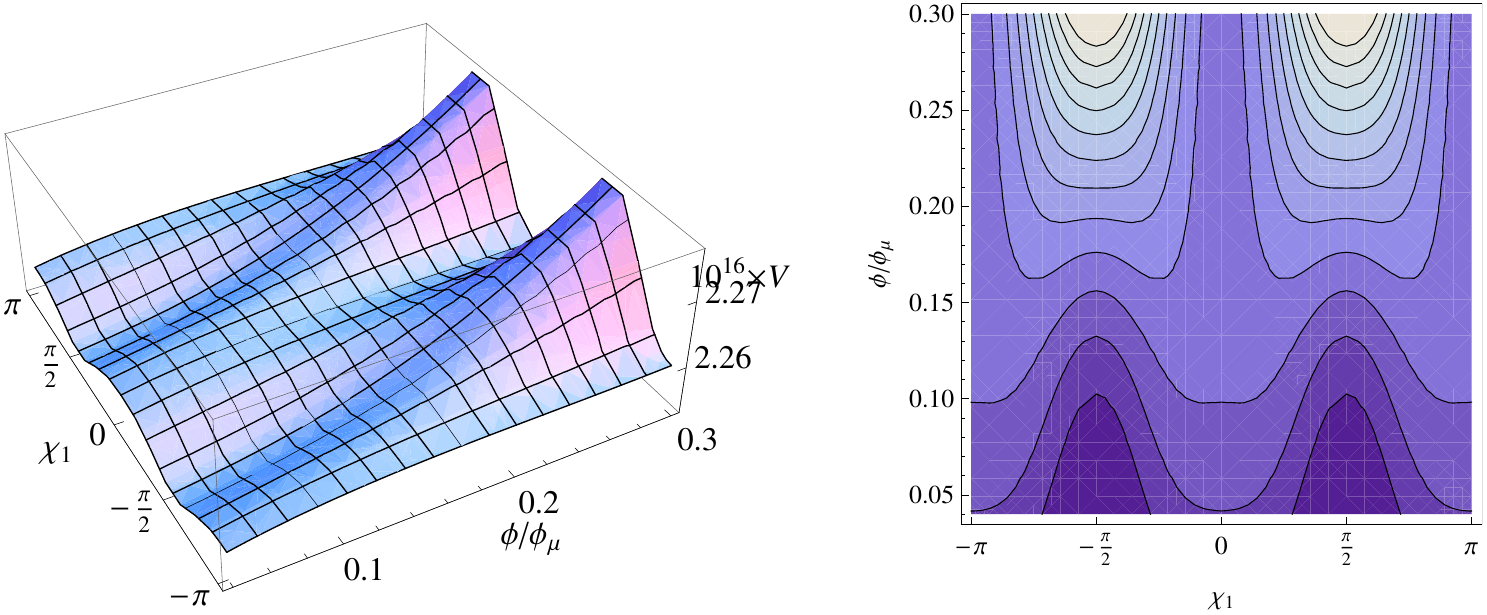}
\caption[Internal space]{The potential as a function of the radial coordinate $\phi$ and a particular choice of angular coordinate $\xi_{1}$ derived in \cite{Chen:2010}. $\phi_{\mu}$ corresponds to the maximal value for $\phi$ in the warped throat. The potential is periodic along the direction $\xi_{1}$ with the period $\pi$. Along the line $\xi_{1} = 0$, the $\xi_{1}$ field is massive when $\phi/\phi_{\mu} > 0.25$ and tachyonic when $\phi/\phi_{\mu} < 0.1$. The figure is taken from \cite{Chen:2010}}{\label{fig:multifieldouyangembedding}}
\end{figure}

In DBI inflation, the scalar fields describe the position of a brane in the bulk. The explicit form of the multi-field potential depends on the details of the geometry of the warped conifold and various effects from the stabilisation of moduli fields. In \cite{Chen:2010}, a potential is derived assuming a specific embedding called the Ouyang embedding. As shown in figure \ref{fig:multifieldouyangembedding}, the potential is similar to a potential in hybrid inflation models in which a field becomes either massive or tachyonic depending on the values of other fields. 

In this potential shown in figure \ref{fig:multifieldouyangembedding}, the inflaton rolls down along the radial direction first. Eventually, it arrives at the transition point where the entropy field becomes tachyonic. Then the inflaton rolls down to the true vacuum along the entropy direction and moves down in the radial direction along the true vacuum. 
In this chapter, we investigate a two-field potential which captures the essential feature of the potential derived in string theory as described above.
We assume the radial field has the form of the potential with a constant sound speed as discussed in subsection \ref{constantsoundspeed} to simplify the calculation. If we define the field $\phi$ as a radial direction and define the field $\chi$ as an angular direction in the warped throat, the two-field potential is given by
\begin{equation}
 V(\phi, \chi)
= \frac{1}{2}\lambda(\chi^2 - \chi^{2}_{0})^2 + g\left(\frac{\chi}{\phi} \right)^2 + \frac{V_{0}}{\phi^{4}}.
\label{waterfallpotential}
\end{equation}
Let us assume that the inflaton starts rolling down in the radial direction $\phi$ with a small deviation from $\chi = 0$. Then, in the early stage when $\chi \ll 1$, the 
potential is effectively a single field potential for $\phi$;
\begin{equation}
 V(\phi, \chi) \sim \frac{1}{2}\lambda \chi^{4}_{0} +  \frac{V_{0}}{\phi^{4}}.
\end{equation}
At this stage, if we assume
\begin{equation}
 \frac{1}{2}\lambda \chi^{4}_{0} \ll \frac{V_{0}}{\phi^{4}},\label{conditionichi}
\end{equation}
the effective potential becomes
\begin{equation}
 V_{\rm{eff},1} \sim \frac{V_{0}}{\phi^{4}}.
\end{equation}
\begin{figure}[t]
\centering
\includegraphics[keepaspectratio=true,height=7cm]{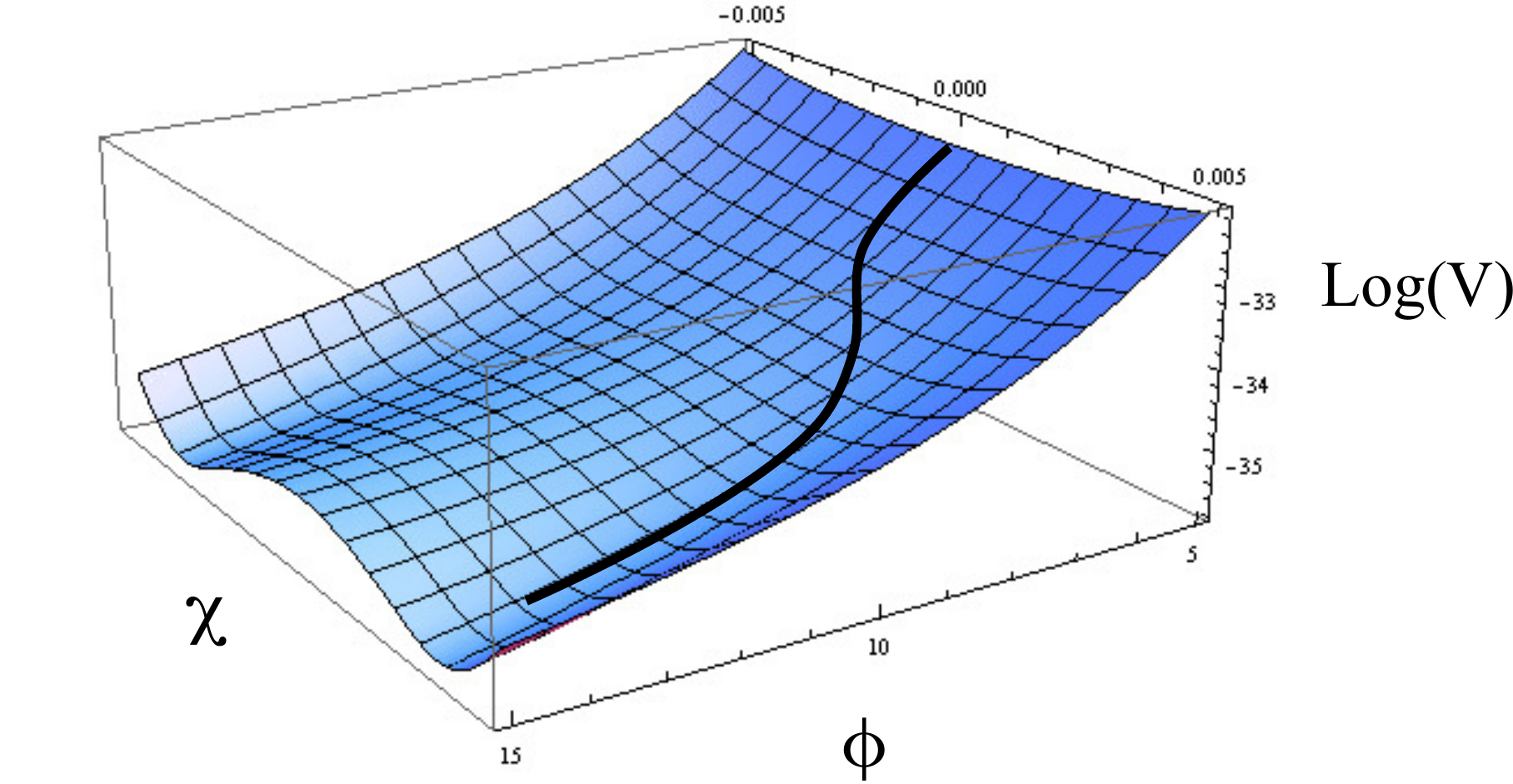}
\caption[Water fall potential]{The potential given by equation (\ref{waterfallpotential}). In order to show the feature of the potential clearly, we plot $\log V(\phi, \chi)$. The parameters are chosen as $\lambda = 3.75 \times 10^{-6}$, $\chi_{0} = 0.004$, $V_{0} = 5 \times 10^{-12}$ and $g = 3 \times 10^{-9}$. For small $\phi$ there is a minimum at $\chi=0$ but for large $\phi$ true vacua appear and the field rolls down to the true vacuum causing the waterfall phase transition.}\label{waterfall}
\end{figure}
Because this is in the same form as the potential (\ref{potentialmizuno}), there is a late-time attractor solution with a constant sound speed given by equation (\ref{attractorsoundspeed}).
As the inflaton rolls down in the radial direction, a waterfall phase transition occurs as shown in figure \ref{waterfall} where the inflaton rolls down to the true minimum of the potential (\ref{waterfallpotential}). This can be seen more clearly by rewriting the potential (\ref{waterfallpotential}) as
\begin{equation}
 V(\phi, \chi) = \frac{1}{2}\lambda \left[\chi^2 - \left(\chi^{2}_{0} - \frac{g}{\lambda \phi^{2}}\right) \right]^2 + V_{0}\left(1 - \frac{g^{2}}{2 \lambda V_{0}} \right)\frac{1}{\phi^{4}} + g\left(\frac{\chi_{0}}{\phi} \right)^2.\label{transformedpotential}
\end{equation}
In this form, $\chi$ appears only in the first term. We can clearly see that $\chi = 0$ is the minimum in the $\chi$ direction when $\phi^{2} < g / (\lambda \chi^{2}_{0})$, while $\chi^{2} = \chi^{2}_{0} - g/(\lambda \phi^{2})$ becomes the minimum in the $\chi$ direction when $\phi^{2} > g / (\lambda \chi^{2}_{0})$. Therefore, $\phi^{2} = g / (\lambda \chi^{2}_{0})$ is the critical transition value for $\phi$. The effective potential in the true vacuum with $\chi^{2} = \chi^{2}_{0} - g/(\lambda \phi^{2})$ is given by 
\begin{equation}
 V(\phi, \chi) \sim V_{0}\left(1 - \frac{g^{2}}{2 \lambda V_{0}} \right)\frac{1}{\phi^{4}} + g\left(\frac{\chi_{0}}{\phi} \right)^2.
\end{equation}
Thus if we assume
\begin{equation}
 \frac{g\left(\chi_{0} / \phi \right)^2}{V_{0}\left(1 - g^{2} / 2 \lambda V_{0} \right) 1/ \phi^{4}} = \frac{g \chi^{2}_{0}}{V_{0}}\frac{\phi^{2}}{1- g^{2} / 2\lambda V_{0}} \ll 1, \label{conditionni}
\end{equation}
the effective potential in the true vacuum becomes
\begin{equation}
 V_{\rm{eff},2} \sim V_{0}\left(1 - \frac{g^{2}}{2 \lambda V_{0}} \right)\frac{1}{\phi^{4}}.
\end{equation}
Again, this is in the form $\sim \phi^{-4}$ and we have a late-time attractor with a constant sound speed
\begin{equation}
 c_{s} = \sqrt{\frac{3}{16f_{0}\tilde{V}_{0}+3}},
\end{equation}
where
\begin{equation}
 \tilde{V}_{0} = V_{0}\left(1 - \frac{g^{2}}{2 \lambda V_{0}} \right).
\end{equation}

\section{Background dynamics}{\label{sec:backtwodbi}}
We first introduce the equations of motion with the phase space variables that was introduced in subsection \ref{phasespacevarispin}. Then, we show our numerical results for the background quantities including the slow-roll parameters. 

\subsection{Background equations}\label{subsec:phasespacegeneralintro}
In this model, the field space metric is given by the simple diagonal form (\ref{simplediagonalform}) where 
\begin{equation}\label{chapterfourmetricfactors}
A_{\phi} = 1,\,\,\,\,\,A_{\chi} = \phi^{2}.
\end{equation}
with the Kronecker delta $\delta_{IJ}$. With this field space metric, X becomes 
\begin{equation}
 X = - \frac{1}{2} G_{IJ} \partial _{\mu} \phi^{I} \partial ^{\mu} \phi^{J} = \frac{1}{2} \left(\dot{\phi}^{2} + \phi^{2} \dot{\chi}^{2} \right),
 \end{equation}
in the homogeneous background. 
The Friedmann equation (\ref{friedmanngeneralderived}) in this model is given by 
\begin{equation}
 3 H^{2} = \frac{1}{f(\phi)} \left(\frac{1}{c_{s}} - 1 \right) + V(\phi_{I}).\label{friedmannhybridchapfive}
\end{equation}
The equations of motion for the fields (\ref{eqofmotionai}) are given by
\begin{equation}
 \ddot{\phi} + 3 H \dot{\phi} - \frac{\dot{c_{s}}}{c_{s}}\dot{\phi} - \phi \dot{\chi}^2 + c_{s} V_{,\phi} - \frac{(1 - c_{s})^2}{2} \frac{f_{,\phi}}{f^2} = 0,\label{phibackgroundeq}
\end{equation}
and
\begin{equation}
\phi^2 \left(\ddot{\chi} + 3 H \dot{\chi} - \frac{\dot{c_{s}}}{c_{s}}\dot{\chi} \right) + 2 \phi \dot{\phi} \dot{\chi} + c_{s} V_{,\chi} = 0.\label{chibackgroundeq}
\end{equation}
The potential $V$ is given by equation (\ref{waterfallpotential}) while the warp factor $f$ is given by equation (\ref{warpfactormizuno}) with $q = 4$ as
\begin{equation}\label{chapterfourwarpfactor}
f(\phi) = f_{0} \phi^{6}.
\end{equation}
Substituting equations (\ref{chapterfourmetricfactors}) and (\ref{chapterfourwarpfactor}) into the equations (\ref{phaseeqofmotiongeneone}) and (\ref{phaseeqofmotiongenetwo}), we obtain
\begin{equation}\label{equationforphichapterfour}
\begin{split}
\dot{x_{\phi}}=&\left(1-x_{\phi}^{2}\right)\left[-\frac{x_{\chi}^{2}}{\sqrt{f_{0}}\phi^{4}} - 3Hx_{\phi} + \sqrt{f_{0}}\phi^{3}c_{s}V_{,\phi} + \frac{6}{\sqrt{f_{0}}\phi^{4}} \left(-1+c_{s}+\frac{x_{\chi}^{2}}{2}\right)\right]\\
&-x_{\phi}x_{\chi}\left[\frac{-2}{\sqrt{f_{0}}\phi^{4}}x_{\phi}x_{\chi} - 3Hx_{\chi} + \sqrt{f_{0}}\phi^{2}c_{s}V_{,\chi} \right],
\end{split}
\end{equation}
\begin{equation}\label{equationforchichapterfour}
\begin{split}
\dot{x_{\phi}}&=\left(1-x_{\chi}^{2}\right)\left[\frac{-2}{\sqrt{f_{0}}\phi^{4}}x_{\phi}x_{\chi} - 3Hx_{\chi} + \sqrt{f_{0}}\phi^{2}c_{s}V_{,\chi} \right]\\
&-x_{\phi}x_{\chi}\left[-\frac{x_{\chi}^{2}}{\sqrt{f_{0}}\phi^{4}} - 3Hx_{\phi} + \sqrt{f_{0}}\phi^{3}c_{s}V_{,\phi} + \frac{6}{\sqrt{f_{0}}\phi^{4}} \left(-1+c_{s}+\frac{x_{\chi}^{2}}{2}\right)\right], 
\end{split}
\end{equation}
where the subscripts $_{,\phi}$ and $_{,\chi}$ denote the derivatives with respect to $\phi$ and $\chi$ and the phase space variables are defined as
\begin{equation}
x_{\phi} = - \sqrt{f_{0}}\phi^{3}\dot{\phi},\,\,\,\,\,x_{\chi} = - \sqrt{f_{0}}\phi^{4}\dot{\chi},
\end{equation}
from equation (\ref{phasespacevari}). In our numerical code, equations (\ref{equationforphichapterfour}) and (\ref{equationforchichapterfour}) are solved instead of equations (\ref{phibackgroundeq}) and (\ref{chibackgroundeq}) for the numerical efficiency. 

\subsection{Numerical results}
As we saw above, the background dynamics is determined by solving equations (\ref{equationforphichapterfour}) and (\ref{equationforchichapterfour}). We choose the parameters as follows; $\lambda = 3.75 \times 10^{-6}$, $\chi_{0} = 0.004$, $V_{0} = 5 \times 10^{-12}$ and $g = 3 \times 10^{-9}$. The warp factor is given by equation (\ref{warpfactormizuno}) with $f_{0} = 1.2 \times 10^{15}$. These parameters are chosen so that all the observables satisfy the constraints obtained by the WMAP observations \cite{Komatsu:2011}. 

\begin{figure}[h]
\centering
\includegraphics[width=14cm]{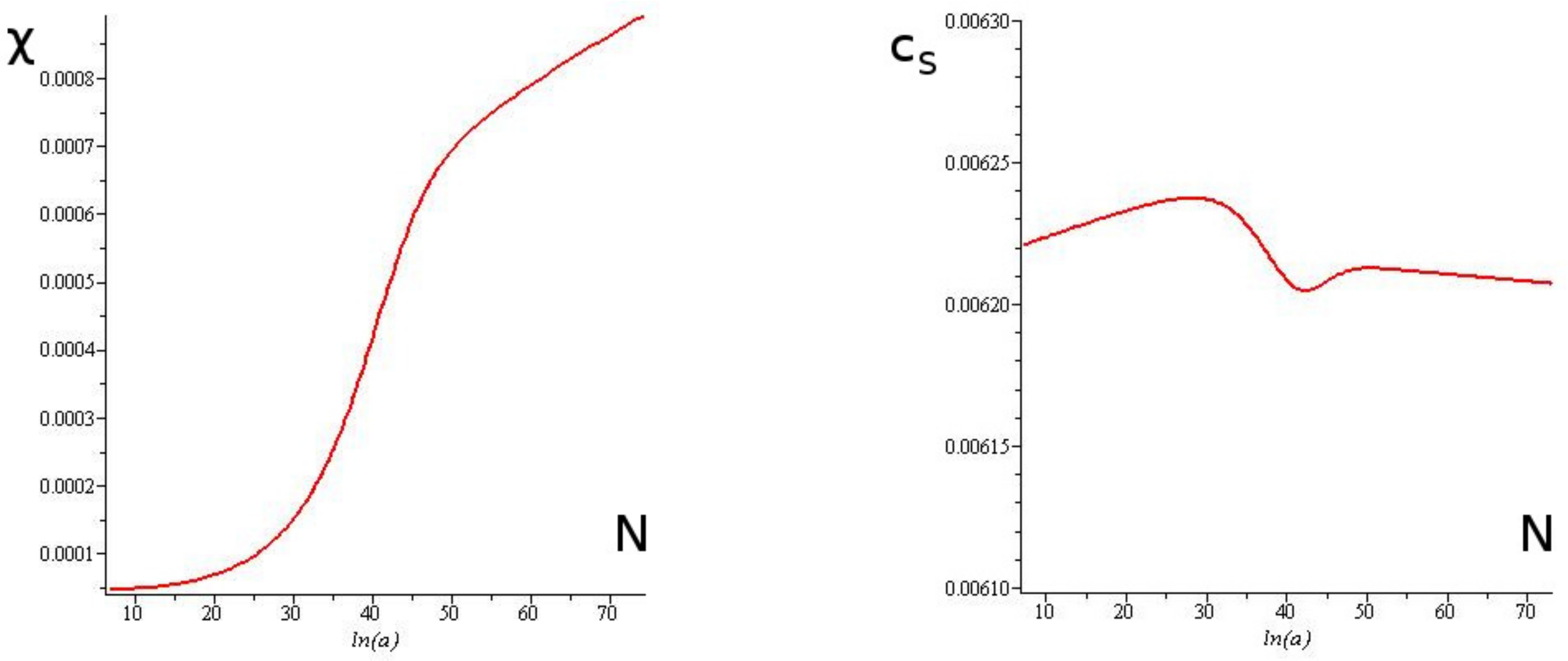}
\caption[The $\chi$ field and the sound speed]{Left: The dynamics of the $\chi$ field. In the early stage ($10 \lesssim N \lesssim 15$), the inflaton rolls down the potential almost in the $\phi$ direction with $\chi \sim 0$. The transition occurs during $20 \lesssim N \lesssim 55$. After $N \sim55$, the inflaton rolls along the true vacuum. Right: The sound speed. The sound speed is almost constant because
it changes very slowly even during the transition.}\label{chisound}
\end{figure}

Firstly, the left panel of Fig.~\ref{chisound} shows the dynamics of the inflaton in the $\chi$ direction. Before the transition happens, the potential has its minimum at $\chi = 0$ in the $\chi$ direction. Therefore, regardless of the initial conditions, the inflaton rolls down the potential and the value of $\chi$ approaches 0 unless the transition occurs while $\chi$ is still large. We use the e-folding number $N=\ln a$
as time. We normalise the e-folding number so that the transition finishes after $N=60$. In this example, $\chi$ is sufficiently small at $N = 10$ and we have an effective single field dynamics until around $N \sim 15$. 
As we expected, the inflaton rolls down to the true vacuum in the transition, which occurs during $20 \lesssim N \lesssim 55$. After the transition ends, it rolls down along the true vacuum. Notice that the true vacuum is not along a constant $\chi$ line but the value of $\chi$ along the true vacuum $\chi_{true}$ is a function 
of $\phi$;
\begin{equation}
\chi_{true} = \sqrt{\chi^{2}_{0} - \frac{g}{\lambda \phi^{2}}},
\end{equation}
which can be obtained from equation (\ref{transformedpotential}).
However, as we see in the right panel of Fig.~\ref{chisound}, the trajectory along the true vacuum curves slowly so that the coupling between the adiabatic mode and the entropic mode can be ignored.
Using the value of the Hubble parameter $H \sim 2.8 \times 10^{-8} M_{P}$ that is obtained with our choice of parameters, we can roughly estimate how many e-folds we need after the sound horizon exit of the mode that is responsible for large scale CMB anisotropies, assuming instant reheating \cite{Lyth:2009}. In our model, it is around 60 e-folds. We can see that the transition ends within 60 e-folds after the sound horizon exit if we consider modes that exit the sound horizon in the effective single field regime ($10 < N < 15$). We assume inflation ends after the transition by some mechanisms such as an annihilation of a D brane with an anti-D brane. 

\begin{figure}[h]
\centering
\includegraphics[width=15cm]{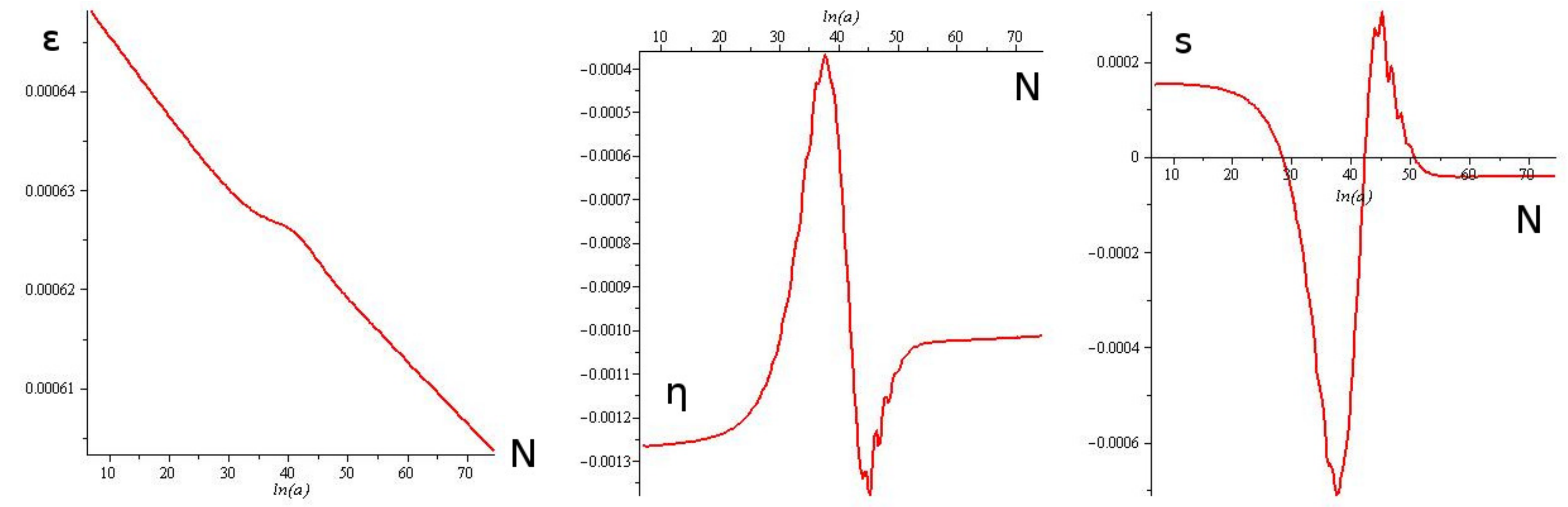}
\caption[Slow-roll parameters]{Left: The slow-roll parameter $\epsilon$. It is always smaller than $1 \times 10^{-3}$ even during the transition.
Middle: The slow-roll parameter $\eta$. Its absolute value is always smaller than $2 \times 10^{-3}$.
Right: The slow-roll parameter $s$. $\lvert s \rvert$ takes the largest value $-7 \times 10^{-4}$ during the transition. However, its absolute value is still much smaller than one.}\label{slowroll}
\end{figure}

The right panel of Fig.~\ref{chisound} shows the sound speed. Before the transition, the sound speed slowly changes. This is because the condition (\ref{conditionichi}) is not fully satisfied. On the other hand, after the transition, we can clearly see that the sound speed is almost constant. The sound speed changes the most during the transition and the slow-roll parameter $s$ takes the largest value during the transition. However, the largest value of $s$ is still around $-7 \times 10^{-4}$. 
As shown in Fig.~\ref{slowroll}, all the slow-roll parameters are always much smaller than unity even during the transition. Therefore, the slow-roll approximation always holds in this model and the sound speed is almost constant even during the transition.

\section{\label{sec:Observables}Prediction of the observables}
In this section, we show the numerical results from solving the equations of motion for the linear perturbations of the scalar fields (\ref{equationofmotionone}) and (\ref{equationofmotiontwo}). We also explain how we set the initial conditions for those equations of motion. Then, we show the numerical results for the observables including the curvature perturbation power spectrum, spectral index for the curvature perturbation and tensor-to-scalar ratio. We use the $\delta$N-formalism to obtain the numerical predictions of the non-Gaussianity parameters. 

\subsection{Initial conditions}
\begin{figure}[h]
\centering
\includegraphics[width=15cm]{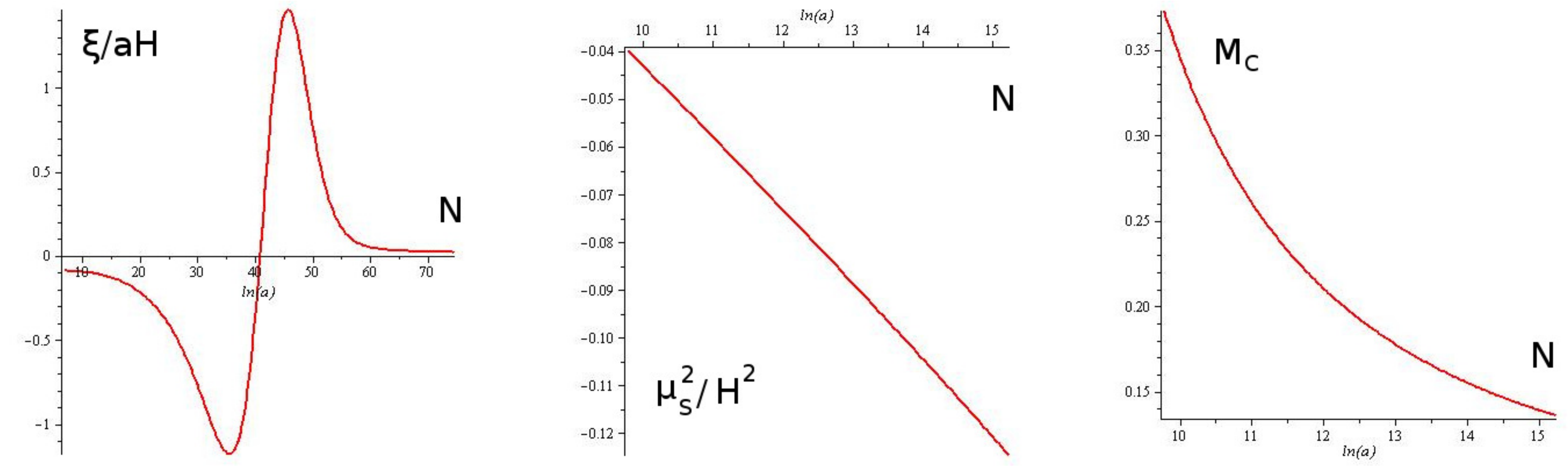}
\caption[The mass and its change]{Left: The coupling $\xi/aH$. Before and after the transition the coupling is small but it becomes large during the transition converting the entropy perturbation into the curvature perturbation. Middle: The mass of the entropy perturbation divided by the Hubble parameter squared. Its absolute value is much smaller than unity around sound horizon exit at $N \sim 10$.
Right: The quantity $M_{c}$ that shows how the mass of the entropy perturbation changes. It is also smaller than unity around sound horizon exit.}\label{massofentropy}
\end{figure}

The initial conditions for equations (\ref{equationofmotionone}) and (\ref{equationofmotiontwo}) are given by equations (\ref{dbisolone}) and (\ref{dbisoltwo}) if the coupling $\xi/a H$ and the slow-roll parameters are much smaller than unity when $\mu^{2}_{s} / H^{2}$ is negligible for the entropy mode as introduced in section \ref{dbilinear}. The slow-roll parameters are much smaller than unity as shown in figure \ref{slowroll}. We show the numerical results for the coupling and the mass of the entropy perturbation to see that we can use those initial conditions in figure \ref{massofentropy}. 

We investigate a mode which exits the sound horizon at $N \sim 10$ where we have an effective single field dynamics. As shown in figure \ref{massofentropy}, the coupling $\xi/aH$ is much smaller than unity around the sound horizon exit. Also, we can see that $\lvert \mu^{2}_{s} / H^{2} \rvert$ is also much smaller than unity around the sound horizon exit at $N \sim 10$. The mass also changes very slowly. If we define 
\begin{equation}
 M_{c} \equiv \frac{\dot{\mu}_{s}}{\mu_{s} H},
\end{equation}
which quantifies how rapidly the mass of the entropy perturbation changes, we can see in Fig.~\ref{massofentropy} that $M_{c}$ is smaller than unity for at least 5 e-folds after the sound horizon exit. Therefore, we can set the initial conditions for equations (\ref{equationofmotionone}) and (\ref{equationofmotiontwo}) by the solutions (\ref{dbisolone}) and (\ref{dbisoltwo}). Note that we set the initial conditions at $N \sim 7$ when the mode which we consider is still well within the sound horizon. 

\subsection{Numerical results}
We treat $v_{\sigma}$ and $v_{s}$ as two independent stochastic variables for the modes well inside the sound horizon as in \cite{Tsujikawa:2003}.
This means that we perform two numerical computations to obtain $\mathcal{P}_{\mathcal{R}}$. In one computation corresponding to the Bunch Davis vacuum state
for $v_{\sigma}$, $v_{s}$ is set to zero to obtain the solution $\mathcal{R}_{1}$ with equation (\ref{curvaturepowerspectrumhorizon}). In another computation corresponding to the Bunch Davies vacuum state for $v_{s}$, $v_{\sigma}$ is set to zero to obtain the solution $\mathcal{R}_{2}$. Then, the curvature power spectrum can be expressed as a sum of two solutions;
\begin{equation}\label{minorcorrectionindependent}
 \mathcal{P}_{\mathcal{R}} = \frac{k^{3}}{2 \pi^{2}} \left(\lvert \mathcal{R}_{1} \rvert^{2} + \lvert \mathcal{R}_{2} \rvert^{2} \right).
\end{equation}
This procedure is applied to all the numerical computations in this paper. Equation (\ref{minorcorrectionindependent}) holds as long as the perturbations are stochastic random variables initially. 

\begin{figure}[h]
\centering
\includegraphics[width=15cm]{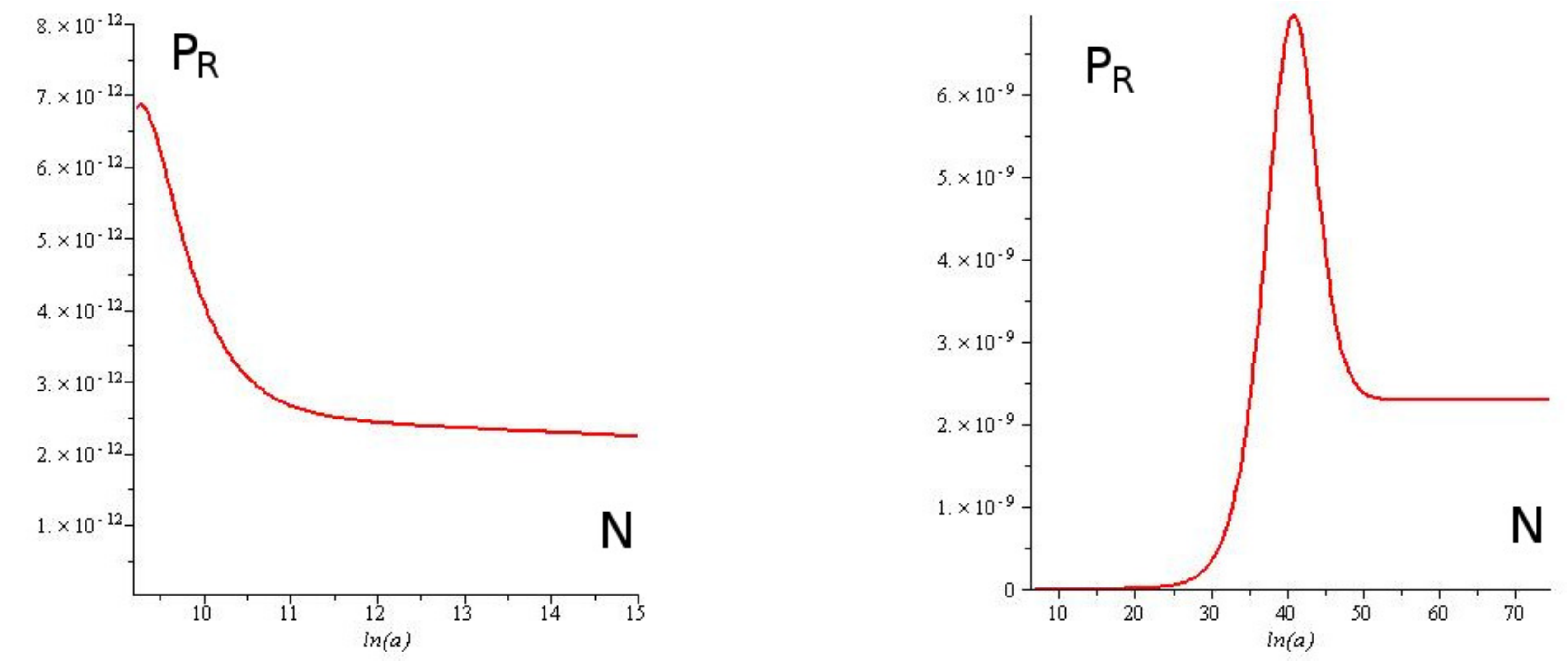}
\caption[The curvature power spectrum]{Left: The power spectrum of the curvature perturbation becomes constant a few e-folds after sound horizon exit which is around $N \sim 10$ because of the effective single field phase \cite{Nalson:2013, Nalson:2013b}. 
Right: We can see that the power spectrum of the curvature perturbation is enhanced during the transition.
Then, it becomes constant again after the transition. Note that both of these figures describe the contribution from the real part of $v_{\sigma k}$. The contribution
from the imaginary part shows the same behaviour. Therefore, it is sufficient to show only these figures in order to know its behaviour because the power spectrum of the curvature perturbation is just a sum of those two contributions.}\label{curvature}
\end{figure}

Figure \ref{curvature} shows the numerical results for the power spectrum of the curvature perturbation. As we can see from the small value of $\epsilon$, the Hubble parameter $H$ changes very slowly. It changes only by a few percent between $N \sim 10$ and $N \sim 70$. Substituting $H \sim 2.8 \times 10^{-8}$, $c_{s} \sim 6.2 \times 10^{-3}$ and $\epsilon = 6.4 \times 10^{-4}$ into equation (\ref{curvaturepowerspectrumhorizon}), we obtain the value of the curvature power spectrum after the sound horizon exit as
\begin{equation}
 \mathcal{P}_{\mathcal{R}_{*}} = \left. \frac{H^{2}}{8 \pi^{2} \epsilon c_{s}} \right\rvert_{*} \sim 2.5 \times 10^{-12}.
\end{equation}
This should coincide with the result of the numerical solution.
In the left panel of Fig.~\ref{curvature}, it is shown that $\mathcal{P}_{\mathcal{R}}$ becomes almost constant soon after sound horizon exit and it is given by $\mathcal{P}_{\mathcal{R}} \sim 2.5 \times 10^{-12}$. 
It does not change significantly until $N \sim 15$ because the dynamics is effectively single field \cite{Nalson:2013, Nalson:2013b}. When the trajectory in the field space curves, the curvature perturbation is sourced by the entropy perturbation and it is enhanced. We can actually see that the power spectrum of the curvature perturbation is enhanced by a factor of $\sim 9 \times 10^{2}$ during the transition in the right panel of figure \ref{curvature}. After the transition, it takes a constant value $\mathcal{P}_{\mathcal{R}} \sim 2.3 \times 10^{-9}$, which is compatible with the constraint (\ref{plancksatteliteconstraintcurvaturespectrum}) by the CMB observation with the Planck satellite. 

Because we can express the final curvature power spectrum as equation (\ref{prmouse}), the enhancement is quantified by the function $\cos^{2}{\Theta}$. In this model, we have $\cos^{2}{\Theta} \sim 1.1 \times 10^{-3}$. Therefore, the curvature perturbation originating from the entropy perturbation dominates the final curvature perturbation. From equation (\ref{chaptertwospectralindexfinal}) combined with the numerical values of the background quantities, we see that the spectral index is indeed determined by the mass of the entropy mode as
\begin{equation}
 n_{s} \sim \left. \frac{2 \mu_{s}^{2}}{3 H^{2}} \right \rvert_{*} + 1 \sim  0.972.
\end{equation}
This is compatible with the Planck satellite observations with the 95 $\%$ CL region \cite{Ade:2013b} even though it is not compatible with the 68 $\%$ CL region (\ref{observedspectralindexandratio}). The tensor-to-scalar ratio is obtained by substituting $\epsilon \sim 6.4 \times 10^{-4}$, $c_{s} \sim 6.2 \times 10^{-3}$ and $\cos^{2} {\Theta} \sim 1.1 \times 10^{-3}$ into equation (\ref{chaptertwofinaltensortoscalar}) as
\begin{equation}\label{chapterfourvalueofthetensortoscalar}
 r = \left. 16 \epsilon c_{s} \right\rvert_{*} \cos^{2} {\Theta} \simeq 7.0 \times 10^{-8}.
\end{equation}
This is compatible with the Planck satellite observations (\ref{observedspectralindexandratio}) as well as the upper bound (\ref{rcondition}) from the string theoretical analysis. 

\subsection{Non-Gaussianity parameters}
\begin{figure}[h]
\centering
\includegraphics[width=15cm]{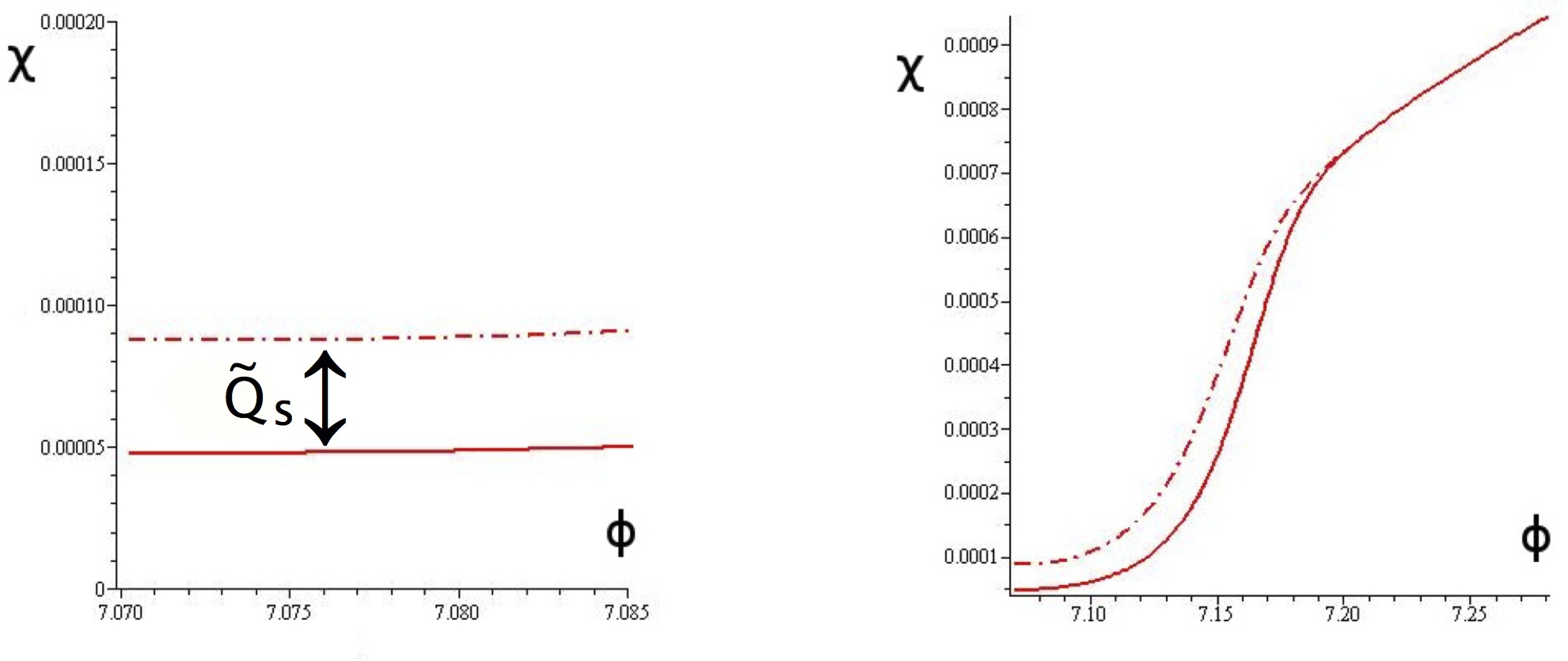}
\caption[$\delta$N in the entropic direction]{Left: Two trajectories
in the field space obtained by perturbing the initial value of the entropy field. 
The solid line describes the unperturbed trajectory
while the dotted line describes the perturbed trajectory.
Right: The transition occurs at different values of $\phi$. The solid line is the unperturbed trajectory
while the dotted line is the perturbed trajectory.}\label{entropyperturbation}
\end{figure}

In order to obtain the numerical values of the non-Gaussianity parameters, we use the $\delta$N-formalism introduced in section \ref{sec:deltaN}. Let us first compute the power spectrum of the curvature perturbation at $t = t_{2}$ after the transition by the $\delta N$-formalism. We take $t_{1}$ in equation (\ref{ndecomposition}) to be well before the transition when the curvature power spectrum is constant after sound horizon exit. 
In this case, as shown in Fig.~\ref{entropyperturbation}, the transition occurs at different $\phi^{I}$ if we perturb the initial field values in the entropic direction. Because taking perturbations in the initial field values changes the trajectory, unlike single field cases, we need to be careful about the definition of the final slice. We define $\phi^{I}$ as the field values along the unperturbed trajectory which correspond to the solid line in figure \ref{entropyperturbation} while we define $^{\rm{p}}\phi^{I}$ as the field values along the perturbed trajectory which correspond to the dotted line in figure \ref{entropyperturbation}. Then, as introduced in section \ref{sec:deltaN}, $\delta$N is given by 
\begin{equation}
 \delta N = \int ^{^{\rm{p}} t_{2}}_{^{\rm{p}} t_{1}} \tilde{H} d \, ^{\rm{p}}t - \int ^{t_{2}}_{t_{1}} H dt,
\label{deltanentropy}
\end{equation}
where the superscript on the left side $^{\rm{p}}$ denotes the quantities with the perturbed initial conditions.
For example, if we perturb the initial field values to the entropic direction, the perturbed initial field values $^{\rm{p}}\phi(^{\rm{p}} t_{1})$ correspond to the position in the field space as
\begin{equation}
 ^{\rm{p}}\phi^{I}(^{\rm{p}} t_{1}) = \phi^{I}(t_{1}) + \tilde{Q}_{s} \tilde{e}_{s}^{I}, \label{entropicperturbedfield}
\end{equation}
with $\tilde{e}_{s}^{I}$ given in equation (\ref{entropyappendixcomponents}). Note that $^{\rm{p}} t_{2}$ in equation (\ref{deltanentropy}) is the time when $^{\rm{p}}\phi^{I}$ takes the same value as $\phi^{I}(t_{2})$ well after the transition. Because both trajectories merge into the attractor solution in the true vacuum after the transition as shown in figure \ref{entropyperturbation}, we take the final uniform density slice so that both unperturbed and perturbed trajectories have the same field values $\phi_{f}$ and $\chi_{f}$ on the final slice. Because we can determine all the phase space variables if we know the values of the fields in the slow-roll case, we have the same $\phi$, $\dot{\phi}$, $\chi$ and $\dot{\chi}$ on the final slices which results in the same $H$ (i.e. uniform density) from equation (\ref{friedmannhybridchapfive}). By using the definition of $\delta$N in equation (\ref{deltanentropy}), we can numerically compute the quantity
\begin{equation}
 N_{, \tilde{s}} = \frac{N\left(\phi^{I}(t_{1}) + \tilde{Q}_{s} \tilde{e}_{s}^{I}  \right) - N\left(\phi^{I}(t_{1}) - \tilde{Q}_{s} \tilde{e}_{s}^{I} \right)}{2 \tilde{Q}_{s}},
\label{nphis}
\end{equation}
where we make $\tilde{Q}_{s}$ sufficiently small so that the value of $N_{, \tilde{s}}$ does not depend on the value of $\tilde{Q}_{s}$. Because the contribution from $N_{, \tilde{\sigma}}$ in equation (\ref{ndecompositionwiththenewbases}) is negligible in our model, we obtain
\begin{equation}
 \mathcal{P}_{\mathcal{R}} = \mathcal{P}_{\zeta} \simeq N_{, \tilde{s}}^{2} \mathcal{P}_{\tilde{Q}_{s}} \rvert _{t=t_{i}},\label{deltanentropic}
\end{equation}
where $\mathcal{P}_{\tilde{Q}_{s}}$ is given by equation (\ref{powerspectrumoffieldsdbiwithkoyamasanbases}). Note that the amplitude of the curvature perturbation on the uniform density hypersurface coincides with the amplitude of the comoving curvature perturbation on super-horizon scales \cite{Bassett:2006}. The numerical result shows that the $\delta$N-formalism successfully predicts the value of the final curvature perturbation $\mathcal{P}_{\mathcal{R}} \sim 2.3 \times 10^{-9}$ within a few percent error in this model as shown in the appendix \ref{app:numericalmethod}. 

In this model, the equilateral non-Gaussianity is obtained numerically using equation (\ref{dbimultiequilateralfinalresult}) as 
\begin{equation}\label{valueoffnlequilateralinchapfour}
f_{\rm{NL}}^{equil} \sim  - 9.5,
\end{equation}
with $c_{s} \sim 6.2 \times 10^{-3}$ and $\cos^{2}{\Theta} \sim 1.1 \times 10^{-3}$. In subsection \ref{subsec:observationalconstraintsdbisingle}, it is shown that the value of $f_{\rm{NL}}^{equil}$ cannot be small enough to be compatible with the Planck satellite observations in single field cases, if the value of the tensor-to-scalar ratio $r$ is smaller than the string theoretical upper bound (\ref{rcondition}) because of the small sound speed. However, this multi-field model satisfies the string theoretical constraint (\ref{rcondition}) as shown in equation (\ref{chapterfourvalueofthetensortoscalar}) with a small value of $f_{\rm{NL}}^{equil}$ as equation (\ref{valueoffnlequilateralinchapfour}) satisfies the Planck satellite constraint (\ref{fnlplanckobservationvaluesixtyeight}). 

Finally, let us show the numerical result of $f_{\rm{NL}}^{local}$ using equation (\ref{chaptertwofinalfnllocalformula}). Numerically, we can compute the second derivative of $N$ with respect to $\tilde{\phi}_{s}$ as
\begin{equation}
 N_{,\tilde{s} \tilde{s}} = \frac{N\left(\phi^{I}(t_{1}) + \tilde{Q}_{s} \tilde{e}_{s}^{I}  \right) - 2 N\left(\phi^{I}(t_{1}) \right) + N\left(\phi^{I}(t_{i}) - \tilde{Q}_{s} \tilde{e}_{s}^{I}  \right)}{\tilde{Q}_{s}^{2}},
\label{secondorderderivative}
\end{equation}
The result of our numerical computation shows that we have
\begin{equation}
 f_{\rm{NL}}^{local} = \frac{5}{6} \frac{N_{,\tilde{s} \tilde{s}}}{N_{,\tilde{s}}^{2}} \sim 40.1,
\end{equation}
in our model (see appendix \ref{app:numericalmethod} for the details of the numerical computations). Note that $f_{\rm{NL}}^{local}$ becomes constant after the transition. This value is excluded with the Planck satellite observations (\ref{fnllocalchapteronefinalvalue}) even though it was compatible with the WMAP observation \cite{Komatsu:2011}
\begin{equation}
 -10 < f_{\rm{NL}}^{local} < 74, 
 \: \: \: \: \: \: \: \: \: \: \: \: \rm{at\:\: 95\% \:\: C.L.}
\end{equation}
which was the strongest constraint when our result was published in Phys. Rev. D. \cite{Kidani:2012}. We can see that the statistical properties of the CMB become powerful tools to exclude inflationary models thanks to the development of the CMB observations. Even though the values of all the observables, except for $f_{\rm{NL}}^{local}$, can be made compatible with the Planck satellite observations, we have not succeeded to find a parameter set with which the values of all the observables including $f_{\rm{NL}}^{local}$ are consistent with the Planck constraints despite the extensive search of parameters. Of course, this does not imply that multi-field DBI inflation models are excluded as our model uses a very specific form of the potential. However, this means that generally it is very hard to find an example of multi field DBI models that satisfy all the observational constraints obtained by Planck satellite observations. 
\chapter{Conclusions and Discussions}

Inflation in the early stage of the Universe is believed to be the most promising model to solve the problems of the standard Big Bang scenario, namely the horizon problem, the flatness problem and the relic problem. Inflation is also successful in generating the fluctuations in the CMB temperature anisotropies naturally as a result of the quantum fluctuations of the inflaton, the scalar field that drives inflation. However, the origin of the inflaton is still unknown. In the DBI inflation models motivated by string theory, the extra dimensional coordinates play the roles of the scalar fields that drive inflation. Therefore, we can identify the geometrical origin of the inflaton in DBI inflation models. 

The CMB observations place strong constraints on inflation models. Among the CMB observables that quantify the statistical properties of the CMB temperature anisotropies, the non-Gaussianities have recently been studied extensively in the literature because the values of the non-Gaussianity parameters including $f_{\rm{NL}}^{equil} $ and $f_{\rm{NL}}^{local} $ were measured to great precision with the Planck satellite. Non-Gaussianities combined with other observables including the curvature perturbation power spectrum, tensor-to-scalar ratio and the spectral index for the scalar perturbation place strong constraints with which many inflation models are excluded. 

DBI inflation is the most plausible model that generates large equilateral non-Gaussianity. However, single field UV DBI models in string theory are strongly disfavoured by the current observations of the spectral index and equilateral non-Gaussianity. It has been shown that these constraints are significantly relaxed in multi-field models if there is a large conversion of the entropy perturbations into the curvature perturbation. 

In chapter \ref{chapter:spinflation}, we studied the DBI inflation model with the simplest two-field potential derived in string theory \cite{Gregory:2012}. When we consider the cases where we have the effective single field dynamics around horizon crossing, the model is approximated with a simple model studied in \cite{Alishahiha:2004}. After the horizon exit, the power spectrum of the curvature perturbation is significantly enhanced if the trajectory of the brane in the field space is bent sharply. Because all the analytic expressions derived in section \ref{sec:observable sinmultifield} are valid with the effective single field dynamics around horizon crossing, we can predict the value of the non-Gaussianity parameters. It has been shown that the model is excluded with the constraints on the power spectrum of the curvature perturbation and $f_{\rm{NL}}^{equil}$ regardless of the model parameters when we take into account the constraint on the volume of the internal space \cite{Baumann:2007}. 

We need further analysis if the coupling between the adiabatic and entropy perturbations is not negligible around horizon crossing. In such cases, we need to calculate the non-Gaussianity performing the full calculations with the in-in formalism. In \cite{Chen:2010b}, they have done similar calculations in quasi-single field inflation where the coupling is not negligible around horizon crossing. They obtained large bispectra whose shape is between the equilateral and local shapes. It would be interesting to apply their method to study the shape of non-Gaussianity in the case with the large coupling to know whether the stringent microphysical constraint can be avoided. 

In chapter \ref{chapter:spinflation}, we study the simplest potential that takes into account the leading order correction to the potential. The shape of the potential can be more complicated  depending on the embbeding of branes in the internal space. For example, in \cite{Chen:2010}, they obtained a potential where a waterfall phase transition connects two different radial trajectories (see figure \ref{fig:multifieldouyangembedding}). In chapter \ref{chapter:hybrid}, for the first time, we quantified the effect of the angular dynamics on observables using a toy model representing this type of the potential. We demonstrated that all the Planck observational constraints can be satisfied, except for the constraint on $f_{\rm{NL}}^{local}$, while obeying the bound on the tensor-to-scalar ratio imposed in string theory models. In general, the large conversion creates large local type non-Gaussianity. In our model, this is indeed the case and we expect that large equilateral non-Gaussianity is generally accompanied by large local type non-Gaussainity in multi-field DBI models. When our result \cite{Kidani:2012} was published, all the WMAP observational constraints \cite{Komatsu:2011} were satisfied with the parameter set introduced in chapter \ref{chapter:hybrid}. However, we have not succeeded to make the value of $f_{\rm{NL}}^{local}$ compatible with the Planck satellite observations despite the extensive parameter search. 

There are a number of extensions of our study in chapter \ref{chapter:hybrid} where we consider a toy two-field model. It would be important to study directly the potentials obtained in string theory taking into account higher order corrections to confirm our results; although it is a challenge to compute the potential when the sound speed is small. The curve in the trajectory in field space, which is essential to evade the strong constraints in string theory and responsible for large local non-Gaussianity, can be caused not only by the potential but also by non-trivial sound speeds \cite{Emery:2012, Emery:2013}. This happens in a model with more than one throat for example where there appear multiple different sound speeds. It would be interesting to compare the two cases to see if one can distinguish between them observationally. Finally, it has been shown that the multi-field effects enhance the equilateral type trispectrum for a given $f_{\rm{NL}}^{equil}$ \cite{Mizuno:2009, Mizuno:2009b} and its shape has been studied in detail \cite{Mizuno:2010, Izumi:2012}. Moreover, it was shown that there appears a particular momentum dependent component whose amplitude is given by $f_{NL}^{local} f_{NL}^{equil}$ \cite{Renaux-Petel:2009}. Thus the trispectrum will provide further tests of multi-field DBI inflation models. 

In this thesis we investigated whether it is possible to realise DBI inflation models that satisfy all the observational constraints while satisfying the microphysical constraint in string theory. We found it rather difficult to find such an example and generally it requires specific forms of potentials and fine-tuned initial conditions. DBI inflation model could leave distinct signatures in the CMB temperature anisotropies once realised because DBI inflation models provide the most plausible mechanism that generates the equilateral type non-Gaussianity. It is important to study further the realisation of this model in string theory. 

\bibliographystyle{JHEP}
    \bibliography{bibliography}

\appendix
\chapter{``in-in" formalism}\label{app:inin}

In this appendix, let us review the in-in formalism which will be used to calculate the bispectra of the curvature perturbation following the derivation of \cite{Weinberg:2005}. Let us consider a general Hamiltonian system with canonical variables $\phi_{a}\left(\textbf{x}, t\right)$ and conjugates $\pi_{a}\left(\textbf{x}, t\right)$ which satisfy the commutation relations
\begin{equation}\label{inincommutation}
\begin{split}
\left[\phi_{a}\left(\textbf{x}, t\right), \pi_{b}\left(\textbf{y}, t\right) \right] &= i \delta_{ab} \delta^{\left(3\right)}\left(\textbf{x} - \textbf{y} \right), \\
\left[\phi_{a}\left(\textbf{x}, t\right), \phi_{b}\left(\textbf{y}, t\right) \right] &= \left[\pi_{a}\left(\textbf{x}, t\right), \pi_{b}\left(\textbf{y}, t\right) \right] = 0,
\end{split}
\end{equation}
where a is an index labelling fields and their spin components, $\delta^{\left(3\right)}$ is the Kronecker delta and $\left[X,Y\right] \equiv XY - YX$ is the anti-commutator in this section. The equations of motion are
\begin{equation}\label{ininequationofmotion}
\begin{split}
\dot{\phi}_{a}\left(\textbf{x}, t\right) &= i \left[H\left(\phi\left(t\right), \pi_{a}\left(t\right) \right),  \phi_{a}\left(\textbf{x}, t\right)\right], \\
\dot{\pi}_{a}\left(\textbf{x}, t\right) &= i \left[H\left(\phi\left(t\right), \pi_{a}\left(t\right) \right),  \pi_{a}\left(\textbf{x}, t\right)\right],
\end{split}
\end{equation}
where Hamiltonian $H$ is a functional of the variables $\phi_{a}\left(\textbf{x}, t\right)$ and conjugates $\pi_{a}\left(\textbf{x}, t\right)$ at fixed time $t$. We expand $\phi_{a}\left(\textbf{x}, t\right)$ and $\pi_{a}\left(\textbf{x}, t\right)$ around the background FRW parts, which are time-dependent c-number solutions $\bar{\phi}_{a}\left(t\right)$ and $\bar{\pi}_{a}\left(t\right)$, as
\begin{equation}
\phi_{a}\left(\textbf{x}, t\right) = \bar{\phi}_{a}\left(t\right) + \delta\phi_{a}\left(\textbf{x}, t\right),\:\:\:\pi_{a}\left(\textbf{x}, t\right) = \bar{\pi}_{a}\left(t\right) + \delta\pi_{a}\left(\textbf{x}, t\right).
\end{equation}
When we expand the Hamiltonian in powers of the perturbations $\delta \phi_{a}\left(\textbf{x}, t\right)$ and $\delta \pi_{a}\left(\textbf{x}, t\right)$ at some time $t$, let us define the sum of the all terms of second and higher order in the perturbations as $\tilde{H}\left(\delta\phi\left(t\right), \delta\pi\left(t\right); t\right)$. Then, using the fact that both the background parts and the perturbation parts of the $\phi_{a}\left(\textbf{x}, t\right)$ and $\pi_{a}\left(\textbf{x}, t\right)$ satisfy equations (\ref{inincommutation}) and (\ref{ininequationofmotion}), after some simple calculations, we can separate the equations of motion for the background parts and for the perturbation parts which are given by
\begin{equation}\label{ininperteqofmotion}
\begin{split}
\delta\dot{\phi}_{a}\left(\textbf{x}, t\right) &= i \left[\tilde{H}\left(\delta\phi\left(t\right), \delta\pi\left(t\right); t\right),  \delta\phi_{a}\left(\textbf{x}, t\right)\right], \\
\delta\dot{\pi}_{a}\left(\textbf{x}, t\right) &= i \left[\tilde{H}\left(\delta\phi\left(t\right), \delta\pi\left(t\right); t\right),  \delta\pi_{a}\left(\textbf{x}, t\right)\right].
\end{split}
\end{equation}
Note that the Hamiltonian only depends on time although it contains the perturbations $\delta \phi_{a}\left(\textbf{x}, t\right)$ and $\delta \pi_{a}\left(\textbf{x}, t\right)$, which are dependent on the space coordinates, because those space coordinates are integrated when we obtain the Hamiltonian from the Hamiltonian density. 

Let us introduce unitary transformations
\begin{equation}\label{inintimeevolutionoperator}
\begin{split}
\delta\phi_{a}\left(t\right) &= U^{-1}\left(t, t_{0}\right) \delta\phi_{a}\left(t_{0}\right) U\left(t, t_{0}\right),\\
\delta\pi_{a}\left(t\right) &= U^{-1}\left(t, t_{0}\right) \delta\pi_{a}\left(t_{0}\right) U\left(t, t_{0}\right),
\end{split}
\end{equation}
where $U\left(t, t_{0}\right)$ is a time evolution operator. Then, by substituting equation (\ref{inintimeevolutionoperator}) into equation (\ref{ininperteqofmotion}), we can see
\begin{equation}\label{inintimeoperatorone}
\begin{split}
\frac{d}{dt}U\left(t, t_{0}\right) &= -i\,\tilde{H}\left(\delta\phi\left(t_{0}\right), \delta\pi_{a}\left(t_{0}\right); t\right)U\left(t, t_{0}\right),\\
\frac{d}{dt}U^{-1}\left(t, t_{0}\right) &= i\,U^{-1}\left(t, t_{0}\right)\tilde{H}\left(\delta\phi\left(t_{0}\right), \delta\pi_{a}\left(t_{0}\right); t\right),
\end{split}
\end{equation}
and the initial conditions
\begin{equation}
U\left(t_{0}, t_{0}\right) = U^{-1}\left(t_{0}, t_{0}\right) = 1.
\end{equation}
Note that we used the relation
\begin{equation}
\tilde{H}\left(\delta\phi\left(t\right), \delta\pi\left(t\right); t\right) = U^{-1}\left(t, t_{0}\right)\tilde{H}\left(\delta\phi\left(t_{0}\right), \delta\pi\left(t_{0}\right); t\right)U\left(t, t_{0}\right),
\end{equation}
and that equations (\ref{inintimeoperatorone}) satisfy the relation $\dot{U}U^{-1}+U\dot{U}^{-1}=0$ which is derived by differentiating the equation $UU^{-1}=1$. To calculate the time evolution operator $U$, let us decompose $\tilde{H}$ into $H_{0}$, which is quadratic in the perturbations, and an interaction term $H_{I}$, which is higher order in the perturbations, as
\begin{equation}\label{ininhubbleexpansion}
\tilde{H}\left(\delta\phi\left(t\right), \delta\pi\left(t\right); t\right) = H_{0}\left(\delta\phi\left(t\right), \delta\pi\left(t\right); t\right) + H_{I}\left(\delta\phi\left(t\right), \delta\pi\left(t\right); t\right),
\end{equation}
and seek to calculate $U$ as a power series in $H_{I}$. Now, we introduce the ``interaction picture". Let us define the fluctuation operators $\delta\phi^{I}_{a}\left(t\right)$ and $\delta\pi^{I}_{a}\left(t\right)$ whose time dependence is generated by $H_{0}$ as
\begin{equation}\label{ininperteqofmotiontwo}
\begin{split}
\delta\dot{\phi}^{I}_{a}\left(\textbf{x}, t\right) &= i \left[H_{0}\left(\delta\phi\left(t\right), \delta\pi\left(t\right); t\right),  \delta\phi^{I}_{a}\left(\textbf{x}, t\right)\right], \\
\delta\dot{\pi}^{I}_{a}\left(\textbf{x}, t\right) &= i \left[H_{0}\left(\delta\phi\left(t\right), \delta\pi\left(t\right); t\right),  \delta\pi^{I}_{a}\left(\textbf{x}, t\right)\right],
\end{split}
\end{equation}
and the initial conditions
\begin{equation}\label{inininteractioninitial}
\delta\phi^{I}_{a}\left(t_{0}\right) = \delta\phi_{a}\left(t_{0}\right),\:\:\:\delta\pi^{I}_{a}\left(t_{0}\right) = \delta\pi_{a}\left(t_{0}\right).
\end{equation}
Because $H_{0}$ is quadratic, the fluctuation operators in the interaction picture are free fields which satisfy the linear wave equations. We can define unitary transformations as
\begin{equation}\label{inintimeevolutionoperatortwo}
\begin{split}
\delta\phi^{I}_{a}\left(t\right) &= U^{-1}_{0}\left(t, t_{0}\right) \delta\phi_{a}\left(t_{0}\right) U_{0}\left(t, t_{0}\right),\\
\delta\pi^{I}_{a}\left(t\right) &= U^{-1}_{0}\left(t, t_{0}\right) \delta\pi_{a}\left(t_{0}\right) U_{0}\left(t, t_{0}\right),
\end{split}
\end{equation}
where $U_{0}\left(t, t_{0}\right)$ is a time evolution operator in the interaction picture, using the initial conditions (\ref{inininteractioninitial}). By substituting equations (\ref{inintimeevolutionoperatortwo}) into equations (\ref{ininperteqofmotiontwo}), we obtain
\begin{equation}\label{inintimeoperatortwo}
\begin{split}
\frac{d}{dt}U_{0}\left(t, t_{0}\right) &= -i\,H_{0}\left(\delta\phi\left(t_{0}\right), \delta\pi\left(t_{0}\right); t\right)U_{0}\left(t, t_{0}\right),\\
\frac{d}{dt}U_{0}^{-1}\left(t, t_{0}\right) &= i\,U^{-1}_{0}\left(t, t_{0}\right)H_{0}\left(\delta\phi\left(t_{0}\right), \delta\pi\left(t_{0}\right); t\right),
\end{split}
\end{equation}
and the initial condition
\begin{equation}
U_{0}\left(t_{0}, t_{0}\right) = U_{0}^{-1}\left(t_{0}, t_{0}\right) = 1.
\end{equation}
Note that we used the relation
\begin{equation}
H_{0}\left(\delta\phi^{I}\left(t\right), \delta\pi^{I}\left(t\right); t\right) = U_{0}^{-1}\left(t, t_{0}\right)\tilde{H}\left(\delta\phi\left(t_{0}\right), \delta\pi_{a}\left(t_{0}\right); t\right)U_{0}\left(t, t_{0}\right). 
\end{equation}
Then, from equations (\ref{inintimeoperatorone}), (\ref{ininhubbleexpansion}) and (\ref{inintimeoperatortwo}), we have
\begin{equation}\label{inintransformationequation}
\begin{split}
\frac{d}{dt}\left\{U^{-1}_{0}\left(t, t_{0}\right)U\left(t, t_{0}\right)\right\} &= - i \, U^{-1}_{0}\left(t, t_{0}\right) H_{I}\left(\delta\phi\left(t_{0}\right), \delta\pi_{a}\left(t_{0}\right); t\right) U\left(t, t_{0}\right)\\
 &= - i \, H_{I}\left(t\right)U^{-1}_{0}\left(t, t_{0}\right)U\left(t, t_{0}\right),
\end{split}
\end{equation}
where we define the interaction Hamiltonian in the interaction picture as
\begin{equation}
\begin{split}
H_{I}\left(t\right) &\equiv U^{-1}_{0}\left(t, t_{0}\right) H_{I}\left(\delta\phi\left(t_{0}\right), \delta\pi_{a}\left(t_{0}\right); t\right) U_{0}\left(t, t_{0}\right) \\
 &= H_{I}\left(\delta\phi^{I}\left(t\right), \delta\pi^{I}\left(t\right); t\right).
\end{split}
\end{equation}
If we define a function $F\left(t,t_{0}\right)$ as
\begin{equation}
U\left(t, t_{0}\right) = U_{0}\left(t, t_{0}\right) F\left(t,t_{0}\right),
\end{equation}
equation (\ref{inintransformationequation}) is rewritten as
\begin{equation}
\frac{d}{dt} F\left(t,t_{0}\right) = - i \, H_{I}\left(t\right) F\left(t,t_{0}\right),\:\:\:F\left(t_{0},t_{0}\right) = 1,
\end{equation}
which can be rewritten as
\begin{equation}\label{ininrecurrence}
F\left(t,t_{0}\right) = 1 - i\,\int^{t}_{t_{0}}H_{I}\left(t_{1}\right)F\left(t_{1},t_{0}\right)dt_{1}.
\end{equation}
By eliminating $F\left(t,t_{0}\right)$ from the right hand side of equation (\ref{ininrecurrence}) by iterative calculations, we can obtain
\begin{equation}\label{ininfsolution}
\begin{split}
F\left(t,t_{0}\right) &= 1 + \left(-i\right)\int^{t}_{t_{0}}H_{I}\left(t_{1}\right)dt_{1} + \left(-i\right)^{2} \int^{t}_{t_{0}} \int^{t_{1}}_{t_{0}}H_{I}\left(t_{1}\right)H_{I}\left(t_{2}\right)dt_{2}dt_{1}+...\\
& = T \exp{\left(-i\,\int^{t}_{t_{0}}H_{I}\left(t_{1}\right)dt_{1}\right)},
\end{split}
\end{equation}
where $T$ indicates that the products of $H_{I}$s in the power series expansion of the exponential are time-ordered. This means that they are written from left to right in decreasing order of the time arguments. With the solution (\ref{ininfsolution}), if we define $Q\left(t\right)$ to be any $\delta\phi\left(\textbf{x},t\right)$, $\delta\pi\left(\textbf{x},t\right)$, or any product of the $\delta\phi\left(\textbf{x},t\right)$s and/or $\delta\pi\left(\textbf{x},t\right)$s all at the same time t but in general with different space coordinates, and $Q^{I}\left(t\right)$ is the same product of $\delta\phi^{I}\left(\textbf{x},t\right)$ and/or $\delta\pi^{I}\left(\textbf{x},t\right)$, it is clear that we have the following relation
\begin{equation}\label{ininresultzero}
\begin{split}
Q\left(t\right) &= F^{-1}\left(t,t_{0}\right)Q^{I}\left(t\right)F\left(t,t_{0}\right)\\
 &=\left\{\bar{T} \exp{\left(i\,\int^{t}_{t_{0}}H_{I}\left(t_{1}\right)dt_{1}\right)}\right\} Q^{I}\left(t\right) \left\{T \exp{\left(-i\,\int^{t}_{t_{0}}H_{I}\left(t_{1}\right)dt_{1}\right)}\right\}.
\end{split}
\end{equation}
Here $\bar{T}$ denotes anti-time-ordering which means that the power series expansion of the exponential is written from left to right in increasing order of the time arguments. From equation (\ref{ininresultzero}), the expectation values of some product $Q\left(t\right)$ of field operators all at the same time $t$ are given by
\begin{equation}\label{ininresultone}
\left<Q\left(t\right)\right> = \left<\left\{\bar{T} \exp{\left(i\,\int^{t}_{t_{0}}H_{I}\left(t_{1}\right)dt_{1}\right)}\right\} Q^{I}\left(t\right) \left\{T \exp{\left(-i\,\int^{t}_{t_{0}}H_{I}\left(t_{1}\right)dt_{1}\right)}\right\}\right>.
\end{equation}
Equation (\ref{ininresultone}) can also be written in the form
\begin{equation}\label{ininresulttwo}
\begin{split}
\left<Q\left(t\right)\right> &= \sum^{\infty}_{N=0}i^{N}\int^{t}_{t_{0}}dt_{N}\int^{t_{N}}_{t_{0}}dt_{N-1}\cdots\int^{t_{3}}_{t_{0}}dt_{2}\int^{t_{2}}_{t_{0}}dt_{1}\\
&\times\left<\left[H_{I}\left(t_{1}\right), \left[H_{I}\left(t_{2}\right), \left[H_{I}\left(t_{3}\right),\cdots[H_{I}\left(t_{N}\right),Q^{I}\left(t\right)]...\right]\right]\right]\right>,
\end{split}
\end{equation}
with the $N=0$ term understood to be $<Q^{I}\left(t\right)>$ where you can take $t_{0}=-\infty$ in cosmology. You can see that equation (\ref{ininresultone}) is equivalent to equation (\ref{ininresulttwo}) for arbitrary t to order N as follows. It is obvious that this is true for the zeroth ($N=0$ in equation (\ref{ininresulttwo})) and first order ($N=1$ in equation (\ref{ininresulttwo})) in $H_{I}$. If we assume that equation (\ref{ininresultone}) is equivalent to equation (\ref{ininresulttwo}) up to order $N-1$ in $H_{I}$, by differentiating these equations, we easily see that the time derivatives of the right hand sides are equal up to order N. 
\chapter{Numerical method}\label{app:numericalmethod}

We introduce the details of the numerical calculations in chapter \ref{chapter:hybrid} here. First, the derivation of the components of the basis vectors is introduced. Then, using those bases, we explain how the value of the curvature perturbation power spectrum is obtained numerically with the $\delta$N formalism. 

Let us first summarise how the adiabatic and entropic bases are defined here. We consider the linear perturbations of the scalar fields defined in equation (\ref{allperturbations}). We decompose the perturbations into the instantaneous adiabatic and entropy perturbations; the adiabatic direction corresponds to the direction of the background field's evolution while the entropy directions are orthogonal to the adiabatic direction as introduced in section \ref{kflationlinear}. For this purpose, we use orthogonal bases $\tilde{e}^{I}_{\rm{n}} \, (n = 1,2...N)$ in field space. 
The orthonormal condition in general multi-field inflation is given by
\begin{equation}
 P_{,X^{I J}} \tilde{e}^{I}_{n} \tilde{e}^{J}_{m} = \delta_{n m},
\label{normalization}
\end{equation}
so that the gradient term $P_{,X^{I J}} \partial_{\rm{i}} Q^{I} \partial^{\rm{i}} Q^{J}$ is diagonalised when we use these bases. Here we assume that $P_{,X^{I J}}$ is
invertible and it can be used as a metric in the field space. The adiabatic vector is
\begin{equation}
 \tilde{e}_{1}^{I} = \frac{\dot{\phi}^{I}}{\sqrt{P_{,X^{J K}} \dot{\phi}^{J} \dot{\phi}^{K}}},
\end{equation}
which satisfies the normalization given by equation (\ref{normalization}). The field perturbations are decomposed in those bases as
\begin{equation}
 Q^{I} = \tilde{Q}_{n} \tilde{e}^{I}_{n}.
\end{equation}
For multi-field DBI inflation, using the relation
\begin{equation}
 P_{,X^{I J}} \dot{\phi}^{I} \dot{\phi}^{J} = c_{s} G_{IJ} \dot{\phi}^{I} \dot{\phi}^{J} + \frac{1-c_{s}^{2}}{2 X c_{s}} G_{IK} G_{JL} \dot{\phi}^{I} \dot{\phi}^{K} \dot{\phi}^{J} \dot{\phi}^{L} = \frac{2 X}{c_{s}},
\end{equation}
we can show that the adiabatic vector is given by
\begin{equation}
 \tilde{e}_{1}^{I} = \frac{\sqrt{c_{s}}}{2X} \dot{\phi}^{I},
\label{adiabaticvector}
\end{equation}
which is the adiabatic vector introduced in equation (\ref{newadiabaticbasis}). This implies that 
\begin{equation}
 G_{IJ} \tilde{e}_{1}^{I} \tilde{e}_{1}^{J} = c_{s},
\end{equation}
\begin{equation}
 P_{,X^{IJ}} = c_{s} G_{IJ} + \frac{1 - c_{s}^{2}}{c_{s}^{2}} G_{IK} G_{JL} \tilde{e}_{1}^{K} \tilde{e}_{1}^{L}.
\label{canonicaldecomposition}
\end{equation}
Substituting equation (\ref{canonicaldecomposition}) into equation (\ref{normalization}), with $m = 1$, we obtain
\begin{equation}
 G_{I J} \tilde{e}_{1}^{I} \tilde{e}_{n} ^{J} = c_{s} \delta_{n 1}.
\label{orthogonal}
\end{equation}
Substituting equations (\ref{canonicaldecomposition}) and (\ref{orthogonal}) into 
equation (\ref{normalization}), we obtain
\begin{equation}
 G_{I J} \tilde{e}_{n}^{I} \tilde{e}_{m} ^{J} = \frac{1}{c_{s}} \delta_{m n} - \frac{1 - c_{s}^{2}}{c_{s}} \delta_{m 1} \delta_{n 1}.
\label{finalnormalization}
\end{equation}
For two-field models with $G_{I J} \left(\phi^{K} \right) = A_{I} \left(\phi^{K} \right) \delta_{I J}$ where $A_{\phi} = 1$ and $A_{\chi} = \phi^{2}$, from equation (\ref{adiabaticvector}), the adiabatic vector is obtained as
\begin{equation}
 \left(\tilde{e}_{\sigma}^{\phi}, \tilde{e}_{\sigma}^{\chi} \right) = \left(\sqrt{c_{s}}\frac{\dot{\phi}}{\sqrt{\dot{\phi}^{2}+\phi^{2}\dot{\chi}^{2}}}, \sqrt{c_{s}}\frac{\dot{\chi}}{\sqrt{\dot{\phi}^{2}+\phi^{2}\dot{\chi}^{2}}} \right).
\end{equation}
From the orthogonal condition (\ref{finalnormalization}), the entropy vector $\tilde{e}_{s}^{I}$ satisfies
\begin{equation}
 G_{IJ} \tilde{e}_{\sigma}^{I} \tilde{e}_{s}^{J} = \tilde{e}_{\sigma}^{\phi} \tilde{e}_{s}^{\phi} + \phi^{2} \tilde{e}_{\sigma}^{\chi}e_{s}^{\chi} = 0,
\end{equation}
\begin{equation}
 G_{IJ} \tilde{e}_{s}^{I} \tilde{e}_{s}^{J} = \left(\tilde{e}_{s}^{\phi}\right)^{2} + \phi^{2}\left( \tilde{e}_{s}^{\chi}\right)^{2} = \frac{1}{c_{s}},
\end{equation}
which leads to 
\begin{equation}\label{entropyappendixcomponents}
 \left( \tilde{e}_{s}^{\phi}, \tilde{e}_{s}^{\chi} \right) = \left(\frac{1}{\sqrt{c_{s}}}\frac{\phi \dot{\chi}}{\sqrt{\dot{\phi}^{2}+\phi^{2}\dot{\chi}^{2}}}, \frac{1}{\sqrt{c_{s}}}\frac{-\dot{\phi}}{\phi\sqrt{\dot{\phi}^{2}+\phi^{2}\dot{\chi}^{2}}} \right).
\end{equation}

Below, we explain how the $\delta$N-formalism is used in the numerical computations using the components of the basis vectors derived above. In single field cases, we just need to perturb the initial conditions along the trajectory. In the numerical computations, it is easy to compute $\phi \left(t_{1} + \delta t \right)$ in single field cases because the trajectory with the perturbed initial conditions is the same as the one with the original initial conditions.

However, in two-field cases, the trajectory with the initial conditions perturbed in the entropic direction is different
from the one with the original initial conditions as shown in figure \ref{entropyperturbation}. Although the perturbed initial values of the fields are defined in equation (\ref{entropicperturbedfield}), once we set the value of $\tilde{Q}_{s}$, we also need to know how to perturb the values of the time derivatives of the fields. This is because we need to set the values of all the phase space variables in order to solve the second order differential equations numerically. From equation (\ref{entropicperturbedfield}), we obtain
\begin{equation}
 \dot{^{\rm{p}} \phi}^{I}\left(^{\rm{p}} t_{1}\right) = \dot{\phi}^{I}(t_{1}) + \dot{\tilde{Q}}_{s} \tilde{e}_{s}^{I} + \tilde{Q}_{s} \dot{\tilde{e}}_{s}^{I}.
\label{timederivativeofpert}
\end{equation}
From equation (\ref{entropyappendixcomponents}), the derivatives of the components of the entropy basis vector are obtained as 
\begin{equation}
 \dot{\tilde{e}}_{s}^{\phi} = \tilde{e}_{s}^{\phi} \left(-\frac{f_{,\phi} \dot{\phi}}{2 f} - \frac{s H}{2} - \frac{\ddot{\sigma}}{\dot{\sigma}} + \frac{\dot{q}(t)}{q(t)} \right),
\label{esphidotrelation}
\end{equation}
and
\begin{equation}
 \dot{\tilde{e}}_{s}^{\chi} = \tilde{e}_{s}^{\chi} \left(-\frac{f_{,\phi} \dot{\phi}}{2 f} - \frac{s H}{2} - \frac{\ddot{\sigma}}{\dot{\sigma}} - \frac{\dot{\phi}}{\phi} + \frac{\dot{p}(t)}{p(t)} \right),
\label{eschidotrelation}
\end{equation}
where $\dot{\sigma}$ is defined in equation (\ref{definitionsincludingsigmadotinchaptertwo}), $s$ is the slow-roll parameter and $\ddot{\sigma} = d \dot{\sigma} / dt$. Here new variables $p(t)$ and $q(t)$ are defined as
\begin{equation}
  p(t) \equiv -\sqrt{f} \dot{\phi}, \: \: \: \: \: q(t) \equiv -\sqrt{f} \phi \dot{\chi},
\end{equation}
so that the sound speed is expressed as
\begin{equation}
 c_{s} = \sqrt{1 - p(t)^{2} - q(t)^{2}}.
\end{equation}
Given that we obtain the numerical values of $\dot{\tilde{e}}_{s}^{\phi}$ and $\dot{\tilde{e}}_{s}^{\chi}$ using equations (\ref{esphidotrelation}) and (\ref{eschidotrelation}), we now know how to perturb all the phase space variables ($\phi$, $\dot{\phi}$, $\chi$, $\dot{\chi}$) from equation (\ref{timederivativeofpert}) if we know the value of $\dot{\tilde{Q}}_{s}$. Because we set the value of $\tilde{Q}_{s}$, we can determine the value of $\dot{\tilde{Q}}_{s}$ if there is a relation between $\tilde{Q}_{s}$ and $\dot{\tilde{Q}}_{s}$. Actually, before the transition, it is possible to obtain the analytic solution for $v_{s}$ and hence the solution for $\tilde{Q}_{s}$ from equation (\ref{relationofvandq}). As mentioned in section \ref{kflationlinear}, the approximations $z''/z \simeq 2 / \tau^{2}$ and $\alpha'' / \alpha \simeq 2 / \tau^{2}$ hold because the slow-roll approximation holds and the coupling $\xi$ is negligible before the transition. Therefore, equation (\ref{equationofmotiontwo}) is approximated as
\begin{equation}
 v_{s}'' + \left(c_{s}^{2} k^{2} - 2 / \tau^{2} + a^{2} \mu_{s}^{2} \right)v_{s}  \simeq 0,
\end{equation}
which can be rewritten as
\begin{equation}
 \frac{d^{2} \tilde{v}_{s}}{d \tilde{\tau} ^{2}} + \frac{1}{\tilde{\tau}} \frac{d \tilde{v}_{s}}{d \tilde{\tau}} + \left(1 - \frac{9/4 - \mu_{s}^{2}/H^{2}}{\tilde{\tau}^{2}} \right)\tilde{v}_{s}  \simeq 0,
\label{besselequation}
\end{equation}
where
\begin{equation}
  \tilde{v}_{s} \equiv \frac{v_{s}}{\sqrt{- \tau}}, \: \: \: \: \: \tilde{\tau} \equiv -c_{s} k \tau.
\end{equation}
Note that we regard $c_{s}$ as a constant because the slow-roll parameter $s$ is much smaller than unity as we showed in figure \ref{slowroll}.
Then, if we approximate $\mu_{s}$ to be a constant, we have the analytic solution for equation (\ref{besselequation}) because it is the Bessel differential equation.
By choosing the Bunch-Davies vacuum initial condition, we obtain
\begin{equation}
 v_{s} = \frac{\sqrt{\pi}}{2} e^{i (\nu_{s} + 1/2) \pi/2} \left(- \tau \right)^{1/2} H_{\nu_{s}}^{(1)}\left(-c_{s} k \tau \right),
\end{equation}
where
\begin{equation}
 \nu_{s} = \sqrt{9/4 - \mu_{s}^{2}/H^{2}},
\end{equation}
and $H_{\nu_{s}}^{(1)}$ is the Hankel function of the first kind. In the super-horizon limit $-c_{s} k \tau << 1$, using the asymptotic form of the Hankel function
\begin{equation}
 H_{\nu_{s}}^{(1)}\left(-c_{s} k \tau \right) \sim - i \frac{\Gamma(\nu_{s})}{\pi} \left(\frac{2}{-c_{s} k \tau} \right)^{\nu_{s}},
\end{equation}
we have
\begin{equation}
 v_{s} = \frac{- i \sqrt{\pi}}{2} e^{i (\nu_{s} + 1/2) \pi/2} \left(- \tau \right)^{1/2} \frac{\Gamma(\nu_{s})}{\pi} \left(\frac{2}{-c_{s} k \tau} \right)^{\nu_{s}},
\end{equation}
and
\begin{equation}
 \tilde{Q}_{s} = \frac{- i \sqrt{\pi} c_{s} H}{2} e^{i (\nu_{s} + 1/2) \pi/2} \left(- \tau \right)^{3/2} \frac{\Gamma(\nu_{s})}{\pi} \left(\frac{2}{-c_{s} k \tau} \right)^{\nu_{s}},
\label{qssuperhorizon}
\end{equation}
where we used the relation $\tau \sim -1/aH$ during slow-roll inflation. From equation (\ref{qssuperhorizon}), we obtain the first derivative of $\tilde{Q}_s$ in terms of 
$\tilde{Q}_s$ as 
\begin{equation}
 \dot{\tilde{Q}}_{s} = \left[s\left(1 - \nu_{s} \right) - \epsilon + \nu_{s} - \frac{3}{2} \right] H \tilde{Q}_{s}.
\label{qdot}
\end{equation}

Let us show the details of the numerical results for obtaining the value of the curvature perturbation power spectrum using the $\delta$N formalism. As shown in figure \ref{chisound}, the transition begins around $N \sim 20$. We take the initial hypersurface around $N \sim 12$ where we can evaluate $P_{\tilde{Q}_{\sigma}}$ with equation (\ref{powerspectrumoffieldsdbiwithkoyamasanbases}) because the trajectory is still effectively a single field one after sound horizon crossing.
On the initial hypersurface, we set the values of all the phase space variables as $\phi(t_{1}) \simeq 7.09, \dot{\phi}(t_{1}) \simeq 8.11 \times 10^{-11}, \chi(t_{1}) \simeq 5.06 \times 10^{-5}, \dot{\chi}(t_{1}) \simeq 3.42 \times 10^{-14}$. If we set $\delta \chi = 10^{-7}$, we obtain
\begin{equation}
 \delta \phi = \frac{e^{\phi}_{s}(t_{1})}{e^{\chi}_{s}(t_{1})} \delta \chi = -2.12 \times 10^{-9},
\label{deltaphi}
\end{equation}
given that $\delta \phi = \tilde{Q}_{s} \tilde{e}^{\phi}_{s}$ and $\delta \chi = \tilde{Q}_{s} \tilde{e}^{\chi}_{s}$ from equation (\ref{entropicperturbedfield}), where $e^{\phi}_{s}(t_{1})$ and $e^{\chi}_{s}(t_{1})$ are obtained numerically. Then, we also have
\begin{equation}
 \tilde{Q}_{s} = \frac{10^{-7}}{\tilde{e}^{\chi}_{s}(t_{1})} = -5.59 \times 10^{-8}.
\label{qsnumerical}
\end{equation}
Using equation (\ref{qdot}), we obtain $\dot{Q}_{s} \simeq -4.29 \times 10^{-17}$. Actually, we can also obtain the first derivative numerically
\begin{equation}
 \dot{\tilde{Q}}_{s} = \left( \frac{\dot{v_{s}}}{v_{s}} + H \left(s - 1 \right) \right) Q_{s} \simeq -3.07 \times 10^{-17},
\label{qsdotnumerical}
\end{equation}
from equation (\ref{relationofvandq}) using the numerical values of $\dot{v}_{s}$ and $v_{s}$. We can see that the values of $\dot{\tilde{Q}}_{s}$ obtained in both ways are the same with around 40 percent error. This error comes from the fact that the solution (\ref{qdot}) is exact only if $\mu_{s}$ is perfectly constant and the slow-roll parameters are zero. Using equations (\ref{timederivativeofpert}), (\ref{qsnumerical}), (\ref{qsdotnumerical}) and the numerical values of $e_{s}^{I}$ and $\dot{e}_{s}^{I}$ obtained by using equations (\ref{entropyappendixcomponents}), (\ref{esphidotrelation}) and (\ref{eschidotrelation}), we obtain the first derivatives of the fields as 
\begin{equation}
 \delta \dot{\phi} \equiv \dot{^{\rm{p}}\phi} \left(^{\rm{p}}t_{1}\right) - \dot{\phi} \left(t_{1}\right) \simeq -1.47 \times 10^{-17},
\label{deltaphidot}
\end{equation}
and
\begin{equation}
 \delta \dot{\chi} \equiv \dot{^{\rm{p}} \chi} (\tilde{t}_{1}) - \dot{\chi}(t_{1}) \simeq 5.35 \times 10^{-17}.
\label{deltachidot}
\end{equation}
Using $\delta \chi = 10^{-7}$ and equations (\ref{deltaphi}), (\ref{deltaphidot}) and (\ref{deltachidot}), we can now perturb the initial conditions. 
Using these initial condition, the first derivative of $N$ with respect to $\tilde{\phi}_s$
is obtained using equations (\ref{deltanentropy}) and (\ref{nphis}) as
\begin{equation}
N_{,\tilde{s}} \simeq 7.82 \times 10^{2}.
\end{equation}
Then, from equation (\ref{deltanentropic}), the power spectrum of the curvature 
perturbation is given by 
\begin{equation}
 P_{\mathcal{R}} = P_{\zeta} \simeq N_{,\tilde{s}}^{2} P_{\tilde{Q}_{s}} \rvert _{t=t_{1}} \simeq 2.29 \times 10^{-9},
\end{equation}
where we used the numerical result for the power spectrum of the entropy perturbation
\begin{equation}
 P_{\tilde{Q}_{s}} \rvert _{t=t_{1}} = \frac{k^{3}}{2 \pi^{2}} \frac{c_{s}^{2} \lvert v_{s k} \rvert^{2}}{a^{2}} \simeq 3.74 \times 10^{-15}.
\end{equation}
This coincides with the value $P_{\mathcal{R}} \sim 2.3 \times 10^{-9}$ with less than one percent error, which is obtained directly by solving the equations of motion for the linear perturbations numerically. This confirms the validity of the $\delta$N formalism in this model. In a similar way, we perturb $N$ and obtain the second order derivative using equation (\ref{secondorderderivative}). It is then possible to compute the local type non-Gaussianity.



\end{document}